\documentclass[a5paper,12pt]{amsbook}
\usepackage[active]{srcltx}
\usepackage[OT2,OT1]{fontenc}
\usepackage{hyperref}
\usepackage[utf8]{inputenc}
\usepackage[russian,spanish]{babel}
\usepackage{enumerate}
\usepackage{appendix}
\usepackage{amsmath, amssymb,stackrel}
\hypersetup{colorlinks=true,linktoc=page,linkcolor=blue,citecolor=red,urlcolor=cyan}
\numberwithin{equation}{section}
\usepackage{feynmp}
\usepackage{feynmp-auto}
\usepackage{chngcntr}
\usepackage{etoolbox}
\usepackage{graphics}
\usepackage{epigraph}
   \usepackage[pdftex]{graphicx}
   \usepackage{epstopdf}
\usepackage{csquotes}
\usepackage{caption}
\usepackage{multirow}
\usepackage[%
    maxbibnames=4,maxcitenames=2,style=authoryear-comp,
    sortcites=true,, doi=false,url=false,
    firstinits=true,isbn=false,hyperref,backend=biber,backref=true]{biblatex}
\DefineBibliographyStrings{spanish}{andothers={\textsl{et al}\adddot}}
\renewbibmacro{in:}{}    
\setquotestyle[american]{english}

\renewcommand{\theequation}{\Roman{chapter}.\arabic{section}.\arabic{equation}}

\AtBeginEnvironment{subappendices}{%
     \chapter*{Anexos}
\counterwithin{figure}{section}
\counterwithin{table}{section}
}
\DeclareNameAlias{sortname}{last-first}
\newtheorem{definicion}{Definici\'on}
\newtheorem{teorema}{Teorema}
\newtheorem{lema}{Lema}

\newtheorem{propiedad}{Propiedad}[definicion]
\renewcommand*{\bibfont}{\footnotesize}
\renewcommand*{\multicitedelim}{\addcomma\space}
\begin{document}
\allowdisplaybreaks[1]
        \renewcommand{\contentsname}{\'Indice}
        \renewcommand{\bibname}{Bibliograf\'ia}
 \renewcommand{\chaptername}{Cap\'itulo}
    \renewcommand{\appendixname}{Ap\'endice}
     \renewcommand{\appendixpagename}{Anexos}
     \renewcommand{\figurename}{Figura}
\normalsize
 \newcommand{\trv}{\text{Tr}_{V_x}}
\newcommand{\iM}{\int_{\mathcal{M}}\,dx\, \sqrt{g}\,}
\newcommand{\dd}{\mathcal{D}}

 \begin{titlepage}
\begin{center}

\vspace{1cm}
\textbf{{ \LARGE  Formalismo de L\'inea de Mundo\\[0.5cm] en \\[0.7cm]
Teor\'ias No Conmutativas}}\\[0.8cm] 

\rule{1\linewidth}{0.9mm} 
\\[2.5cm]
\huge

\includegraphics[width=0.6\textwidth]{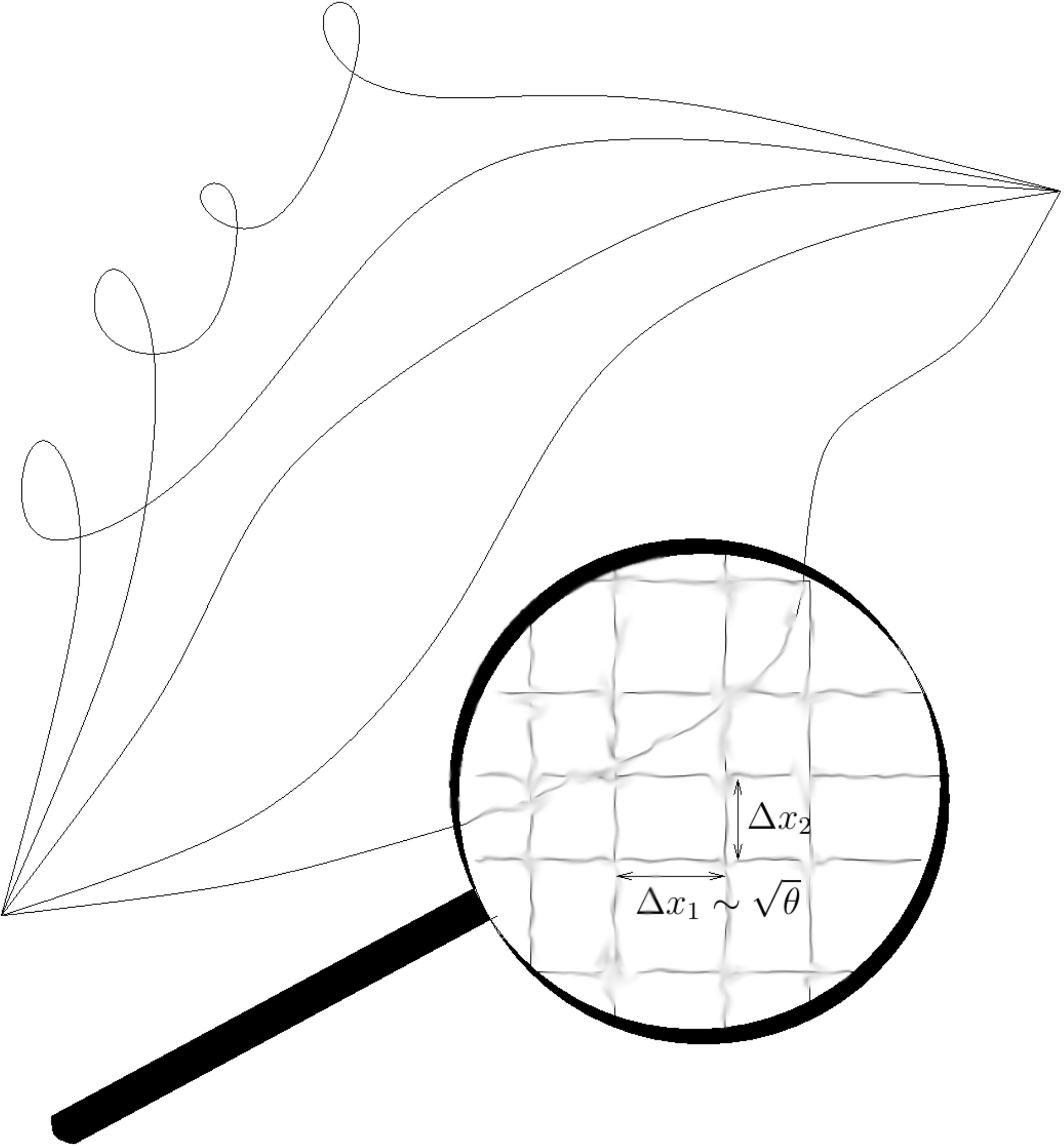}~\\[4.5cm]

\Large

\texttt{S.A. \textsc{Franchino Vi\~nas}}
\vfill


\end{center}
\end{titlepage}
\begin{titlepage}
\begin{center}

\includegraphics[width=0.40\textwidth]{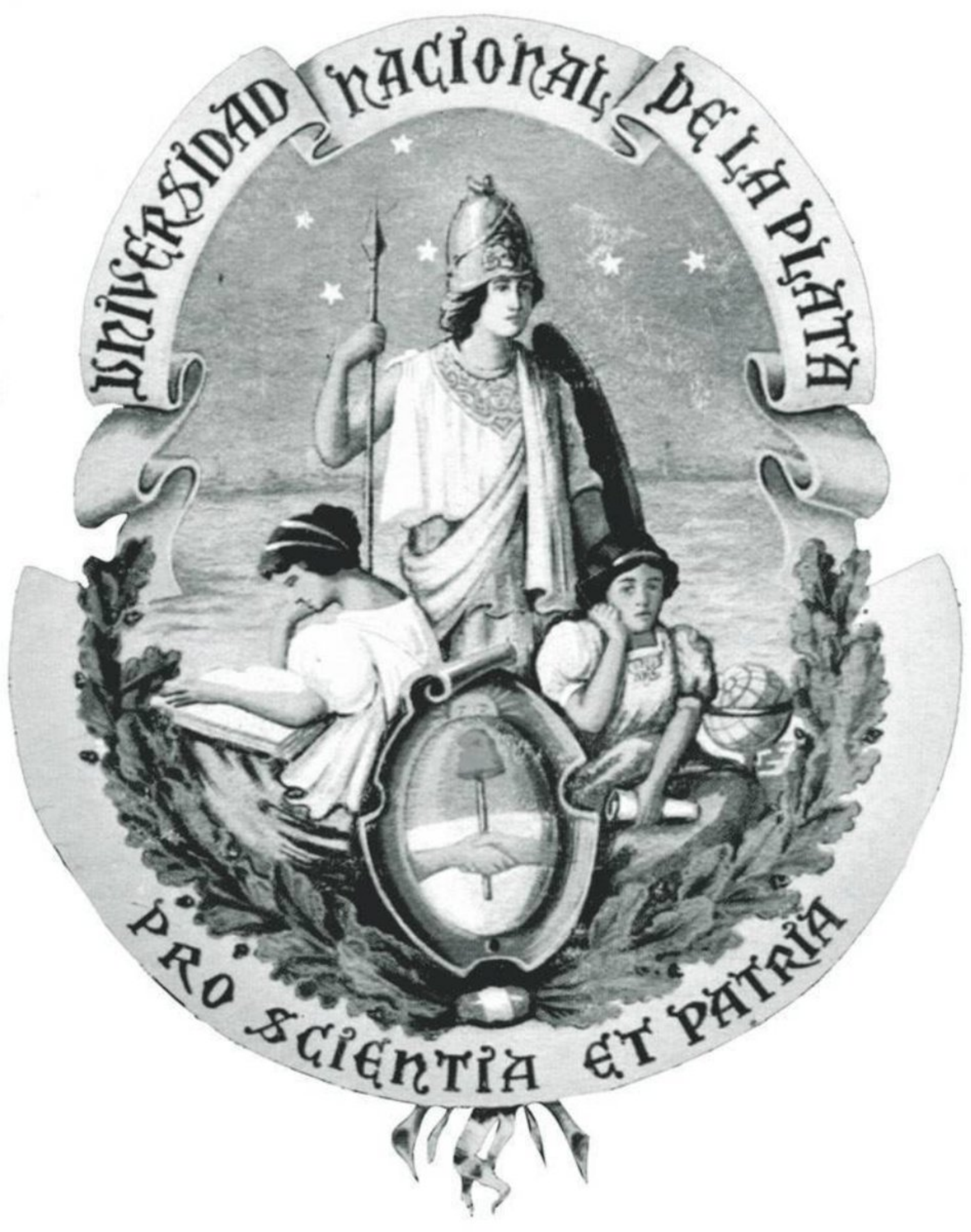}~\\[0.8cm]

\textsc{\large Universidad Nacional de La Plata}\\[.2cm]
Facultad de Ciencias Exactas - Departamento de F\'isica\\[1.1cm]

\textsc{\large Tesis presentada para optar al grado de Doctor de la Facultad de Ciencias Exactas}\\[0.1cm]

\rule{\linewidth}{0.5mm}\\[-0.3cm]
\rule{\linewidth}{0.5mm}\\[0.4cm]
{ \Large \bfseries Formalismo de L\'inea de Mundo\\[0.3cm] en \\[0.3cm]
Teor\'ias No Conmutativas\\[0.4cm] }

\rule{1\linewidth}{0.5mm} \\[0.8cm]
\huge

\Large

Sebasti\'an A. \textsc{Franchino Vi\~nas}\\[0.3cm]

\large
Director: Prof. Dr. Pablo A. \textsc{Gonz\'alez Pisani.}\\[0.1cm]
Codirector: Prof. Dr. Horacio A. \textsc{Falomir.}

\vfill
{\normalsize 15 de Septiembre del 2015}

\end{center}
\end{titlepage}

\begin{titlepage}
 
\end{titlepage}

\null\newpage
\noindent\rule{\linewidth}{0.5mm} \\[0.0cm]
\begin{flushleft} 
\normalsize
\emph{Tesis para optar al grado de}\\[0.3cm]
{Doctor de la Facultad de Ciencias Exactas}\\[0.3cm]
\emph{de la Universidad Nacional de La Plata.}\\[1cm]

\normalsize
\emph{Autor:}\\[0.1cm]
Lic. Sebasti\'an A. \textsc{Franchino Vi\~nas.}\\[0.5cm]
\emph{Director:}\\[0.1cm]
Prof. Dr. Pablo A. \textsc{Gonz\'alez Pisani.}\\[0.5cm]
\emph{Codirector:}\\[0.1cm]
Prof. Dr. Horacio A. \textsc{Falomir.}\\[0.5cm]
\emph{Lugar de Trabajo:}\\[0.1cm]
Instituto de F\'isica La Plata, CONICET y \\
Departamento de F\'isica, Facultad de Ciencias Exactas, Universidad~Nacional~de~La~Plata.\\[0.5cm]
\emph{Miembros del Jurado Evaluador:}\\[0.1cm]
Prof. Dr. Mauricio  \textsc{Leston}\\
Prof. Dr. Gerardo  \textsc{Rossini}\\
Prof. Dr. Dmitri  \textsc{Vassilevich} 
\vspace{1cm}
\begin{center}
\rule{0.4\linewidth}{0.5mm}
\end{center}
\vspace{1cm}

Este trabajo fue financiado por una beca de posgrado del CONICET.\\[0.5cm]
\end{flushleft}

\newpage~

\newpage~

\vspace{0.2\textheight}
\normalsize
\begin{flushright}
\begin{minipage}{0.6\textwidth}
 \begin{flushright}
 \emph{A quienes han convergido en nuestra sangre, y a los que \'esta engendrar\'a.}
\end{flushright}
\end{minipage}
\end{flushright}

\newpage~
\newpage~
\vspace{0.0\textheight}

 \begin{flushright}
\begin{minipage}{0.6\textwidth}
\normalsize\emph{Io cominciai: \guillemotleft Poeta che mi guidi,}\\[0.1cm]
\emph{guarda la mia virt\`u sell \`e possente,}\\[0.1cm] 
\emph{prima che lalto passo tu mi fidi (\ldots)\guillemotright}\\[0.3cm]
\emph{\guillemotleft (\ldots) 
Or va chun sol volere \`e dambedue:\\[0.1cm]
tu duca, tu segnore e tu maestro\guillemotright.\\[0.1cm]
Cos\`i li dissi; e poi che mosso fue,\\[0.3cm]
intrai per lo cammino alto e silvestro.}\\[0.3cm]

\begin{flushright}
\textsc{Dante Alighieri},\\ \emph{La Divina Commedia}, Inferno Canto II.
\end{flushright}
\end{minipage}
 
\end{flushright}


\normalsize
\tableofcontents
\chapter*{Resumen}
\normalsize La unificaci\'on de las fuerzas descriptas en el modelo est\'andar y gravitatoria en una teor\'ia de la gravedad cu\'antica es quiz\'as el problema m\'as importante, desde el punto de vista te\'orico, a resolver por la f\'isica actual. La teor\'ia cu\'antica de campos (TCC) no conmutativa (NC) se ha establecido en los \'ultimos a\~nos como un posible modelo efectivo de la gravedad cu\'antica debido, fundamentalmente, a las benignas propiedades de renormalizaci\'on que han demostrado tener algunos modelos dentro de este marco. 

El objetivo de esta tesis es la implementaci\'on de las t\'ecnicas  del formalismo de l\'inea de mundo (FLM), las cuales ya han demostrado su eficacia en el estudio de las TCC usuales, al c\'alculo de cantidades a un bucle de diversos modelos de TCC NC. El principal resultado consiste en el an\'alisis a un bucle del  modelo de Grosse-Wulkenhaar utilizando el FLM. Asimismo, mostramos c\'omo adaptar el FLM a modelos no conmutativos de campos escalares autointeractuantes en el plano Moyal. En el camino de esta adaptaci\'on, encontramos los desarrollos (para tiempo propio peque\~no) del n\'ucleo de calor de operadores con potenciales singulares y no locales; estos resultan de inter\'es tanto por su posible aplicaci\'on f\'isica como por su contenido matem\'atico.


\chapter*{Lista de Abreviaturas}

\setlength{\tabcolsep}{15pt}

\begin{tabular}{ll}
AE & Acci\'on Efectiva.\\
ET & Espaciotiempo.\\
FE & Funciones espectrales.\\
FG & Funcional Generatriz.\\
FLM & Formalismo de L\'inea de Mundo,\\
& \hspace{0.5cm}del ingl\'es \emph{Worldline Formalism}.\\
IdC & Integral de Camino.\\
IR & Infrarrojo.\\
GW & Grosse-Wulkenhaar,\\
&\hspace{0.5cm}en referencia al modelo.\\
LS & Langmann-Szabo,\\
& \hspace{0.5cm}en referencia a la simetr\'ia.\\ 
NC & No Conmutativo/a/os/as.\\
NdC & N\'ucleo de Calor, del ingl\'es, \emph{Heatkernel}.\\
QCD & \emph{Quantum Chromodynamics},\\
& \hspace{0.5cm}en ingl\'es, Cromodin\'amica Cu\'antica.\\
QED & \emph{Quantum Electrodynamics},\\
& \hspace{0.5cm}en ingl\'es, Electrodin\'amica Cu\'antica.\\
SDW & Seeley-DeWitt,\\
&\hspace{0.5cm}en referencia a los coeficientes.\\
TCC & Teor\'ia Cu\'antica de Campos.\\
t.f. & T\'erminos finitos.\\ 
UV & Ultravioleta.
\end{tabular}


\normalsize \chapter{Introducci\'on}\label{INTRO}
\setlength\epigraphwidth{6.5cm}
\epigraph{\itshape (...) da{\ss} die Quantentheorie nicht nur die Maxwellsche Electrodynamik, sondern auch die neue Gravi\-ta\-tions\-theo\-rie wird modifizieren m\"ussen.}{-- \textsc{A. Einstein}, \textit{N\"aherungsweise Inte\-gra\-tion der Feldgleichungen der Gravitation (1916).}}

\section{Sobre el problema de la gravitaci\'on cu\'antica}\label{INTRO.gc}
La f\'isica del siglo XX estuvo marcada por el triunfo de dos revolucionarias teor\'ias, a decir, aquellas anunciadas por las dos peque\~nas nubes que Lord Kelvin, en un discurso de 1900, ve\'ia en el di\'afano cielo de la f\'isica\footnote{Lord Kelvin hac\'ia referencia a dos fen\'omenos f\'isicos que eran por ese entonces incomprendidos:  la dependencia del calor espec\'ifico de gases y s\'olidos con la temperatura, y los resultados del experimento de Michelson-Morley que refutaban la existencia del \'eter.}: la relatividad general y la teor\'ia cu\'antica de campos (TCC). 

El modelo est\'andar, una TCC, ha sido m\'as que exitoso en la descripci\'on de fen\'omenos microsc\'opicos que involucran la f\'isica de part\'iculas y en los cuales los efectos gravitatorios pueden despreciarse. Para ello, como es sabido, fue necesario lidiar con los infinitos que plagaban la teor\'ia y parec\'ian tornarla inutilizable. 
Como arquetipo de este \'exito suele tomarse la precisa determinaci\'on de la inversa de la constante de estructura fina $\alpha$; como se explica en \textcite[p\'ags. 196--198]{Peskin:1995ev} esta determinaci\'on involucra c\'alculos a un orden de cuatro bucles en la teor\'ia de la electrodin\'amica cu\'antica (QED), y experimentos que se basan en estudiar, entre otras cantidades, el momento magn\'etico an\'omalo del electr\'on ($\alpha^{-1}=137,035\, 992\, 35_{73}$), la estructura hiperfina del muonio ($\alpha^{-1}=137,035\,994_{18}$) o el efecto Hall cu\'antico ($\alpha^{-1}=137,035\, 997\, 9_{32}$). Por m\'as que una indeterminaci\'on de menos de una parte en mil millones parecer\'ia definitiva, la comunidad contin\'ua hoy en d\'ia intentando correr los 
l\'imites tanto en los c\'alculos como en las mediciones \parencite{Aoyama:2014sxa}.

En el otro extremo, la teor\'ia general de la relatividad ha demostrado su validez explicando la f\'isica de los grandes cuerpos, para los cuales las propiedades cu\'anticas pueden ser dejadas de lado. Ha descripto los ya cl\'asicos fen\'omenos de precesi\'on del perihelio de mercurio y de deflecci\'on de la luz por el sol, y ha superado su puesta a prueba por modernos experimentos sobre el retardo en el tiempo de viaje de la luz \parencite{Bertotti:2003rm} o el efecto geod\'etico \parencite{Everitt:2011hp}.

Vale entonces preguntarse: ?`qu\'e es lo que sucede cuando ninguna de las dos teor\'ias puede ser obviada? Simplemente\ldots \phantom{i}no lo sabemos. Para colmo de males, parecer\'ia ser que no poseemos a disposici\'on datos experimentales de una tal situaci\'on. Esto ha motivado a algunos investigadores a sugerir que este problema, el cual llamaremos de la gravedad cu\'antica, es un problema m\'as adecuado al campo de la filosof\'ia que al de la f\'isica. 

Tenemos empero una certeza: no podr\'iamos explicar esos fe\-n\'o\-menos utilizando las actuales teor\'ias. No tenemos la m\'as remota idea de qu\'e suceder\'ia al hacer colisionar dos part\'iculas con energ\'ias del orden de la escala de energ\'ia de Planck\footnote{La energ\'ia de Planck es aquella \'unica que se obtiene a partir de las constantes fundamentales de ambas teor\'ias: la velocidad de la luz en el vac\'io $c$, la constante de Planck reducida $\hbar$ y la constante de gravitaci\'on universal de Newton  $G$.} $E_P=\sqrt{\frac{c^5\hbar}{G}}\sim10^{28}\,\text{eV}$. Por separado, las dos teor\'ias arrojan resultados incongruentes. Lo que es peor a\'un, al intentar vincularlas se encuentran inconsistencias l\'ogicas o matem\'aticas que nos previenen de realizar predicciones cuantitativas. Esto era de esperarse desde el momento en que uno comprende c\'omo se construye el espaciotiempo (ET) en ambas teor\'ias: en TCC un punto puede localizarse s\'olo utilizando una part\'icula de prueba de masa 
infinita \parencite{Salecker:1957be}, mientras que 
en relatividad general se precisa de 
part\'iculas de masa nula.

En todo caso, queda en claro que estas teor\'ias tienen cierto rango de validez, relacionado con la energ\'ia $E$ de los elementos a estudiar. ?`Ser\'a que existe una teor\'ia que engloba ambas, y que en el l\'imite adecuado (tal vez $E\ll E_p$) se reduce a ellas? Entre las diversas opciones que ofrece la f\'isica actual, la teor\'ia de cuerdas es la m\'as aceptada por la comunidad cient\'ifica como posible teor\'ia del todo.  Allende las discusiones sobre su poder predictivo y dos novedosas caracter\'isticas a\'un no observadas (la supersimetr\'ia y las seis dimensiones adicionales al ET), es cierto que el c\'alculo de cantidades f\'isicas involucra un pesado formalismo matem\'atico. 

En estas condiciones, seguramente ser\'a de ayuda disponer de otros modelos que, conservando la base de las teor\'ias de cuerdas, simplifiquen la obtenci\'on de resultados num\'ericos. En b\'usqueda de estas nuevas alternativas es plausible suponer que, vista la estrecha relaci\'on entre la gravedad y la geometr\'ia, una teor\'ia de la gravedad cu\'antica debe introducir una cuantizaci\'on del espacio ordinario. En otras palabras, es de esperar que el \'algebra de las coordenadas se torne no conmutativo (NC).

\section[Breve rese\~na hist\'orica de la TCC NC]{Breve rese\~na hist\'orica de la  teor\'ia cu\'antica de campos no conmutativa}\label{INTRO.resena}

La primera referencia que encontramos acerca de la partici\'on del espacio en celdas se remonta a 1930 y corresponde a \textcite{Heisenberg:1930}, quien busca	 una forma de regularizar las divergencias en la autoenerg\'ia del electr\'on.  En cambio, debemos transportarnos hasta los \'ultimos a\~nos de la d\'ecada de 1940 para dar con el pionero en la construcci\'on de una TCC sobre un ET NC, \textcite{Snyder:1946qz}. En su trabajo, contempor\'aneo y rival de la incipiente renormalizaci\'on perturbativa de Tomonaga, Schwinger, Feynman y \textcite{Dyson:1949ha}, abriga la esperanza de que la longitud m\'inima $\ell$, inducida en el espacio por la no conmutatividad de las coordenadas, podr\'ia funcionar tambi\'en como un par\'ametro regulador de divergencias, en este caso las ultravioletas (UV) de las TCC. 

Para comprender esa idea, tomemos como ejemplo la teor\'ia $\lambda\phi^4$ definida sobre un ET eucl\'ideo. La predicci\'on de magnitudes observables implica, al orden de un bucle, el c\'alculo de las contribuciones de los diagramas tipo renacuajo y tipo pez, las cuales son respectivamente
\begin{fmffile}{tadpole}
\begin{align}
\begin{split}
\begin{gathered}
{\begin{fmfgraph}(50,50)
\fmfpen{thin}
   \fmfleft{i1}
 \fmfright{o1}
   \fmf{plain,tension=1}{i1,v1}
\fmf{plain,left}{v1,v1}
 \fmf{plain,tension=1}{v1,o1}
\fmfdot{v1}
  \end{fmfgraph}}
\end{gathered}
\propto& \int d^4p \frac{1}{p^2+m^2},\\
\begin{gathered}
 {\begin{fmfgraph}(50,50)
\fmfpen{thin}
   \fmfleft{i1,i2}
 \fmfright{o1,o2}
   \fmf{plain,tension=1}{i1,v1}
\fmf{plain,tension=1}{i2,v1}
\fmf{plain,right=1,tension=1}{v1,v2,v1}
 \fmf{plain,tension=1}{v2,o1}
\fmf{plain,tension=1}{v2,o2}
\fmfdot{v1,v2}
  \end{fmfgraph}}
\end{gathered}\propto
&\int d^4p \frac{1}{p^2+m^2} \frac{1}{(p-q)^2+m^2}.
\end{split}\label{intro:diagramas.planares}
\end{align}
\end{fmffile}Estas integrales sobre todo el espacio son ciertamente divergentes, debido al comportamiento de los integrandos para grandes valores del impulso (regi\'on UV). Sin embargo, si la integraci\'on fuera limitada, digamos, a la regi\'on $p^2<\Lambda^2\sim\ell^{-2}$, la integral resultar\'ia convergente.

Pese a los esfuerzos de \textcite{Snyder:1947nq} por obtener una teor\'ia del campo electromagn\'etico en un espacio NC, la idea no tuvo en la comunidad f\'isica la repercusi\'on de la que s\'i goz\'o la teor\'ia de renormalizaci\'on perturbativa. 
No fue sino hasta los \'ultimos a\~nos de la d\'ecada de 1980 que las semillas sembradas por Snyder finalmente germinaron. En esos a\~nos, luego de varios trabajos que dieron lugar a la formalizaci\'on matem\'atica de la rama que dio a llamarse geometr\'ia no conmutativa \parencite{Connes:1994}, \textcite{DuboisViolette:1988vq} propusieron utilizar este formalismo para estudiar teor\'ias de campos cl\'asicos de gauge; poco despu\'es, la aplicaci\'on al modelo standard fue analizada por \textcite{Connes:1990qp}. En su forma final \parencite{Chamseddine:1996zu, Connes:2006qv}, la idea corresponde a estudiar la traza de cierto operador de Dirac sobre una geometr\'ia no conmutativa, que conduce tanto a la parte fermi\'onica como bos\'onica de la acci\'on cl\'asica del modelo est\'andar. 

Casi a la par, en 1990, \textcite{Filk:1990hp} comenz\'o a examinar el desarrollo perturbativo de un modelo de TCC NC para un campo escalar autointeractuante sobre el plano no conmutativo. La formulaci\'on en este espacio, definido por la relaci\'on de conmutaci\'on 
$$i\,[\hat{x}^{\mu},\hat{x}^{\nu}]=2\,\Theta^{\mu\nu}\in\mathbb{R}$$
entre los operadores coordenada,  supuso dos avances:

\renewcommand{\labelenumi}{\arabic{enumi})}

\begin{figure}[ht]
\begin{minipage}{.9\textwidth}
\centering
\begin{fmffile}{noplanar}
{\begin{fmfgraph}(100,100)
\fmfpen{thick}
   \fmfleft{i1}
 \fmfright{o1}
\fmf{plain,tension=1}{i1,v1}
\fmf{plain,tension=1}{v1,v2}
\fmf{plain,tension=1}{v2,v3}
 \fmf{plain,tension=4.5}{v3,o1}
\fmfdot{v1}
\fmffreeze
\fmf{plain,left,tension=0.3}{v1,v2}
\fmf{plain,right,tension=-0.3}{v1,v2}
\fmf{plain,rubout,tension=1.5}{v1,v3} 
  \end{fmfgraph}}
\end{fmffile}
\caption{\small Diagrama de renacuajo no planar.\label{fig:noplanar}}
\end{minipage}
\end{figure}
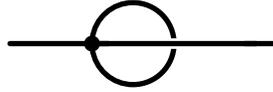

\begin{enumerate}
 \item existe una correspondencia un\'ivoca entre el \'algebra de funciones de operadores posici\'on con el cual trabaja y el \'algebra de funciones con un producto no conmutativo (Moyal o $\star$), el cual a fines pr\'acticos suele definirse en la forma
\begin{align}
(f\star g)(x):=e^{-i\,\partial_{\mu}^{f}\Theta^{\mu\nu}\partial^{g}_{\nu}}\,f(x)g(x),
\end{align}
donde $\partial^f_{\mu}$ representa el operador $\partial_{\mu}$ que act\'ua sobre $f(x)$.
Este hecho hab\'ia sido ya apreciado por \textcite{Groenewold:1946kp,Moyal:1949sk}, quienes inspeccionaron la analog\'ia, propuesta por \textcite{Weyl:1927,Wigner:1932},  entre una teor\'ia estad\'istica sobre el espacio de fases y la mec\'anica cu\'antica usual\footnote{La conexi\'on entre la mec\'anica cu\'antica usual en una dimensi\'on y el plano no conmutativo es evidente luego de observar los conmutadores involucrados.}. En otras palabras, esta correspondencia significa que, a trav\'es de un mapeo entre operadores y funciones, multiplicar funciones de operadores  es id\'entico a utilizar el producto Moyal ($\star$) entre funciones de variable real y evaluar el resultado en los operadores 
\begin{align}
f(\hat{x})\cdot g(\hat{x})=(f\star g)(\hat{x});
\end{align}                                                                                                                                                                                                                                                                                                                                                                                                         
\item en el espacio dual a las coordenadas (en el sentido de Fourier), el producto Moyal se reduce a la introducci\'on de una fase $e^{i\sum_{\mu,\nu}p_{\mu}\Theta^{\mu\nu} p_{\nu}}$ en las reglas de Feynman para el v\'ertice, donde $p_{\mu}$ son los momentos asociados a las patas de dicho v\'ertice.  Como consecuencia,  surgen a nivel diagram\'atico dos grupos: el de los diagramas planares y el de aquellos no planares. Para los primeros, las fases ligadas a momentos internos se cancelan y la estructura de divergencias de la contribuci\'on es id\'entica a la de los diagramas de la teor\'ia conmutativa subyacente ($\Theta^{\mu\nu}=0$). Para una teor\'ia $\lambda\phi^4_{\star}$, este es el caso de los diagramas de la ecuaci\'on \eqref{intro:diagramas.planares}. Los no planares, en cambio, poseen una fase que alentadoramente los vuelve m\'as convergentes. Un ejemplo, siempre en la teor\'ia $\lambda\phi^4_{\star}$, es el diagrama de la Figura \ref{fig:noplanar}; a causa del ordenamiento de las patas resulta 
diverso al diagrama de 
renacuajo usual.
\end{enumerate}

A mediados de la d\'ecada de 1990 una serie de trabajos hizo hincapi\'e en los problemas de colapso gravitacional que surgen en un proceso ideal de medici\'on de coordenadas al intentar conjugar la TCC y la relatividad general \parencite{Ahluwalia:1993dd, Doplicher:1994tu}. Su punto de vista novedoso consisti\'o en explotar estos inconvenientes para motivar relaciones de incerteza entre las coordenadas, requiriendo justamente que un proceso de medici\'on no pueda derivar en la creaci\'on de un agujero negro. Esto correspond\'ia a seguir un camino en cierto sentido contrario al que era utilizado en ese momento, visto que eran las teor\'ias propuestas de gravedad cu\'antica las que suger\'ian un comportamiento granular a altas energ\'ias \parencite{Ashtekar:1993yb}. 

Por otro lado, luego de un lustro \textcite{Seiberg:1999vs} mostraron que ciertas teor\'ias cuerdas se reducen a TCC NC en cierto r\'egimen de bajas energ\'ias, reforzando la idea esbozada en la secci\'on anterior de que podr\'ian ser utilizadas para estudiar (fenomenol\'ogicamente) la gravedad cu\'antica. Como respuesta a este resultado, se gener\'o una oleada de publicaciones sobre el tema; tan solo unos pocos meses despu\'es, apareci\'o entre ellas una con un resultado sumamente desesperanzador: las TCC NC, en vez de curar las divergencias usuales, padec\'ian de un nuevo inconveniente llamado mezcla ultravioleta-infrarroja (UV-IR) \parencite{Minwalla:1999px}. Si bien los diagramas que muestran este comportamiento son divergentes UV en la teor\'ia conmutativa y se vuelven convergentes UV en la teor\'ia NC, su inclusi\'on en diagramas de orden mayor genera divergencias IR que parecieran ser no renormalizables.

El punto crucial para solucionar este problema, de acuerdo a Grosse y Wulkenhaar (GW), fue el an\'alisis de las simetr\'ias cl\'asicas del potencial $\lambda\phi^4_{\star}$ \parencite{Grosse:2004yu}. En efecto, este potencial resulta invariante frente a las transformaciones de dualidad de Langmann-Szabo (LS), mediante las cuales coordenadas y momentos son intercambiados \parencite{Langmann:2002cc}. Con este argumento fundamentaron que la funci\'on de dos puntos posiblemente tendr\'ia contribuciones divergentes que respetaban dicha dualidad; ergo, para que el modelo resultara renormalizable el propagador tambi\'en deb\'ia respetar esa simetr\'ia. La soluci\'on m\'as simple, correspondiente a agregar un potencial arm\'onico a los t\'erminos del lagrangiano, fue la adoptada por GW. En este caso, la originalidad de GW corresponde no tanto a la introducci\'on de este t\'ermino harm\'onico, cuyo surgimiento se pod\'ia vislumbrar en el trabajo de \textcite{Kempf:1994qp}, sino a la demostraci\'on de la 
renormalizabilidad 
perturbativa del 
modelo a todo orden.

Desde entonces, el modelo de GW se ha mostrado poseedor de ciertas propiedades m\'as que interesantes.  Comencemos mencionando que, a diferencia de lo que parece ocurrir para el modelo conmutativo $\lambda\phi^4$, no sufre del problema del polo de Landau. Este inconveniente, que afecta tambi\'en a la QED \parencite{Landau:1956zr} y puso en jaque a las TCC promediando la d\'ecada de 1950, puede ser planteado de la siguiente manera: la constante de acoplamiento desnuda $\lambda$, aquella con la cual se comienza el proceso de renormalizaci\'on, adquiere una dependencia en la constante renormalizada $\lambda_R$, un par\'ametro medible, y la escala de energ\'ia asociada a un par\'ametro de corte $\Lambda$. Verbigracia, luego de resolver las ecuaciones del grupo de renormalizaci\'on a orden dominante en el desarrollo perturbativo del modelo $\lambda\phi^4$  en un ET de dimensi\'on cuatro, se encuentra que
\begin{align}\label{intro.pololandau}
 \lambda= \frac{\lambda_R}{1-\beta_2\lambda_R\log(\frac{\Lambda}{m})},
\end{align}
donde $\beta_2>0$ es una constante que depende de la teor\'ia en cuesti\'on\footnote{En QED se observa un comportamiento an\'alogo identificando la constante de acoplamiento con el cuadrado de la carga del electr\'on.}. Para $\lambda_R>0$, al incrementar el valor del par\'ametro de corte nos encontramos con un valor para el cual la constante $\lambda$ diverge; dado que para obtener esta f\'ormula se hab\'ia supuesto la peque\~nez de la constante $\lambda$ a fin de obtener desarrollos perturbativos, el modelo resulta inconsistente. Esto sucede siempre y cuando $\lambda_R\neq0$; caso contrario, logramos eliminar el polo y eludir la divergencia a fuerza de obtener una constante de acoplamiento $\lambda_R$ id\'enticamente nula, a saber, la teor\'ia de un campo libre. Esto suele ligarse al fen\'omeno de trivialidad, equivalente a la anulaci\'on de las funciones de m\'as de dos puntos de la teor\'ia o a su posible descripci\'on en t\'erminos de campos libres. 
Ha sido demostrado que dicha trivializaci\'on acaece para el modelo $\lambda\phi^4$ en dimensiones $d>4$ \parencite{Aizenman:1981zz}, mientras que algunos estudios anal\'iticos \parencite{Frohlich:1982tw} y num\'ericos \parencite{Suslov:2008ca,Wolff:2009ke} sugieren lo mismo para $d=4$.


  Una forma elegante de salvar estas dificultades result\'o ser, para muchos, el descubrimiento de la libertad asint\'otica   para las teor\'ias de Yang-Mills en la d\'ecada de 1970 \parencite{tHooft:1972zz, Gross:1973id, Politzer:1973fx}. El modelo de GW, como hemos dicho previamente, burla el problema del polo de Landau, pero no exhibiendo una libertad asint\'otica. A fines del 2006, \textcite{Disertori:2006nq} analizaron el flujo del grupo de renormalizaci\'on utilizando una identidad similar a la de Ward en QED. Sorpresivamente, aunque no tanto dados los resultados previos al orden de uno y tres  bucles \parencite{Grosse:2004by, Disertori:2006uy},  la funci\'on beta se anulaba a todo orden perturbativo en el punto autodual de LS y demostraban as\'i que la constante de acoplamiento pose\'ia una cota superior: el modelo estaba en consecuencia dotado de lo que se denomina seguridad asint\'otica.

Recientemente \textcite{Grosse:2012uv} han demostrado que cierto l\'imite altamente no conmutativo de su modelo puede ser resuelto, es decir, es posible conocer todas las funciones de correlaci\'on, cuanto menos formalmente. Adem\'as, estos resultados permiten analizar el modelo de GW en el marco de las TCC constructivas, teor\'ias que surgieron a mediados del siglo XX como respuesta a la falta de rigurosidad matem\'atica que aquejaba a las TCC, proponiendo una formulaci\'on axiom\'atica de estas \'ultimas basada en principios f\'isicos incuestionables, ya sea en el espacio minkowskiano \parencite{Wightman:1956zz} o eucl\'ideo \parencite{Schwinger:1959zz}. Los indicios expuestos al respecto en las publicaciones  de \textcite{Grosse:2014nza,Grosse:2014lxa} son promitentes.

Para finalizar, no debemos olvidar referirnos a las posibles evidencias experimentales. Aseverar que no disponemos de datos experimentales para evaluar la validez de las teor\'ias no conmutativas nos parece apresurado. En este sentido, numerosos modelos y datos experimentales han
sido analizados con el fin de obtener cotas inferiores para la escala de energ\'ia $\Lambda_{NC}$ de no conmutatividad. Entre aquellos que
involucran procesos de altas energ\'ias, del orden del TeV, podemos mencionar el estudio del proceso de recalentamiento posterior a la
inflaci\'on (\textcite{Horvat:2011wh}, $\Lambda_{NC}\gtrsim10^{-6}\,E_{planck}$), de la interacci\'on entre neutrinos y electrones
(\textcite{Horvat:2011iv}, $\Lambda_{NC}\gtrsim10^{-16}\,E_{planck}$) y de \textit{Gamma Ray Bursts} (\textcite{Nemiroff:2011fk},
$\Lambda_{NC}\gtrsim E_{planck}$).  Tambi\'en ha sido dispensada atenci\'on a la investigaci\'on de procesos de bajas energ\'ias, del orden del
eV, centrada fundamentalmente en el surgimiento de un t\'ermino magn\'etico que tiene implicancias en la f\'isica de part\'iculas y \'atomos (\textcite{Adorno:2011me}, $\Lambda_{NC}\gtrsim10^{-12}\,E_{planck}$ y \textcite{Mocioiu:2001nz},
$\Lambda_{NC}\gtrsim10^{-5}\,E_{planck}$). 

En todo caso, es probable que la posibilidad de falsar estas teor\'ias est\'e m\'as cerca de lo que
intuimos \parencite{AmelinoCamelia:2008qg}. Est\'e de acuerdo o no con esta afirmaci\'on, acompa\~nenos en el siguiente razonamiento:
consideremos rayos c\'osmicos de protones ultra energ\'eticos, con energ\'ias $E\sim 10^{20}\,\text{eV}$. Coincidimos en que $\frac{E}{E_p}\sim10^{-8}$, lo cual sugiere dificultades para detectar posibles desviaciones de los datos respecto a los de un modelo conmutativo. Sin embargo, ?`qu\'e impide que eventualmente las correcciones 
puedan involucrar la masa $m_p$ del prot\'on, ser de la forma $\frac{E^2}{E_P\,m_p}\sim 10^3$ y quiz\'as observables? 

\section{Formalismo de L\'inea de Mundo}
 
Los m\'etodos funcionales han demostrado su potencialidad desde la concepci\'on  de las integrales de camino por parte de \textcite{Feynman:1948ur}. Basta, por ejemplo, notar la simpleza y claridad que goza la rederivaci\'on de las reglas de Feynman a trav\'es de estas t\'ecnicas. Sin lugar a dudas, uno de sus principales sucesos ha sido en el marco de las teor\'ias de Yang-Mills, donde permite implementar elegantemente la invarianza de gauge en el c\'alculo perturbativo de diversas cantidades (diagramas de Feynman, identidades de Slavnov-Taylor, etc.).

No obstante, su \'exito no se reduce a la obtenci\'on de resultados perturbativos, sino que resulta fundamental en el an\'alisis de cantidades no perturbativas. Podemos citar como ejemplo el estudio de solitones e instantones \parencite{tHooft:1976fv,tHooft:1976up,Shifman:1999mv}, o de teor\'ias de gauge supersim\'etricas no abelianas \parencite{Vandoren:2008xg}. Para no correr el riesgo de aburrir al lector a trav\'es de una larga enumeraci\'on de aplicaciones que van incluso m\'as all\'a del campo de la f\'isica, lo remitimos a la bibliograf\'ia especializada \parencite{Bastianelli:2006rx,Kleinert:2004ev}.

Por su parte, el formalismo de l\'inea de mundo (FLM) es un m\'etodo que involucra el c\'alculo de integrales funcionales en mec\'anica cu\'antica para el c\'alculo de cantidades en TCC \parencite{Schubert:2001he}. En consecuencia, esta t\'ecnica hereda las bondades conceptuales y pr\'acticas de los m\'etodos funcionales. 

Al rastrear los usos del FLM podemos llegar hasta el trabajo de \textcite{Feynman:1950ir}, en cuyo ap\'endice se ve un atisbo de la t\'ecnica: utiliz\'o integrales de camino en la primera cuantizaci\'on como m\'etodo para encontrar soluciones de la ecuaci\'on de Klein-Gordon. A nuestro entender, la primera aplicaci\'on del FLM propiamente dicho corresponde al estudio de la producci\'on de pares electr\'on-positr\'on ante la presencia de campos el\'ectricos externos peque\~nos \parencite{Affleck:1981bma}. Cerca en el tiempo, \textcite{AlvarezGaume:1983ig} consideraron dos part\'iculas, una bos\'onica y una fermi\'onica, para calcular anomal\'ias gravitacionales en teor\'ias de campos fermi\'onicos en interacci\'on con un campo gravitatorio d\'ebil.

M\'as adelante, t\'ecnicas inspiradas en integrales funcionales sobre hojas de mundo en teor\'ias de cuerdas fueron utilizadas por \textcite{Bern:1987tw,Bern:1991aq}, con el objeto de calcular correcciones de un bucle a acciones efectivas de campos escalares en presencia de campos de gauge de fondo. Estos resultados fueron luego  elegantemente derivados por \textcite{Strassler:1992zr} en t\'erminos de integrales de camino para part\'iculas, a saber lo que hoy en d\'ia conocemos como FLM. 

Entre las diversas aplicaciones exitosas del FLM que desde entonces se han realizado, podemos mencionar su uso al an\'alisis, al orden de un bucle, del acoplamiento de un fondo gravitacional externo con campos cu\'anticos de diverso spin  \parencite{Bastianelli:2002fv,Bastianelli:2002qw,Bastianelli:2004zp,Bastianelli:2005vk,Bastianelli:2007pv,Bastianelli:2008nm} y campos escalares en variedades chatas con borde  \parencite{Bastianelli:2006hq,Bastianelli:2007jr,Bastianelli:2008vh,Bastianelli:2009mw}; a ordenes superiores en el n\'umero de bucles ha permitido el estudio de teor\'ias escalares \parencite{Sato:1998sf}, espinores en QED \parencite{Schubert:1996jj} y teor\'ias de Yang-Mills \parencite{Sato:1999xy}. 
Siempre en el marco de QCD, resulta de gran ayuda, por ejemplo, al realizar c\'alculos de amplitudes con $n$ fotones o gluones \parencite{Ahmadiniaz:2012ie}.

Asimismo, cabe destacar que  las integrales de camino resultan apropiadas para la realizaci\'on de simulaciones computacionales de modelos de TCC, tomando  o no un ET discretizado \parencite{Dunne:2009zz}. Combinadas con
los m\'etodos de Monte Carlo en lo que ha recibido el nombre de \emph{worldline numerics}, ha posibilitado el  c\'alculo de energ\'ias de Casimir en diversas geometr\'ias y el estudio de su dependencia con la 
temperatura \parencite{Gies:2003cv,Gies:2006bt,Gies:2006cq,Gies:2006xe,Gies:2008zz,Weber:2010kc,Weber:2010xv}, como as\'i tambi\'en el de la producci\'on de pares ante la presencia de campos electromagn\'eticos \parencite{Gies:2005bz}. Al encarar este \'ultimo problema, \textcite{Dunne:2006ff} mostraron la posibilidad de realizar c\'alculos anal\'iticos no perturbativos considerando instantones en la l\'inea de mundo.

\section{Objetivos y estructura de esta tesis}\label{INTRO.objetivos}
Hemos ya planteado algunos de los problemas que surgen al intentar formular una teor\'ia de la gravedad cu\'antica y sugerido que su abordaje desde una posible teor\'ia efectiva, la TCC NC,  puede resultar de ayuda. El prop\'osito global de la presente tesis es entonces abordar ciertos modelos de TCC NC a trav\'es del prisma del FLM, con el objeto de calcular diversas cantidades f\'isicas al orden de un bucle; como se podr\'a observar de los resultados, la eficiencia que los m\'etodos funcionales ofrecen en la TCC usual se conserva en el caso NC.  La comparaci\'on de este tipo de correcciones con datos experimentales podr\'ia llevar luego a acotar los par\'ametros de no conmutatividad $\Theta^{\mu\nu}$. Pasemos ahora a detallar la estructura de esta tesis, desglosando los objetivos particulares de cada cap\'itulo. 

En el cap\'itulo \ref{DET} introducimos conceptos y cantidades b\'asicas de TCC. Este cap\'itulo est\'a concebido no para ofrecer un tratamiento acabado del tema (que el lector puede encontrar refiri\'endose a la rica literatura existente) sino como un resumen que, actuando a modo de recordatorio y fijando notaciones a trav\'es de ejemplos, dote de cierto grado de independencia a la tesis. De este modo introducimos someramente las cantidades y t\'ecnicas que ser\'an de inter\'es en nuestros c\'alculos posteriores: la acci\'on efectiva (escrita perturbativamente en t\'ermino de funciones espectrales), la t\'ecnica de renormalizaci\'on y la energ\'ia de Casimir.

En el decurso de ese cap\'itulo se puede adivinar el rol primordial que tienen las funciones espectrales (FE), en especial el n\'ucleo de calor (NdC), en el desarrollo de este trabajo. Por este motivo, creemos conveniente realizar una exposici\'on formal de las mismas a lo largo del cap\'itulo \ref{FE}, evitando demostraciones que podr\'ian desviar la atenci\'on de los objetivos de la tesis. 

M\'as adelante, en el cap\'itulo \ref{FLM}, presentamos el FLM para el caso m\'as sencillo, el de un campo cu\'antico escalar autointeractuante. Como resultado intermedio obtenemos el desarrollo asint\'otico, para tiempo propio peque\~no, del NdC del operador de fluctuaciones cu\'anticas, suponiendo ciertas condiciones de regularidad sobre el potencial.  A modo de ejemplo, computamos al orden de un bucle la acci\'on efectiva del modelo $\lambda\phi^4$, para luego analizar su renormalizaci\'on a id\'entico orden. En forma suplementaria y finalizando este cap\'itulo dedicado a las teor\'ias conmutativas, consideramos un modelo con un potencial delta de Dirac de fondo; para este potencial singular, mostramos como calcular la energ\'ia efectiva. Estos \'ultimos resultados, de inter\'es en diversos modelos que proponen densidades de carga localizadas en superficies, corresponden al trabajo de \textcite{Vinas:2010ix}.

El cap\'itulo \ref{NC} est\'a dedicado a la generalizaci\'on del FLM a TCC NC en las cuales la no conmutatividad entre las coordenadas est\'a dada por par\'ametros constantes. Una vez hecho un breve comentario sobre como implementar la no conmutatividad haciendo uso del producto Moyal, el modelo 
$\lambda\phi^3_{\star}$  muestra como novedad que el operador de fluctuaciones cu\'anticas puede poseer t\'erminos no locales. La clave en nuestra implementaci\'on del FLM es, como era de esperar a partir del \'algebra de operadores posici\'on y momento, trabajar con integrales de camino en el espacio de fases. El resultado principal, contenido en el trabajo publicado en conjunto con \textcite{Bonezzi:2012vr}, es una f\'ormula maestra para el desarrollo, a tiempo propio peque\~no, del NdC de operadores con potenciales no locales.  Adem\'as, tomamos diversos modelos no conmutativos para 
ejemplificar las peculiaridades que en general ofrecen; entre ellos, el modelo del disco no conmutativo, estudiado junto a \textcite{Falomir:2013vaa}.

Prosiguiendo con el estudio de las TCC NC, el cap\'itulo \ref{GW} detalla la aplicaci\'on del FLM al modelo de GW. En este caso, el potencial de fondo arm\'onico sugiere una ligera adaptaci\'on de las t\'ecnicas del cap\'itulo previo; una vez realizados esos retoques, el estudio de la renormalizaci\'on del modelo al orden de un bucle es inmediato. Dos casos especiales, el de la no conmutatividad degenerada y el del oscilador anisotr\'opico, merecen particular atenci\'on. Este cap\'itulo est\'a basado en el trabajo realizado por \textcite{Vinas:2014exa}.

Por \'ultimo, el cap\'itulo \ref{CONC} contiene las conclusiones de este trabajo y ciertas ideas que, habiendo quedado parcial o totalmente en el tintero, podr\'ian dar frutos en un futuro cercano.

A modo de resumen, enumeramos a continuaci\'on la lista de las publicaciones que contienen los resultados originales de esta tesis:\\[0.1cm]
\begin{itemize}
\item   S.~A.~Franchino Vi\~nas and P.~A.~G.~Pisani,
  ``Semi-transparent Boundary Conditions in the Worldline Formalism,''
  J.\ Phys.\ A {\bf 44} (2011) 295401
  [arXiv:\-1012.2883 [hep-th]].\\[0.1cm]
\item R.~Bonezzi, O.~Corradini, S.~A.~Franchino Vi\~nas and P.~A.~G.~Pisani,
  ``Worldline approach to noncommutative field theory,''
  J.\ Phys.\ A {\bf 45} (2012) 405401  [arXiv:1204.1013 [hep-th]].\\[0.1cm]
\item H.~Falomir, S.~A.~Franchino Viñas, P.~A.~G.~Pisani and F.~Vega,
  ``Boundaries in the Moyal plane,''
  JHEP {\bf 1312} (2013) 024
  [arXiv:1307.4464 [hep-th]].\\[0.1cm]
\item S.~Franchino Viñas and P.~Pisani,
  ``Worldline approach to the Grosse-Wulkenhaar model,''
  JHEP {\bf 1411} (2014) 087
  [arXiv:1406.7336 [hep-th]].\\[0.1cm]
\end{itemize}

A esta lista corresponde agregar dos trabajos relacionados que se encuentran actualmente en preparaci\'on:\\[0.1cm]
\begin{itemize}
 \item S.~Franchino Viñas y P.~Pisani, ``Thermodynamics in the Noncommutative Disc'';\\[0.1cm]
 \item S.~Franchino ~Viñas y  P.~Pisani, ``Zeta function and Casimir energy for two parallel plates in Euclidean Moyal Space''.\\[0.1cm]
\end{itemize}


\chapter{Aspectos generales de teor\'ia cu\'antica de campos}\label{DET}
\setlength\epigraphwidth{5cm}
\epigraph{\itshape Al mismo r\'io entras y no entras, pues eres y no eres.}{-- \textsc{Her\'aclito}.}

Tal como hemos explicado en la Introducci\'on, esta tesis se propone el estudio de un tipo particular de TCC: el de las teor\'ias NC. Las cantidades que habremos de calcular, no obstante poseer\'an ciertas peculiaridades inherentes a la no conmutatividad, han sido definidas en general para cualquier TCC. En esta secci\'on intentaremos proveer de un breve resumen o ayuda memoria que facilite la lectura del resto del trabajo.

Primero y principal, introduciremos el concepto de acci\'on efectiva (AE), la contraparte cu\'antica de la acci\'on a nivel cl\'asico. Pr\'acticamente sin riesgo a equivocarnos, podemos aseverar que esa fue su raz\'on de ser: construir una cantidad que poseyendo toda la informaci\'on cu\'antica del modelo responda a los principios variacionales t\'ipicos de la formulaci\'on cl\'asica. La acci\'on efectiva, concebida como una funcional del valor de expectaci\'on de vac\'io del campo cu\'antico, no es m\'as que una modificaci\'on de la acci\'on cl\'asica, en el sentido que contiene las correcciones necesarias de los diagramas de Feynman de la teor\'ia y que su m\'inimo variacional corresponde al valor de expectaci\'on del campo en el estado de vac\'io. Por este motivo es usual emplear un desarrollo en potencias de $\hbar$ para la acci\'on efectiva, cuyo primer t\'ermino corresponde justamente a la acci\'on cl\'asica. El segundo t\'ermino puede ser escrito utilizando funciones espectrales (FE); a ellas 
dedicaremos el pr\'
oximo cap\'itulo. Orden a orden en $\hbar$, permite 
observar si las simetr\'ias que pose\'ia el sistema a nivel cl\'asico se mantienen o son rotas por las correcciones cu\'anticas, dando lugar a las denominadas anomal\'ias cu\'anticas.

Por cuanto al examinar los t\'erminos de dicho desarrollo nos encontramos con expresiones divergentes, nos es  necesario presentar las ideas de regularizaci\'on que permiten dar sentido f\'isico a estas teor\'ias. Veremos que las t\'ecnicas de renormalizaci\'on, confirmadas a mediados del siglo pasado por el \'exito de la QED, son adaptables al estudio de la acci\'on efectiva a trav\'es de FE. 

Para finalizar, consideraremos el efecto Casimir, una manifestaci\'on observable de la energ\'ia de vac\'io que, como era de prever, precisa ser regularizada. Las aplicaciones de este fen\'omeno son diversas; desde el punto de vista de la f\'isica aplicada, su comprensi\'on se ha vuelto vital desde la incursi\'on tecnol\'ogica en la escala nanom\'etrica.

\section{La acci\'on efectiva}\label{DET.eff1}
Para definir la AE, consideraremos el ejemplo m\'as simple. Tomemos un campo cu\'antico $\varphi$  escalar y real sobre un espacio de base (digamos $\mathbb{R}^d$) descripto por la acci\'on $S[\varphi]$; la funci\'on de partici\'on $Z[J]$ y la funcional generatriz $W[J]$, son definidas de manera que dependen de la fuente cl\'asica $J(x)$ a trav\'es de la integral funcional
\begin{align}\label{DET:zeta-W}
  Z[J]:=e^{-\frac1\hbar\,W[J]}:=\mathcal{N}\int \mathcal{D}\varphi\ e^{-\frac1\hbar\,S[\varphi]+\frac1\hbar\,\int dx\,J\varphi}\,.
\end{align}
Aquellos no familiarizados con integrales de camino son referidos a  \textcite{Bastianelli:2006rx,Kleinert:2004ev}  para su tratamiento formal. A grandes rasgos, ser\'a suficiente su interpretaci\'on f\'isica como integral sobre todas las configuraciones del campo posibles en el espacio. La constante $\mathcal{N}$ suele ser elegida de manera que\footnote{Esta  normalizaci\'on implica que $Z[J]$ ser\'a la funcional generatriz de las funciones de Green de la teor\'ia. A su vez, en t\'erminos de diagramas de Feynman s\'olo ser\'a necesario considerar aquellos que no tienen subdiagramas de vac\'io, i.e. no poseen subdiagramas sin patas externas.} $Z[J=0]=1$:
\begin{align}\label{DET:normalizacion}
 \mathcal{N}^{-1}=\int \mathcal{D}\varphi\ e^{-\frac1\hbar\,S[\varphi]}\,.
\end{align}

Luego, introducimos el campo medio (o campo cl\'asico) $\phi(x)$ en presencia de dicha fuente $J(x)$ a trav\'es de la relaci\'on
\begin{align}
\begin{split}
  \phi(x)&:=\frac{\mathcal{N}}{Z[J]}\ \int \mathcal{D}\varphi
  \ e^{-\frac1\hbar\,S[\varphi]+\frac1\hbar\,\int J\varphi}\, \varphi(x)\\
  &=-\frac{\delta W[J]}{\delta J(x)}\,.
\end{split}\end{align}
Por este motivo se dice que los campos $J(x)$ y $\phi(x)$ son campos conjugados; adem\'as, esto nos permite considerar la transformada de Legendre de la funcional generatriz $W[J]$ en funci\'on del campo medio $\phi$
\begin{align}\label{DET:eff.def}
  \Gamma[\phi]:=\left\lbrace W[J]+\int J\phi\right\rbrace_{J=J[\phi]}\,.
\end{align}
La funci\'on $\Gamma[\phi]$, recibe el nombre de acci\'on efectiva. Ciertamente, a partir de esta definici\'on vale la igualdad
\begin{align}\label{DET:derGamma/derphi=j}
  \frac{\delta \Gamma[\phi]}{\delta\phi(x)}=J(x).
\end{align}
Al tomar $J(x)\equiv0$ obtenemos el principio variacional al que hemos hecho referencia anteriormente. Adicionalmente, notemos que los valores de expectaci\'on de vac\'io\footnote{Estos valores de expectaci\'on de vac\'io corresponden a campos temporalmente ordenados.} de productos del campo cu\'antico $\varphi$ 
pueden escribirse en t\'erminos de derivadas funcionales de $Z[J]$ evaluadas en $J\equiv0$:
\begin{align}\label{DET:derivadaW}
\begin{split}
 \langle \varphi(x_1)\cdots\varphi(x_n)\rangle:&=\int \mathcal{D}\varphi\ e^{-\frac1\hbar\,S[\varphi]} \varphi(x_1)\cdots\varphi(x_n)\\
&=\left.\frac{\delta^n Z[J]}{\delta J(x_1)\cdots\delta J(x_n)}\right\lvert_{J\equiv0}.
\end{split}
\end{align}
A partir de ellos, la f\'ormula de reducci\'on de \textcite{Lehmann:1954rq} permite determinar las amplitudes de dispersi\'on de cualquier proceso. Esto, sumado a las propiedades de la transformada de Legendre que permiten reescribir las derivadas de \eqref{DET:derivadaW} en t\'erminos de derivadas funcionales de $\Gamma[\phi]$ respecto a la variable $\phi$, implica que la informaci\'on de todo proceso de dispersi\'on est\'a contenida en la AE. 

En general obtener una expresi\'on cerrada para la AE es una ardua tarea; en su lugar, demostraremos en la pr\'oxima secci\'on c\'omo es m\'as sencillo obtener un desarrollo en potencias de $\hbar$.

\section{Desarrollo perturbativo de la acci\'on efectiva}\label{DET.eff2}
Hemos visto que la AE es la transformada de Legendre de la FG $W[J]$, mientras que en la definici\'on de esta \'ultima entran en juego integrales funcionales de exponenciales. En base a esto, es de esperar que el m\'etodo de descenso empinado permita obtener un desarrollo perturbativo de la AE. De acuerdo a este m\'etodo, buscando el desarrollo en serie de potencias de $Z[J]$ conviene considerar la acci\'on alrededor de la configuraci\'on cl\'asica $\phi_0(x)$, definida como aquella que minimiza la acci\'on $S[\phi]$; en otras palabras, $\phi_0(x)$ satisface la ecuaci\'on cl\'asica de movimiento
\begin{align}\label{DET:configclasica}
 \frac{\delta S}{\delta \varphi(x)}[\phi_0]=J(x).
\end{align}
En ausencia de fuentes externas, i.e. $J(x)=0$, la configuraci\'on cl\'asica minimiza la acci\'on $S[\phi]$ mientras que el campo $\phi(x)$, que representa el valor de expectaci\'on de vac\'io del campo, minimiza la funcional $\Gamma[\phi]$ (ver \eqref{DET:derGamma/derphi=j}). Como hab\'iamos adelantado, la acci\'on efectiva resulta contener, en este sentido, los efectos cu\'anticos del campo $\varphi$.

Retomando la definici\'on original \eqref{DET:zeta-W}, realizamos la traslaci\'on del campo cu\'antico $\varphi$ en una cantidad igual a la configuraci\'on cl\'asica $\phi_0(x)$ para llegar a la expresi\'on
\begin{align}\label{DET:zetadesarrollada}
  Z[J]=\mathcal{N}e^{-\frac1\hbar\,S[\phi_0]+\frac1\hbar\,\int J\phi_0}
  \int \mathcal{D}\varphi
  \ e^{-\frac1\hbar
  \int (\delta S_{\phi_0}-J)\,\varphi-\tfrac1{2\hbar}\iint \delta^2S_{\phi_0}\,\varphi\varphi+\ldots},
\end{align}
donde $\delta S_{\phi_0}$ denota la funci\'on obtenida como la derivada funcional $\delta S[\varphi]/\delta \varphi(x)$ evaluada en la configuraci\'on cl\'asica $\phi_0(x)$. An\'alogamente, el n\'ucleo $\delta^2S_{\phi_0}$ es la variaci\'on segunda $\frac{\delta^2 S[\varphi]}{\delta\varphi(x)\delta\varphi(x')}$ evaluada en $\phi_0(x)$. Por fortuna, la traslaci\'on en los campos no introduce una modificaci\'on en la medida de integraci\'on de la integral funcional. 

La ecuaci\'on \eqref{DET:zetadesarrollada} puede simplificarse recordando la definici\'on \eqref{DET:configclasica} de la configuraci\'on cl\'asica; m\'as a\'un, cambiando la escala del campo\footnote{Esta transformaci\'on acarrea un jacobiano constante que se cancela al realizar id\'entico cambio en los campos de la integral funcional que define la constante de normalizaci\'on $\mathcal{N'}^{-1}=\int \mathcal{D}\varphi\ e^{-\frac1\hbar\,S[\sqrt{\hbar}\varphi]}$.} $\varphi\rightarrow \sqrt{\hbar}\varphi$ se puede entrever que el m\'etodo de descenso empinado coincide en este caso con un desarrollo en potencias de $\hbar$:
\begin{align}\label{DET:zeta.predesarollo}
  Z[J]=\mathcal{N}'\,e^{-\frac1\hbar\,S[\phi_0]+\frac1\hbar\,\int J\phi_0}
  \int \mathcal{D}\varphi
  \ e^{-\tfrac{1}{2}\iint \delta^2S_{\phi_0}\,\varphi\varphi}\,e^{O(\sqrt{\hbar})}.
\end{align}
Justamente, el reescaleo nos permite proponer, expandiendo  el factor $e^{O(\sqrt{\hbar})}$ y teniendo en cuenta que el teorema de Wick constri\~ne los t\'erminos a poseer un n\'umero par de campos $\varphi$, un desarrollo perturbativo para $W[J]$  de la forma
\begin{align}
 W[J]=\sum_{n=0}^{\infty} \hbar^nW_n[J].
\end{align}
Combinando las expresiones \eqref{DET:zeta-W} y \eqref{DET:zeta.predesarollo}, es inmediato que el resultado al m\'as bajo orden es la transformada de Legendre de la acci\'on cl\'asica,
\begin{align}
W_0[J]=S[\phi_0(x)]-\int J(x)\phi_0(x).
\end{align}

Por otra parte, la integral funcional de los t\'erminos cuadr\'aticos en el campo $\varphi$ da la siguiente contribuci\'on:
\begin{align}\label{DET:w1}
\begin{split}
 e^{-W_1[J]}&=\mathcal{N}'_0\int\mathcal{D}\varphi \,e^{-\tfrac{1}{2}\iint \varphi(x)\delta^2S_{\phi_0} \varphi(x)}\\
 &=\text{Det}^{-1/2}\left(G^{-1}A\right),
\end{split}
\end{align}
donde $A$ es el operador de fluctuaciones cu\'anticas definido a trav\'es del n\'ucleo $\delta^2S_{\phi_0}$, el operador $G^{-1}$ es el inverso de $(-\partial^2+m^2)$ y  $\mathcal{N}'_0$ es la contribuci\'on a orden $\hbar^0$ de la constante de normalizaci\'on $\mathcal{N}'$. En este punto es preciso realizar un par de comentarios. Primero, $G$ tiene su origen en la elecci\'on realizada para $\mathcal{N}$, expresada en \eqref{DET:normalizacion}; un an\'alisis perturbativo en $\hbar$ para $\mathcal{N}'$ demuestra la validez de \eqref{DET:w1}. En forma concreta, estamos suponiendo que la acci\'on consta \'unicamente de un t\'ermino cin\'etico, uno de masa y uno de  potencial autointeractuante con potencias de $\varphi$ mayores a tres:
\begin{align}
 S[\varphi]=\int \varphi\, (-\partial^2+m^2)\varphi+V(\varphi).
\end{align}

Segundo, el determinante funcional $\text{Det}(X)$ est\'a definido como el producto de los autovalores del operador $X$, a saber,   
\begin{align}\label{DET:detfunc}
 \text{Det}(X):=\prod_{i=1}^{\infty}\lambda_i\,,
\end{align}
siendo $\lambda_{i\in\mathbb{N}}$, los autovalores del operador $X$.
El resultado \eqref{DET:w1} puede entenderse de la siguiente manera, suponiendo que el operador $\delta^2S_{\phi_0}$ admite una base  de autovectores $\psi_{i\in\mathbb{N}}$: todo campo puede expandirse en esta base como 
\begin{align}
\varphi(x)=\sum_{i\in\mathbb{N}} a_i\psi_i 
\end{align}
y, por consiguiente, realizar la integral sobre toda configuraci\'on equivale a sumar las contribuciones de campos con componentes arbitarias en la base elegida
\begin{align}
 \int\mathcal{D}\varphi e^{-\tfrac{1}{2}\int \varphi(x)\delta^2S_{\phi_0} \varphi(x)}=\int \prod_{i=1}^{\infty} da_i e^{-\tfrac{1}{2}\sum_{i=1}^{\infty} \lambda_i a_i^2}.
\end{align}
Salvo para algunas excepciones en las que s\'olo hay un n\'umero finito de autofunciones o los autovalores tienden a acumularse alrededor de la unidad, el determinante funcional \eqref{DET:detfunc} es una expresi\'on divergente; estas divergencias se corresponden con las que se educen haciendo uso de los diagramas de Feynman en un desarrollo perturbativo. 

Dejando moment\'aneamente de lado esas divergencias, recordemos que para calcular la acci\'on efectiva debemos primero obtener una expresi\'on para el campo $\phi(x)$ conjugado a la fuente $J(x)$
\begin{align}\label{DET:phi-phi0}
 \begin{split}
\phi(x)&=-\frac{\delta W[J]}{\delta J(x)}\\
 &=\phi_0(x)+\int \left\{\frac{\delta S}{\delta \phi}[\phi_0]-J(x)\right\} \frac{\delta \phi_0}{\delta J(x)}+O(\hbar)\\[0.2cm]
 &= \phi_0(x)+\delta \phi(x),
\end{split}
\end{align}
con $\delta \phi(x)$ de orden $\hbar$, puesto que el t\'ermino entre llaves es nulo para configuraciones cl\'asicas. 

Por su parte, tomando la definici\'on \eqref{DET:eff.def} y utilizando la relaci\'on \eqref{DET:phi-phi0}, la acci\'on efectiva resulta  
\begin{align}\label{DET:gamma.casidesarrollo}
  \begin{split}
\Gamma[\phi]&=\left\lbrace S[\phi(x)-\delta \phi(x)]-\int J(x)(\phi(x)-\delta \phi(x))\right.\\
  &\hspace{2.0cm}\left.-\hbar\log\text{Det}^{-1/2}\left(G^{-1}A\right)+O(\hbar^2)+\int J(x)\phi(x)\right\rbrace_{J=J[\phi]}\\[0.1cm]
  &=S[\phi(x)]-\int \left\{\frac{\delta S}{\delta \phi}[\phi_0]-J[\phi_0]\right\}\delta \phi(x)\\[0.2cm]
&\hspace{4cm}-\hbar\log\text{Det}^{-1/2}\left(G^{-1}A\right)+O(\hbar^2).
  \end{split}
\end{align}
En la \'ultima l\'inea hemos utilizado el hecho que $\delta \phi$ es orden $\hbar$ para reemplazar $\phi$ por la configuraci\'on cl\'asica $\phi_0$ en el t\'ermino entre las llaves. Basta por \'ultimo reconocer la ecuaci\'on de movimiento \eqref{DET:configclasica} de la configuraci\'on cl\'asica  en dicho t\'ermino de \eqref{DET:gamma.casidesarrollo} para obtener el desarrollo de la acci\'on efectiva al orden de un bucle\footnote{Puede demostrarse que el desarrollo en potencias de $\hbar$ corresponde diagram\'aticamente con el desarrollo en el n\'umero de bucles del diagrama.}
\begin{align}\label{DET:efectiva.logdet}
\Gamma[\phi]&=S[\phi(x)]+\frac{\hbar}{2}\log\text{Det}\left(G^{-1}A\right)+O(\hbar^2).
 \end{align}

M\'as a\'un, si acaso $B=G^{-1}A$ fuera una matriz diagonalizable con autovalores no nulos, podr\'iamos utilizar las propiedades del logaritmo para escribir 
\begin{align}
\begin{split}
\log\det B&=\log\left(\prod_{i=1}^n \lambda_i\right)\\
&=\text{tr}\log B,
\end{split}
\end{align}
llamando $\lambda_i$ a los autovalores de la matriz $B$. Esta propiedad se puede generalizar para cualquier matriz no singular e incluso para operadores invertibles que difieren del operador identidad en $K$, un operador de traza finita \parencite{Kontsevich:1994nc}. Dando por cierta esta igualdad entre\footnote{Las may\'usculas en $\text{Tr}$ y $\text{Det}$ son utilizadas para la traza y el determinante de operadores sobre espacios de dimensi\'on infinita.} $\text{log}\,\text{Det}$ y $\text{Tr}\,\text{log}$, la correcci\'on de un bucle $\Gamma_{1-\text{\rm bucle}}[\phi]$ a la acci\'on efectiva es\footnote{Fuera de esta secci\'on utilizaremos las unidades en las cuales $\hbar$ puede ser tomada como la unidad.}
\begin{align}\label{DET:efectiva0}
 \Gamma_{1-\text{\rm bucle}}[\phi]
 &=\frac{\hbar}{2}\text{Tr}\,\text{log}\, \left(G^{-1}A\right),
\end{align}
Si adem\'as utilizamos la f\'ormula v\'alida para operadores C y D definidos positivos
\begin{align}
 -\log\left(\frac{C}{D}\right)=\int_0^{\infty}\frac{d\beta}{\beta} \left(e^{-\beta C}-e^{-\beta D}\right),
\end{align}
e intercambiamos el signo integral con la traza, obtenemos el resultado final
\begin{align}\label{DET:efffinal}
 \Gamma_{1-\text{\rm bucle}}[\phi]=-\frac{\hbar}{2}\int_0^{\infty}\frac{d\beta}{\beta} \left(\text{Tr}\, e^{-\beta A}-\text{Tr}\,e^{-\beta G}\right).
\end{align}
Gracias a esta expresi\'on, podemos relacionar las correcciones cu\'anticas a la acci\'on efectiva de orden un bucle con la funci\'on espectral $K_X(\beta)$, llamada traza del NdC del operador $X$ y definida como
\begin{align}\label{DET:heat-ker}
K_X(\beta):=\text{Tr}(e^{-\beta X}).
\end{align}
El estudio de este tipo de funciones, definidas a trav\'es del espectro de un operador, pertenece a la rama de las matem\'aticas denominada geometr\'ia espectral \parencite{Kirsten:2000xc}. Siendo su motivaci\'on el estudio de \eqref{DET:efffinal}, en el cap\'itulo siguiente enunciaremos y formalizaremos algunas de las propiedades generales que las FE han demostrado poseer. 

Antes, en la siguiente secci\'on, mostraremos c\'omo el m\'etodo de renormalizaci\'on permite regularizar las divergencias que impregnan las FE y hasta este momento hemos omitido.

\section{Sobre la renormalizaci\'on}\label{DET.renor}
Es sabido que las TCC se encuentran plagadas de divergencias que deben ser eliminadas para dar sentido f\'isico a los resultados. En el caso de un desarrollo diagram\'atico en el espacio de momentos, tal cual hemos bosquejado en la secci\'on \ref{INTRO.resena}, las divergencias se presentan en general en la integraci\'on de grandes o peque\~nos momentos, y por este motivo reciben el nombre de divergencias UV  o IR.

La f\'ormula \eqref{DET:efffinal} para la AE no es la excepci\'on, visto que arrastra las divergencias en la definici\'on del determinante funcional del operador de fluctuaciones cu\'anticas. Actualmente, la forma aceptada de curar estas divergencias es haciendo uso del proceso de renormalizaci\'on, propuesta inicialmente por \textcite{Tomonaga:1948zz,Schwinger:1948yj,Feynman:1950ir,Dyson:1949ha}. En esta secci\'on, afrontaremos la renormalizaci\'on de la acci\'on efectiva al orden de un bucle haciendo uso de las propiedades del NdC. Un m\'etodo relacionado y m\'as frecuente en la literatura es el de la renormalizaci\'on a partir de la definici\'on de otra funci\'on espectral, la funci\'on zeta \parencite{Vassilevich:2003xt}; estas t\'ecnicas fueron concebidas por \textcite{Dowker:1975tf,Hawking:1977} y fueron r\'apidamente difundidas debido a la simpleza con la que permiten encarar el estudio de campos definidos sobre  variedades curvas. En el anexo \ref{DET.zeta} estableceremos 
la conexi\'on entre 
ambos m\'etodos. 

En el lenguaje moderno, la renormalizaci\'on esquem\'aticamente consta de los siguientes pasos: primero se regularizan las contribuciones divergentes a la AE a trav\'es de la introducci\'on de un par\'ametro $\epsilon$, de forma que en el l\'imite $\epsilon\rightarrow \epsilon_0$ la expresi\'on regularizada concuerde formalmente con el resultado original. Luego, se agrupan en la AE todos los t\'erminos que tienen la misma dependencia en los campos, dando como resultado un lagrangiano con nuevas constantes de acoplamiento (que llamaremos  renormalizadas), definidas en t\'ermino de las originales (desnudas) y del par\'ametro $\epsilon$. 

La forma de determinar las constantes renormalizadas es a trav\'es de mediciones experimentales, si corresponden a t\'erminos que estaban en el lagrangiano original o son divergentes en el l\'imite $\epsilon\rightarrow\epsilon_0$; el resto de las constantes, las surgidas de contribuciones finitas, queda definido por las anteriores. Esto motiva la siguiente clasificaci\'on: si el n\'umero de constantes a determinar experimentalmente crece sin l\'imite al aumentar el n\'umero de bucles, el modelo se llama no renormalizable; si es finito pero con infinitas contribuciones divergentes es renormalizable; finalmente, en el caso restante estamos frente a un modelo superrenormalizable.

Para analizar c\'omo se realiza la renormalizaci\'on utilizando el NdC, comencemos notando que a partir de \eqref{DET:efffinal} se comprende intuitivamente que las contribuciones diagram\'aticas con part\'iculas virtuales muy energ\'eticas corresponden a grandes autovalores del operador de fluctuaciones $A$; por ende,  adquieren importancia para valores peque\~nos del par\'ametro $\beta$. En otras palabras, las divergencias UV deber\'ian manifestarse como un mal comportamiento del integrando para valores de $\beta$ cercanos a cero. Por el contrario, posibles problemas a bajas energ\'ias ser\'ian relevantes en el l\'imite en el cual $\beta$ es indefinidamente grande; usualmente, estos inconvenientes se ver\'an controlados por la presencia de un t\'ermino de masa en los operadores $A$ y $G$, que confiere al integrando un decrecimiento exponencial $e^{-\beta m^2}$.

Con el fin de establecer estas conclusiones en forma precisa, usaremos un resultado que formalizaremos en el pr\'oximo cap\'itulo, referente al n\'ucleo de calor: bajo ciertas condiciones, es posible encontrar un desarrollo asint\'otico 
\begin{align}\label{DET:desa.hk}
 \text{Tr}\,e^{-\beta A}\sim\sum_{n=0}^\infty a_n(A)\,\beta^{(n-d)/2},\qquad \text{para}\;\;\beta\downarrow0,
\end{align}
donde $d$ es la dimensi\'on del espacio sobre el cual se encuentra definido el campo y las cantidades $a_n(A)$ se denominan coeficientes de Seeley-DeWitt (SDW). Reemplazando en \eqref{DET:efffinal}, resulta patente que los primeros t\'erminos de este desarrollo no ser\'an bien comportados al integrar valores peque\~nos de la variable $\beta$. 

Buscando regularizar dichas cantidades introducimos un par\'ametro $\Lambda$, llamado par\'ametro de corte UV, que posee unidades de momento y en cuyo l\'imite formal $\Lambda\rightarrow\infty$ reobtenemos la expresi\'on original de la AE. Haciendo expl\'icito el t\'ermino de masa a trav\'es de la definici\'on $A=:B+m^2$ y usando \eqref{DET:desa.hk}, la contribuci\'on del operador de fluctuaciones cu\'anticas $A$ a la AE resulta
\begin{align}\label{DET:eff.regular}
 \begin{split}
\Gamma_{1-\text{\rm bucle},A}[\phi]&=-\frac{1}{2}\int_{\Lambda^{-2}}^{\infty}d\beta\,\frac{e^{-\beta m^2}}{\beta} \text{Tr}\, e^{-\beta B}\\
  &=-\frac{1}{2}\,\sum_{n=0}^\infty m^{d-n}\,\Gamma\left((n-d)/2,\tfrac{m^2}{\Lambda^2}\right)\ a_n(B)\,,
  \end{split}
\end{align}
escrito en t\'erminos de la funci\'on gamma incompleta $\Gamma(\cdot,\cdot)$ \parencite{A-S}. De acuerdo al desarrollo de esta funci\'on para peque\~nos valores de su segundo argumento, dado por 
\begin{align}
 \Gamma(s,a)\sim \int_{a}^1 x^{s-1}+\ldots\,,
\end{align}
la expresi\'on \eqref{DET:eff.regular} posee un n\'umero finito de t\'erminos divergentes, a saber, aquellos para los cuales $0\leq n\leq d$. Como hemos establecido previamente, estos deben ser eliminados redefiniendo las constantes de acoplamiento en la acci\'on original: el lagrangiano, si es que a\'un no los ten\'ia,  pasar\'a a poseer t\'erminos de la forma de $a_0(B),a_1(B),a_2(B),\ldots,a_{d}(B)$. 

Este proceso no es \'unico, visto que las constantes renormalizadas s\'olo est\'an definidas a menos de t\'erminos finitos (t.f.); por este motivo es preciso fijar una regla en la elecci\'on de dichas constantes. Una de las m\'as simples a la hora de realizar c\'alculos es la prescripci\'on de m\'inima sustracci\'on, la cual establece que la constante renormalizada  se construye a partir de la constante desnuda agregando s\'olo la parte divergente de las contribuciones. Por ej., de acuerdo a esta prescripci\'on, $\Gamma(0,\frac{m^2}{\Lambda^2})\approx -\log(\frac{m^2}{\Lambda^2})-\gamma+\cdots$, donde $\gamma$ es la constante de Euler, aportar\'ia s\'olo el t\'ermino logar\'itmico a la renormalizaci\'on.

Habiendo cancelado las divergencias UV, resta adicionar a la acci\'on original los t\'erminos finitos en el l\'imite $\Lambda\rightarrow\infty$,
\begin{align}\label{DET:hk.renor.n>d}
-\frac{1}{2} \sum_{n=d+1}^\infty m^{d-n}\,\Gamma\left((n-d)/2\right)\ a_n(B)\,.
\end{align}

En este momento podr\'ia surgir la sensaci\'on de que s\'olo hemos escondido el problema de las divergencias debajo de la alfombra. Una interpretaci\'on que despeja estas dudas la ofrecen las ideas del grupo de renormalizaci\'on de Wilson \parencite{Wilson:1973jj,Polchinski:1983gv}. Desde este punto de vista el par\'ametro de corte $\Lambda$ se vincula a la escala de energ\'ia fundamental a la cual el modelo est\'a definido; dicho en otros t\'erminos, el modelo involucra s\'olo los modos del campo con energ\'ia menores a $\Lambda$, lo cual limita los eventos para los cuales podr\'an realizarse predicciones. M\'as a\'un, la dependencia de las constantes de acoplamiento con la escala no debe ser entendida como un mero paso intermedio antes de tomar el l\'imite $\Lambda\rightarrow\infty$, sino que es una consecuencia natural, inherente a la incorporaci\'on de informaci\'on sobre los modos del campo con grandes impulsos. Para analizar esta dependencia se define para cada constante de acoplamiento desnuda 
$\lambda$ la funci\'on
\begin{align}
\beta_{\lambda}=\Lambda\left.\frac{\partial \lambda}{\partial \Lambda}\right\rvert_{\lambda_R},
\end{align}
donde la derivada debe realizarse a valores fijos de la constante renormalizada $\lambda_R$. La funci\'on $\beta$, como veremos en las secciones \ref{FLM.phi4} y \ref{GW.FLM}, brinda informaci\'on valiosa sobre la validez del an\'alisis perturbativo de la teor\'ia a diversas escalas de energ\'ia.  

En resumen, hemos mostrado c\'omo es posible aplicar las t\'ecnicas espectrales a la renormalizaci\'on; en la siguiente secci\'on, veremos  que tambi\'en ofrecen un m\'etodo eficiente para el c\'alculo de energ\'ias de Casimir.

\section{El efecto Casimir}\label{DET.casimir}
En 1948, Hendrik Casimir public\'o un trabajo \parencite{Casimir:1948dh} en el cual estudiaba la interacci\'on entre dos placas met\'alicas debido a la presencia de un campo electromagn\'etico en el marco de la segunda cuantizaci\'on  o, visto en perspectiva, la influencia de las condiciones externas en una teor\'ia cu\'antica de campos. La peculiaridad de este efecto, que vendr\'a llamado efecto Casimir, radica en su manifestaci\'on macrosc\'opica pese a su origen cu\'antico: dadas dos placas de $1\,\text{cm}^2$ de \'area, separadas por una distancia de $1\,\mu\text{m}$, la fuerza de interacci\'on es del orden de $10^{-7}\,\text{N}$. Asimismo, su fuerte dependencia en la forma de las placas lo hace un excelente candidato para aplicaciones nanotecnol\'ogicas \parencite{Bellucci:2009hh}, sin mencionar su uso en modelos de cromodin\'amica cu\'antica \parencite{Bordag:1996ma,Elizalde:1997hx}. Los experimentos m\'as precisos dise\~nados para medir la fuerza de Casimir, son una serie de trabajos realizados por 
\textcite{Mohideen:1998iz,Chang:2012fh}; 
se basan en el uso de microscopios de fuerza at\'omica y sus resultados poseen un 1\% de error experimental. Para un an\'alisis detallado sobre el efecto Casimir el lector puede referirse a los trabajos de \textcite{Plunien:1986ca,Bordag:2001qi,Milton:2001yy}.	

El razonamiento propuesto originalmente por Casimir es sencillo: comencemos considerando una cavidad c\'ubica formada por paredes met\'alicas de arista $L$. El campo electromagn\'etico, cu\'anticamente, se comporta como un conjunto de infinitos osciladores arm\'onicos con vectores de onda $\mathbf{k}_{\mathbf{n}}$ y frecuencias $\omega_{\mathbf{n}}$, relacionados por la velocidad de la luz $c$:
\begin{align}
\omega_{\mathbf{n}}=c\,\mathbf{k}_{\mathbf{n}}=\frac{\pi c}{L}\mathbf{n}\,,\qquad \mathbf{n}=(n_1,n_2,n_3)\in\mathbb{N}^3.
\end{align}
De este modo la menor energ\'ia posible para un estado (la energ\'ia $E_0$ de vac\'io) queda escrita como la suma sobre las frecuencias\footnote{Casimir, al trabajar con el campo electromagn\'etico, agrega un factor 2 a la expresi\'on \eqref{DET:casimir}, correspondiente a las posibles polarizaciones  del campo
.}
\begin{align}\label{DET:casimir}
E_0= \frac{1}{2}\sum_{\mathbf{n}\in\mathbb{N}^3}\hbar\, \omega_{\mathbf{n}} =\frac{1}{2}\sum_{\mathbf{n}\in\mathbb{N}^3} \frac{\pi\hbar c}{L}\sqrt{n_1^2+n_2^2+n_3^2}.
\end{align}

En el marco de la TCC, procediendo de acuerdo a los lineamientos de la cuantizaci\'on can\'onica, es costumbre obtener una expansi\'on del operador de campo en la base de funciones que minimizan la acci\'on, en conjunto con operadores de creaci\'on  $a_{\mathbf{k}}^{\dagger}$ y destrucci\'on $a_{\mathbf{k}}$. Notemos que tanto la base como los operadores est\'an rotulados con un \'indice\footnote{Este \'indice puede tener componentes tanto continuas como discretas dependiendo de la geometr\'ia del problema.} $\mathbf{k}$. De esta manera el operador hamiltoniano, para un campo escalar de masa $m$, toma la forma
\begin{align}
 \hat{H}=\frac{1}{2}\sum_{\mathbf{k}} \omega_{\mathbf{k}} \left(a^{\dagger}_{\mathbf{k}}a_{\mathbf{k}}+\frac{1 }{2}\right),
\end{align}
donde $\omega_{\mathbf{k}}$ son las energ\'ias de los modos, mientras que $a_{\mathbf{k}}$  y $a_{\mathbf{k}}^{\dagger}$ satisfacen las reglas de conmutaci\'on bos\'onicas
\begin{align}
 [a_{\mathbf{k}},a^{\dagger}_{\mathbf{k'}}]=\delta_{{\mathbf{k}}{\mathbf{k'}}},\qquad [a_{\mathbf{k}}^{\dagger},a^{\dagger}_{\mathbf{k'}}]=[a_{\mathbf{k}},a_{\mathbf{k'}}]=0,\qquad \forall\,{\mathbf{k}},{\mathbf{k'}}.
\end{align}
Definiendo el estado de vac\'io $\vert 0\rangle$ como aquel para el cual $a_k\vert 0\rangle=0$, el valor de expectaci\'on del hamiltoniano resulta ser en este caso la energ\'ia de vac\'io\footnote{La energ\'ia de vac\'io no puede ser eliminada introduciendo el orden normal de los campos en un espacio gen\'erico \parencite{Birrell:1982ix}, al menos no si se desea analizar qu\'e sucede al modificar lentamente la geometr\'ia del problema. Tomando por ejemplo un campo escalar sobre un intervalo de longitud $L$ en cuyos extremos satisface condiciones de contorno Dirichlet, s\'i podr\'ia elegirse la energ\'ia de vac\'io como nula para un dado $L$, pero fijado este valor dejar\'ia de serlo para longitudes $L'\neq L$.}:
\begin{align}
 E_0=\langle 0\vert \hat{H}\vert 0\rangle.
\end{align}

Recapitulando, la suma \eqref{DET:casimir} es evidentemente divergente y carece de sentido. No obstante, esta energ\'ia de vac\'io no es una cantidad medible. La genialidad de Casimir consisti\'o en encontrar una expresi\'on finita para una cantidad medible, la fuerza de interacci\'on entre las placas: con ese fin, analiz\'o el caso de dos placas paralelas de \'area infinita  y separadas por una distancia $a$. Esto es an\'alogo a tomar el l\'imite $L\rightarrow \infty$ en las dimensiones de las placas consideradas, lo cual transforma las sumas correspondientes en integrales
\begin{align}
 \mathcal{E}:=\frac{E_0}{L^2}=\frac{\hbar c}{2(2\pi)^2}\sum_{n}\int_{\mathbb{R}^2} \sqrt{\frac{\pi^2n^2}{a^2}+k^2}\, d^2k .
\end{align}
La variaci\'on $\frac{\delta \mathcal{E}}{\delta a}$ de la densidad de energ\'ia por unidad de \'area de las placas al modificar la distancia de separaci\'on $a$ es por tanto la fuerza por unidad de \'area a la que se encuentran sometidas estas placas para que el sistema se encuentre en equilibrio.

En este punto Casimir propuso introducir una funci\'on $f(k)$ que regulariza la expresi\'on, es decir, una funci\'on tal que resulta convergente la integral
\begin{align}
 \sum_{n}\int_{\mathbb{R}^2}\sqrt{\frac{\pi^2n^2}{a^2}+k^2}\,f(k)\, d^2k .
\end{align}

Hoy en d\'ia es frecuente la utilizaci\'on de t\'ecnicas similares, como la de la funci\'on zeta: consideraremos la integral 
dependiente del par\'ametro\footnote{Esta integral es b\'asicamente la funci\'on $\zeta$, definida en el anexo \ref{DET.zeta}, del operador derivada segunda con condiciones de contorno Dirichlet en dos planos paralelos. Por supuesto, de aqu\'i deriva el nombre de la t\'ecnica.} $s\in\mathbb{C}$ 
\begin{align}
\mathcal{E}(s)=\frac{\hbar c}{2(2\pi)^2\,\mu}\sum_{n}\int_{\mathbb{R}^2} \left(\frac{\pi^2n^2a^{-2}+k^2}{\mu^2}\right)^{-s}\, d^2k,
\end{align}
y la calcularemos en la regi\'on en la que se encuentre bien definida; posteriormente extenderemos anal\'iticamente el resultado al valor de inter\'es $s=-1/2$. En esta expresi\'on, el par\'ametro $\mu$ posee unidades de masa y f\'isicamente debe ser introducido para lidiar con las unidades del integrando. Haciendo uso de la representaci\'on de tiempo propio de \textcite{Schwinger:1951nm} para la potencia y la funci\'on Zeta de Riemann \parencite{A-S} $\zeta_R(\cdot)$, la expresi\'on para la densidad de energ\'ia, v\'alida en la regi\'on $\text{Re}\, s>2$, es
\begin{align}\label{DET:casimir.s}
\begin{split}
\mathcal{E}(s)&=\frac{\hbar c }{2\Gamma(s)}\,\mu^{2s-1}\sum_n\int_{\mathbb{R}^2} \frac{d^2k}{(2\pi)}\int_0^{\infty} \frac{dt}{t}\, t^{s}\, e^{-t(k^2+n^2\pi^2/a^2)}\\
&=-\frac{\pi^{1/2-2s}}{16}\,\hbar c\, \mu^{2s-1}a^{2(s-1)}\, \Gamma(s-1)\, \zeta_R\big(2(s-1)\big).
\end{split}
\end{align}
Si observamos detenidamente, en \eqref{DET:casimir.s} ha surgido naturalmente el NdC de la parte espacial del operador de fluctuaciones cu\'anticas para el campo escalar libre entre dos placas. Esto no es una casualidad sino una consecuencia de la estrecha relaci\'on existente entre el NdC y la funci\'on zeta, relaci\'on estudiada con m\'as detalle en el anexo \ref{DET.zeta}.

Retornando al c\'alculo de la energ\'ia de Casimir, la extensi\'on anal\'itica al punto de inter\'es $s=-1/2$ es finita y arroja el valor
\begin{align}
\mathcal{E}(-1/2)=-\frac{\pi^2}{1440}\frac{\hbar c}{a^3}.
\end{align}
Por otra parte, es inmediata la obtenci\'on de la fuerza por unidad de superficie 
\begin{align}
 \mathcal{F}=-\frac{\partial \mathcal{E}}{\partial a}=-\frac{\pi^2}{480} \frac{\hbar c}{a^4}.
\end{align}
Este, salvo por el factor 2 de la polarizaci\'on del campo, es el resultado original de Casimir. 

Una cantidad que guarda estrecha relaci\'on con la energ\'ia de Casimir, al punto tal de ser confundidas,  es la energ\'ia efectiva $E_{\rm eff}$. A continuaci\'on intentaremos despejar esta confusi\'on.

\subsection{La energ\'ia efectiva}\label{DET.energiaeff}
La energ\'ia efectiva $E_{\rm eff}$ est\'a definida a trav\'es de la contribuci\'on de un bucle $\Gamma_{1-\text{\rm bucle}}$ a la acci\'on efectiva como \parencite{Blau:1988kv}
\begin{align}\label{DET:energia.efectiva}
 E_{\rm eff}:=\frac{\Gamma_{1-\text{\rm bucle}}[\phi=0]}{T},
\end{align}
siendo $T$ el tama\~no temporal del universo.  Para mostrar los argumentos que motivan la introducci\'on de esta cantidad precisaremos algunos resultados previos sobre la AE. 

Suponiendo que la AE es una funcional regular, admite un desarrollo de la forma
\begin{align}\label{DET:eff.desarrollo.regular}
 \Gamma[\phi]=\sum_{n=0}^{\infty} \frac{1}{n!}\int 	dx_1\cdots dx_n\, \Gamma^{(n)}(x_1,\ldots,x_n) \phi(x_1)\ldots\phi(x_n),
\end{align}
donde\footnote{Estas funciones $\Gamma^{(n)}$ resultan ser las funciones propias de v\'ertice, es decir, aquellas que contienen la informaci\'on proveniente de los diagramas de Feynman irreducibles de 1 part\'icula \parencite{Itzykson}. Ergo, la acci\'on efectiva es la funcional generatriz de las funciones propias de v\'ertice.} $\Gamma^{(n)}(x_1,\ldots,x_n)=\frac{\delta^n\Gamma}{\delta\phi(x_1)\ldots\phi(x_n)}$. Otro desarrollo posible de la acci\'on efectiva es en el n\'umero de derivadas del campo o, si se prefiere, en potencias de las variables de momento:
\begin{align}\label{DET:eff.desarrollo.derivadas}
 \Gamma[\phi]=\int V_{\rm eff}(\phi(x))+\frac{1}{2}Z(\phi(x)) (\partial\phi(x))^2+ O(\partial\phi(x))^4.
\end{align}

Igualando las expresiones \eqref{DET:eff.desarrollo.regular} y \eqref{DET:eff.desarrollo.derivadas}, el potencial efectivo $V_{\rm eff}$ puede ser escrito en t\'ermino de las funciones $\Gamma^{(n)}$ como
\begin{align}
 V_{\rm eff}(\phi(x))=\sum_{n=0}^{\infty}\frac{1}{n!}\left(\int dx_2\cdots dx_n\, {\Gamma}^{(n)}(x,x_2\ldots,x_n)\right)\phi^n(x).
\end{align}
Es posible demostrar que para una teor\'ia invariante traslacional, el t\'ermino entre par\'entesis resulta ser independiente de la variable $x$. M\'as a\'un, tomando el potencial efectivo sobre configuraciones que toman un valor $\phi$ constante en el espacio, se puede obtener la relaci\'on 
\begin{align}\label{DET:derivadas.Veff}
 \left.\left(\frac{d}{d\phi}\right)^nV_{\rm eff}(\phi)\right\rvert_{\phi=0}=\int dx_2\cdots dx_n {\Gamma}^{(n)}(x,x_2\ldots,x_n).
\end{align}
En particular y en tono con lo expuesto a continuaci\'on de \eqref{DET:derGamma/derphi=j}, resulta  que el potencial efectivo posee un extremo para el valor medio de vac\'io $\phi=\phi_{\text{m\'in}}$ tal que:
\begin{align}
\left. \frac{\delta \Gamma[\phi]}{\delta\phi(x)}\right\rvert_{\phi=\phi_{\text{m\'in}}}=\Bigl.J[\phi](x)\Bigr\rvert_{\phi=\phi_{\text{m\'in}}}=0=V'_{\rm eff}(\phi_{\text{m\'in}}).
\end{align}

El resultado general establece que se puede relacionar el potencial efectivo $V_{\rm eff}(\phi)$ con el valor m\'inimo que la densidad Hamiltoniana toma sobre estados para los cuales el valor medio del campo cu\'antico es $\langle \varphi(x)\rangle =\phi$, un valor constante; de aqu\'i la importancia que reviste \eqref{DET:derivadas.Veff}. El valor medio de vac\'io $\phi=\phi_{\text{m\'in}}$ es especial ya que corresponde al m\'inimo del potencial efectivo y, por ende, al m\'inimo de la densidad Hamiltoniana sobre estados cuyo valor medio es independiente de las coordenadas \parencite{Coleman:1988}. Verbigracia, $\phi_{\text{m\'in}}$ ser\'ia el estado de vac\'io ante la hip\'otesis de invariancia traslacional; sin p\'erdida de generalidad podr\'iamos adem\'as suponer $\phi_{\text{m\'in}}=0$, pues para el campo $\phi'(x)$, definido de acuerdo a
\begin{align}
 \phi(x)= \phi'(x)+\phi_{\text{m\'in}},
\end{align}
la configuraci\'on de vac\'io es la trivial. Hechas estas aclaraciones, la definici\'on \eqref{DET:energia.efectiva} equivale a la contribuci\'on de orden\footnote{La contribuci\'on $\hbar^{0}$ a $V_{\rm eff}(\phi_{\text{m\'in}})$ siempre puede ser elegida como nula.} $\hbar$ a $V_{\rm eff}(\phi_{\text{m\'in}})$.

Finalmente, es posible demostrar que para universos ultraest\'aticos la diferencia entre la energ\'ia de Casimir y la energ\'ia efectiva es un t\'ermino que depende de un coeficiente del n\'ucleo de calor o, por transitividad, de la geometr\'ia del problema \parencite{Elizalde:2012zza,Blau:1988kv}.

\begin{subappendices}
   \renewcommand{\theequation}{\Roman{chapter}.\Alph{section}.\arabic{equation}}

 \section[Renormalizaci\'on \textit{alla} zeta y \textit{alla} n\'ucleo de calor]{Relaci\'on entre la renormalizaci\'on \textit{alla} zeta y \textit{alla} n\'ucleo de calor}\label{DET.zeta}
 
Dado un operador $A$ que act\'ua sobre funciones de $\mathbb{R}^d$ y con autovalores $\lambda_n$ se define su funci\'on espectral zeta en la forma
 \begin{align}\label{DET:zeta}
\begin{split} 
\zeta_A(s):&=\text{Tr}\; A^{-s}\\
 &=\sum_n \lambda_n^{-s}.
 \end{split}
 \end{align}
En este anexo no nos centraremos en estudiar las propiedades de esta funci\'on; ser\'a suficiente notar que, bajo hip\'otesis bastante generales, la suma converge eligiendo $\mathrm{Re}(s)$ suficientemente grande. Luego, puede realizarse la extensi\'on anal\'itica de la funci\'on resultante al resto del plano complejo $s$. 
 
 Conviene mencionar antes de proseguir que el NdC y la funci\'on zeta se encuentran relacionados a trav\'es de la transformada de Mellin
 \begin{align}
  \zeta_A(s)=\frac{1}{\Gamma(s)}\int_0^{\infty} d\tau\, \tau^{s}\,K_A(\tau),
 \end{align}
lo cual se puede verificar formalmente utilizando la definici\'on de ambos en t\'erminos de los autovalores del operador $A$. A la inversa, se puede representar el n\'ucleo de calor en t\'erminos de la funci\'on zeta utilizando la integral 
\begin{align}
 K_A(\tau)=\frac{1}{2\pi i} \oint ds\,  t^{-s}\,\Gamma(s) \zeta_A(s)
\end{align}
sobre una curva cerrada que encierra todos los polos del integrando.
 
Consideremos ahora la derivada de la funci\'on zeta evaluada en cero; de acuerdo a la definici\'on \eqref{DET:zeta} vale
\begin{align}\label{DET:zeta.derivada}
\begin{split}
-\zeta'_A(0):
 &=\log \left(\prod_n\lambda_n\right).
 \end{split}
\end{align}
Esta ecuaci\'on motiva, dados un operador $A$ y un par\'ametro de escala $\mu$, la definici\'on del determinante del operador en t\'erminos de la funci\'on zeta de acuerdo a la expresi\'on
\begin{align}
  \frac{1}{2}\log{{\rm Det}\,\mu^{-1}A}:=-\frac{1}{2}\zeta_A'(0)-\log{\mu}\,\zeta_A(0)\,.
\end{align}

En pos de analizar su valor en $s=0$, podemos obtener un desarrollo para la funci\'on zeta de $A$ combinando la transformada de Mellin con el desarrollo asint\'otico \eqref{DET:desa.hk} del NdC de $B:=A-m^2$:
\begin{align}\label{DET:zeta.desarrollo}
  \zeta_A(s)
  &=\frac{1}{\Gamma(s)}\int_0^\infty d\tau\,\tau^{s-1}\,e^{-\tau m^2}
  \,\sum_{n=0}^\infty a_n(B)\,\tau^{(n-d)/2}
  \nonumber\\
  &=m^{-2s}\,\sum_{n=0}^\infty m^{d-n}
  \,\frac{\Gamma(s+(n-d)/2)}{\Gamma(s)}\ a_n(B)\,.
\end{align}
Si $d$ es impar ninguno de los factores de \eqref{DET:zeta.desarrollo} posee polos, de manera que el t\'ermino $1/\Gamma(s)$ anula todas las contribuciones\footnote{Esta aseveraci\'on es v\'alida bajo la hip\'otesis de que el espacio sobre el cual est\'a definido el operador $A$ no posee bordes (en esta ocasi\'on $\mathbb{R}^d$)	, pues entonces los coeficientes $a_n(B)$ se anulan si $n$ es impar.}. Consecuentemente, para dimensiones impares $\zeta_A(0)=0$ y el determinante resulta independiente del par\'ametro de escala $\mu$. En el caso de dimensiones pares, las divergencias de los t\'erminos para los cuales $0\leq n\leq d$ son compensadas por el factor $1/\Gamma(s)$, por lo que las contribuciones a la funci\'on zeta para $s=0$ son
\begin{align}\label{DET:zetacero}
  \zeta_A(0)
  =\sum_{n=0}^{d} m^{d-n}
  \,\frac{(-1)^{(d-n)/2}}{((d-n)/2)!}\ a_n(B)\,.
\end{align}

En definitiva, la dependencia en la escala arbitraria $\mu$ es proporcional a los coeficientes $a_0(A),a_1(A),\ldots,a_{d}(A)$ y puede ser eliminada de las cantidades f\'isicas renormalizando las constantes de acoplamiento adecuadas. Este es el an\'alogo a las contribuciones divergentes en \eqref{DET:eff.regular}, encontradas mediante el m\'etodo del n\'ucleo de calor.

A continuaci\'on, podemos hacer un an\'alisis similar para la derivada de la funci\'on zeta, la cual,  derivando formalmente \eqref{DET:zeta.desarrollo}, se puede escribir en t\'erminos de la funci\'on digamma  $\psi(z):=\Gamma'(z)/\Gamma(z)$ \parencite{A-S}:
\begin{multline}\label{DET:zetaprima}
  -\zeta_A'(s)
  =\log{m^2}\,\zeta_A(s)\\
  -m^{-2s}\,\sum_{n=0}^\infty m^{d-n}
  \,\frac{\Gamma(s+(n-d)/2)}{\Gamma(s)}
  \left\{\psi(s+(n-d)/2)-\psi(s)\right\}
  \ a_n(B) \,.
\end{multline}
El primer t\'ermino es proporcional a la funci\'on zeta y por ende no precisa de un nuevo an\'alisis. El restante, conviene analizarlo tomando para $n$ los siguientes intervalos: si $0\leq n\leq d$, hay contribuciones finitas ya sea porque $n$ tiene la paridad de $d$ y la funci\'on $\Gamma(s+(n-d)/2)$ posee un polo, a la vez que la diferencia $\psi(s+(n-d)/2)-\psi(s)$ es anal\'itica en $s=0$, o porque $d$ tiene la restante paridad, y sucede lo contrario. En todo caso, obtenemos contribuciones adicionales finitas que en el proceso de renormalizaci\'on deben ser a\~nadidas a las de \eqref{DET:zetacero}. Por su parte, el t\'ermino $n=d$  se anula independientemente de la dimensi\'on $d$ con la cual trabajemos. En cuanto se refiere a los restantes t\'erminos, i.e. aquellos con $n> d$, originan las contribuciones a la AE
\begin{align}
  -\frac{1}{2}\sum_{n= d+1}^\infty \frac{\Gamma((n-d)/2)}{m^{n-d}}\,a_{n}(B)\,,
\end{align}
coincidentes con las encontradas en \eqref{DET:hk.renor.n>d} gracias al m\'etodo del NdC.
\end{subappendices}


\chapter{Sobre las funciones espectrales}\label{FE}
\setlength\epigraphwidth{8cm}
\epigraph{\itshape\small La filosofia \`e scritta in questo grandissimo libro che continuamente ci sta aperto innanzi a gli occhi (io dico l'universo), ma non si pu\`o intendere se prima non s'impara a intender la lingua, e conoscer i caratteri, ne' quali \`e scritto. Egli \`e scritto in lingua matematica, e i caratteri son triangoli, cerchi, ed altre figure geometriche, senza i quali mezzi è impossibile a intenderne umanamente parola; senza questi \`e un aggirarsi vanamente per un oscuro laberinto. }{-- \textsc{Galileo Galilei}, \textit{Il Saggiatore}, Cap. VI.}

El NdC responde al problema cl\'asico de la difusi\'on del calor, modelado a trav\'es de ecuaciones diferenciales en derivadas parciales \parencite{Hadamard,Courant}. En el contexto de las teor\'ias cu\'anticas, la primera referencia que encontramos sobre su uso corresponde a \textcite{Fock:1937dy}, quien lo utiliz\'o en el c\'alculo de funciones de Green. M\'as cerca en el tiempo, quiz\'as el trabajo m\'as famoso en el que ha sido empleado es en la construcci\'on de una teor\'ia covariante de la gravedad cu\'antica por parte de \textcite{DeWitt:1967yk}. Por cierto, en el cap\'itulo precedente hemos visto, para el caso sencillo de un campo escalar en TCC, c\'omo  surgen con naturalidad las FE de operadores diferenciales en la realizaci\'on de c\'alculos perturbativos, cfr. \eqref{DET:efffinal}.

Detr\'as de las FE existe un formalismo matem\'atico que ha dado lugar a la creaci\'on de una entera rama de las matem\'aticas llamada geometr\'ia espectral. Esta rama investiga la relaci\'on entre la geometr\'ia de una variedad y los autovalores de operadores sobre ella definidos; esto se suele resumir en la metaf\'orica pregunta formulada por  \textcite{Kac:1966xd}: \emph{``Can we hear the shape of a drum?''}\footnote{``?`Podemos escuchar la forma de un tambor?'', en ingl\'es.}. 

El puntapi\'e inicial para la concepci\'on de la geometr\'ia espectral fue dado a principios del siglo XX por \textcite{Weyl:1911}, quien demostr\'o que el comportamiento asint\'otico de los autovalores del operador Laplaciano, con condiciones de contorno Dirichlet en el borde de una regi\'on compacta suave, permite determinar el volumen de esta \'ultima. Seguidamente se observ\'o que tanto el NdC como la funci\'on zeta del Laplaciano contienen  gran cantidad de informaci\'on referida a la variedad sobre la cual se encuentran definidos \parencite{Carleman,Minakshisundaram:1949xg}. Los resultados para operadores diferenciales m\'as generales que el Laplaciano fueron acu\~nados por Seeley en una serie de trabajos  publicados a fines de de la d\'ecada de 1960
\parencite{Seeley:1967,Seeley:1967an,Seeley:1969re}.

Este cap\'itulo tiene como prop\'osito dar sustento a las t\'ecnicas espectrales que consideraremos en el resto de esta tesis. Para ello, enunciaremos algunos resultados de la geometr\'ia espectral, dedic\'andonos especialmente al NdC de cierta clase de operadores diferenciales. Como paso previo, deberemos establecer ciertas propiedades de los operadores en cuesti\'on \parencite{Gilkey:1995mj,Wloka}.

\section{Nociones generales sobre operadores diferenciales}
Antes de definir qu\'e es lo que entendemos por operador diferencial en derivadas parciales, conviene introducir la notaci\'on de multi-\'indices, la cual facilitar\'a la notaci\'on de este cap\'itulo.
\begin{definicion}
 Un multi-\'indice $\alpha$ es una $m$-upla $\alpha=\left(\alpha_1,\ldots,\alpha_m\right)$ de enteros no negativos $\alpha_j$.
\end{definicion}
\begin{definicion}
 Dado un multi-\'indice $\alpha$ definimos
  \begin{align}
  \begin{split}
 \lvert\alpha\rvert&:=\alpha_1+\ldots+\alpha_m,\\
 \alpha!&:=\alpha_1!\,\cdots\alpha_m!,\\
 x^{\alpha}&:=x_1^{\alpha_1}\cdots x_m^{\alpha_m},\\
 D^{\alpha}_x&:=(-i)^{\lvert\alpha\rvert}\,\left(\frac{\partial}{\partial x_1}\right)^{\alpha_1}\cdots \left(\frac{\partial}{\partial x_m}\right)^{\alpha_m}.
 \end{split}
 \end{align}
\end{definicion}

Haciendo uso de dicha notaci\'on, podemos definir en forma compacta la clase de operadores sobre los que centraremos nuestra atenci\'on.
\begin{definicion}
 Un operador lineal $P:\left[\mathcal{C}^{\infty}(\mathbb{R}^d)\right]^{k}\rightarrow\left[\mathcal{C}^{\infty}(\mathbb{R}^d)\right]^{k}$ que mapea el espacio de vectores k-dimensionales de funciones suaves definidas sobre $\mathbb{R}^d$ en s\'i mismo es un operador en derivadas parciales de orden $b$ si tiene una expresi\'on polin\'omica
 \begin{align}\label{FE:opdif}
  P:=p(x,D):=\sum_{\lvert\alpha\rvert\leq b}a_{\alpha}(x) D_x^{\alpha},
 \end{align}
donde los coeficientes $a_{\alpha}(x)\in\mathcal{C}^{\infty}:\mathbb{R}^d\rightarrow\mathbb{C}^{k\times k}$ son funciones suaves que toman valores en un espacio matricial complejo de dimensi\'on $k$.
\end{definicion}
A este tipo de operadores  corresponden  $G=-\partial^2+m^2$, definido luego de \eqref{DET:w1} y coincidente con el operador laplaciano en el caso $m=0$, y, por lo general, el operador $\delta_\phi^2S$ introducido en \eqref{DET:zetadesarrollada}.
De acuerdo a la definici\'on \eqref{FE:opdif}, la acci\'on de un operador diferencial $P$ sobre una funci\'on $f(x)\in\mathcal{S}$, el espacio de Schwartz\footnote{El espacio de Schwartz $\mathcal{S}$ es el conjunto de funciones $f(x)$ suaves definidas sobre $\mathbb{R}^d$, tales que para multi-\'indices $\alpha,\beta$ cualesquiera existen constantes $C_{\alpha,\beta}$ que satisfacen la relaci\'on $\lvert x^{\alpha}D^{\beta}_xf\rvert\leq C_{\alpha,\beta}$. La transformada de Fourier es un isomorfismo en este espacio.}, es
\begin{align}\label{FE:op.transfor}
\begin{split} P\,f(x)&=\int e^{i x \xi}\, p(x,\xi)\, \tilde{f}(\xi)\, d\tilde{\xi}\\ 
&=\int e^{i(x-y)\xi}\,p(x,\xi)\, f(y)\, {dy} {d\tilde{\xi}}.
\end{split}\end{align}
En esta expresi\'on hemos utilizado la transformada de Fourier $\tilde{f}(p)$ de la funci\'on $f(x)$, escrita de acuerdo a la convenci\'on
\begin{equation}
\begin{split}    \tilde{f}(p)&=\int e^{-ipx}\,f(x)\, dx\\
    &=\int e^{-i(p-q)x}\,\tilde{f}(q)\, dx d\tilde{q},
\end{split}\end{equation}
en la cual\footnote{Usaremos la siguiente notaci\'on: una tide sobre una funci\'on representa su transformada de fourier $\mathcal{F}$, mientras que una tilde en un diferencial implica un factor $(2\pi)^{-d}$.} $d\tilde{q}:=dq/(2\pi)^d$. La ec. \eqref{FE:op.transfor} motiva la siguiente definici\'on, de la que a su vez resultan inmediatas dos propiedades caracter\'isticas de un operador diferencial de orden $b$:
\begin{definicion}
 El s\'imbolo $\sigma(P):\mathbb{R}^{2d}\rightarrow\mathbb{C}^{k\times k}$ de un operador diferencial $P=p(x,D)$ est\'a definido como
 \begin{align}
  \sigma(P)(x,\xi):=p(x,\xi).
 \end{align}
Por otro lado, el s\'imbolo principal $\sigma_L(P)$ consta de los t\'erminos de mayor orden como funci\'on en $\xi$ de $\sigma(P)$:
 \begin{align}
  \sigma_L(P)(x,\xi):=\sum_{\lvert\alpha\rvert=b} a_{\alpha}(x)\xi^{\alpha}.
 \end{align}
  \end{definicion}

\begin{propiedad}
 El s\'imbolo $\sigma(P)(x,\xi)$  es un polinomio de orden $b$ en la variable dual $\xi$.
\end{propiedad}
\begin{propiedad}
 El s\'imbolo principal $\sigma_L(P)(x,\xi)$ es un polinomio homog\'eneo de grado $b$ en $\xi$.
\end{propiedad}

Los resultados que estableci\'o R.T. Seeley se refieren a un conjunto m\'as restringido que el de los operadores lineales en derivadas parciales. Las restricciones de las siguientes definiciones bastar\'an para poder enunciar el lema fundamental de esta secci\'on sobre operadores diferenciales y el teorema sobre su NdC.

\begin{definicion}
 Un operador diferencial $P$ se dice el\'iptico sii su s\'imbolo principal $\sigma_L(P)$ es invertible para $\lvert\xi\rvert=1$ o, equivalentemente, sii $\det \sigma_L(P)(x,\xi)\neq0$ para $\lvert\xi\rvert=1$.
\end{definicion}

\begin{definicion}
 Se dice que un operador diferencial $P$ posee un s\'imbolo principal definido positivo sii su s\'imbolo principal puede ser elegido para $\xi\neq0$ como una matriz herm\'itica definida positiva (con autovalores $\lambda>0$).
\end{definicion}

Es preciso realizar algunos comentarios. Primero, la importancia del car\'acter el\'iptico radica en que todo operador diferencial el\'iptico $P$, en caso de que exista, posee un operador adjunto $P^{\dagger}$ que resulta ser tambi\'en un operador diferencial. Por otro lado, a diferencia de lo que podr\'ia parecer a primera vista, el espectro de un operador definido positivo no es necesariamente no negativo, aunque s\'i posee una cota inferior.

Todas las definiciones y propiedades precedentes son f\'acilmente generalizables al caso de un fibrado vectorial $V$ suave, definido sobre $\mathcal{M}$, una variedad Riemaniana   $d$-dimensional, suave y compacta, que supondremos sin borde. Para ello, ser\'a necesario suponer que $V$ posee una dada m\'etrica Riemanniana compatible, de forma que el espacio $L^2(V)$ de funciones de cuadrado integrable Lebesgue sobre la variedad est\'e definido en forma invariante. Salvo que explicitemos lo contrario, todo aquello que enunciaremos en el resto del cap\'itulo corresponder\'a a este caso.
 
Finalmente, podemos establecer el principal resultado de esta secci\'on, expresado en el\footnote{Para su demostraci\'on, ver el trabajo de \textcite{Gilkey:1995mj}, cap\'itulo 1.}
\begin{lema}\label{FE,lema}
 Sea $P:C^{\infty}(V)\rightarrow C^{\infty}(V)$ un operador diferencial el\'iptico y autoadjunto de orden $b>0$:
 \begin{enumerate}[a)]
  \item se puede encontrar una base ortonormal completa de autovectores $\{\phi_{n\in\mathbb{N}}\}$ para $L^2(V)$, con autovalores $\lambda_n\in\mathbb{R}$;
  \item los autovectores $\phi_n$ son suaves y se cumple que $$\lim_{n\rightarrow\infty} \lvert\lambda_n\rvert=\infty;$$
  \item si ordenamos los autovalores en orden creciente $\lvert\lambda_1\rvert\leq\lvert\lambda_2\rvert\leq\cdots$, entonces existen constantes $C,\delta>0$ tales que $\lvert\lambda_n\rvert\geq C\,n^{\delta}$ si $n$ es grande.
 \end{enumerate}

\end{lema}

\section{El n\'ucleo de calor}
La traza del NdC, funci\'on espectral que en \eqref{DET:heat-ker} hemos presentado, puede ser analizada con formalidad en el contexto de la teor\'ia de operadores diferenciales el\'ipticos. El punto de partida es, indudablemente, la generalizaci\'on del afamado problema de la difusi\'on de calor estudiado por J. Fourier en el siglo XIX. Efectivamente, consideremos~la

\begin{definicion}
Dado un operador diferencial $P$ el\'iptico de orden $b>0$, autoadjunto y con s\'imbolo principal definido positivo sobre un fibrado vectorial \!\footnote{Recordemos que supondremos de aqu\'i en adelante que el fibrado satisfar\'a las hip\'otesis detalladas en la secci\'on anterior. Utilizaremos tambi\'en id\'entica notaci\'on.} $V$ con variedad de base $\mathcal{M}$, la ecuaci\'on de calor asociada es
\begin{align}\label{FE:ecuacion-calor}
 \left(\frac{d}{dt}+P\right) f(x,t)=0,\qquad \text{para}\;\;x\in\mathcal{M}  \text{ y } t\geq0,
\end{align}
con la condici\'on inicial
\begin{align}
 f(x,0)=f(x).
\end{align}
\end{definicion}

La soluci\'on a \eqref{FE:ecuacion-calor} es, formalmente, $f(x,t)=e^{-tP}f(x)$. En forma m\'as precisa, de acuerdo a los resultados del Lema \ref{FE,lema}, existe una base de autofunciones $\phi_{n\in\mathbb{N}}$ del operador $P$ en t\'erminos de la cual podemos escribir la  serie de Fourier generalizada $f(x,t)=\sum_n a_n(t) \phi_n(x)$. En consecuencia, resulta inmediato que la soluci\'on a la ecuaci\'on de calor puede escribirse como
\begin{align}
 f(x,t)&=\sum_n e^{-\lambda_n t}\, (\phi_n,f)\,\phi_n(x)\\
 &=\int_{\mathcal{M}}\,K(t,x,y)\, f(y)\,\sqrt{g}\,dy  ,
\end{align}
donde hemos introducido la medida invariante sobre la variedad base $\mathcal{M}$ utilizando su m\'etrica $g$ y hemos definido (formalmente) el NdC
\begin{align}
 K(t,x,y)=\sum_n e^{-\lambda_n t}\,\phi_n(x) \otimes\, \phi_n^{\dagger}(y).
\end{align}
De estas ecuaciones, deducimos que el operador $e^{-tP}$ es un operador integral cuyo n\'ucleo resulta ser $K(t,x,y)$. Este n\'ucleo tiene la propiedad de ser una funci\'on infinitamente suave de las variables $(t,x,y)$ para $t>0$, por lo que se pueden justificar todos los manejos formales realizados precedentemente.

Entre otras caracter\'isticas del n\'ucleo $K(t,x,y)$, enunciaremos el resultado principal de este cap\'itulo, resumido en el siguiente\footnote{Para su demostraci\'on, ver  el trabajo de \textcite{Gilkey:1995mj}, cap\'itulo 1.}
\begin{teorema}\label{FE,teor}
 Sea $P$ un operador en derivadas parciales de orden $b>0$, el\'iptico, autoadjunto y definido positivo sobre un fibrado vectorial $V$. Entonces, si $K(t,x,y)$ es el n\'ucleo de $e^{-tP}$, vale el desarrollo asint\'otico de su diagonal
\begin{align}\label{FE:desarrollo.heat-local}
  K(t,x,x)\sim \sum_{n=0}^{\infty} t^{\frac{n-d}{b}}\,e_n(x,P) ,\qquad \text{para}\;\;\; t\!\downarrow\!0.
\end{align}
 Adem\'as, las cantidades $e_n(x,P)$ dependen de un n\'umero finito de coeficientes del s\'imbolo $p(x,\xi)$ pensado como polinomio en $\xi$, y resultan ser invariantes frente a cualesquiera elecciones del sistema de coordenadas en $\mathcal{M}$ y la trivializaci\'on local de $V$. M\'as a\'un, si $n$ es impar resulta que\footnote{Seguidamente veremos que para variedades con borde no vale esta propiedad.} $e_n(x,P)=0$.
\end{teorema}

A partir de \eqref{FE:desarrollo.heat-local} se puede obtener el resultado correspondiente a la traza del n\'ucleo de calor, la cual en el caso de operadores con s\'imbolo principal escalar es
\begin{align}\label{FE:primerdesarrollo.hk}
\begin{split}
 \text{Tr}_{L_2}e^{-tP}&=\int_{\mathcal{M}}\text{Tr}_{V_x} K(t,x,x)\,\sqrt{g}\, dx \\
 &\sim \sum_{n=0}^{\infty} t^{\frac{n-d}{b}}\iM a_n(x,P) ,\qquad \text{para}\;\;\; t\!\downarrow\!0\\
 &\sim \sum_{n=0}^{\infty} t^{\frac{n-d}{b}} a_n(P) ,\qquad \text{para}\;\;\; t\!\downarrow\!0.
\end{split}
\end{align}
Los coeficientes $a_n(x,P)$ reciben el nombre de invariantes escalares locales de Seeley-DeWitt (SDW) del desarrollo asint\'otico del n\'ucleo de calor y est\'an definidos como la traza $Tr_{V_x}$ de $e_n(x,P)$ sobre la fibra $V$ situada en el punto $x$
\begin{align}
 a_n(x,P)=\text{Tr}_{V_x} e_n(x,P).
\end{align}
En tanto, sus integrales $a_n(P)=\int_{\mathcal{M}}a_n(x,P)$ sobre la variedad se denominan escalares invariantes integrados de SDW o simplemente coeficientes de SDW. Ambos heredan las propiedades de los coeficientes $e_n(x,P)$; entre otras, pueden ser calculados en cualquier sistema de coordenadas y en forma relativa a cualquier trivializaci\'on local como expresiones combinatorias de un n\'umero finito de coeficientes del s\'imbolo $p(x,\xi)$ de $P$ pensado como polinomio en $\xi$. Tambi\'en es cierto que $a_n(x,P)=0$ si $n$ es impar.

Es frecuente utilizar una funci\'on $f(x)$ suave definida sobre $\mathcal{M}$ para definir el desarrollo asint\'otico
\begin{align}\label{FE:desarrollo.heat-local.f}
\text{Tr}_{L_2}(f\,e^{-tP})\sim \sum_{n=0}^{\infty} t^{\frac{n-d}{b}}\,a_n(f,P) ,\qquad \text{para}\;\;\; t\!\downarrow\!0.
\end{align}
El uso de esta funci\'on resultar\'a de ayuda al analizar resultados similares sobre variedades no compactas.

Para finalizar, conviene mencionar que estos resultados pueden ser generalizados a variedades con borde, donde se imponen ciertas condiciones de contorno a trav\'es del operador $\mathcal{B}$. En la gran mayor\'ia de los casos de inter\'es f\'isico, evitando condiciones de contorno no locales \parencite{Grubb:1995}, se obtiene un desarrollo 
\begin{align}\label{FE:heatkernel.desarrollo}
    {\rm Tr}\left(f\,e^{-t A}\right)\sim \sum_{n=0}^{\infty} t^{\frac{n-d}{b}} a_n(f,A,\mathcal{B}) ,\qquad \text{para}\;\;\; t\!\downarrow\!0.
\end{align}
Los coeficientes $a_n(P,\mathcal{B})$ poseen ahora contribuciones tanto de volumen como de borde de invariantes locales
\begin{multline}
a_n(f,P,\mathcal{B})=\iM \, f(x)\,a_n(x,P)\\
+\sum_{j=1}^{n} \int_{\partial\mathcal{M}}dy\, \sqrt{h} \,f^{(j)}(y)\, a_{n,j}(y,P,\mathcal{B})\,,
\end{multline}
simbolizando con $f^{(j)}(y)$ la j-\'esima derivada en la coordenada normal al borde de la funci\'on regularizadora $f(x)$. Bajo estas hip\'otesis, los coeficientes impares no son necesariamente nulos.

\subsection{El n\'ucleo de calor para operadores generalizados de Laplace}
Para ejemplificar el teorema de la secci\'on previa, podemos analizar el caso de un operador $D$ de segundo orden del tipo de Laplace
\begin{align}\label{FE:laplaciano.generico}
 D=-\left(g^{\mu\nu}\nabla_{\mu}\nabla_{\nu}+E\right),
\end{align}
donde $g^{\mu\nu}$ es la inversa del tensor m\'etrico sobre $\mathcal{M}$, $\nabla_{\mu}$ es la derivada covariante que contiene tanto la conexi\'on de Christoffel, $$\Gamma_{\mu\nu}^{\sigma}=\frac{1}{2}g^{\sigma\rho}\left(\partial_{\mu}g_{\nu\rho}+\partial_{\nu}g_{\mu\rho}-\partial_{\rho}g_{\mu\nu}\right),$$ como la de spin ($\omega_{\mu}$), y $E$ es un endomorfismo en $V$. Por cuanto el c\'alculo del desarrollo asint\'otico del n\'ucleo de calor involucra tan s\'olo el producto de invariantes locales, conviene identificarlos; en este ejemplo, dichos t\'erminos pueden armarse utilizando el endomorfismo $E$, la m\'etrica $g^{\mu\nu}$, la derivada covariante $\nabla_{\mu}$ y la funci\'on suave $f$ que introducimos al tomar la traza. Igualmente, deben considerarse sus derivados: la curvatura de Riemann
\begin{equation}
{R^\mu}_{\nu\rho\sigma}:=\partial_\sigma \Gamma_{\nu\rho}^\mu -
\partial_\rho \Gamma_{\nu\sigma}^\mu +
\Gamma_{\nu\rho}^\lambda \Gamma_{\lambda\sigma}^\mu -
\Gamma_{\nu\sigma}^\lambda \Gamma_{\lambda\rho}^\mu, 
\label{Riemann}
\end{equation}
el tensor de Ricci $R_{\mu\nu}:={R^\sigma}_{\mu\nu\sigma}$, la curvatura escalar $R:=R^\mu_\mu$ y el tensor de campo $\Omega_{\mu\nu}$ de la conexi\'on $\omega_{\mu}$:
\begin{equation}
\Omega_{\mu\nu}=\partial_\mu\omega_\nu - \partial_\nu\omega_\mu
+\omega_\mu\omega_\nu -\omega_\nu\omega_\mu \, .
\label{Omega}
\end{equation}

Un m\'etodo veloz para determinar la forma de los t\'erminos de los coeficientes $a_n(f,D)$ consiste en su an\'alisis dimensional. Siendo cierto que $t$ es un par\'ametro con dimensi\'on de longitud al cuadrado ($\left[t\right]=l^2$) y que la diagonal del NdC es adimensional, $a_n(x,D)$ debe tener unidades $\left[a_n(x,D)\right]=l^{2(d-n)/2}$. Partiendo del hecho que cada una de las posibles contribuciones tiene dimensiones de $l^{-p}$, con $p\geq0$, el n\'umero de candidatos a contribuir a un dado coeficiente resulta finito, tal y como lo establece el Teorema  \ref{FE,teor}. Por ejemplo, el primer t\'ermino debe ser del tipo
\begin{align}
a_{0}(f,D)=\iM\,\trv\{\alpha_0 f\}
\end{align}
con $\alpha_0$ un n\'umero a determinar; por fortuna, bastar\'a hacerlo para un dado operador sobre un cierto fibrado vectorial, visto que no depende de la elecci\'on de estos. En el anexo  \ref{FE.app.calculo} resumimos una de las t\'ecnicas utilizadas para el c\'alculo de coeficientes, que consiste en tomar el caso especial $\mathcal{M}=\mathbb{R}^d$.

\begin{subappendices}
  \renewcommand{\theequation}{\Roman{chapter}.\Alph{section}.\arabic{equation}}

\section{C\'alculo de los primeros coeficientes de Seeley-DeWitt para la variedad de base \texorpdfstring{$\mathbb{R}^d$}{R**d}}\label{FE.app.calculo}
En este anexo, estudiaremos el desarrollo asint\'otico del n\'ucleo de calor correspondiente al operador diferencial introducido en \eqref{FE:laplaciano.generico}
\begin{align}
 D=-\left(g^{\mu\nu}\nabla_{\mu}\nabla_{\nu}+E\right),
\end{align}
definido sobre un espacio vectorial $V$ con variedad de base $\mathcal{M}=\mathbb{R}^d$, una variedad no compacta. En este caso, es necesario suponer ciertas condiciones de decrecimiento en el infinito sobre $E$ y la funci\'on $f(x)$, con el fin de obtener un n\'ucleo de calor 
\begin{align}
K(f;t)&={\rm Tr}_{L^2} (fe^{-tD})
\end{align}
bien comportado. Dada por v\'alida esta suposici\'on, podemos calcular la traza utilizando la base de ondas planas\footnote{Esta base resulta conveniente para $\mathbb{R}^d$. En general, podr\'a utilizarse este m\'etodo para otras variedades eligiendo la base m\'as adecuada al problema.} y la traza $\text{Tr}_{V_x}$ sobre la fibra en el punto $x$ de la variedad:
\begin{align}
\begin{split}\label{FE:heatk.Rm}
K(f;t)&=\int dx \int 
d\tilde{k}\, e^{-ikx}\,\trv \{f(x) e^{-tD} e^{ikx}\} \\
&=\int dx \int d\tilde{k}\, \trv\{ f(x) 
e^{-tk^2}e^{t(2ik^{\mu}\nabla_{\mu}+\nabla^2+E)}\}\,.
\end{split}
\end{align}
Conviene notar, en primer lugar, que en la \'ultima l\'inea de esta ecuaci\'on el operador diferencial $\nabla_{\mu}$ act\'ua solo sobre $E$. En segundo lugar, que hemos separado la exponencial en dos t\'erminos, con la intenci\'on de desarrollar en serie aquel que contiene derivadas; el resultado que obtendr\'iamos ser\'ia un polinomio en $t$ y en $k$ donde, debido a la presencia de la exponencial $e^{-tk^2}$, cada factor $k$ ser\'ia del orden de $\sqrt{t}$ luego de realizar la integral correspondiente.

Siguiendo ese plan y  utilizando la notaci\'on $\nabla_{\mu_1}\cdots\nabla_{\mu_n}f=f_{;\mu_1\cdots\mu_n}$, el resultado que se obtiene para el desarrollo del n\'ucleo de calor  es
\begin{align}\label{FE:heat.Rm1}
\begin{split}
K(f;t)=\frac 1{(4\pi t)^{d/2}} &\int dx \trv \Biggl\{ f(x)
\Biggl(1 +t E \\
&+ t^2 \left(
\frac 12 E^2 +\frac 16 E_{;\mu\mu}+
\frac 1{12} \Omega_{\mu \nu}\Omega^{\mu \nu} \right) +O(t^3)\Biggr)
\Biggr\}\,.
\end{split}\end{align}
De esta ecuaci\'on se pueden leer los primeros coeficientes del desarrollo
\begin{align}
\begin{split}
a_{0}(f,D)&=\frac{1}{(4\pi)^{d/2}}\int_{\mathbb{R}^m}\, \trv\{f\},\\
a_2(f,D)&=\frac{1}{(4\pi)^{d/2}}\int_{\mathbb{R}^m}\, \trv\{f\,E\},\\
a_4(f,D)&=\frac{1}{12\,(4\pi)^{d/2}}\int_{\mathbb{R}^m}\, \trv\{f(2\,E_{;kk}+6\,E^2+\Omega_{ij}\Omega_{ij})\}.\label{FE:a024.Rm}
\end{split}
\end{align}
Dado que $\mathbb{R}^d$ es una variedad plana, si tom\'aramos una variedad Riemanniana arbitraria deber\'iamos agregar a estos coeficientes contribuciones provenientes de t\'erminos que involucraran el tensor de curvatura.

\end{subappendices}


\chapter{Formalismo de l\'inea de mundo}\label{FLM}
\setlength\epigraphwidth{6.5cm}
\epigraph{\itshape Un'ansia inconsueta da qualche tempo si accende in me alla sera, e non \`e pi\`u rimpianto delle gioie lasciate, come accadeva nei primi tempi del viaggio; piuttosto \`e l'impazienza di conoscere le terre ignote a cui mi dirigo. }{-- \textsc{D. Buzzatti}, \textit{I Sette Messaggeri.}}

\normalsize
El an\'alisis perturbativo de cantidades f\'isicas en TCC suele portar indefectiblemente a la construcci\'on y posterior c\'alculo de diagramas de Feynman. El principal problema que esto acarrea es de c\'omputo: hay un veloz incremento en el n\'umero de diagramas relevantes al aumentar el orden de la perturbaci\'on en las correcciones cu\'anticas o el n\'umero de part\'iculas involucradas en el proceso de dispersi\'on. No obstante, en muchas teor\'ias los diversos diagramas aportan contribuciones que se cancelan entre ellas y dan lugar a un resultado final simplificado. 

El FLM es un m\'etodo eficiente para realizar el c\'alculo de acciones efectivas, amplitudes de dispersi\'on y anomal\'ias en TCC que explica algunas de esas cancelaciones y muestra varias ventajas conceptuales y pr\'acticas respecto al m\'etodo diagram\'atico \parencite{Schubert:2001he,Bastianelli:2006rx}. En la secci\'on \ref{INTRO.objetivos} hemos ya mencionado diversas aplicaciones en las cuales ha demostrado su eficiencia.

A grandes rasgos, las t\'ecnicas del FLM pueden ser utilizadas para estudiar el NdC de diversos operadores diferenciales y, en consecuencia, permiten el an\'alisis del determinante funcional del operador diferencial que describe las fluctuaciones cu\'anticas en TCC. 
El operador en cuesti\'on, el cual se obtiene tomando la variaci\'on segunda de la acci\'on cl\'asica, en general se trata de un operador diferencial local cuyos coeficientes de Seeley-De Witt (SDW), coeficientes de la expansi\'on asint\'otica (para peque\~nos valores del tiempo propio) de la traza del NdC del operador, han sido en muchos casos estudiados.  

En particular, como veremos en este cap\'itulo para el caso de campos escalares, el FLM nos permite determinar, a trav\'es del c\'alculo intermedio de coeficientes de SDW, cantidades a 1 bucle en TCC utilizando las integrales de camino de R. Feynman. Tan pronto hayamos introducido el formalismo, consideraremos el caso general de potenciales regulares. Esto, sumado a los m\'etodos desarrollados en la  secci\'on \ref{DET.renor} para la renormalizaci\'on implementada en t\'erminos de t\'ecnicas espectrales, permitir\'a concentrarnos en el an\'alisis de la renormalizabilidad del modelo $\lambda\phi^4$.

Para concluir, mostraremos c\'omo el FLM puede ser utilizado incluso en el estudio de potenciales singulares tipo delta de Dirac. Esta clase de potenciales se asocia a bordes semitransparentes, los cuales imponen particulares condiciones de contorno a los campos. Como es usual, es posible investigar la dependencia de la energ\'ia efectiva (cfr. \ref{DET.energiaeff}) ante geometr\'ias con este tipo de bordes; el ejemplo del que nos ocuparemos es la configuraci\'on de dos placas paralelas.

Vale decir que todas las t\'ecnicas desarrolladas a lo largo de este cap\'itulo, junto a los ejemplos que analizamos, tienen como objetivo preparar el recorrido para la generalizaci\'on del FLM a TCC no conmutativas, teor\'ias que son propuestas como modelos efectivos de la gravedad cu\'antica. A ello nos dedicaremos a partir del cap\'itulo \ref{NC}.

\section{Formalismo de l\'inea de mundo para campos escalares}
Consideremos un campo escalar $\varphi$ real, masivo y autointeractuante a trav\'es de un potencial $U(\varphi)$, definido sobre un espacio eucl\'ideo de $d$ dimensiones. Como hemos visto en la secci\'on \ref{DET.eff2}, la correcci\'on a un bucle de la AE  est\'a dada por\footnote{La ecuaci\'on \eqref{DET:efffinal} posee adem\'as un t\'ermino que corresponde al n\'ucleo de calor de una part\'icula libre, proveniente de la normalizaci\'on de la funcional generatriz $Z[J])$. Por econom\'ia en la escritura,  salvo menci\'on contraria, omitiremos este t\'ermino.} 
\begin{align}\label{DET:eff.campoescalar}
 \Gamma_{1-\text{\rm bucle}}[\phi]=-\frac{1}{2}\int_0^{\infty}\frac{d\beta}{\beta}\, \text{Tr}\, e^{-\beta A},
\end{align}
la integral de la traza del NdC de $A=-\Delta +m^2+U''(\phi)$, el operador de fluctuaciones cu\'anticas evaluado en el valor de expectaci\'on de vac\'io $\langle \varphi \rangle= \phi$. Esta traza puede ser escrita expl\'icitamente sobre el espacio de coordenadas $x$ utilizando la notaci\'on de Dirac, resultando
\begin{align}\label{DET:efectiva}
 \Gamma_{1-\text{\rm bucle}}[\phi]&=-\frac{1}{2}\int_0^{\infty}\frac{d\beta}{\beta} \int d^{d}x\, \langle x\lvert\, e^{-\beta (-\Delta +m^2+U''(\phi))}\,\lvert x\rangle.
\end{align}

Llegados a este punto, podemos notar un aspecto clave del FLM: el integrando de \eqref{DET:efectiva} puede ser asociado a una amplitud de transici\'on en mec\'anica cu\'antica no relativista. En efecto, supongamos que estuvieramos estudiando el comportamiento de una part\'icula de masa $m'$ cuya evoluci\'on fuera dictada por el hamiltoniano H, correspondiente a un potencial $V$ independiente del tiempo. En ese caso, la amplitud de transici\'on $K(x',x'',it)$ desde la posici\'on $x'$ hasta  $x''$, en un intervalo de tiempo $t$, estar\'ia dada por
\begin{align}\label{DET:transicion}
\begin{split}K(x',x'',it)&:= \langle x''\lvert e^{-itH} \lvert x'\rangle\\[0.2cm]
&=\int_{x(0)=x'}^{x(t)=x''} \mathcal{D}x(\tau)\; e^{i\int_0^td\tau \left(\frac{m'}{2}\dot{x}^2-V(x(\tau))\right)},
\end{split}\end{align}
al expresarla como una integral de camino de Feynman\footnote{Si el hamiltoniano estuviera definido a valores matriciales uno deber\'ia tener la precauci\'on de ordenar temporalmente el integrando de la IdC  para obtener los resultados correctos \parencite{Bastianelli:1992ct}. En los casos que analizaremos ser\'a irrelevante la introducci\'on de dicho ordenamiento.}. Comparando las expresiones \eqref{DET:efectiva} y \eqref{DET:transicion}, es inmediato realizar las correspondencias
\begin{align}
\begin{split}
\left\{
 \begin{array}{ccl}
H&\rightarrow &p^2 +m^2+U''(\phi(x))\\
 t&\rightarrow &-i\beta\\
 m'&\rightarrow &\frac{1}{2}\\
 x',x''&\rightarrow &x
\end{array}\right.
\end{split}
\end{align}

En resumen, notando que el exponente en la integral de camino \eqref{DET:transicion} es el lagrangiano de la part\'icula imaginaria, la contribuci\'on de un bucle a la AE puede ser escrita como la f\'ormula maestra del FLM,
\begin{align}
 \Gamma_{1-\text{\rm bucle}}[\phi]&=-\frac{1}{2}\int_0^{\infty}\frac{d\beta}{\beta} \int d^{d}x \int_{x(0)=x}^{x(\beta)=x} \!\!\mathcal{D}x(\tau)\; e^{-\int_0^\beta d\tau \left(\frac{1}{4}\dot{x}^2+m^2+U''(\phi(x(\tau)))\right)}.\label{DET:efectivafinal}
\end{align}
Equivalentemente, la suma sobre todas las trayectorias cerradas puede ser tenida en cuenta eligiendo en la integral de caminos la medida asociada a trayectorias peri\'odicas\footnote{El c\'alculo de la traza del NdC involucra una integral sobre el ET; ya sea sobre el punto de intersecci\'on de las trayectorias, en el caso de condiciones Dirichlet, o sobre el modo cero (o modo centro de masa) de $-\partial_t^2$, para caminos peri\'odicos, el resultado es el mismo. Sin embargo, en geometr\'ias curvas han sido encontradas discrepancias entre ambos enfoques al momento de calcular cantidades locales; el problema radica en que las coordenadas de centro de masa pueden no ser covariantes \parencite{Schalm:1998ix}.}:
\begin{align}
 \Gamma_{1-\text{\rm bucle}}[\phi]&=-\frac{1}{2}\int_0^{\infty}\frac{d\beta}{\beta}  \int_{x(0)=x(\beta)} \mathcal{D}x(\tau)\; e^{-\int_0^\beta d\tau \left(\frac{1}{4}\dot{x}^2+m^2+U''(\phi(x(\tau)))\right)}.\label{DET:efectivafinal2}
\end{align} 
Las ventajas de utilizar \eqref{DET:efectivafinal} o \eqref{DET:efectivafinal2} respecto a la otra var\'ian con el modelo en consideraci\'on. En lo sucesivo veremos c\'omo aplicar concretamente estas f\'ormulas a diversos ejemplos.

\section[Coeficientes de SDW para un potencial regular]{Coeficientes de Seeley-DeWitt para un potencial regular}\label{FLM.regular}
Dado un operador del tipo laplaciano $A=-\Delta +V(x)$ en $\mathbb{R}^d$, el cual podr\'ia asociarse a las fluctuaciones cu\'anticas de un campo escalar interactuante con el potencial de fondo $V(x)$ o autointeractuante con un potencial $U(\varphi(x))$ tal que $\delta^2_{\phi}U=V(x)$, la traza de su NdC puede ser escrita, utilizando el FLM, como
\begin{align}
\begin{split} \text{Tr} \,e^{-\beta A}&=\int d^{d}x\, K(x,x,\beta)\\
&=\int d^{d}x \int_{x(0)=x(\beta)=x} \mathcal{D}'x(\tau)\; e^{-\int_0^\beta d\tau \left(\frac{1}{4}\dot{x}^2+V(x(\tau))\right)}\label{FLM:regular0}
 .\end{split}
\end{align}
A modo de primer ejemplo sobre la aplicaci\'on de (\ref{FLM:regular0}), determinaremos los coeficientes de SDW para el caso de un potencial $ V(x)$ al que requeriremos ciertas condiciones de regularidad. 

Para ello, centraremos en un primer momento nuestra atenci\'on sobre el NdC. Visto que obtener los coeficientes de SDW es al fin y al cabo obtener un desarrollo v\'alido para valores peque\~nos del tiempo propio $\beta$, conviene realizar como primer paso un reescaleo en la variable de integraci\'on temporal, de forma que
\begin{align}
K(x,x',\beta)&=\int_{x(0)=x}^{x(1)=x'} \mathcal{D}x(\tau)\; e^{-\int_0^1 d\tau \left(\frac{1}{4\beta}\dot{x}^2+\beta \,V(x(\tau))\right)}.\label{FLM:regularheat}
\end{align}
El cambio de escala acarrea la aparici\'on de un factor dependiente de $\beta$ en la medida de la integral de camino. Desconocer la forma precisa de este factor no traer\'a mayores inconvenientes pues la normalizaci\'on ser\'a posteriormente determinada comparando el resultado con el NdC de una part\'icula libre, cuyo valor es conocido\footnote{Otro camino posible es analizar el efecto de la modificaci\'on de la escala en la medida de integraci\'on.}.

De  (\ref{FLM:regularheat}) resulta inmediato que el desarrollo para peque\~nos tiempos $\beta$ ser\'a com\-pa\-ti\-ble con una expansi\'on en potencias de V, en cuanto, pensada como una TCC en 0+1 dimensiones, el propagador es proporcional a $\beta$, la constante de acoplamiento del potencial es justamente el tiempo propio, y el desarrollo en potencias de $V$ no es m\'as que el desarrollo en el n\'umero de bucles:
\begin{align}
K(x,x',\beta)&=\int_{x(0)=x}^{x(1)=x'}  \mathcal{D}x(\tau)\; e^{-\int_0^1 d\tau \frac{1}{4\beta}\dot{x}^2}\sum_{n=0}^{\infty}\frac{(-\beta)^n}{n!}\left(\int_0^{1}d\tau\, V(x(\tau))\right)^n.\label{FLM:regular1}
\end{align}

Una forma natural de proseguir es expandiendo $x(\tau)$ alrededor de la trayectoria cl\'asica (correspondiente a $V\equiv0$) e introduciendo las fluctuaciones cu\'anticas $\varphi(\tau)\in\mathbb{R}^d$ a trav\'es de sus componentes $\varphi_{{\mu}}(\tau)$:
\begin{equation}
 x_{\mu}(\tau)=:\tau (x'-x)_{\mu}+x_{\mu} +\varphi_{\mu}(\tau).
\end{equation}
Recalquemos un aspecto que pronto ser\'a clave: a diferencia de los caminos $x(\tau)$, las fluctuaciones $\varphi(\tau)$ satisfacen condiciones Dirichlet en $\tau=0$ y $\tau=1$. Realizando esta traslaci\'on en la IdC, integral cuya medida de integraci\'on no se ve afectada, obtenemos
\begin{align}
\begin{split}
{K}(x,x',\beta)&=\\
&\hspace{-1cm}=e^{\textstyle-\frac{(x-x')^2}{4\beta}}\sum_{n=0}^{\infty}\frac{(-\beta)^n}{n!}
\int\limits_{\substack{\varphi(0)=0}}^{\varphi(1)=0} \mathcal{D}\varphi(\tau)
\;e^{-\int^{1}_0 dt'\,\dot{\varphi}^2(t')/\!4\beta}\;
\\
&\hspace{-1cm}\hspace{5cm}\times\left[\int^1_0 dt\, V\left((x-x')\,t+x'+\varphi(t)\right)\right]^n\\&\\
&\hspace{-1cm}=e^{\textstyle-\frac{(x-x')^2}{4\beta}}\sum_{n=0}^{\infty}\frac{(-\beta)^n}{n!}
\left\langle\left(\int^1_0 dt\, V\left((x-x')\,t+x'+\varphi(t)\right)\right)^n\right\rangle_{\!D}\\[0.2cm]
&\hspace{-1cm}\hspace{5cm}\times \int\limits_{\substack{\varphi(0)=0}}^{
\varphi(1)=0} \mathcal{D}\varphi(\tau)
\;e^{-\int^{1}_0 dt'\,\dot{\varphi}^2(t')/\!4\beta}\ .\end{split}
\label{FLM:regular.promedio}
\end{align}
En la \'ultima ecuaci\'on hemos utilizado la siguiente notaci\'on, tomada prestada de la mec\'anica estad\'istica, para el valor medio de una funci\'on arbitraria de los campos $\varphi(\tau)$ con una medida gaussiana dada por el cociente de integrales de caminos
\begin{equation}
\left\langle\ f[\varphi(t)] \right\rangle_{\!D}:=\frac{\int\limits_{\substack{\varphi(0)=\varphi(1)=0}} \mathcal{D}\varphi(\tau)
\;e^{-\int^{1}_0 dt'\,\dot{\varphi}^2(t')/\!4\beta}\ f[\varphi(t)]}{\int\limits_{\substack{\varphi(0)=\varphi(1)=0}} \mathcal{D}\varphi(\tau)
\;e^{-\int^{1}_0 dt'\,\dot{\varphi}^2(t')/\!4\beta}\ }\ ,
\end{equation}
De la definici\'on resulta trivial la igualdad $\langle 1\rangle=1$.

Este es un punto adecuado para dedicarnos a un aspecto que hemos dejado de lado hasta el momento: la normalizaci\'on de la integral de caminos. Notemos que, al tomar $V\equiv 0$, la expresi\'on (\ref{FLM:regular.promedio}) se reduce al NdC $K_0(x,x',\beta)$ de una part\'icula libre en la primera cuantizaci\'on, tambi\'en asociado al problema del flujo del calor. Su forma exacta puede obtenerse aprovechando su interpretaci\'on como amplitud de transici\'on e introduciendo la relaci\'on de completitud en la base de estados de momento definido, ya que en ella el hamiltoniano asociado resulta diagonal; el resultado es
\begin{align}\label{FLM:nucleolibre}
    \begin{split}
{K}_0(x,x',\beta)
    &=\frac{e^{\textstyle-\frac{(x-x')^2}{4\beta}}}{(4\pi\beta)^{d/2}}\ .
\end{split}
\end{align}
Una comparaci\'on directa de \eqref{FLM:regular.promedio} con \eqref{FLM:nucleolibre} nos permite elegir la normalizaci\'on 
\begin{align}\label{FLM:norm}\int\limits_{\substack{\varphi(0)=0}}^{\varphi(1)=0}
\mathcal{D}\varphi(\tau)
\;e^{-\int^{1}_0 dt'\,\dot{\varphi}^2(t')/\!4\beta}:=
\frac{1}{(4\pi\beta)^{d/2}}\ .
\end{align}

Luego de esta breve digresi\'on sobre la notaci\'on y la normalizaci\'on de la integral de camino, podemos proseguir con el estudio de \eqref{FLM:regular.promedio}. Una de las alternativas para explotar las condiciones de regularidad del potencial $V$,  es realizar un desarrollo en t\'erminos de las fluctuaciones $\varphi$ y los incrementos en la posici\'on $(x-x')$. En tal caso, las contribuciones de los t\'erminos correspondientes a $n=0,1,2$ en (\ref{FLM:regular.promedio}) son\footnote{La elecci\'on de los t\'erminos que figuran en \eqref{FLM:desarrollo1} puede parecer a primera vista arbitraria; no obstante, responde a un desarrollo a orden $\beta^2$. Para convencerse, basta observar que ante la presencia del factor exponencial, la variable $x^{\mu}$ es del orden de $\sqrt{\beta}$ al ser integrada.}
\begin{align}
\label{FLM:desarrollo1}
\begin{split}
\\[-0.4cm]
{K}&(x,x',\beta)=\\[0.1cm]
=&\frac{e^{\textstyle-\frac{(x-x')^2}{4\,\beta}}}{(4\pi\beta)^{d/2}}\left\lbrace\langle1\rangle_{\!D}-\beta\,V(x')\int_0^1\,dt\,\langle1\rangle_{\!D}\right.\\
&\hspace{0cm}-\left.\beta\,\partial_{\mu}V(x')
\int_0^1\,dt\, \,\left[(x-x')^{\mu}\,t\,\langle1\rangle_{\!D}
+\langle\varphi^{\mu}(t)\rangle_{\!D}\right]\right.
\\
&\hspace{0.5cm}-\left.\frac{\beta}{2}\,\partial^{2}_{{\mu}\nu}V(x')\int_0^1\,dt\, \,\langle\left[(x-x')\,t
+\varphi(t)\right]^{\mu}\left[(x-x')\,t
+\varphi(t)\right]^{\nu}\rangle_{\!D}\right.
\\
&\hspace{4.3cm}-\left.\ldots+
\frac{\beta^2}{2}\,V^2(x')\int_0^1\int_0^1\,dt\,dt' \langle1\rangle_{\!D}+\ldots\right\rbrace.
\end{split}\end{align}

Es claro que hemos reducido el problema al c\'alculo de funciones de transici\'on de $n$ puntos, a saber, valores medios de expectaci\'on del producto de $n$ campos $\varphi$. Como hemos aprendido de TCC, una forma elegante de tratar con estas cantidades es a trav\'es de la funcional generatriz (FG) $Z_D[j]$, definida por la ecuaci\'on
\begin{align}\label{FLM:zetafunction}
    \begin{split} 
    Z_D[j]&:=\frac{\int\limits_{\substack{\varphi(0)=0}}^{\varphi(1)=0}
    \mathcal{D}\varphi(\tau) \;e^{-\int^{1}_0 dt'\,
    \left\{\dot{\varphi}^2(t')/\!4\beta
    -j(t')\varphi(t')\right\}}}{\int\limits_{\substack{\varphi(0)=0}}^{\varphi(1)=0}
    \mathcal{D}\varphi(\tau) \;e^{-\int^{1}_0 dt'\,
    \left\{\dot{\varphi}^2(t')/\!4\beta
    \right\}}}
    \\[0.2cm]
    &=e^{ \beta\int_0^1
    dt\, j(t)\left((-\partial^2)^{-1}_Dj\right)(t)}.\
\end{split}
\end{align}
Para utilizar esta expresi\'on, obtenida completando cuadrados en la funcional cuadr\'atica en $\varphi$, debemos encontrar el operador integral $(\partial^2)^{-1}_D$, el cual act\'ua sobre cada componente del vector fuente $j$ como el inverso del operador de Sturm-Liouville $\partial^2$ en una dimensi\'on con condiciones de contorno Dirichlet; la soluci\'on a este problema es harto conocida y est\'a dada por  la funci\'on de Green sim\'etrica 
\begin{align}\label{FLM:greenfunc}
    g(t,t')    
    =\left\lbrace \begin{array}{ll}
t(1-t')\qquad& {\rm si\ }t<t'\\
t'(1-t)\qquad& {\rm si\ }t>t'\
                  \end{array}\right.\ .
\end{align}
Al respecto conviene hacer una aclaraci\'on: ?`qu\'e habr\'ia sucedido en caso de que hubi\'eramos trabajado con la IdC en t\'erminos de trayectorias con condiciones de contorno peri\'odicas en lugar de las tipo Dirichlet? El operador $(\partial^2)^{-1}_{per}$ involucrado  habr\'ia tenido modos cero (las trayectorias constantes) y no habr\'ia sido invertible. Por supuesto, ello se podr\'ia remendar separando el modo cero, una alternativa un poco m\'as laboriosa que la elegida.

Retornando a la expresi\'on \eqref{FLM:zetafunction} para la FG, su ventaja operativa consiste en que, derivadas funcionales por medio, permite conocer cualquier funci\'on de transici\'on de $n$ puntos computando
\begin{align}
\langle \varphi_{i_1}(t_1)\ldots\varphi_{i_n}(t_n)\rangle_{\!D}:
&=\left.\frac{\delta}{\delta j_{{\mu}_1}(t_1)}\,\ldots\,\frac{\delta}{\delta j_{{\mu}_n}(t_n)}
    \ Z_D[j]\right|_{j\equiv 0}.\label{FLM:npoint}
\end{align}
A modo de ejemplo, reuniendo las ecuaciones (\ref{FLM:npoint}) y (\ref{FLM:zetafunction}), podemos calcular las cantidades
\begin{align}\label{FLM:prop}
\begin{split}\langle
    \varphi_{\mu}(t)
    \rangle_{\!D}
    &=0\ ,
\\
\langle
    \varphi_{\mu}(t)\varphi_{\nu}(t')
    \rangle_{\!D}
    &=\delta_{{\mu}{\nu}}\,2\beta\,\left\lbrace \begin{array}{ll}
t(1-t')\qquad& {\rm si\ }t<t'\\
t'(1-t)\qquad& {\rm si\ }t>t'\
                  \end{array}\right. ,
    \end{split}
\end{align}
utilizando la delta de Kronecker  $\delta_{\mu\nu}$. De esto resulta inmediato reconocer que el c\'alculo de los valores medios de la ecuaci\'on (\ref{FLM:desarrollo1}), si bien puede resultar trabajoso, es directo. Es as\'i que, vali\'endonos de \eqref{FLM:prop}, encontramos el siguiente resultado para la expansi\'on del NdC en t\'erminos de los incrementos $\delta_{\mu}:=x_{\mu}-x'_{\mu}$ y el tiempo propio $\beta$:
\begin{align}\label{FLM:desarrollofinal}
{K}(x,x',\beta)&=\\
   \nonumber &\hspace{-2cm}=\frac{e^{\textstyle-\frac{\delta^2}{4\,\beta}}}{\left(4\pi\beta\right)^{d/2}}\ \Biggl\lbrace
    1-\beta \, V
    \mp\frac{\beta}{2}\,\partial_{\mu}V\cdot\delta^{\mu}
    \mbox{}
    +\frac{\beta^2}{2}\,\left(-\frac{1}{3}\triangle V+V^2\right)\Biggr.
    \\
   \nonumber &\hspace{-1.5cm}-\frac{\beta}{6}\,\partial^2_{{\mu}{\nu}}V\cdot\delta^{\mu}\delta^{\nu}
    \pm\frac{\beta^2}{2}\,\left(-\frac{1}{6}\partial_{\mu}\triangle V+V\partial_{\mu}V\right)\cdot \delta^{\mu}\mp\frac{\beta}{24}\,\partial^3_{{\mu}{\nu}{\rho}}V\cdot\delta^{\mu}\delta^{\nu}\delta^{\rho}
    \\
    \nonumber &\hspace{-0.cm}
    +\frac{\beta^3}{6}\,\left(-\frac{1}{10}\triangle\triangle V+V\triangle V+\frac{1}{2}\partial^{\mu}V\partial_{\mu}V-V^3\right)
    \\
    \nonumber &\hspace{1.0cm}+\Biggl.
    \frac{\beta^2}{2}\,\left(-\frac{1}{20}\partial^2_{{\mu}{\nu}}\triangle V+\frac{1}{3}V\partial^2_{{\mu}{\nu}}V
    +\frac{1}{4}\partial_{\mu}V\partial_{\nu}V\right)\cdot\delta^{\mu}\delta^{\nu}\\
\nonumber &\hspace{5.0cm}-\frac{\beta}{120}\,\partial^4_{{\mu}{\nu}{\rho}{\sigma}}V\cdot\delta^{\mu}\delta^{\nu}\delta^{\rho}\delta^{\sigma}+\ldots\Biggr\rbrace \ .
\end{align}
En esta expresi\'on, los signos superior e inferior corresponden al caso en el que el desarrollo del potencial es alrededor del punto inicial ($x'$) o final ($x$), respectivamente. Por otro lado, el prefactor exponencial implica que al realizar la integral del NdC  en la variable $x$, cada factor $\delta_i$ dar\'a una contribuci\'on de orden $\beta^{1/2}$. Como consecuencia, los puntos en la expresi\'on (\ref{FLM:desarrollofinal}) indican contribuciones de orden $\beta^{7/2}$.

M\'as all\'a que el NdC posea informaci\'on adicional tanto a nivel matem\'atico como f\'isico, la ecuaci\'on (\ref{FLM:regular0}) requiere \'unicamente el desarrollo para tiempo propio peque\~no de la traza del NdC. Tomando lo que se conoce como la diagonal del NdC, es decir $x_{\mu}=x'_{\mu}$ o $\delta_{{\mu}}=0$ en (\ref{FLM:desarrollofinal}), llegamos a la expresi\'on 
\begin{align}
\nonumber\\[-0.4cm]
\label{FLM:desarrollodiagonal}
{K}&(x,x,\beta)=\\
\nonumber    &=\frac{1}{\left(4\pi\beta\right)^{d/2}}\ \Biggl\lbrace
    1-\beta \, V
    +\frac{\beta^2}{2}\,\left(-\frac{1}{3}\triangle V+V^2\right)+\Biggr.
    \displaybreak\\
\nonumber    &\hspace{1cm}\mbox{}\Biggl.
    +\frac{\beta^3}{6}\,\left(-\frac{1}{10}\triangle\triangle V+V\triangle V+\frac{1}{2}\partial_{\mu}V\partial^{\mu}V-V^3\right)
    +O(\beta^4)\Biggr\rbrace\ ,
\end{align}
 donde el potencial y sus derivadas est\'an evaluados en $x$. Finalmente, de acuerdo a las expresiones (\ref{FE:primerdesarrollo.hk}) y (\ref{FLM:regular0}), basta solo integrar (\ref{FLM:desarrollodiagonal}) en $x\in\mathbb{R}^{d}$ para obtener los primeros coeficientes de SDW
\begin{align}\label{FLM:sdwregular}
\begin{split}
a_0&=\frac{1}{(4\pi)^{d/2}}\int_{\mathbb{R}^{d}} 1\ ,\\
a_2&=-\frac{1}{(4\pi)^{d/2}}\int_{\mathbb{R}^{d}} V(x)\ ,\\
a_4&=\frac{1}{6\,(4\pi)^{d/2}}\int_{\mathbb{R}^{d}}\Bigl[-\triangle V\!\left(x\right)+3\,V^2\!\left(x\right)\Bigr]\ ,\\
a_6&=\frac{1}{60\,(4\pi)^{d/2}}\int_{\mathbb{R}^{d}}
\Bigl[-\triangle\triangle V(x)+
10\,V(x)\,\triangle V(x)\\
&\hspace{4cm}+5\,\partial_{\mu}V(x)\,\partial^{\mu}V(x)-10\,V^3(x)\Bigr]\ ,\\
a_1&=a_3=a_5=0\ .
\end{split}
\end{align}
A modo de verificaci\'on, constatamos que este desarrollo cumple con los teoremas detallados en la secci\'on \ref{FE}: los coeficientes son integrales de productos del potencial y sus derivadas, y, puesto que la variedad de base $\mathbb{R}^{d}$ no posee borde, el desarrollo asint\'otico de la traza del NdC contiene solamente potencias enteras de $\beta$. Adicionalmente, est\'a de acuerdo con \eqref{FE:a024.Rm} si identificamos $E\equiv-V$, $\Omega_{ij}=0$ y $\nabla_{\mu}\equiv\partial_{\mu}$.

\section{Renormalizaci\'on a un bucle del modelo \texorpdfstring{$\lambda\varphi^4$}{lambda phi**4}}\label{FLM.phi4}

Como aplicaci\'on de los resultados obtenidos en la secci\'on anterior a un modelo f\'isico sencillo y de importancia, consideraremos la renormalizaci\'on  de un campo cu\'antico $\varphi$, escalar, real y masivo, cuyo comportamiento en un espacio eucl\'ideo de $d$ dimensiones se encuentra regido por la densidad Lagrangiana
\begin{align}
 \mathcal{L}=\frac{1}{2}(\partial \varphi)^2+\frac{m^2}{2}\varphi^2+\frac{\lambda}{4!}\varphi^4.
\end{align}
 
En el cap\'itulo \ref{DET} hemos demostrado que la correcci\'on de un bucle a la AE y la posterior renormalizaci\'on de la teor\'ia est\'a determinada por el operador de fluctuaciones cu\'anticas $A$ o, para ser precisos, por la integral de Schwinger \eqref{DET:efffinal} del NdC de $A$, la cual en cierto desarrollo involucra los coeficientes de SDW. El c\'alculo de $A=B+m^2$ en este modelo arroja el resultado
\begin{align}
 A=\delta^2 S=-\partial^2+m^2+\frac{\lambda}{2}\phi^2,
\end{align}
el cual permite identificar el potencial\footnote{La presencia de la masa contribuye un factor $e^{-\beta m^2}$ que puede ser factorizado en el c\'alculo del n\'ucleo de calor.} $V(x)=\frac{\lambda}{2}\phi^2(x)$, dependiente del campo medio $\phi(x)$, y utilizar la f\'ormula \eqref{FLM:sdwregular} para la determinaci\'on de los primeros coeficientes de SDW de $B$:
\begin{align}\label{FLM:phi4.coef}
\begin{split}
a_0&=\frac{1}{(4\pi)^{d/2}}\int_{\mathbb{R}^{d}} 1,\\
a_2&=-\frac{\lambda}{2\,(4\pi)^{d/2}}\int_{\mathbb{R}^{d}} \phi^2(x),\\
a_4&=\frac{\lambda^2}{8\,(4\pi)^{d/2}}\int_{\mathbb{R}^{d}}\phi^4\!\left(x\right),\\
a_6&=\frac{\lambda^2}{48\,(4\pi)^{d/2}}\int_{\mathbb{R}^{d}}
\left(\phi^2\,\triangle \phi^2-\lambda\,\phi^6(x)\right)\ .
\end{split}
\end{align}

Recordemos que la ecuaci\'on \eqref{DET:eff.regular} muestra que, si las hubiere, las contribuciones divergentes a la AE de orden un bucle son un n\'umero finito y provienen de los primeros t\'erminos del desarrollo del NdC para tiempo propio peque\~no. Tomemos por caso $d=4$: los t\'erminos divergentes de la AE son tres, y su dependencia funcional en el campo cl\'asico $\phi$ est\'a dada por los coeficientes $a_n$, con $n=0,2,4$. Debido a la naturaleza de las divergencias resulta conveniente estudiar cada una de las contribuciones por separado.

La primera, aquella proporcional a $a_0$, se trata de una contribuci\'on de vac\'io, puesto que no involucra el potencial. Si agreg\'aramos el t\'ermino asociado al propagador $G^{-1}$, tal y como lo dicta \eqref{DET:efectiva0}, esta contribuci\'on no se presentar\'ia.

Por otro lado, el t\'ermino correspondiente a $a_2$ tiene una dependencia funcional en $\phi$ id\'entica a la del t\'ermino de masa. En t\'erminos de diagramas de Feynman, est\'a asociado al diagrama de renacuajo, cfr. \eqref{intro:diagramas.planares}. Efectivamente, al orden de un bucle y seg\'un \eqref{DET:eff.regular}, el t\'ermino de la AE cuadr\'atico en los campos  es\footnote{Dado que todos los resultados enunciados a continuaci\'on corresponden al orden de un bucle, omitiremos la inclusi\'on del sub\'indice ``$1-\text{\rm bucle}$'' en lo que resta de la secci\'on.}
\begin{align}
 \Gamma^{(2)}=\int dx\,\left(m^2+\frac{\lambda}{32\pi^2}\Gamma\left(-1,\frac{m^2}{\Lambda^2}\right)\right)\phi^2(x).
\end{align}
Si utilizamos la prescripci\'on de m\'inima sustracci\'on para renormalizar, explicado en la secci\'on \ref{DET.renor}, encontramos la masa renormalizada
\begin{align}
 m^2_R= m^2\left[1+\frac{\lambda}{32\pi^2}\left(\frac{\Lambda^2}{m^2}+\log\left(\frac{m^2}{\Lambda^2}\right)\right)
 \right],
\end{align}
puesto que el desarrollo de $\Gamma\left(-1,x\right)$  para peque\~nos valores de $x$ resulta
\begin{align}
 \Gamma\left(-1,x\right)=\frac{1}{x}+\log x +O(x^0).
\end{align}

Por \'ultimo, a\'un nos resta analizar la contribuci\'on divergente que involucra al coeficiente $a_4$. Esta es proporcional a la potencia cuarta de $\phi$, al igual que el potencial de interacci\'on; reuniendo ambos t\'erminos la funci\'on de cuatro puntos $\Gamma^{(4)}$, puede escribirse como 
\begin{align}
 \Gamma^{(4)}=\frac{\lambda}{4!}\,\int dx\,\left(1-\frac{3}{32\pi^2}\,\lambda\,\Gamma\left(0,\frac{m^2}{\Lambda^2}\right)\right) \, \phi^4(x) .
 \end{align}
Empleando nuevamente la prescripci\'on de m\'inima sustracci\'on, encontramos que la constante de acoplamiento  renormalizada $\lambda_R$ posee el desarrollo 
\begin{align}\label{FLM:lambdaR}
 \lambda_R=\lambda\,\left(1+\frac{3}{32\pi^2}\,\lambda\,\log\frac{m^2}{\Lambda^2}
 +\cdots\right)
\end{align}

A partir de \eqref{FLM:lambdaR} se puede calcular la funci\'on $\beta$ de la constante de acoplamiento (desnuda), con el fin de estudiar su comportamiento con la escala de energ\'ia $\Lambda$:
\begin{align}\label{FLM:funcionbeta}
 \beta(\lambda)=\Lambda\left.\partial_{\Lambda}\lambda\right\rvert_{\lambda_R}=\frac{3}{16\pi^2}\lambda^2.
\end{align}
Por cuanto $\lambda^2>0$, la constante de acoplamiento resulta ser una funci\'on creciente de $\Lambda$, al menos al orden de un bucle. Si este fuera su comportamiento a todo orden podr\'ia acarrear problemas en el desarrollo perturbativo en potencias de $\lambda$ que hemos propuesto para la AE. En particular, se puede resolver la ecuaci\'on diferencial \eqref{FLM:funcionbeta} para obtener
\begin{align}\label{FLM:lambda-Lambda}
 \lambda= \frac{\lambda_0}{1-\frac{3}{16\pi^2}\lambda_0\log(\frac{\Lambda}{\Lambda_0})},
\end{align}
donde $\lambda_0$  es el valor de la constante de acoplamiento para una escala de energ\'ia $\Lambda_0$. Esta f\'ormula muestra que, partiendo de un valor peque\~no para la constante de acoplamiento, el flujo es tal que al aumentar la escala de energ\'ia el denominador se anula y la expresi\'on diverge para una escala de energ\'ia
\begin{align}
\Lambda=\Lambda_0e^{\frac{16\pi^2}{3\lambda_0}}.
\end{align}

Este inconveniente es llamado polo o fantasma de Landau, en reconocimiento al f\'isico ruso cuyo grupo de trabajo encontr\'o un comportamiento similar en la teor\'ia de la electrodin\'amica cu\'antica. Un problema relacionado es el de la trivialidad cu\'antica: para eliminar la inconsistencia de este polo, la \'unica soluci\'on posible es la elecci\'on $\lambda_0=0$, la teor\'ia trivial sin potencial. Si bien estos c\'alculos corresponden solo al orden de un bucle, hay varios resultados anal\'iticos y num\'ericos a \'ordenes superiores que sugieren la presencia de este comportamiento en el modelo $\lambda\phi^4$ en 4 dimensiones \parencite{Frohlich:1982tw,Suslov:2008ca,Wolff:2009ke}.

Con estas consideraciones damos por concluido el estudio de potenciales regulares. En la pr\'oxima secci\'on, veremos c\'omo generalizar el m\'etodo al caso en el que el potencial es una delta de Dirac. Siguiendo ciegamente el mismo procedimiento, al intentar hacer el desarrollo del potencial alrededor del punto inicial $x'$ en la ecuaci\'on (\ref{FLM:desarrollo1}), obtendr\'iamos el producto mal definido de deltas de Dirac y sus derivadas evaluadas todas en el mismo punto $x'$. Afortunadamente, encontraremos una forma de ladear este problema.

\section[NdC para un potencial tipo delta de Dirac]{N\'ucleo de calor para un potencial tipo delta de Dirac}\label{FLM.delta}

Vistas las aplicaciones exitosas del FLM a diversos problemas en TCC, existe un creciente inter\'es en extender estas t\'ecnicas 	para estudiar la influencia de diferentes condiciones externas en la AE de campos cu\'anticos. Recientemente se ha comenzado a construir una generalizaci\'on sistem\'atica del FLM en variedades con borde, comenzando con el c\'alculo del desarrollo asint\'otico del NdC de un campo escalar en diversas variedades chatas con borde \parencite{Bastianelli:2006hq,Bastianelli:2007jr,Bastianelli:2008vh,Bastianelli:2009mw}. Por otro lado, la delta de Dirac con soporte en una superficie de codimensi\'on uno es un problema matem\'atico bien definido  \parencite{Albe} y  ha sido utilizada para modelar diversos sistemas f\'isicos. De todos ellos, se mencionar\'a sobre el comienzo del pr\'oximo par\'agrafo su aplicaci\'on al estudio de l\'aminas de plasma \parencite{Bordag:2005by}. Como motivaci\'on adicional, investigaciones m\'as recientes  \parencite{Falomir:2001iw,Falomir:2003vw,
Falomir:2005xh,Kirsten:2008wu} 
han establecido que en presencia de singularidades en el espacio-tiempo o en los coeficientes del operador diferencial la estructura del desarrollo asint\'{o}tico del heat-kernel cambia sustancialmente.

En esta secci\'on mostraremos que el FLM puede ser aplicado al estudio de campos escalares con condiciones de pegado sobre una superficie de codimensi\'on uno, condiciones que son impuestas por un potencial tipo delta de Dirac con soporte en dicha superficie y  reciben el nombre de borde semitransparentes. Estos resultados nos permitir\'an analizar, en la secci\'on \ref{FLM.delta.cas}, la energ\'ia efectiva y la fuerza de Casimir en una configuraci\'on de placas paralelas con acoplamientos peque\~nos.

Para comenzar, analizaremos el desarrollo asint\'otico de la traza del NdC de un operador $A_{\gamma}$ tipo Schr\"odinger, cuyo potencial contiene una delta de Dirac con soporte en el hiperplano $x_1=0$ de $\mathbb{R}^{d+1}$, i.e.
\begin{align}\label{FLM:operadordelta}
 A_{\gamma}=-\Delta+V(x)+\gamma\delta(x_1).
\end{align}
Supondremos adem\'as que $V(x)$ es un potencial regular y $\gamma\in\mathbb{R}^+$. Por convenci\'on, llamaremos  $y\in\mathbb{R}^d$ a las coordenadas sobre el hiperplano $x_1=0$. Este operador describe las fluctuaciones cu\'anticas de un campo escalar en un espacio eucl\'ideo $d+1$ dimensional y en interacci\'on con un potencial de fondo que es la suma de uno regular ($V$) y uno altamente localizado (la delta de Dirac). 

De acuerdo a la ecuaci\'on (\ref{FLM:regularheat}) la diagonal del NdC de $A_{\gamma}$ se puede escribir como 
\begin{multline}
{K_\gamma}(x,x,\beta)=\\
=
\int\limits_{\substack{x(0)=x\\x(1)=x}} \mathcal{D}x(\tau)
\;e^{\textstyle-\int^{1}_0 dt'\,\left[\dot{x}^2(t')/\!4\beta+\beta\, V\left(x(t')\right)+\beta\,\gamma\,\delta[x_1(t')]\right]}\ .
\end{multline}
Dado que ya hemos mostrado c\'omo se puede obtener el desarrollo para un potencial regular en el apartado \ref{FLM.regular}, no nos ocuparemos del factor que involucra $V$ y desarrollaremos en serie s\'olo la exponencial de la parte singular, para obtener
\begin{multline}\label{FLM:pathdelta}
K_\gamma(x,x,\beta)-K_0(x,x,\beta)=\sum_{n=1}^{\infty}\frac{(-\beta\,\gamma)^n}{n!}\int\limits^{1}_0\cdots\int\limits^{1}_0 dt_1 \cdots dt_n\,
\\
\times  \int\limits_{\substack{x(0)=x\\x(1)=x}} \mathcal{D}x(\tau)
\;e^{\textstyle-\int^{1}_0 dt'\,\left[\dot{x}^2(t')/\!4\beta+\beta\, V\left(x(t')\right)\right]}\,\delta \left[x_1(t_1)\right]\cdots\,\delta\left[x_1(t_n)\right]\ .
\end{multline}
En esta expresi\'on hemos sustra\'ido la contribuci\'on de $K_0(x,x,\beta)$, el NdC correspondiente a $\gamma=0$, para el cual se pueden utilizar los resultados precedentes.

Llegados a este punto, es claro que no se puede realizar la expansi\'on de las deltas de Dirac alrededor de la posici\'on inicial: en caso de hacerlo obtendr\'iamos un producto de deltas y sus derivadas todas con el mismo soporte. En cambio, s\'i podemos interpretar estas integrales de camino como amplitudes de transici\'on para una part\'icula bajo la acci\'on del potencial regular $V$, sometida a las restricciones que imponen las deltas.

En otras palabras, la presencia de las deltas en el $n$-\'esimo t\'ermino de la ec. (\ref{FLM:pathdelta}) implica que a la integral de caminos s\'olo contribuir\'an aquellas trayectorias cerradas que comienzan y terminan en el punto $x$, y adicionalmente tocan la superficie $x_1=0$ (el soporte de la funci\'on delta) en cada uno de los tiempos $t_1,\cdots,t_n$. En consecuencia, cada una de las contribuciones con $n$ deltas puede ser escrita como el producto de $n+1$ funciones de transici\'on que tengan en cuenta las condiciones impuestas sobre la coordenada $x_1$:
\begin{align}\label{FLM:eq.hkreg+delta}
K_\gamma(x,x,\beta)-K_0(x,x,\beta)&=\\
\nonumber &\hspace{-4.cm}=\sum_{n=1}^{\infty}\left(-\beta\,\gamma\right)^n\int\limits^{1}_0\cdots\int\limits^{t_3}_0\int\limits^{t_2}_0\,dt_1\,dt_2\,\ldots dt_n
\\
\nonumber &\hspace{-3cm}
\times
\ \int\limits_{\substack{y(0)=y}}^{y(1)=y} \mathcal{D}y(\tau)
\;e^{\textstyle-\int^{1}_{0} dt'\,\dot{y}^2(t')/\!4\beta}
\,
\\
\nonumber  &\hspace{-3.0cm}\times\,
\int\limits_{\substack{x_1(t_n)=0}}^{
x_1(1)=x_1} \mathcal{D}x_1(\tau)
\;e^{\textstyle-\int^{1}_{t_n} dt'\,\left[\dot{x_1}^2(t')/\!4\beta+\beta\, V\left(x(t')\right)\right]}\, 
\\
\nonumber  &\hspace{-3.cm}\times\,
\prod_{i=1}^{n-1}\int\limits_{\substack{x_1(t_i)=0}}^{x_1(t_{i+1})=0}
 \mathcal{D}x_1(\tau)
\;e^{\textstyle-\int^{t_{i+1}}_{t_i} dt'\,\left[\dot{x_1}^2(t')/\!4\beta+\beta\, V\left(x(t')\right)\right]}
\,\\
\nonumber  &\hspace{-3.cm}\times\,
\int\limits_{\substack{x_1(0)=x_1}}^{x_1(t_{1})=0} \mathcal{D}x_1(\tau)
\;e^{\textstyle-\int^{t_{1}}_{0} dt'\,\left[\dot{x_1}^2(t')/\!4\beta+\beta\, V\left(x(t')\right)\right]}
\ .
\end{align}
Si consideramos una dada trayectoria $y(\tau)$ en las coordenadas paralelas al soporte de la delta, se puede observar de (\ref{FLM:eq.hkreg+delta}) que cada una de las integrales de camino en la variable $x_1$ corresponde a un NdC en un espacio unidimensional con un potencial regular. Por consiguiente, no obstante la dependencia temporal que genera la trayectoria $y(\tau)$ en el argumento de los potenciales, estos NdC  pueden ser calculados siguiendo un lineamiento an\'alogo al detallado en la secci\'on \ref{FLM.regular}. Posteriormente, la integral en $x\in\mathbb{R}^{d+1}$ proporcionar\'ia el desarrollo deseado para la traza del NdC.

No obstante el m\'etodo explicado carece de vericuetos, el c\'alculo resulta ser m\'as eficiente realizando una peque\~na variaci\'on: a\~nadiremos factores en (\ref{FLM:pathdelta}) para trabajar con deltas que tengan soporte no en un hiperplano sino en un punto. Consecuentemente, introducimos por cada funci\'on $\delta(x_1(t_i))$ en la f\'ormula  (\ref{FLM:pathdelta}) la integral en $y_0^{(i)}\in\mathbb{R}^d$ de una funci\'on delta $d$	-dimensional en las coordenadas $y$, es decir, un factor 
$$1=\int_{\mathbb{R}^d} dy_0^{(i)}\,\delta^{(d)}(y(t_i)-y_0^{(i)}).$$
Intercambiando las integrales en estas nuevas variables con las integrales de camino, podemos reinterpretar a las funciones delta como restricciones sobre los caminos, los cuales deber\'an pasar por el punto $x^{(i)}:=x(t_i)=(0,y_0^{(i)})$ al tiempo $t_i$:
\begin{multline}\label{FLM:eq.hkreg+delta2}
K_\gamma(x,x,\beta)-K_0(x,x,\beta)=\sum_{n=1}^{\infty}\left(-\beta\,\gamma\right)^n
\int\limits^{1}_0\cdots\int\limits^{t_3}_0\int\limits^{t_2}_0\,dt_1\,dt_2\,\ldots dt_n\,
\\
\times\int_{\mathbb{R}^d}\ldots \int_{\mathbb{R}^d}dy^{(1)}\ldots dy^{(n)}
\,
K(x,x^{(n)},\beta(1- t_n))\,
\times\,
\ldots\,\times\,
K(x^{(1)},x,\beta t_1)\ .
\end{multline}
El c\'alculo ahora resulta inmediato, puesto que cada uno de los NdC que componen (\ref{FLM:eq.hkreg+delta2}) puede desarrollarse de acuerdo a (\ref{FLM:desarrollofinal}).

Antes de obtener el resultado final vale la pena hacer el siguiente comentario a una posible objeci\'on. Es sabido que una part\'icula en un espacio de $d+1$ dimensiones bajo la acci\'on de un potencial tipo delta con soporte en un punto da lugar a un hamiltoniano tipo Schr\"odinger mal definido, al cual corresponden las funciones de transici\'on de la expresi\'on (\ref{FLM:eq.hkreg+delta2}). Esto es consistente con el hecho de que en ausencia de las integrales en $y^{(j)}$ las integrales temporales resultan divergentes.

Retomando el camino que nos llev\'o a la f\'ormula (\ref{FLM:eq.hkreg+delta2}), luego de reemplazar los NdC por los desarrollos escritos seg\'un (\ref{FLM:desarrollofinal}) es posible integrar finalmente las variables $x\in\mathbb{R}^{d+1}$ para obtener el desarrollo asint\'otico para $\beta$ peque\~no de la traza del NdC de~$A_{\gamma}$,
\begin{align}\label{FLM:desarrollodeltafinal}
\begin{split}
\\[-0.4cm]
{\rm Tr}\,e^{-\beta A_\gamma}-{\rm Tr}\,e^{-\beta A_0}&=\\[0.1cm]
&\hspace{-4cm}=\frac{\gamma}{(4\pi\beta)^{(d+1)/2}}\int\,d^{d}y\,
\Biggl\{
-\beta+\frac{\sqrt{\pi}}{4}\,\gamma\,\beta^{3/2}+\left(V-\frac{1}{6}\,\gamma^2\right)\beta^2\Biggr.\\
&\hspace{-4cm}\hspace{0.5cm}\Biggl.-\frac{\sqrt{\pi}}{32}\,\gamma\left(8V-\gamma^2\right)\beta^{5/2}+
\frac{1}{6}\left(\triangle V-3V^2+\gamma^2\,V\right)\beta^3+O(\beta^{7/2})
\Biggr\}\ .
\end{split}
\end{align}
En esta expresi\'on $A_0$ denota el operador de Schr\"odinger regular dado por (\ref{FLM:operadordelta}) para el valor $\gamma=0$ y, tal como hemos mencionado anteriormente, el desarrollo de la traza de su NdC  ha sido obtenido en la secci\'on \ref{FLM.regular}.

Del desarrollo (\ref{FLM:desarrollodeltafinal}), podemos inmediatamente leer las correcciones $\Delta a_n$ que reciben los primeros coeficientes $a_n$ de SDW  debido a la presencia de un potencial tipo delta de Dirac sumado al potencial regular. An\'alogamente a lo que sucede en los casos en que la variedad de base posee un borde, la presencia del potencial tipo delta implica la aparici\'on de t\'erminos de borde en el desarrollo de la traza del NdC. Las correcciones $\Delta a_n$ a los coeficientes del caso regular, los cuales est\'an dados por (\ref{FLM:sdwregular}) para $n$ par y son nulos para $n$ impar, se pueden escribir como
\begin{align}\label{FLM:eq.5.coef.delta}
\begin{split}
\Delta a_{0}&=-\frac{\gamma}{(4\pi)^{(d+1)/2}}\int d^dy\, 1\ ,
\\
\Delta a_{1}&=\frac{\sqrt{\pi}}{4 (4\pi)^{(d+1)/2}}\,\gamma^2 \int d^dy\, 1\ ,
\\
\Delta a_{2}&=\frac{\gamma}{(4\pi)^{(d+1)/2}}\int d^dy\, \left(V(0,y)-\frac{1}{6}\,\gamma^2\right)\ ,
\\
\Delta a_{3}&=-\frac{\sqrt{\pi}}{32(4\pi)^{(d+1)/2}}\,\int d^dy\, \gamma^2\left(8V(0,y)-\gamma^2\right)\ ,
\\
\Delta a_4&=\frac{1}{6(4\pi)^{(d+1)/2}}\,\gamma\int d^dy\,\left(\triangle V(0,y)-3V^2(0,y)+\gamma^2\,V(0,y)\right)\ .
\end{split}\end{align}
Los primeros cuatro coeficientes escritos en (\ref{FLM:eq.5.coef.delta}) pueden ser obtenidos como un caso particular de los resultados de  \textcite{Bordag:1999ed}.

\subsection{Fuerza de Casimir para condiciones de borde semitransparentes}\label{FLM.delta.cas}

Un aspecto interesante a analizar es la influencia de las condiciones de contorno semitransparentes sobre los campos en TCC. En este par\'agrafo determinaremos, utilizando el FLM, la energ\'ia efectiva y la correspondiente fuerza de Casimir\footnote{Es com\'un dar a esta fuerza, derivada de la energ\'ia efectiva, el nombre de fuerza de Casimir, el mismo que se utiliza para la obtenida a partir de la energ\'ia de Casimir.} para la configuraci\'on de dos l\'aminas paralelas debido a las oscilaciones cu\'anticas de un campo escalar que se encuentra d\'ebilmente acoplado a dos potenciales tipo delta. Estos resultados pueden ser comparados con los obtenidos por \textcite{Bordag:1992cm}, quienes trabajaron con una expresi\'on de la funci\'on de Green.

Cabe mencionar que no es la primera vez que el FLM se aplica al c\'alculo de energ\'ias efectivas y de Casimir:  sobre todo en combinaci\'on con m\'etodos de Monte Carlo ha permitido el estudio de diversas geometr\'ias \parencite{Gies:2003cv,Gies:2006bt,Gies:2006cq,Gies:2006xe}, como as\'i tambi\'en el de interesantes comportamientos de la energ\'ia de Casimir con la temperatura  \parencite{Gies:2008zz,Weber:2010kc,Weber:2010xv}. 

Por su parte, las condiciones de borde semitransparentes recientemente han sido utilizadas para cuantizar los modos transversales el\'ectricos del campo electromagn\'etico en la vecindad de densidades de carga localizadas \parencite{Barton1,Barton2,Barton3,Bordag:2005qv,Bordag:2005by,Bordag:2006kx,Bordag:2007zz,Bordag:2008rc,Bordag:2009zzc}. Podemos citar, a modo de ejemplo, el caso de las mol\'eculas grandes de carbono: la densidad electr\'onica de estas mol\'eculas puede ser considerada como una l\'amina infinitamente delgada de plasma que impone condiciones de borde semitransparentes a los modos de oscilaci\'on cu\'anticos del campo electromagn\'etico. Para peque\~nos valores de carga neta y corriente sobre la l\'amina, el problema se reduce a resolver la ecuaci\'on de onda con un potencial tipo delta de Dirac y su derivada con soporte en la l\'amina. 

Consideremos entonces un campo escalar masivo $\varphi(x)$, definido en $x\in\mathbb{R}^{d+1}$, y que interact\'ua con deltas de Dirac que imponen condiciones de contorno semitransparente en los hiperplanos $\lvert x_1\lvert=L/2$ de acuerdo a la siguiente acci\'on:
\begin{multline}\label{FLM:cas.accion}
    S[\phi]=\frac{1}{2}\int_{\mathbb{R}^{d+1}}(\partial\varphi(x))^2+
    m^2\varphi^2(x)\\
    +\gamma\left[\delta(x_1+L/2)+\delta(x_1-L/2)\right]\varphi^2(x)\ .
\end{multline}
De las $d+1$ coordenadas una corresponde al tiempo eucl\'ideo, otra a la espacial que define la posici\'on de las placas ($x_1$) y las restantes a las espaciales paralelas a dichas placas.

En este caso, el operador de fluctuaciones cu\'anticas posee como potencial la suma de dos deltas de Dirac. Sirvi\'endonos de (\ref{FLM:regularheat}), la f\'ormula de partida del FLM, luego de factorizar las contribuciones del tiempo eucl\'ideo y de las $d-1$ coordenadas paralelas a las placas encontramos que el NdC en la diagonal est\'a dado por
\begin{multline}\label{FLM:cas.heat}
{K}(x,x,\beta)=\frac{1}{\textstyle{(4\pi\beta)^{d/2}}}\ e^{-\beta m^2}\,\int\limits_{\substack{x_1(0)=x_1}}^{x_1(1)=x_1}
\mathcal{D}x_1(\tau)
\;e^{\textstyle-\int^{1}_0 dt'\,\dot{x_1}^2(t')/\!4\beta}\\[0.2cm]
\times
e^{\textstyle-\int^{1}_0 dt'\beta\, \gamma\,\delta \left[x_1(t')+L/2\right]+\beta\, \gamma\,\delta \left[x_1(t')-L/2\right]}\ .
\end{multline}
Si tomamos como escala de referencia la distancia $L$ entre las placas para definir las dimensiones de las constantes $m$, $\beta$ y $\gamma$, podemos considerar el caso de acoplamiento peque\~no $L\gamma\ll 1$ y desarrollar consecuentemente  (\ref{FLM:cas.heat}):
\begin{align}\label{FLM:cas.heat.pequeno}
\begin{split}
{K}(x,x,\beta)&=\frac{1}{\textstyle{(4\pi\beta)^{d/2}}}\ e^{-\beta m^2}
\int\limits_{\substack{x_1(0)=x_1}}^{
x_1(1)=x_1}
 \mathcal{D}x_1(\tau)
\;e^{\textstyle-\int^{1}_0 dt'\,\dot{x_1}^2(t')/\!4\beta}\;
\\[0.1cm]
&\hspace{-1.6cm}\times\left\lbrace1-\beta\,\gamma\int^1_0 dt\, \delta\left[x_1(t)+L/2\right]-
\beta\,\gamma \int^1_0 dt\, \delta\left[x_1(t)-L/2\right]+\right.
\\[0.2cm]
&\hspace{0.cm}+\left.\mbox{}\beta^2\,\gamma^2 \int^1_0\int^1_0 ds\, dt\, \delta\left[x_1(s)+L/2\right]\,\delta\left[x_1(t)-L/2\right]+\ldots\right\rbrace\ .
\end{split}
\end{align}

En esta expresi\'on, el primer t\'ermino corresponde a la densidad de energ\'ia constante del espacio vac\'io. El segundo y el tercero, en cambio, contribuyen a la autoenerg\'ia de cada placa y ciertamente son independientes de la distancia $L$ entre ellas. Es por ello que la contribuci\'on principal a la energ\'ia de interacci\'on entre las l\'aminas viene dada por el cuarto t\'ermino y es orden $\gamma^2$. Llamando $K_L(x,x,\beta)$ a esta contribuci\'on, podemos seguir mutatis mutandis el m\'etodo de la secci\'on \ref{FLM.delta} para deshacernos de las deltas: 
\begin{multline}\label{FLM:cas.interaccion}
\hspace{-0.2cm}\int_{\mathbb{R}^{{d+1}}}dx\,K_{L}(x,x,\beta):=\frac{2\textstyle{T\,A_{d-1}}}{\textstyle{(4\pi\beta)^{d/2}}}\ e^{-\beta m^2}\,\beta^2\,\gamma^2
\,\int_{\mathbb{R}}dx_1\,\int^1_0\int^t_0 ds\, dt\,
\\[0.3cm]
\hspace{0.3cm}\times\,K_0(-L/2,x,\beta s)\times K_0(L/2,-L/2,\beta (t-s)) \times K_0(x,L/2,\beta (1-t))\ .\\[-0.4cm]
\end{multline}
Las amplitudes de transici\'on $K_0$ pertenecen a part\'iculas libres, pues el campo no interact\'ua m\'as que con las deltas. Su expresi\'on cerrada la hemos escrito  en \eqref{FLM:nucleolibre} y nos permite calcular
\begin{align}
\begin{split}
\int_{\mathbb{R}^{{d+1}}}dx\,K_{L}(x,x,\beta)&=\\
&\hspace{-3.5cm}=\frac{\textstyle{T\,A_{d-1}}}{\textstyle{(4\pi\beta)^{d/2}}}\ e^{-\beta m^2}2\beta^2\,\gamma^2\,\int_{\mathbb{R}}dx_1\\
&\hspace{-2.5cm}\times
\,
\int\limits^{1}_0\int\limits^{t}_0 ds\,dt\ \frac{\textstyle{e^{-(x_1+L/2)^2/\!4\beta s}}}{\textstyle{\sqrt{4\pi\beta\,s}}}\,\frac{\textstyle{e^{-L^2/\!4\beta (t-s)}}}{\textstyle{\sqrt{4\pi\beta\,(t-s)}}}\,\frac{\textstyle{e^{-(x_1-L/2)^2/\!4\beta (1-t)}}}{\textstyle{\sqrt{4\pi\beta\,(1-t)}}}
\\
&\hspace{-3.5cm}=\frac{\textstyle{T\,A_{d-1}}}{\textstyle{(4\,\pi\,\beta)^{d/2}}}\,\frac{\textstyle{\beta\,\gamma^2}}{\textstyle{4}}
\,e^{-\beta\,m^2}\,\text{erfc}(L/\sqrt{\beta})\ .
\end{split}
\end{align}
En este resultado, T representa la longitud infinita del intervalo de tiempo eucl\'ideo, $A_{d-1}$ es la superficie tambi\'en infinita de las l\'aminas y erfc($\cdot$) es la funci\'on error complementaria.

Teniendo en cuenta que la energ\'ia efectiva por unidad de \'area de las l\'aminas, de acuerdo a lo expuesto en la secci\'on \ref{DET.energiaeff}, est\'a definida utilizando la correcci\'on $\Gamma_{1-\text{\rm bucle}}$ de un bucle a la AE como 
$$E_{\rm eff}:=\Gamma_{1-\text{\rm bucle}}/T\,A_{d-1},$$ 
y recordando la expresi\'on (\ref{DET:efectivafinal}) para $\Gamma_{1-\text{\rm bucle}}$, la densidad de energ\'ia de interacci\'on entre las placas por unidad de \'area debida a las oscilaciones de vac\'io del campo escalar es
\begin{align}\label{FLM:cas.energia}
   \begin{split} E_{\rm eff} &=-\frac{1}{T\,A_{d-1}}\frac{1}{2}\int_0^\infty\frac{d\beta}{\beta}
    \int_{\mathbb{R}^{{d+1}}}dx\,K_{L}(x,x,\beta)\\
    &=-\frac{\textstyle{\gamma^2}}{\textstyle{4}}
\,\int_0^\infty d\beta\frac{\textstyle{1}}{\textstyle{(8\,\pi\,\beta)^{d/2}}}\,e^{-\beta\,m^2}\,\text{erfc}(L/\sqrt{\beta}).
\end{split}
\end{align}

En tanto, la presi\'on de Casimir $p$,  a saber la fuerza de Casimir por unidad de \'area paralela a las placas, puede determinarse como la derivada respecto a la distancia de separaci\'on $L$ entre ellas,
\begin{equation}\label{FLM:cas.press}
    p:=-\frac{dE_{\rm eff}}{dL}\ .
\end{equation}
Reuniendo las ecuaci\'ones (\ref{FLM:cas.interaccion}), (\ref{FLM:cas.energia}) y (\ref{FLM:cas.press}) arribamos a la siguiente expresi\'on para la primera contribuci\'on en $\gamma$ a la presi\'on de Casimir:
\begin{equation}\label{FLM:cas.pressweak}
p=-\frac{\gamma^2}{(4\pi)^{(d+1)/2}}\,(m/L)^{d/2-1/2}\,K_{d/2-1/2}(2mL)\ .
\end{equation}

Resulta de inter\'es analizar los casos l\'imite de (\ref{FLM:cas.pressweak}) como funci\'on de la variable adimensional $m\,L$. Consideremos primero la presi\'on de Casimir cuando las placas est\'an muy distanciadas o, an\'alogamente, tomemos el l\'imite $m\,L\gg1$ en (\ref{FLM:cas.pressweak}) para obtener el comportamiento 
\begin{equation}\label{FLM:cas.pressweak2}
    p\simeq-\frac{\gamma^2}{2^{d+2}\pi^{d/2}}\,\frac{m^{d/2-1}}{L^{d/2}}\,e^{-2mL}\ .
\end{equation}
Esta expresi\'on muestra que la presi\'on de Casimir est\'a suprimida, para grandes separaciones y campos muy masivos, por un factor exponencial que depende de la variable $m\,L$.

Por \'ultimo, estudiemos el l\'imite de masa peque\~na de la f\'ormula (\ref{FLM:cas.pressweak}). Dado $m\,L\ll1$, resulta que la presi\'on de Casimir puede aproximarse, de acuerdo a la dimensi\'on $d$ del espacio, por
\begin{equation}\label{FLM:cas.pressweak3}
    p\simeq
    \left\{
    \begin{array}{lcr}
    \frac{\textstyle\gamma^2}{\textstyle4\pi}\,\log{(Lm)}&&{\rm si\ }d=1\ ,\\\mbox{}\\
    -\frac{\textstyle\gamma^2}{\textstyle(4\pi)^{(d+1)/2}}
    \,\frac{\textstyle\Gamma((d-1)/2)}{\textstyle2}\,\frac{\textstyle1}{\textstyle L^{d-1}}&&{\rm si\ }d\geq 2\ .
    \end{array}
    \right.
\end{equation}
La primera l\'inea de la expresi\'on (\ref{FLM:cas.pressweak3}) es consistente con el hecho de que en 1+1 dimensiones (d=1) un campo sin masa genera una energ\'ia de Casimir que resulta no ser anal\'itica en $\gamma=0$  \parencite{Milton:2004vy}. Este resultado adem\'as coincide, para $d=3$, con el que se obtiene de tomar el l\'imite de acomplamiento d\'ebil en las expresiones de \textcite{Bordag:2005by}.

Pese a que podr\'iamos continuar exponiendo otras aplicaciones del FLM en el marco de la TCC usual, creemos que estos ejemplos otorgan una idea lo suficientemente acabada del mismo como para motivar su generalizaci\'on a TCC no conmutativas en los siguientes cap\'itulos.


\chapter[FLM en la  TCC NC]{Formalismo de l\'inea de mundo en la teor\'ia cu\'antica de campos no conmutativa}\label{NC}
\setlength\epigraphwidth{6.5cm}
\epigraph{\itshape Lo supieron los arduos alumnos de Pit\'agoras:\\
los astros y los hombres vuelven c\'iclicamente;\\
los \'atomos fatales repetir\'an la urgente\\
Afrodita de oro, los tebanos, las \'agoras.}{-- \textsc{J. L. Borges}, \textit{La noche c\'iclica.}}

En el decurso del cap\'itulo \ref{FLM}, hemos visto c\'omo los coeficientes de SDW del operador de fluctuaciones cu\'anticas, y en consecuencia las correcciones de un bucle a la AE, pueden ser obtenidos para una variedad de modelos en TCC. Dicho operador es, usualmente, un operador diferencial local; esa localidad se ve tambi\'en reflejada en  los coeficientes de SDW asociados, los cuales resultan ser integrales de invariantes locales y pueden ser expresados en t\'erminos de productos del potencial y sus derivadas. En este marco, la renormalizaci\'on de la teor\'ia puede ser llevada a cabo regularizando los factores divergentes que acompa\~nan a estos coeficientes, cfr. \eqref{DET:eff.regular}.

Al considerar las TCC NC, los resultados mencionados precedentemente sufren modificaciones. Cualitativamente, esto puede entenderse notando que las TCC NC a\~naden una idea que puede ser considerada el ingrediente principal de una teor\'ia de la gravitaci\'on cu\'antica: una escala de longitud m\'inima \parencite{Douglas:2001ba,Szabo:2001kg}. La presencia de esta escala m\'inima, a su vez, genera una estructura granular del espacio a peque\~nas distancias y da lugar a efectos no locales. Este es el motivo por el cual el operador de fluctuaciones cu\'anticas es, en general, un operador no local. Los coeficientes de SDW correspondientes a este tipo de operadores no locales, adem\'as de ser no locales, presentan algunas propiedades peculiares; un ej. es la mezcla UV-IR \parencite{Minwalla:1999px}, relacionada con la existencia de divergencias infrarrojas incluso en el caso de campos masivos. 

A lo largo de este cap\'itulo utilizaremos las t\'ecnicas del FLM para obtener una descripci\'on sistem\'atica de los coeficientes de SDW para este tipo de operadores no locales. Como veremos en la secci\'on \ref{NC.tccnc}, este tipo de operadores ser\'a  relevante a la hora de estudiar la cuantizaci\'on de campos escalares no conmutativos autointeractuantes en espacios deformados con un producto no conmutativo llamado Moyal. Posteriormente, en las secciones \ref{NC.intcaminofase} y \ref{NC.fg} veremos c\'omo la no localidad de los operadores sugiere naturalmente la implementaci\'on de integrales de camino en el espacio de fases. Estos resultados nos permitir\'an obtener una f\'ormula magistral en la secci\'on \ref{NC.FLM} para los coeficientes de SDW, f\'ormula con la que analizaremos diversos modelos: partiendo de un caso general con potenciales que multiplican en forma Moyal a derecha e izquierda, continuaremos con el estudio de otra geometr\'ia (el toro no conmutativo, secci\'on \ref{NC.toro}) y el 
modelo 
$\lambda\,\phi_{\star}^4$ 
en el espacio eucl\'ideo 
Moyal (secci\'on \ref{NC.phi4}). En este \'ultimo caso analizaremos como renormalizar la teor\'ia al orden de un bucle.

Con esta base seremos capaces de analizar el modelo de Grosse-Wulkenhaar en el cap\'itulo \ref{GW}. Este es un modelo $\lambda\,\phi_{\star}^4$ al que se le ha modificado el propagador mediante la adici\'on de un t\'ermino harm\'onico \parencite{Grosse:2003nw,Grosse:2004yu}. La importancia que ha tomado este modelo en los \'ultimos a\~nos reside en sus interesantes propiedades: es renormalizable a todo orden en teor\'ia de perturbaciones, el flujo de renormalizaci\'on posee un punto fijo en el que los par\'ametros son finitos y no nulos, y podr\'ia ser construido axiom\'aticamente. 

\section[TCC NC en el ET eucl\'ideo Moyal]{Teor\'ia cu\'antica de campos no conmutativa en el espaciotiempo eucl\'ideo Moyal}\label{NC.tccnc}
El cap\'itulo \ref{INTRO} intenta fundamentar el problema de la gravedad cu\'antica. \emph{Grosso modo}, lo podemos resumir mencionando que tanto la TCC como la teor\'ia de la relatividad general tienen una gran capacidad predictiva, cualitativa y cuantitativa, para los fen\'omenos en los que una de las dos puede ser dejada de lado. En contraste, para procesos que involucren ambas ocurren ciertas contradicciones te\'oricas que a\'un no han sido zanjadas. Como complicaci\'on adicional, pareciera ser 
que no hay medidas experimentales que correspondan a ese tipo de procesos.

En este marco surgen las teor\'ias no conmutativas, asi\'endose de uno de los puntos de discordancia entre la relatividad y la cu\'antica: mientras en la primera teor\'ia los puntos en el espacio pueden ser muestreados utilizando una prueba de masa nula, en la ulterior es necesario el uso de part\'iculas de energ\'ia (masa) infinita. Esto sugiere que en una teor\'ia de gravedad cu\'antica el espacio no podr\'ia ser cl\'asico. Una posibilidad a seguir es generalizar la relaci\'on de conmutaci\'on entre coordenadas y momentos, introduciendo una matriz $\Theta$ de no conmutatividad para los $d$ operadores de posici\'on $\hat{x}_{{\mu}}$; dicha matriz ser\'a antisim\'etrica, de dimensi\'on $d\times d$ y tendr\'a componentes $\Theta^{{\mu}{\nu}}\in\mathbb{R}$ tales que
\begin{equation}\label{NC:conmutador}
    [\hat{x}^{\mu},\hat{x}^{\nu}]=2i\,\Theta^{{\mu}{\nu}}\, .
\end{equation}
Como hemos mencionado en el cap\'itulo \ref{INTRO}, las constantes caracter\'isticas de la gravitaci\'on y la TCC, sugieren que $\Theta^{{\mu}{\nu}}\sim \ell_p$, donde la longitud de Planck toma el valor $\ell_p=\hbar/E_p c\approx10^{-35}\text{m}$.

Las implicancias de esta regla de conmutaci\'on son inmediatas: al igual que acaece en mec\'anica cu\'antica usual, donde es imposible encontrar autoestados simult\'aneos de los operadores conjugados de posici\'on y momento, en esta situaci\'on no existen estados cuya posici\'on pueda determinarse con certeza en todas sus componentes (si $\Theta\neq0$). M\'as a\'un, existir\'a una relaci\'on de incerteza de Heisenberg generalizada, la cual invoca la imagen de un espaciotiempo granulado en celdas de dimensi\'on igual a alg\'un par\'ametro caracter\'istico de $\Theta$. La geometr\'ia resultante, llamada ET eucl\'ideo Moyal, es un ejemplo de geometr\'ia no conmutativa. El caso general ha sido estudiado rigurosamente en la publicaci\'on de \textcite{Connes:1994,GraciaBondia:2001tr}, mientras que una motivaci\'on f\'isica puede encontrarse en el trabajo de \textcite{Landi:1997sh}; un punto crucial en ese contexto es que el conmutador \eqref{NC:conmutador} no tiene por qu\'e ser una matriz constante, sino que 
puede adquirir una dependencia en las coordenadas.

El lector avezado se estar\'a preguntando a esta altura como implentar estas ideas de no conmutatividad en una TCC; los campos devendr\'ian funciones de operadores en lo que parecer\'ia un intrincado camino hacia la obtenci\'on de observables. Un enfoque alternativo para el ET eucl\'ideo Moyal est\'a inspirado en el trabajo de \textcite{Weyl:1927}. La idea, que originalmente correspond\'ia a un espacio de fases bidimensional\footnote{En el trascurso de esta secci\'on veremos que el ET Moyal eucl\'ideo para dimensiones $d>2$ puede ser escrito como suma de espacios uni- y bidimensionales conmutantes entre s\'i.}, se puede 
resumir en la siguiente proposici\'on: el producto de operadores no es m\'as que un producto NC. 

En efecto, utilizando el mapeo $\Omega$ de Weyl, para cada funci\'on $f(x)$ es posible definir un operador $\hat{f}$ en la forma 
\begin{align}\label{NC:weyl.x1x2}
 \hat{f}:=\Omega(f):=\int d\tilde{p}\, \tilde{f}(p)\, e^{ip \hat{x}},
\end{align}
donde $\hat{x}$ es el vector cuyas componentes son los operadores de posici\'on y $\tilde{f}(p)$ es la transformada de Fourier de $f(x)$. Al momento de multiplicar dos operadores $\hat{f}$ y $\hat{g}$, la utilizaci\'on de \eqref{NC:weyl.x1x2} conduce a
\begin{align}
\begin{split}
 \hat{f}\hat{g}&=\int d\tilde{p}d\tilde{q}\, \tilde{f}(p)\, e^{ip\hat{x}}  \tilde{g}(q)\, e^{iq \hat{x}}\\
 &=\int d\tilde{p}d\tilde{q}\, \tilde{f}(p)\tilde{g}(q)\,e^{ip \Theta q}\,e^{i(p+q) \hat{x}}\\
 &=\int d\tilde{p}\left(\int d\tilde{q}\, \tilde{f}(p-q)\tilde{g}(q)\,e^{ip \Theta q}\right)\,e^{ip \hat{x}},
\end{split}
\end{align}
dado que para agrupar las exponenciales en la segunda l\'inea hay que tener en cuenta la relaci\'on de conmutaci\'on \eqref{NC:conmutador} y la f\'ormula de Baker-Campbell-Hausdorff. Por simple observaci\'on vemos que la funci\'on $ \Omega^{-1}(\hat{f}\hat{g})$ asociada al producto de operadores puede ser escrita en t\'erminos de un producto $\star$ no local:
\begin{align}\begin{split}
 \Omega^{-1}(\hat{f}\hat{g})&=\int d\tilde{p}d\tilde{q}\, \tilde{f}(p-q)\tilde{g}(q)\,e^{ip \Theta q}\,e^{ipx}\\
 &=: (f\star g)(x).
\end{split}\end{align}
A id\'entico resultado arribamos si implementamos el mapeo inverso al de Weyl, el mapeo de \textcite{Wigner:1932}; para $d=2$, la regla de conmutaci\'on $[\hat{x}_1,\hat{x}_2]=2\,i\theta$ y los autovectores $\lvert x_1\rangle$ de  $\hat{x}_1$ nos permiten escribir 
\begin{align}
 \Omega^{-1}(\hat{f}):=f(x_1,x_2)=2\int_{-\infty}^{\infty} dy\,e^{-ix_2y/\theta}\,\langle x_1+y\rvert \hat{f}\lvert x_1-y\rangle,
\end{align}

Una forma alternativa de escritura para el producto $\star$, tomando dos funciones escalares $\phi$ y $\psi$ que dependen de $x\in\mathbb{R}^d$, es
\begin{equation}\label{NC:moyal}
    (\phi\star \psi)(x)=e^{i\,\partial^{^\phi}\Theta\partial^{^\psi}}\,\phi(x)\psi(x)\,,
\end{equation}
donde $\partial^{\phi}$ y $\partial^{\psi}$ denotan el operador gradiente que act\'ua sobre $\phi$ y $\psi$ respectivamente. La expresi\'on $\partial^{^\phi}\Theta\partial^{^\psi}$ representa $\partial^{^\phi}_{{\mu}}\Theta^{{\mu}{\nu}}\partial^{^\psi}_{{\nu}}$: hemos omitido los \'indices de suma para facilitar la escritura, algo que haremos de aqu\'i en adelante salvo que pueda dar lugar a confusi\'on. Observando \eqref{NC:moyal} se vuelve patente que el producto $\star$ es una deformaci\'on del producto usual, en el sentido que para par\'ametros de no conmutatividad peque\~nos se recupera este \'ultimo. Aquellos interesados en las condiciones bajo las cuales la exponencial en la expresi\'on (\ref{NC:moyal}) est\'a bien definida pueden referirse a la publicaci\'on de  \textcite{Estrada:1989da}.

De entre las diversas propiedades del producto Moyal, resultan fundamentales la no conmutatividad y la asociatividad; ellas pueden demostrarse partiendo de la f\'ormula (\ref{NC:moyal}) y empleando la antisimetr\'ia de la matriz $\Theta$. Adem\'as, podemos verificar que, dotadas de este producto, las coordenadas poseen el conmutador
\begin{equation}\label{NC:conmutador2}
    [x^{\mu},x^{\nu}]_{\star}:=x^{\mu}\star x^{\nu}-x^{\nu}\star x^{\mu}=2i\,\Theta^{{\mu}{\nu}}\, ,
\end{equation}
en concordancia con \eqref{NC:conmutador}. Asimismo, la antisimetr\'ia de $\Theta$ garantiza que su rango sea par y, en el caso general, se pueda descomponer 
$$\mathbb{R}^d=\mathbb{R}^{c}\oplus\mathbb{R}^{2b}$$
 eligiendo coordenadas $x=(\bar{x},\hat{x})$ con $\bar{x}\in\mathbb{R}^{c}$ conmutantes y $\hat{x}\in\mathbb{R}^{2b}$ no conmutantes. Por ende, siempre es posible escribir 
\begin{equation}\label{NC:theta}
    \Theta=\mathbf{0}_c\oplus \Xi\,,
\end{equation}
donde $\mathbf{0}_c$ es la matriz nula en $\mathbb{R}^{c}$ y $\Xi$ es una matriz antisim\'etrica no degenerada en $\mathbb{R}^{2b}$. En adici\'on,  la matriz $\Xi$ puede expresarse como la suma directa de $b$ matrices de dimensi\'on $2\times 2$ de la forma
\begin{align}
 \begin{pmatrix}
\phantom{-}0 &\theta\\
-\theta & 0  
 \end{pmatrix}.
\end{align}
De este modo, podemos interpretar al ET no conmutativo $\mathbb{R}^{2b}$ como dividido en $b$ planos no conmutativos, los cuales poseen dimensi\'on dos, conmutan entre s\'i y quedan definidos por un par\'ametro real de no conmutatividad. Cuando estos par\'ametros no sean iguales todos entre s\'i, llamaremos $\theta_{\mu=1\,\cdots,d}$ a la \'unica entrada no nula de la $\mu$-\'esima fila de modo tal que, por ejemplo, el plano formado por las coordenadas $x_{1}$ y $x_{2}$ est\'a caracterizado por el par\'ametro $\theta_1=-\theta_{2}$.

Conviene, antes de estudiar un modelo de TCC NC, mencionar otras propiedades sobre el producto Moyal que ser\'an de ayuda:
\begin{itemize}
\item bajo el signo integral se puede intercambiar el producto Moyal por el producto usual, es decir
\begin{align}\label{NC:moyal.cuadrat.int}
 \int \phi(x)\star\psi(x)=\int \phi(x)\psi(x);
\end{align}
\item como consecuencia de la anterior proposici\'on, es c\'iclico bajo el signo integral; en otras palabras, dadas tres funciones $\phi_1(x)$, $\phi_2(x)$ y $\phi_3(x)$ vale
\begin{align}
 \int \phi_1(x)\star\phi_2(x)\star\phi_3(x)=\int \phi_3(x)\star\phi_1(x)\star\phi_2(x);
\end{align}
\item su transformada de Fourier
se puede escribir en la forma
\begin{equation}\label{NC:moyal-fourier}
    \widetilde{\phi\star \psi}(p)=\int d\tilde{q}\,e^{-i p\Theta q}\,\tilde{\phi}(p-q)\tilde{\psi}(q)\,.
\end{equation}
\end{itemize}

\subsection{Modelo \texorpdfstring{$\lambda\phi_{\star}^3$}{lambda phi**3} en el espaciotiempo eucl\'ideo Moyal}
 Por regla general, para construir un dado modelo en el ET eucl\'ideo Moyal tomaremos la acci\'on de un modelo conmutativo y reemplazaremos el producto usual por el Moyal. De acuerdo a lo expuesto en la secci\'on anterior, esto se corresponder\'ia con una deformaci\'on de la acci\'on original, la cual en un desarrollo a primer orden en los par\'ametros de no conmutatividad agregar\'ia t\'erminos proporcionales a $\Theta^{{\mu}{\nu}}$.

Para familiarizarnos con las TCC NC, tomemos un modelo simple, el de un campo escalar con una autointeracci\'on c\'ubica. Teniendo en cuenta que en los t\'erminos cuadr\'aticos de la acci\'on es irrelevante la inclusi\'on o no del producto Moyal, vid. \eqref{NC:moyal.cuadrat.int}, el lagrangiano de este modelo es
\begin{equation}\label{NC:lag.phi3}
    \mathcal{L}=\frac{1}{2}(\partial\varphi)^2+\frac{m^2}{2}\varphi^2+\frac{\lambda}{3!}\varphi^3_\star\,,
\end{equation}
donde hemos utilizado la notaci\'on $\varphi^3_\star:=\varphi\star\varphi\star\varphi$.

Tengamos presente que la contribuci\'on de un bucle a la AE est\'a dictada por \eqref{DET:efectiva.logdet} en funci\'on del operador de fluctuaciones cu\'anticas, el cual se computa como\footnote{La ecuaci\'on \eqref{DET:efectiva.logdet} posee asimismo un t\'ermino que corresponde al n\'ucleo de calor de una part\'icula libre, proveniente de la normalizaci\'on de la funcional generatriz $Z[J])$. Dado que es independiente del modelo, no ser\'a de inter\'es sino hasta el momento de la renormalizaci\'on.}  la variaci\'on segunda de la acci\'on con respecto al campo cu\'antico $\varphi$ evaluada en el campo cl\'asico $\phi$. En el caso conmutativo, i.e. $\Theta=0$, esta correcci\'on de un bucle es
\begin{equation}
   \Gamma_{1-\text{\rm bucle}}^{C}=\frac{1}{2}\log{\rm Det}\left\{-\partial^2+m^2+\lambda\,\phi(x)\right\}\, .
\end{equation}
Conforme a los teoremas enunciados en el cap\'itulo \ref{FE}, la traza del NdC de  este operador de Schr\"odinger regular, posee un desarrollo asint\'otico en potencias del tiempo propio de la forma de la ecuaci\'on (\ref{FE:heatkernel.desarrollo}).

Por otro lado, en el caso no conmutativo ($\Theta\neq0$), el c\'alculo del operador de fluctuaciones cu\'anticas es un poco m\'as delicado debido a la presencia de productos Moyal. Un an\'alisis detallado muestra que la contribuci\'on de un bucle $\Gamma_{1-\text{\rm bucle}}^{NC}$ a la AE correspondiente al lagrangiano (\ref{NC:lag.phi3}) es
\begin{equation}\label{NC:nc-operator}
    \Gamma_{1-\text{\rm bucle}}^{NC}=\frac{1}{2}\log{\rm
      Det}\left\{-\partial^2+m^2+\frac{\lambda}{2}\,L(\phi)+\frac{\lambda}{2}R(\phi)\right\}\, .
\end{equation}
A diferencia de lo que suced\'ia en el caso conmutativo, la no conmutatividad del producto da origen a dos t\'erminos diversos en el potencial: $L(\phi)$ es un operador cuya acci\'on sobre una funci\'on $\psi(x)$ est\'a definida como $L(\phi)\psi(x):=(\phi\star\psi)(x)$ mientras que $R(\phi)\psi(x):=(\psi\star\phi)(x)$. En otras palabras, $L(\phi(x))$ y $R(\phi(x))$ corresponden a multiplicaci\'on Moyal a izquierda o a derecha por $\phi$. Estos operadores no locales pueden tambi\'en ser escritos utilizando el llamado corrimiento de Bopp (Bopp's shift) para el producto Moyal:
\begin{align}\label{NC:boppshift}
\begin{split}
    L(\phi)\psi(x)=\phi(x+i\Theta \partial)\psi(x)\,,\\
    R(\phi)\psi(x)=\phi(x-i\Theta \partial)\psi(x)\, .
\end{split}
\end{align}

Como corolario de las expresiones (\ref{NC:lag.phi3}) y (\ref{NC:boppshift}), una posibilidad para regularizar la AE en el modelo $\lambda\phi_{\star}^3$ es estudiar la traza del NdC  del operador herm\'itico\footnote{La hermiticidad est\'a garantizada por la aparici\'on de los t\'erminos multiplicativos tanto a derecha como a izquierda.} y no local
\begin{equation}
    -\partial^2+m^2+\frac{\lambda}{2}\,\phi(x+i\Theta\partial)+\frac{\lambda}{2}\,\phi(x-i\Theta\partial)\,.
\end{equation}

En algunas ocasiones, como en los ejemplos que tratamos en los  Anexos \ref{NC.pot.central} y \ref{NC.disco}, podr\'ia resultar relativamente simple el estudio directo del espectro del operador de fluctuaciones cu\'anticas. En el caso general, nos encontraremos frente a potenciales $l_{1,2}(x)$ y $r_{1,2}(x)$ arbitrarios que forman parte de un operador en la forma
\begin{equation}\label{NC:operator}
-\partial^2+L(l_1(x))+R(r_1(x))+L(l_2(x))R(r_2(x)).
\end{equation}
Sus propiedades pueden ser radicalmente opuestas a las de los operadores con los que estamos habituados a trabajar, derivados de modelos conmutativos. Tambi\'en en los Anexos \ref{NC.pot.central} y \ref{NC.disco} se pueden observar algunas de estas curiosidades.  

Si la traza del NdC de estos peculiares operadores de fluctuaciones cu\'anticas poseen o no un desarrollo de alg\'un tipo que permita estudiar la AE y, en caso afirmativo, cu\'al es su forma, son preguntas que no podemos contestar utilizando los teoremas del cap\'itulo \ref{FE}. Una de las claves para obtener una respuesta, es notar que en la expresi\'on (\ref{NC:operator}), teniendo en mente el corrimiento de Bopp, la variable $x$ y el operador $\partial_{x}$ parecen estar en un pie de igualdad. Esto no suced\'ia en el caso conmutativo, en el cual el operador $\partial_{x}$, vinculado al operador momento en una primera cuantizaci\'on, s\'olo formaba parte  del t\'ermino cin\'etico. Es por este motivo que dedicaremos dos secciones al c\'alculo de IdC en el espacio de fases antes de intentar dar una respuesta a esta interrogante.

\section{Integrales de camino en el espacio de fases}\label{NC.intcaminofase}
Consideremos un operador $H$ que posee un potencial no local formado por $l(x)$ y $r(x)$, funciones de $x\in\mathbb{R}^d$ que multiplican en forma Moyal a izquierda y a derecha respectivamente:
\begin{equation}\label{NC:H}
    H=-\partial^2+L(l)\,R(r)\,.
\end{equation}
Notemos que esta expresi\'on no posee problemas de ordenamiento dado que, debido a la asociatividad del producto Moyal, $L(l)$ y $R(r)$ conmutan entre s\'i. Haciendo uso del corrimiento de Bopp podemos por lo tanto interpretar este operador como la representaci\'on, en el espacio de coordenadas, de un hamiltoniano que en el marco de la primera cuantizaci\'on ser\'ia
\begin{equation}\label{NC:H.cuantico}
    H=p^2+l(x-\Theta p)\,r(x+\Theta p)\,.
\end{equation}
Vale la pena mencionar que, en concordancia con la aseveraci\'on que sigue a la f\'ormula (\ref{NC:H}), los operadores $x_{\pm}^{\mu}=(x^{\mu}\pm\Theta^{{\mu}{\nu}}p_{\nu})$ conmutan entre s\'i y por consiguiente tampoco en la expresi\'on (\ref{NC:H.cuantico}) hay problemas de ordenamiento. 

A la hora de construir la representaci\'on de la amplitud de transici\'on asociada al operador (\ref{NC:H.cuantico}) en t\'erminos de IdC, es empero conveniente reescribir el hamiltoniano de acuerdo al orden de Weyl. Comencemos por definir este ordenamiento: se dice que un operador $A(x,p)$ en el espacio de fases est\'a ordenado de acuerdo a Weyl cuando est\'a escrito de tal manera que $A(x,p)=A_S(x,p)+\Delta A\equiv A_W(x,p)$, donde $A_S(x,p)$ involucra productos sim\'etricos en $x$ y $p$ y $\Delta A$ incluye todos los t\'erminos que resultan de realizar eventuales conmutaciones entre $x$ y $p$, necesarias para reordenar $A(x,p)$ en su forma sim\'etrica. Por ejemplo, el producto $xp  =(xp)_S +\frac12[x,p] = (xp)_S +\frac{i\hbar}{2}\equiv (xp)_W$, con $(xp)_S=\frac{1}{2}(xp+px)$. Un tratamiento preciso sobre este ordenamiento puede ser encontrado en los anexos B y C del libro de \textcite{Bastianelli:2006rx} dedicado a IdC.

En el caso que nos ata\~ne, por cuanto el potencial $l(x-\Theta p)\,r(x+\Theta p)$ mezcla coordenadas y momentos, la expresi\'on \eqref{NC:H.cuantico} no est\'a \emph{a priori} ordenada sim\'etricamente. No obstante, empleando el desarrollo de Taylor para expandir las funciones $l$ y $r$, se puede demostrar que el operador (\ref{NC:H.cuantico}) s\'olo involucra productos sim\'etricos en $x$ y $p$; esto es id\'entico a decir que los t\'erminos que involucran productos entre $x$ y $p$  pueden ser ordenados sim\'etricamente sin introducir t\'erminos adicionales provenientes de conmutadores. Analicemos por ejemplo el producto de las contribuciones lineales de los desarrollos de Taylor de $l$ y $r$, el cual puede ser reescrito como
\begin{align}
  \label{NC:weyl}
\begin{split}
x_-^{\mu} x_+^{\nu} &= (x^{\mu}
  x^{\nu})_S-\Theta^{{\mu}{\rho}}\Theta^{{\nu}{\sigma}} (p_{\rho} p_{{\sigma}})_S -\Theta^{{\mu}{\rho}} p_{\rho}
  x^{{\nu}}+\Theta^{{\nu}{\sigma}} x^{\mu} p_{\sigma}\\
 &= (x_-^{\mu} x_+^{\nu})_S -\frac12 \Theta^{{\mu}{\rho}} [p_{\rho}, x^{\nu}]+\frac{1}{2}
 \Theta^{{\nu}{\sigma}} [x^{\mu},p_{\sigma}]\\
& = (x_-^{\mu} x_+^{\nu})_S,
\end{split}
\end{align}
gracias a que la matriz $\Theta^{{\mu}{\nu}}$ es antisim\'etrica. No es dif\'icil convencerse de que esta propiedad de $\Theta^{{\mu}{\nu}}$, sumada a la relaci\'on de conmutaci\'on $[x_-^{\mu},x_+^{\nu}]=0$  y a la simetr\'ia de los coeficientes de los desarrollos de Taylor de $l$ y $r$, implica que las contribuciones del producto $l(x_-)\,r(x_+)$ son sim\'etricas: $[x_-^{\mu},x_+^{\nu}]=0$ permite reordenar un factor $x_+^{\mu}$ (\'o $x_-^{\nu}$ ) entre varios $x_-^{\nu}$ (respectivamente $x_+^{\mu}$), mientras que la simetr\'ia de los desarrollos de Taylor hace posible la simetrizaci\'on de los productos entre diversas variables $x_+^{\mu}$ (respectivamente $x_-^{\nu}$).

Una vez que nos hemos cerciorado de que el potencial est\'a ordenado en la forma de Weyl, podemos hacer uso de la regla del punto intermedio  \parencite{Bastianelli:2006rx,DeBoer:1995hv} y escribir para el NdC del operador (\ref{NC:H.cuantico}) la siguiente representaci\'on en t\'erminos de IdC:
\begin{multline}\label{NC:pathint}
\langle x+z\vert e^{-\beta H}\vert x \rangle
        = \int_{x(0)=x}^{x(t)=x+z}\mathcal{D}x(t)\mathcal{D}p(t)\,e^{-\int_0^{\beta}
          dt\,\left\{p^2(t)-ip(t)\dot{x}(t)\right\}} \\
\times e^{-\int_0^{\beta}
          dt\,l(x(t)-\Theta p(t))\,r(x(t)+\Theta p(t))}\,.
\end{multline}
En esta expresi\'on $\beta>0$ y $x(t)$, $p(t)$ representan trayectorias en el espacio de fases $\mathbb{R}^{2d}$. Es importante recalcar que la IdC debe realizarse sobre trayectorias $x(t)$ que satisfacen las condiciones de contorno $x(0)=x$ y $x(\beta)=x+z$, mientras que las trayectorias $p(t)$ son arbitrarias. 

La integral \eqref{NC:pathint} se puede trabajar siguiendo los mismos pasos que se suelen realizar en el caso conmutativo. En primera instancia, reemplazamos la integral en las trayectorias $x(t)$ por una integral sobre las perturbaciones $q(t):=x(t)-x_{cl}(t)$ alrededor del camino cl\'asico para una part\'icula libre, $x_{cl}(t)=z\,t/\beta+x$. Posteriormente, efectuamos un reescaleo de todas las variables con dimensi\'on, utilizando el tiempo propio $\beta$ para volverlas adimensionales: $t\rightarrow \beta t$, $q\rightarrow \sqrt{\beta}q$, $p\rightarrow
p/\sqrt{\beta}$. De esta manera (\ref{NC:pathint}) se convierte en
\begin{multline}{\label{NC:pathint2}}
        \langle x+z\vert e^{-\beta H}\vert x \rangle
        = \beta^{-d/2}\int_{q(0)=0}^{q(1)=0}\!\!\!\mathcal{D}q\mathcal{D}p\,e^{-\int_0^{1} dt\,\left\{p^2-ip\dot{q}\right\}}\\[0.1cm]
        \times\,
        e^{i\frac{z}{\sqrt{\beta}}\int_0^1dt\,p}e^{-\beta\int_0^{1}
          dt\,l(x+tz+\sqrt{\beta}q-\Theta p/\sqrt{\beta})
        \,r(x+tz+\sqrt{\beta}q+\Theta p/\sqrt{\beta})}\, .       
\end{multline}

Tambi\'en en forma semejante a lo realizado en el caso conmutativo, con el fin de facilitar la escritura de las pr\'oximas ecuaciones, conviene definir el valor medio de una funcional $f[q(t),p(t)]$  utilizando la medida gaussiana de la integral de camino en el espacio de fases y normalizando de manera tal que $\langle 1\rangle_{\!D}=1$:
\begin{equation}
    \left\langle
      f[q(t),p(t)]\right\rangle_{\!D}:=\frac{\int_{q(0)=0}^{q(1)=0}\mathcal{D}q\mathcal{D}p\,e^{-\int_0^{1}
      dt\,\left\{p^2-ip\dot{q}\right\}} f[q(t),p(t)]}{\int_{q(0)=0}^{q(1)=0}\mathcal{D}q\mathcal{D}p\,e^{-\int_0^{1}
      dt\,\left\{p^2-ip\dot{q}\right\}}}.
\end{equation}
A modo de ejemplo, la f\'ormula (\ref{NC:pathint2}) se puede escribir como 
\begin{multline}
        \langle x+z\vert e^{-\beta H}\vert x \rangle
        = \frac1{(4\pi\beta)^{d/2}}\left\langle
          e^{i\frac{z}{\sqrt{\beta}}\int_0^1dt\,p}\times\right.\\
          \left.\times e^{-\beta\int_0^{1}
            dt\,l(x+tz+\sqrt{\beta}q-\Theta p/\sqrt{\beta})
        \,r(x+tz+\sqrt{\beta}q+\Theta p/\sqrt{\beta})}\right\rangle_{\!D}\,.
\end{multline}

Llegado este punto podemos utilizar el desarrollo en serie de McLaurin de la exponencial del t\'ermino potencial, para luego interpretar que todos los potenciales han sido trasladados del punto $x$ en las cantidades adecuadas. Empleando el gradiente $\partial$ como generador de traslaciones obtenemos 
\begin{multline}{\label{NC:pathint3}}
        \langle x+z\vert e^{-\beta H}\vert x \rangle
        =\frac1{(4\pi\beta)^{d/2}}\sum_{n=0}^\infty \frac{(-\beta)^n}{n!}\int_0^1dt_1\ldots\int_0^1dt_n\times\\ \times\,
        \left\langle e^{i\frac{z}{\sqrt{\beta}}\int_0^1dt\,p}
        e^{\sum_{i=1}^n[t_iz+\sqrt{\beta}q(t_i)-\Theta p(t_i)/\sqrt{\beta}]\partial^l_i+
        [t_iz+\sqrt{\beta}q(t_i)+\Theta
        p(t_i)/\sqrt{\beta}]\partial^r_i}\right\rangle_{\!D}\times\\ \left.\times\,
        l(x_1)\ldots l(x_n)\,r(x_1)\ldots r(x_n)\right|_{x}\,,
\end{multline}
donde $\partial^l_i$ y $\partial^r_i$ corresponden a los gradientes de $l(x_i)$ y $r(x_i)$ respectivamente\footnote{En este caso, los sub\'indices $i$ no hacen referencia a componentes espaciotemporales sino al i-\'esimo punto $x_i\in\mathbb{R}^d$.}. Tal y como est\'a indicado, al final del c\'alculo todas las variables $x_j$ deben ser evaluadas en $x$. Si observamos detenidamente (\ref{NC:pathint3}), notamos que las variables $q(t)$ y $p(t)$ aparecen linealmente en la exponencial del valor medio y, consecuentemente, para hacer la reinterpretaci\'on
\begin{multline}{\label{NC:pathint4}}
        \langle x+z\vert e^{-\beta H}\vert x \rangle
        =\frac1{(4\pi\beta)^{d/2}}\sum_{n=0}^\infty \frac{(-\beta)^n}{n!}\int_0^1dt_1\ldots\int_0^1dt_n\times\\
        \left.\times\,
        e^{\sum_{i=1}^n t_iz(\partial^l_i+\partial^r_i)}\left\langle
        e^{\int_0^1 dt \left(p\, k_n +q\, j_n\right)}
        \right\rangle_{\!D}
        l(x_1)\ldots l(x_n)\,r(x_1)\ldots r(x_n)\right|_{x}
\end{multline}
basta con reconocer la forma apropiada de las fuentes $k_n(t)$ y $j_n(t)$:
\begin{align}\label{NC:fuentes}
\begin{split}
k_n(t)&=\frac{iz}{\sqrt{\beta}}+\frac{\Theta}{\sqrt{\beta}}\sum_{i=1}^{n}\delta(t-t_i)(\partial^l_i-\partial^r_i)\,,\\
j_n(t)&=\sqrt{\beta}\,\sum_{i=1}^{n}\delta(t-t_i)(\partial^l_i+\partial^r_i)\, .
\end{split}
\end{align}
A fin de cuentas, hemos reducido el c\'alculo de IdC de operadores no locales a la determinaci\'on de la funcional generatriz
$$Z_D[k,j]:=\left\langle e^{\int_0^1 dt \left(p\, k +q\, j \right)}\right\rangle_{\!D}$$
en el espacio de fases. La pr\'oxima secci\'on la dedicaremos al c\'alculo de esta FG para fuentes cualesquiera $k(t)$ y $j(t)$.

\section{La funcional generatriz en el espacio de fases}\label{NC.fg}
La FG en el espacio de fases, funcional a partir de la cual se pueden obtener los valores de expectaci\'on de potencias arbitrarias $\langle p^n q^m\rangle$, est\'a definida para fuentes arbitrarias $k(t)$ y $j(t)$ como
\begin{align}\label{NC:generatriz}
\begin{split}
Z_D[k,j]:
&=\frac{\int^{q(1)=0}_{q(0)=0}\mathcal{D}q\mathcal{D}p\,e^{-\int_0^{1} dt\,\left(p^2-ip\,\dot{q}\right)}
e^{\int_0^{1} dt\,(p\, k +q\, j)}}{\int^{q(1)=0}_{q(0)=0}\mathcal{D}q\mathcal{D}p\,e^{-\int_0^{1} dt\,\left(p^2-ip\,\dot{q}\right)}}\\[1mm]
&=\frac{\int^{q(1)=0}_{q(0)=0}\mathcal{D}P\ e^{-\frac{1}{2}\int_0^{1} dt\,P^t A P
+\int_0^1dt\,P^tK}}{\int^{q(1)=0}_{q(0)=0}\mathcal{D}P\ e^{-\int_0^{1} dt\,P^t A P}}.
\end{split}\end{align}
En la segunda l\'inea hemos introducido los vectores $P$ y $K$, y el operador $A$, definido sobre aquellos caminos que cumplen condiciones de contorno tipo Dirichlet en $q$, $q(0)=q(1)=0$:
\begin{align}
    \label{NC:propagator}
P:=\left(\begin{array}{c}p(t)\\q(t)\end{array}\right),\qquad
    K:=\left(\begin{array}{c}k(t)\\j(t)\end{array}\right),\qquad A:=
\begin{pmatrix}
2&-i\partial_t\\
i\partial_t&0
\end{pmatrix}.
\end{align}
De este modo se hace patente que el argumento de la exponencial es una funcional cuadr\'atica en los caminos $p$ y $q$, con la particularidad de que $A$ es un operador invertible teniendo en cuenta las condiciones de contorno mencionadas. Completando cuadrados y recordando la normalizaci\'on elegida para la FG, encontramos que 
\begin{equation}\label{NC:generatriz2}
Z_{ D}[k,j]=e^{\frac{1}{2}\int_0^1 dt\, K^t A^{-1} K}\,.
\end{equation}

Por ende, para determinar la FG, precisamos el n\'ucleo $A^{-1}(t,t')$, correspondiente al operador inverso de $A$. Su c\'alculo no ofrece mayores inconvenientes y arroja como resultado
\begin{equation}
A^{-1}(t,t')=
\begin{pmatrix}
\frac{1}{2}&\frac{i}{2}\left[h(t,t')+f(t,t')\right]\\
\frac{i}{2}\left[h(t,t')-f(t,t')\right]&2g(t,t')
\end{pmatrix}\,,
\end{equation}
vali\'endonos de las funciones auxiliares
\begin{align}
\begin{split}
h(t,t'):&=1-t-t'\,,\\
f(t,t'):&=t-t'-\epsilon(t-t')\,,\\
g(t,t'):&=t(1-t')H(t'-t)+t'(1-t)H(t-t')\,.
\end{split}
\end{align}
En estas definiciones hemos hecho uso de la funci\'on signo $\epsilon(\cdot)$, cuyo valor es $\pm 1$ si su argumento es respectivamente positivo o negativo, y la funci\'on $H(\cdot)$ de Heaviside. Notemos que $g(t,t')$ no es m\'as que la funci\'on de Green del operador $(-\partial_t^2)_D$ con condiciones de contorno  tipo Dirichlet.

Como un caso particular, podemos utilizar los valores de las fuentes (\ref{NC:fuentes}) para computar el valor medio que figura en la ec. (\ref{NC:pathint4}),
\begin{equation}\label{NC:generatrizfinal}
    \left\langle e^{\int_0^1 dt \left(p\, k_n +q\, j_n \right)}\right\rangle_{\!D}=
    e^{-\frac{z^2}{4\beta}+\frac{iz}{2\beta}\Theta\sum_{i=1}^n(\partial^l_i-\partial^r_i)}
    e^{\triangle_n}\,,
\end{equation}
en t\'erminos del operador $\triangle_n$ definido en la forma
\begin{align}
\begin{split}
 \label{NC:dn}
    \triangle_n:&=\sum_{i,j=1}^n\left[\beta
      g(t_i,t_j)(\partial^l_i+\partial^r_i)(\partial_j^l+\partial_j^r)-
      \frac{1}{4\beta}(\partial^l_i-\partial^r_i)\Theta^2(\partial_j^l-\partial_j^r)\right.
      \\&\hspace{3cm}\left.\mbox{}
      -\frac{i}{2}f(t_i,t_j)(\partial_i^l\Theta\partial_j^l-\partial_i^r\Theta\partial_j^r)
      -ih(t_i,t_j)\partial_i^l\Theta\partial_j^r\right]\,.
\end{split}
\end{align}
Por supuesto, podemos verificar que tomando $\Theta=0$ recobramos el resultado conmutativo; efectivamente, el \'unico t\'ermino de $\Delta_n$ que sobrevive en este l\'imite es el asociado a $g(t,t')$, la funci\'on de Green del operador $(-\partial_t^2)_D$.

\section{Formalismo de l\'inea de mundo en el espacio de fases}\label{NC.FLM}
En el decurso de las dos secciones precedentes, hemos establecido varios resultados referentes a IdC en el espacio de fases. Ellos nos ser\'an de gran utilidad para implementar el FLM en el espacio de fases, teniendo en cuenta que, seg\'un hemos visto en el cap\'itulo \ref{FLM}, este formalismo permite  la utilizaci\'on de IdC para el c\'alculo de NdC. De esta manera, intentaremos encontrar en esta secci\'on qu\'e tipos de desarrollo admite la traza del NdC de operadores no locales de la forma (\ref{NC:H.cuantico}). 

Iniciemos considerando la traza del NdC de un operador no local $H$ multiplicado en forma Moyal $\bar{\star}$ con una funci\'on regular $f$:
\begin{equation}\label{NC:tr}
    {\rm Tr}\left(f(x)\,\bar{\star}\, e^{-\beta H}\right)=\int_{\mathbb{R}^d}
    dx\,\left.\langle x+z\vert e^{-\beta H}\vert x \rangle\right|_{z=-i\bar{\Theta}\partial^f}f(x)\,.
\end{equation}
Por supuesto, seguimos la notaci\'on que ya hemos utilizado anteriormente, seg\'un la cual $\partial^{f}$ denota el gradiente que act\'ua s\'olo sobre la funci\'on regularizadora $f$, y la variable $z$ debe ser formalmente reemplazada por el operador $-i\bar{\Theta}\partial^f$. La demostraci\'on de \eqref{NC:tr} es inmediata si se introduce en la parte izquierda la descomposici\'on espectral de la unidad en t\'erminos de autoestados del operador posici\'on y se recuerda que el operador $\partial$ es el generador de las traslaciones:
$$e^{-i\bar\Theta \partial^{f}\partial} \langle x| =\langle x-i\bar\Theta \partial^{f} |\ .$$
Cabe aclarar que el producto $\bar{\star}$ se encuentra definido en t\'erminos de una nueva matriz antisim\'etrica $\bar{\Theta}$ que, en principio, es diferente de $\Theta$. Esto nos permitir\'a considerar luego los casos en los que la funci\'on regularizadora act\'ua multiplicando en forma conmutativa ($\bar{\Theta}=0$), Moyal a izquierda ($\bar{\Theta}=\Theta$) o Moyal a derecha 
($\bar{\Theta}=-\Theta$). 

El procedimiento a esta altura es seguramente evidente; para calcular \eqref{NC:tr} podemos utilizar la expresi\'on (\ref{NC:pathint4}) para la amplitud de transici\'on entre dos puntos arbitrarios, recordando que el valor medio involucrado es (\ref{NC:generatrizfinal}). De esta forma obtenemos la f\'ormula maestra
\begin{multline}{\label{NC:formulamaestra}}
        {\rm Tr}\left(f(x)\,\bar{\star}\, e^{-\beta H}\right)
        =\frac{1}{(4\pi\beta)^{d/2}}\sum_{n=0}^\infty (-\beta)^n
        \int_{\mathbb{R}^d}dx\,f(x)\,e^{\frac{1}{4\beta}\sum_{i,j=1}^n D_iD_j}
        \times\\
        \left.\times\,
        \int_0^1 dt_1\ldots\int_0^{t_{n-1}}dt_n
        \ e^{i\triangle^{NC}_n+\beta\triangle^{C}_n}
        l(x_1)\ldots l(x_n)\,r(x_1)\ldots r(x_n)
        \right|_{x}\,,
\end{multline}
donde, por motivos que ser\'an explicados a continuaci\'on, hemos definido los siguientes operadores diferenciales 
\begin{align}\label{NC:operadoresdif}
    \begin{split}
    \triangle^{C}_n:&=\sum_{i,j=1}^n g(t_i,t_j)(\partial^l_i+\partial^r_i)(\partial^l_j+\partial^r_j)\,,
    \\[1mm]
    D_i:&=\left(\Theta-\bar{\Theta}\right)\partial^l_i-\left(\Theta+\bar{\Theta}\right)\partial^r_i\,,
\end{split}
\displaybreak 
\\[1mm]
\nonumber \triangle^{NC}_n:&=
    \sum_{i< j=1}^n\Big\{\left[\partial^l_i\Theta\partial^l_j-\partial^r_i\Theta\partial^r_j\right]
     -(1-t_i-t_j)(\partial^l_i\Theta\partial^r_j-\partial^r_i\Theta\partial^l_j)
\\
\nonumber    
    \mbox{} &\hspace{0.cm}+ (t_i-t_j)\left[\partial^l_i(-\Theta+\bar{\Theta})\partial^l_j+\partial^r_i(\Theta+\bar{\Theta})
\partial^r_j+
\partial^l_i\bar\Theta\partial^r_j+\partial^r_i\bar\Theta\partial^l_j\right]\Big\}
    \\
\nonumber
    &\hspace{1.cm}-\sum_{i=1}^n (1-2t_i)\partial^l_i\Theta\partial^r_i\,.
\end{align}
Conviene notar que en la deducci\'on de (\ref{NC:formulamaestra}) hemos utilizado la simetr\'ia del integrando con respecto a permutaciones de las variables $t_i$ y hemos integrado por partes para reemplazar derivadas sobre la funci\'on regular $f$ por derivadas sobre los potenciales $l_i$ y $r_i$, es decir $\partial^f\rightarrow-\sum_{i=1}^n(\partial^l_i+\partial^r_i)$. Esta es la raz\'on por la cual el caso sin funci\'on regular puede ser entendido como $f\equiv1$ y adicionalmente, para hacer patente que las derivadas sobre $f$ se anulan, $\bar{\Theta}=0$.

Antes de analizar en detalle los diversos casos que de acuerdo al valor de $\bar{\Theta}$ engloba la expresi\'on (\ref{NC:formulamaestra}), conviene realizar algunos comentarios generales. Como veremos seguidamente, los coeficientes de SDW para el caso conmutativo est\'an enteramente determinados por la acci\'on del operador $\triangle^C_n$, dado que $\triangle_n^{NC}$ y $D_i$ se anulan para $\bar{\Theta}=\Theta=0$. Por otro lado, en el caso que $r(x)\equiv 1$ (o $l(x)\equiv 1$) y la funci\'on regularizadora multiplica en forma Moyal a izquierda (derecha, respectivamente), los operadores $D_i$ se anulan y el \'unico t\'ermino no nulo en $\triangle_n^{NC}$ es el primero $\partial^l_i\Theta\partial^l_j$ ($-\partial^r_i\Theta\partial^r_j$, respectivamente) que reemplaza todos los productos conmutativos por productos Moyal a izquierda (derecha, respectivamente). 

Por otro lado, mostraremos que el factor $e^{1/\beta\sum_{i,j} D_iD_j}$ es responsable de la presencia de coeficientes de SDW no locales en el sentido Moyal, los cuales a su vez corresponden a contribuciones de diagramas no planares y pueden conducir a la mezcla UV-IR. En lo sucesivo investigaremos la influencia de la elecci\'on de los valores de $\Theta$ y $\bar{\Theta}$ sobre (\ref{NC:formulamaestra}).

\subsection{Caso conmutativo}
Examinemos primeramente el caso conmutativo, ya que nos servir\'a como comparaci\'on en los restantes. Debemos por lo tanto fijar $\bar{\Theta}=\Theta=0$ en la f\'ormula maestra \eqref{NC:formulamaestra} para la traza del NdC. En adici\'on podemos elegir, por ej., $r(x)\equiv 1$. Como ya hemos mencionado, los operadores $D_i$ y $\triangle_n^{NC}$ se anulan, de forma que el NdC queda resumido en
\begin{multline}\label{NC:caso.conmu}
        {\rm Tr}\left(f(x)\,\cdot\, e^{-\beta H}\right)
        =\frac{1}{(4\pi\beta)^{d/2}}\sum_{n=0}^\infty (-\beta)^n
        \int_{\mathbb{R}^d}dx\,f(x)
        \times\\
        \left.\times\,
        \int_0^1 dt_1\int_0^{t_1}dt_2\ldots\int_0^{t_{n-1}}dt_n
        \ e^{\beta\sum_{i,j=1}^n g(t_i,t_j)\partial_i\partial_j}
        \ l(x_1)\ldots l(x_n)
        \right|_{x}\,.
\end{multline}
Esta f\'ormula muestra que los coeficientes de SDW pueden ser escritos como integrales del producto de la funci\'on regularizadora y del potencial y sus derivadas. 
En a\~nadidura, recalquemos que respetando el orden que inducen los \'indices de las variables $t_i$ y $x_j$ en (\ref{NC:caso.conmu}), el resultado es correcto incluso para aquellos casos en los que $l(x)$ es un potencial a valores matriciales. Ciertamente, para ello el producto de dos potenciales $l(x)$ debe ser entendido como el producto matricial.

\subsection{Caso no conmutativo con \texorpdfstring{$\bar{\Theta}=0$}{theta=0}}\label{NC.caso.smearingconmu}
En este par\'agrafo, trabajaremos  con $\Theta\neq0$, suponiendo adem\'as que la funci\'on regularizadora multiplica bajo el producto conmutativo ($\bar{\Theta}=0$). Bajo estas condiciones, la expresi\'on (\ref{NC:formulamaestra}) se reduce a
\begin{eqnarray}{\label{NC:caso.smearingconmu}}
        {\rm Tr}\left(f(x)\,\cdot\, e^{-\beta H}\right)
        =\frac{1}{(4\pi\beta)^{d/2}}\sum_{n=0}^\infty (-\beta)^n
        \int_{\mathbb{R}^d}dx\,f(x)\,
        \times\\
        \nonumber\left.\times\,
        \int_0^1 dt_1\int_0^{t_1}dt_2\ldots\int_0^{t_{n-1}}dt_n
        \ e^{\triangle_n}
        l(x_1)\ldots l(x_n)\,r(x_1)\ldots r(x_n)
        \right|_{x}\,.
\end{eqnarray}
Es posible demostrar que algunos de los coeficientes de SDW que se derivan de esta f\'ormula son no locales, incluso en el sentido Moyal \parencite{Vassilevich:2003yz}. Para este fin nos bastar\'a considerar un potencial que involucre s\'olo multiplicaci\'on Moyal a izquierda. En efecto, fijando $r(x)\equiv 1$, los primeros t\'erminos de la serie en (\ref{NC:caso.smearingconmu}) son
\begin{equation}\label{NC:caso.smearingconmu1}
        {\rm Tr}\left(f(x)\,\cdot\, e^{-\beta H}\right)
        =\frac{1}{(4\pi\beta)^{d/2}}
        \int_{\mathbb{R}^d}dx\,f(x)\left(1-\beta\,
        e^{-\frac{1}{4\beta}\partial^l\Theta^2\partial^l}
        l(x)+\ldots\right)\,.
\end{equation}
El primer t\'ermino coincide con el t\'ermino principal del caso conmutativo, el llamado t\'ermino de volumen. El segundo, por otro lado, puede ser reescrito en una forma adecuada para un desarrollo en potencias de $\beta$,
\begin{multline}\label{NC:caso.smearingconmu2}
        -\frac{\beta}{(4\pi\beta)^{d/2}}
        \int_{\mathbb{R}^c}\!d\bar{x}\int_{\mathbb{R}^{2b}}\!d\hat{x}\,f(\bar{x},\hat{x})\int_{\mathbb{R}^{2b}}
        d\hat{y}
        \frac{\beta^{b}}{\pi^{b}\ {\rm det}\,\Xi}
        \ e^{\beta(\hat{x}-\hat{y})\Xi^{-2}(\hat{x}-\hat{y})}
        \ l(\bar{x},\hat{y}),
\end{multline}
para lo cual ha sido necesario descomponer $\Theta$ en su parte singular y no singular seg\'un (\ref{NC:theta}). De esta expresi\'on se puede ver que la presencia de una funci\'on regularizadora multiplicando en la forma usual genera una contribuci\'on al coeficiente $a_{b+1}$ que es no local incluso en el sentido Moyal y toma la forma
\begin{align}
a_{b+1} =-\frac{1}{(4\pi)^{d/2}}\frac{1}{\pi^b\ {\rm det}\,\Xi}
        \int_{\mathbb{R}^c}d\bar{x}
        \left(\int_{\mathbb{R}^{2b}}d\hat{x}\,f(\bar{x},\hat{x})\int_{\mathbb{R}^{2b}}d\hat{y}\,l(\bar{x},\hat{y})\right)+\ldots\,.
\end{align}
Vale la pena resaltar que este efecto ha sido obtenido considerando $r(x)\equiv1$, es decir, no es una consecuencia de la mezcla de productos Moyal actuando a izquierda y a derecha en el potencial.

\subsection{Caso con todos los productos Moyal a izquierda (o a derecha).}
Procederemos ahora a ilustrar el caso en el que la traza del NdC involucra \'unicamente productos Moyal a izquierda ($r(x)\equiv 1$ y $\bar{\Theta}=\Theta$) o a derecha ($l(x)\equiv 1$ y $\bar{\Theta}=-\Theta$). Bajo cualquiera de estas  suposiciones los operadores $D_i$ se anulan, mientras que $\triangle_n^{NC}$ toma la forma
\begin{equation}\label{NC:dfolr}
    \triangle^{NC}_n:=
    \pm\sum_{i< j=1}^n\partial_i\Theta\partial_j\,,
\end{equation}
donde el signo superior (inferior) corresponde a la elecci\'on de los productos Moyal todos a izquierda (derecha) y las derivadas act\'uan consecuentemente sobre $l(x_i)$ ($r(x_i)$). La \'unica contribuci\'on no conmutativa a los coeficientes de SDW es, por esta raz\'on, el factor $e^{i\triangle_{n}^{NC}}$; a su vez, esta fase es exactamente igual a la que surgir\'ia en la multiplicaci\'on Moyal de $n$ funciones de acuerdo con la definici\'on (\ref{NC:moyal}). Teniendo en cuenta este hecho, la traza del NdC puede ser reescrita como
\begin{eqnarray}\label{NC:caso.nomix}
        {\rm Tr}\left(f(x)\star e^{-\beta H}\right)
        =\frac{1}{(4\pi\beta)^{d/2}}\sum_{n=0}^\infty (-\beta)^n
        \int_{\mathbb{R}^d}dx\,f(x)
        \\
        \nonumber\left.\times\,
        \int_0^1 dt_1\int_0^{t_1}dt_2\ldots\int_0^{t_{n-1}}dt_n
        \ e^{\beta\sum_{i,j=1}^n g(t_i,t_j)\partial_i\partial_j}
        \ l(x_1)\star \cdots \star l(x_n)
        \right|_{x}\,.
\end{eqnarray}

Concluimos entonces que los coeficientes de SDW pueden obtenerse en esta ocasi\'on a partir de los coeficientes del caso conmutativo para un potencial a valores matriciales, sencillamente reemplazando todos los productos matriciales (espacialmente locales) con productos Moyal a izquierda (derecha). A orden $\beta^3$ obtenemos el siguiente desarrollo:
\begin{align}
\begin{split}{\rm Tr}\left(f(x)\star e^{-\beta H}\right)
        &=\frac{1}{(4\pi\beta)^{d/2}} \int_{\mathbb{R}^d}dx\,f(x) 
\\
    &\hspace{-2.5cm}\times \Biggl[1-\beta l(x) +\beta^2 \left( \frac12 l_\star^2(x) -\frac16\partial^2 l(x)\right)+\beta^3 \Biggl( -\frac1{60}\partial^4 l(x)
\\
        &\hspace{-2.5cm}+\frac1{12} \Big(\partial^2l \star l(x) +l\star \partial^2 l(x) +\partial l\star \partial l(x)\Big)-\frac1{3!} l_\star^3(x)\Biggr) 
        +\cdots\Biggr].
 \end{split}
\end{align}

En resumen, siempre que haya productos Moyal s\'olo  a izquierda (o a derecha), los coeficientes de SDW son locales en el sentido Moyal \parencite{Vassilevich:2003yz,Gayral:2004ww}.

\subsection{Caso general}
La  opci\'on restante es considerar la presencia de ambas funciones $l(x)$ y $r(x)$ en el potencial del operador (\ref{NC:H.cuantico}), donde por simplicidad tomaremos $f(x)\equiv1$. La f\'ormula (\ref{NC:formulamaestra}) se reduce en este caso a
\begin{multline}\label{NC:caso.lrmix}
        {\rm Tr}\left(e^{-\beta H}\right)
        =\frac{1}{(4\pi\beta)^{d/2}}\sum_{n=0}^\infty (-\beta)^n
        \int_{\mathbb{R}^d}dx\,
        \\
        \left.\times\,
        \int_0^1 dt_1\int_0^{t_1}dt_2\ldots\int_0^{t_{n-1}}dt_n
        \ e^{\triangle_n}
        l(x_1)\ldots l(x_n)\,r(x_1)\ldots r(x_n)
        \right|_{x}\,,
\end{multline}
expresi\'on similar a la obtenida en la secci\'on \ref{NC.caso.smearingconmu} tanto en aspecto como en implicancias. Con la intenci\'on de fundamentar esta aseveraci\'on, analicemos (\ref{NC:caso.lrmix}) t\'ermino a t\'ermino. Para $n=0$, obtenemos la habitual contribuci\'on conmutativa de volumen. Si tomamos $n=1$,  el resultado est\'a dado por
\begin{eqnarray}\label{NC:caso.lrmixn=1}
        -\frac{\beta}{(4\pi\beta)^{d/2}}
        \int_{\mathbb{R}^d}dx\,
        r(x)e^{-\frac{1}{\beta}\partial^l\Theta^2\partial^l}l(x)\,,
\end{eqnarray}
donde la integraci\'on por partes nos ha permitido hacer el reemplazo $\partial^r\rightarrow-\partial^l$. Notemos que la expresi\'on (\ref{NC:caso.lrmixn=1}) puede ser obtenida de (\ref{NC:caso.smearingconmu1}) reemplazando $r(x)$ con $f(x)$. Como en esa ocasi\'on, es conveniente utilizar una expresi\'on de la que se pueda leer el desarrollo en potencias de $\beta$; utilizando siempre la notaci\'on introducida en la secci\'on \ref{NC.tccnc} para variables conmutantes y no conmutantes, obtenemos 
\begin{multline}\label{NC:caso.lrmix.prefinal}
       -\frac{1}{(4\pi\beta)^{d/2}}\frac{\beta^{b+1}}{(4\pi)^{b}\ {\rm det}\,\Xi}
        \\
        \times \int_{\mathbb{R}^c}d\bar{x}\int_{\mathbb{R}^{2b}}d\hat{x}\,r(\bar{x},\hat{x})\int_{\mathbb{R}^{2b}}d\hat{y}\ e^{\frac{\beta}{4}(\hat{x}-\hat{y})\Xi^{-2}(\hat{x}-\hat{y})}
        \ l(\bar{x},\hat{y}).
\end{multline}
Sus contribuciones a los coeficientes $a_{b+1}$ y $a_{b+2}$ son no locales incluso en el sentido Moyal, an\'alogas a las observadas en el par\'agrafo  \ref{NC.caso.smearingconmu}:
\begin{align}\label{NC:caso.lrmix.final}
 \begin{split} 
        a_{b+1}&= -\frac{1}{(4\pi)^{d/2+b}\ {\rm det}\,\Xi}
        \int_{\mathbb{R}^c}d\bar{x}
        \left.
        \left(\int_{\mathbb{R}^{2b}}d\hat{x}\,r(\bar{x},\hat{x})\int_{\mathbb{R}^{2b}}d\hat{y}\,l(\bar{x},\hat{y})\right)+\cdots\right.\,,\\
        a_{b+2}&=-\frac{(\Xi^{-2})_{{\mu}{\nu}}}{4(4\pi)^{d/2+b} {\rm det}\,\Xi}
        \int_{\mathbb{R}^c}d\bar{x}\left\{
        \int_{\mathbb{R}^{2b}}d\hat{x}\,\hat{x}^{\mu}\hat{x}^{\nu}r(\bar{x},\hat{x})
        \cdot
        \int_{\mathbb{R}^{2b}}d\hat{y}\,l(\bar{x},\hat{y})
        \right.\\ 
        &\hspace{3.5cm}+ \int_{\mathbb{R}^{2b}}d\hat{x}\,r(\bar{x},\hat{x})
        \cdot
        \int_{\mathbb{R}^{2b}}d\hat{y}\,\hat{y}^{\mu}\hat{y}^{\nu}l(\bar{x},\hat{y})
        \\
        &\hspace{3.7cm}-\left.2\int_{\mathbb{R}^{2b}}d\hat{x}\,\hat{x}^{\mu}r(\bar{x},\hat{x})
        \int_{\mathbb{R}^{2b}}d\hat{y}\,\hat{y}^{\nu}l(\bar{x},\hat{y})
        +\ldots
        \right\}\,.
 \end{split}
\end{align}
Estas contribuciones no locales a los coeficientes $a_{b+1}$ y $a_{b+2}$, lineales en el producto $r(\hat{x})\,l(\hat{y})$ y mencionadas en la obra de \textcite{Vassilevich:2005vk}, podr\'ian afectar la renormalizabilidad de una TCC cuyo operador de fluctuaciones cu\'anticas fuera $H$ en el caso\footnote{Por cierto, tomaremos un potencial particular y demostraremos en la secci\'on \ref{NC.phi4}, a partir de \eqref{NC:caso.lrmixn=1}, que estos inconvenientes y la llamada mezcla UV-IR son dos caras de la misma moneda \parencite{Gayral:2004cu}.} $d/2-b\geq1$.

A continuaci\'on, utilizaremos los resultados que hemos establecido en esta secci\'on para estudiar primero el NdC de un operador sobre el toro NC y luego el modelo de un campo escalar con una autointeracci\'on cu\'artica en el ET eucl\'ideo Moyal.

\section{Toro no conmutativo}\label{NC.toro}
A diferencia de todos los operadores con los que hemos trabajado hasta este momento, los cuales han estado definidos sobre el ET eucl\'ideo Moyal, en esta secci\'on consideraremos un operador de la forma (\ref{NC:H.cuantico}) sobre el toro no conmutativo  $T^d_{\star}$ de $d$ dimensiones, definido como en el trabajo de \textcite{Gayral:2006vd}. 
De acuerdo a la notaci\'on establecida anteriormente, consideraremos coordenadas $x=(\bar{x},\hat{x})$ sobre el toro $T^d_{\star}$ que pueden ser separadas en componentes conmutantes $\bar{x}\in T^c$ y no conmutantes $\hat{x}\in T^{2b}$, con $d=c+2b$. Adem\'as, tomaremos $0\leq x_{\mu} \leq L_{\mu}$.

Teniendo presente que el toro es isomorfo a una regi\'on de $\mathbb{R}^d$ a la cual se le imponen condiciones de periodicidad en los bordes, puede verificarse que el NdC $\langle y|e^{-\beta
  H}|x\rangle_{T^d_{\star}}$ del operador $H$ en el toro puede ser escrito como la suma de infinitas amplitudes de transici\'on 
$\langle y+t_k|e^{-\beta H}|x\rangle_{\mathbb{R}^d}$
 calculadas en todo el espacio $\mathbb{R}^d$,
\begin{equation}\label{NC:toro}
    \langle y|e^{-\beta H}|x\rangle_{T^d_{\star}}=\sum_{k\in \mathbb{Z}^d}\langle y+t_k|e^{-\beta H}|x\rangle_{\mathbb{R}^d},
\end{equation}
donde $t_k=(L_1 k_1,\ldots,L_d
k_d)$ y $k=(k_1,\ldots,k_d)\in\mathbb{Z}^d$.
Haciendo un ligero abuso de notaci\'on, hemos llamado $H$ a dos operadores diferentes: al de la parte izquierda, definido sobre el toro, y a su extensi\'on peri\'odica en $\mathbb{R}^d$, el de la derecha; de m\'as est\'a decir que los potenciales $l(x)$ y $r(x)$ tambi\'en deben ser extendidos peri\'odicamente.

De esta manera, hemos reducido el problema al ya resuelto de encontrar amplitudes de transici\'on en $\mathbb{R}^d$. Para encontrar un desarrollo de \eqref{NC:toro} en potencias del tiempo propio  basta unir nuevamente las expresiones obtenidas para la amplitud de transici\'on y la FG en teor\'ias NC, cfr. (\ref{NC:pathint4}) y (\ref{NC:generatrizfinal}), recordando la periodicidad del potencial. Con este m\'etodo obtenemos para la contribuci\'on de volumen, correspondiente a $n=0$ en (\ref{NC:pathint4}), el valor
\begin{align}\begin{split}\label{NC:toro.primer}
    \frac{1}{(4\pi\beta)^{d/2}}\,\prod_{i=1}^d L_i\sum_{k\in \mathbb{Z}}
    e^{-\frac{L_i^2}{4\beta} k^2}&\sim \frac{1}{(4\pi\beta)^{d/2}}\,L_1\ldots L_d\\
    &=\frac{V_{T^d_{\star}}}{(4\pi\beta)^{d/2}}\,.
\end{split}
\end{align}
Efectivamente, todos los t\'erminos con $k\neq0$  son exponencialmente decrecientes y no aportan al desarrollo asint\'otico para $\beta$ peque\~no. Este resultado coincide con el del caso conmutativo.

Por otro lado, la contribuci\'on al NdC del toro que se obtiene de tomar el t\'ermino con $n=1$ en (\ref{NC:pathint4}), utilizando la periodicidad tanto de $l(x)$ como de $r(x)$ para reemplazar $\partial^r$ por $-\partial^l$, es
\begin{equation}\label{NC:n1}
    -\frac{\beta}{(4\pi\beta)^{d/2}}\sum_{k\in \mathbb{Z}^d}\int_{T^d}dx\,r(x)
    e^{-\frac{1}{\beta}\left(-i\Theta\partial+\frac{1}{2}t_k\right)^2}
    l(x)\,.
\end{equation}
La convergencia de esta serie no ha podido ser establecida en general, debido a que $\left(-i\Theta\partial+\frac{1}{2}t_k\right)$ podr\'ia anularse o estar indefinidamente cerca de hacerlo. Para evitar este problema, ha sido analizado por \textcite{Gayral:2006vd} el caso en el que los elementos de la matriz $\Theta$ satisfacen cierta condici\'on Diofantina: bajo esta hip\'otesis la cantidad $\left\lvert-i\Theta\partial+\frac{1}{2}t_k\right\rvert$ supera siempre una cantidad no nula dada, ocasionando que \eqref{NC:n1} se comporte como una exponencial $e^{-\frac{a}{\beta}}$ con $a>0$ fijo y no contribuya al desarrollo asint\'otico.  

\section{Modelo \texorpdfstring{${\lambda\varphi^4_{\star}}$}{lambda phi**4}}\label{NC.phi4}
Pasemos ahora a considerar, luego de haber establecido los resultados matem\'aticos generales, un modelo sencillo que podr\'ia tener implicancias al analizar el modelo est\'andar.
Al igual que en el caso conmutativo, un caso simple de estudiar resulta ser el de un campo escalar y real $\varphi$ con una autointeracci\'on cu\'artica. Nos centraremos en esta secci\'on en el c\'alculo de los coeficientes de SDW que nos permitir\'an analizar las correcciones de un bucle al propagador de este campo definido sobre el espacio eucl\'ideo Moyal $\mathbb{R}^d_{\star}$. Su acci\'on se escribe como la generalizaci\'on $\star$ de la correspondiente acci\'on conmutativa,
\begin{equation}\label{NC:phi4.L}
    \mathcal{L}=\frac{1}{2}(\partial\varphi)^2+\frac{m^2}{2}\varphi^2+\frac{\lambda}{4!}\varphi^4_\star\,,
\end{equation}
donde por supuesto $\varphi_{\star}^4:=\varphi\star\varphi\star\varphi\star\varphi$. 

Recordemos que la contribuci\'on de un bucle $\Gamma_{1-\text{\rm bucle}}$ a la AE, como hemos explicado en la secci\'on \ref{DET.eff2}, puede ser obtenida en t\'erminos del operador que se obtiene al realizar la variaci\'on segunda de (\ref{NC:phi4.L}). Utilizando el m\'etodo del tiempo propio de Schwinger, $\Gamma_{1-\text{\rm bucle}}$ puede ser escrita como
\begin{align}\label{NC:phi4.effactapp}
\begin{split}    \Gamma_{1-\text{\rm bucle}} =&\frac{1}{2}\log{\rm Det}\left\{-\partial^2+m^2+\frac{\lambda}{3!}\,[L(\phi_\star^2)+R(\phi_\star^2)
    +L(\phi)R(\phi)]\right\}\\ 
    =&-\frac{1}{2}\int_{\Lambda^{-2}}^\infty \frac{d\beta}{\beta}\,e^{-\beta m^2}\,{\rm Tr}\,e^{-\beta
    \left\{-\partial^2+\frac{\lambda}{3!}\,\left[ L(\phi_\star^2)+R(\phi_\star^2)
    +L(\phi)R(\phi)\right] \right\}}\,.
\end{split}
\end{align}
Notemos que  para controlar posibles divergencias ultravioletas, hemos introducido un par\'ametro de corte $\Lambda$ como regularizador. Asimismo, tal y como suced\'ia en el modelo $\lambda\phi^3_{\star}$, cfr. \eqref{NC:nc-operator}, hay contribuciones al potencial que involucran productos Moyal a derecha y a izquierda.

El recorrido a partir de este momento  es seguramente claro: la traza contenida en (\ref{NC:phi4.effactapp}) puede ser reemplazada por la f\'ormula (\ref{NC:formulamaestra}). Dado que estamos interesados en estudiar el propagador, el cual se obtiene de los t\'erminos cuadr\'aticos en el campo de la AE, s\'olo ser\'a necesario tener en cuenta los t\'erminos de (\ref{NC:phi4.effactapp}) lineales en $\lambda$. Frente a este razonamiento, el problema se torna m\'as sencillo de lo que nos podr\'iamos haber imaginado en un primer momento, por cuanto podremos utilizar la f\'ormula (\ref{NC:formulamaestra}) para obtener las contribuciones de cada uno de los t\'erminos $L(\phi_\star^2)$, $R(\phi_\star^2)$ y $L(\phi)R(\phi)$  por separado. Reemplazando $f(x)\equiv 1$, $\bar{\Theta}=0$ y $r(x)	\equiv 1$ en el t\'ermino correspondiente a $n=1$ de (\ref{NC:formulamaestra}), obtenemos la contribuci\'on de $L(\phi_\star^2)$ a la AE. Semejantemente se puede obtener el aporte de $R(\phi_\star^2)$, con la diferencia que en 
esta ocasi\'on debe fijarse $l(x)\equiv 1$. Eventualmente, ambas 
contribuciones resultan iguales entre s\'i y su suma es
\begin{multline}\label{NC:phi4.propapp}
  \int_{\Lambda^{-2}}^\infty d\beta\,e^{-\beta m^2}\,\frac{1}{(4\pi \beta)^{d/2}}
  \int_{\mathbb{R}^d}dx\,\frac{\lambda}{3!}\phi_\star^2(x)=
\\[0.1cm]
  =  \frac{\lambda}{3!}\frac{m^{d-2}}{(2\pi)^{d/2}}\,\Gamma(1-d/2,m^2/\Lambda^2)\int_{\mathbb{R}^d}\phi^2\,,
\end{multline}
donde $\Gamma(\cdot,\cdot)$ es la funci\'on gama incompleta \parencite{A-S}. 

Si intent\'aramos deshacernos del par\'ametro de corte tomando el l\'imite $\Lambda\rightarrow\infty$ en \eqref{NC:phi4.propapp}, encontrar\'iamos que la expresi\'on diverger\'ia como $\log{\Lambda}$, para $d=2$, o como $\Lambda^{d-2}$, para $d>2$. Esto no deber\'ia sorprendernos, ya que estas contribuciones no dependen de los par\'ametros de no conmutatividad: corresponden a las denominadas contribuciones planares y son id\'enticos a las que se obtienen en el caso conmutativo. En conformidad con ello, la divergencia podr\'ia ser eliminada siguiendo el proceso usual de renormalizaci\'on, el cual consiste en la absorci\'on de la parte divergente en una redefinici\'on de la masa. 

La contribuci\'on restante, asociada al t\'ermino $L(\phi)R(\phi)$, introduce novedades con respecto al caso conmutativo. Reemplazando $f(x)\equiv 1$, $\bar{\Theta}\equiv 0$ y $\partial^l=-\partial^r$ en el t\'ermino correspondiente a $n=1$ de (\ref{NC:formulamaestra}) obtenemos
\begin{multline}\label{NC:phi4.propappmixed}
  \frac{1}{2}\int_{\Lambda^{-2}}^\infty d\beta\,e^{-\beta m^2}\,\frac{1}{(4\pi \beta)^{d/2}}
  \int_{\mathbb{R}^d}dx\,\frac{\lambda}{3!}\phi(x)e^{-\frac{1}{\beta}\partial\Theta^2\partial}\phi(x)=
\\[0.1cm]
  =\frac{\lambda}{12}(16\,\pi^3)^{-d/2}m^{d-2}\,\int d^c\bar{p}\,d^{2b}\hat{p}
  \ \tilde\phi^*(\bar{p},\hat{p})\tilde\phi(\bar{p},\hat{p})
  \cdot\Sigma_{NP}(\hat{p})
  \,,
\end{multline}
donde, por cuanto en el espacio transformado de Fourier la acci\'on del producto Moyal se simplifica notablemente, en la f\'ormula (\ref{NC:phi4.propappmixed}) hemos escrito la contribuci\'on a la AE en t\'erminos de la transformada de Fourier del campo $\phi(\tilde{p},\hat{p})$ y las variables $\bar{p}$ y $\hat{p}$, duales de Fourier de las coordenadas conmutantes $\bar{x}$ y no conmmutantes $\hat{x}$ introducidas en (\ref{NC:theta}). Asimismo, hemos definido 
\begin{equation}\label{NC:phi4.bess}
  \Sigma_{NP}(\hat{p}):=\int_0^\infty \frac{d\beta}{\beta^{d/2}}\,e^{-\beta-\frac{m^2}{\beta}|\Xi \hat{p}|^2}=
  2(m|\Xi \hat{p}|)^{1-d/2}K_{d/2-1}(2m|\Xi \hat{p}|)\,
\end{equation}
en t\'erminos de $K_{d/2-1}(\cdot)$ la funci\'on de Bessel modificada \parencite{A-S}. 

El punto clave a notar en \eqref{NC:phi4.propappmixed} es que, gracias al t\'ermino asociado a la matriz no degenerada $\Xi$, la integral en (\ref{NC:phi4.bess}) converge para $\beta\rightarrow0$. En otras palabras, la no conmutatividad regulariza en este caso las divergencias ultravioletas al orden de un bucle y el par\'ametro de corte $\Lambda$ puede ser removido sin inconvenientes. Como era de 
esperar, se puede demostrar que la expresi\'on (\ref{NC:phi4.propappmixed}) corresponde al resultado que arrojan los diagramas de un bucle no planares.

Continuando el an\'alisis de la expresi\'on (\ref{NC:phi4.propappmixed}), podemos escribir en forma precisa el desarrollo de $ \Sigma_{NP}(\hat{p})$ para peque\~nos momentos $\hat{p}$  como
\renewcommand{\arraystretch}{1.8}
\begin{align}\label{NC:phi4.desarrollosigma}
    \begin{split}
    \Sigma_{NP}(\hat{p})=\left\lbrace\begin{array}{ll}
\renewcommand{\arraystretch}{1.7}\!\!\!\!\!\begin{array}{l}
(d/2-2)!(m|\Xi\hat{p}|)^{2-d} \,(1+O(|\hat{p}|^2))\\ \hspace{1.5cm} +\frac{2(-1)^{d/2}}{(d/2-1)!}\log{(m|\Xi\hat{p}|)}\,\left(1+O(|\hat{p}|^2)\right)\,
\end{array}\!\!,& d>2\\
\renewcommand{\arraystretch}{1.5}
-2\left[\log{m|\Xi\hat{p}|+\gamma}\right]
    (1+O(|\hat{p}|^2))\,,& d=2,
    \end{array}\right.
    \end{split}
\renewcommand{\arraystretch}{1.}
\end{align}
donde $\gamma$ es la constante de Euler. El resultado de (\ref{NC:phi4.desarrollosigma}) evaluado en $d=4$ y $\Theta=\Xi$ ($c=0$) corresponde a la contribuci\'on a la AE calculada por \textcite{Minwalla:1999px} considerando diagramas no planares.

Por otro lado, la convergencia de la integral \eqref{NC:phi4.propappmixed} en los impulsos depende del n\'umero de coordenadas conmutativas. En efecto, $\Sigma_{NP}(\hat{p})$ es del orden de $|\hat{p}|^{-d+2}$ para impulsos peque\~nos si suponemos que $d$ es mayor a dos; ergo, siempre que no requiramos a los campos un comportamiento especial en el origen, el integrando crece como $|\hat{p}|^{1-c}$ para impulsos $\hat{p}$ peque\~nos, donde $c=d-2b$ es el n\'umero de coordenadas conmutativas. Si este n\'umero $c$ iguala o excede a dos, esta contribuci\'on de un bucle a la AE se vuelve divergente IR. 

An\'alogas conclusiones se pueden obtener a partir de la expresi\'on \eqref{NC:caso.lrmix.prefinal}, cuyo t\'ermino principal en un desarrollo en potencias de $\beta$ es
\begin{equation}\label{NC:phi4.mix}
        -\frac{\lambda}{3!\,(4\pi\beta)^{d/2}}\cdot\frac{\beta^{b+1}}{(4\pi)^b\ {\rm det}\,\Xi}
        \int_{\mathbb{R}^c}d\bar{x}
        \left(\int_{\mathbb{R}^{2b}}d\hat{x}\,\phi(\bar{x},\hat{x})\right)
        \left(\int_{\mathbb{R}^{2b}}d\hat{y}\,\phi(\bar{x},\hat{y})\right)\,.
\end{equation}
Si introduj\'esemos este resultado en la \'ultima l\'inea de la f\'ormula (\ref{NC:phi4.effactapp}), el integrando se comportar\'ia como $\beta^{-c/2}$ para peque\~nos $\beta$ y la integral diverger\'ia al tomar el l\'imite $\Lambda\rightarrow\infty$ en  caso de que $c$ fuera mayor o igual a dos. Esta divergencia UV es evidentemente no local, visto que no lo es siquiera en sentido Moyal, y por ende no puede ser eliminada con un t\'ermino local en el lagrangiano.

Ergo, resta establecer qu\'e valores pueden ser tomados por $c$, el n\'umero de coordenadas conmutativas, en una teor\'ia con significado f\'isico. Para ello, podemos seguir el siguiente razonamiento: es ya sabido que la matriz $S$ de teor\'ias con tiempos no conmutativos resulta ser no unitaria\  \parencite{Gomis:2000zz,Seiberg:2000gc,Chaichian:2000ia}. Si proponemos entonces esta teor\'ia en un ET de dimensi\'on par, digamos $d=4$, el principio de unitariedad de la matriz $S$, sumado a la antisimetr\'ia de la matriz de no conmutatividad $\Theta$, impone que al menos dos coordenadas deben ser conmutativas ($c\geq2$). Dando esto por v\'alido, tal y como hemos visto, algunos diagramas no planares generan divergencias que no pueden ser removidas a trav\'es de la redefinici\'on de los par\'ametros de la teor\'ia \parencite{Gayral:2004cu}; en el mejor de los casos, habr\'ia que agregar t\'erminos que tuvieran la forma no local de aquellos divergentes, y aguardar la renormalizabilidad de 
la 
teor\'ia resultante.

Finalmente, para un valor gen\'erico $c$ el tipo de divergencias IR encontrado en (\ref{NC:phi4.desarrollosigma}) aparece tarde o temprano al considerar diagramas con un n\'umero creciente de bucles. Puesto que hemos eliminado las divergencias UV  a costas de encontrarnos con otras IR, el problema ha sido bautizado en la literatura \parencite{Minwalla:1999px} como mezcla ultravioleta-infrarroja (UV-IR). 
Este resultado  muestra que para ciertas teor\'ias NC, la integraci\'on de variables internas de momento en el c\'alculo de diagramas de Feynamn puede generar divergencias para valores peque\~nos de las variables externas de momento, incluso para campos masivos. Se foment\'o as\'i la creencia en la comunidad de que la renormalizabilidad de todas estas teor\'ias estar\'ia comprometida. Sin embargo, el modelo de Grosse-Wulkenhaar, al cual nos abocaremos en el pr\'oximo cap\'itulo, ech\'o por tierra estas ideas. 

\begin{subappendices}
  \renewcommand{\theequation}{\Roman{chapter}.\Alph{section}.\arabic{equation}}

\section[TCC NC con potenciales centrales]{Teor\'ia cu\'antica de campos no conmutativa  con potenciales centrales}\label{NC.pot.central}
En este anexo estudiaremos el operador de fluctuaciones cu\'anticas para un campo escalar real $\varphi(t,x)$, definido sobre un espacio descripto por las variables $x_0\in\mathbb{R}$ y $x=(x_1,x_2)\in\mathbb{R}^2$, el tiempo Minkowskiano y las coordenadas en el plano Moyal respectivamente, que interact\'ua con un potencial central de fondo\footnote{Utilizaremos \'indices latinos para referirnos a las componentes espaciales.} $V(r^2=x_ix^i)$. En este caso no haremos uso de los resultados obtenidos en el trascurso del cap\'itulo haciendo uso del FLM; la intenci\'on es mostrar algunas de las particularidades que poseen los operadores no locales con los que se trabaja en TCC NC. La acci\'on que consideraremos es
\begin{equation}\label{NC:central.action}
    S[\varphi]=\frac12\int_{\mathbb{R}\times\mathbb{R}^2}dx\,\left\{(\partial_0 \varphi)^2-(\partial_i \varphi)^2-m^2\,\varphi^2-V\star\varphi\star\varphi\right\}\,.
\end{equation} 
 
Como ya hemos explicado, el producto Moyal permite implementar la no conmutatividad de los operadores coordenada en el espacio usual de funciones de las coordenadas. Asimismo, la propiedad c\'iclica del producto $\star$ implica que es irrelevante la posici\'on de $V$ respecto a $\varphi$ en el t\'ermino de potencial de \eqref{NC:central.action}, evitando posibles ambig\"uedades en su definici\'on. 

De acuerdo a los resultados perturbativos establecidos en \eqref{DET:efectiva.logdet}, la AE correspondiente a \eqref{NC:central.action}, al orden de un bucle,  es 
 \begin{align}
  \Gamma[\phi]=S[\phi]+\frac{1}{2}\log\text{Det}\left\lbrace\delta^2_{\phi}S\right\rbrace.
 \end{align}
 El operador de fluctuaciones cu\'anticas $\delta^2_{\phi}S$ est\'a definido como la variaci\'on segunda de la acci\'on evaluada en el campo medio $\phi$; tal y como suced\'ia en los dem\'as ejemplos NC, la variaci\'on del t\'ermino de potencial debe ser llevada a cabo cuidadosamente por la presencia del producto $\star$ y genera m\'as de una contribuci\'on. En concreto,
 \begin{equation}\label{NC:central.olqf}
    \delta^2_{\phi}S:=-\partial_0^2-m^2-A\,,
\end{equation}
donde por practicidad hemos introducido el operador A, relacionado \'unicamente a las variables espaciales y poseedor de un espectro que determinaremos en breve:
\begin{equation}\label{NC:central.opa}
    A:=-\partial^2_i+\frac12\, V(x_i^+x_i^+)+\frac12\, V(x_i^-x_i^-)\, .
\end{equation}
 Los operadores $x^{\pm}_i=(x_i\mp i\Theta^{ij}\partial_j)$ en \eqref{NC:central.opa}, resultantes del corrimiento de Bopp, son an\'alogos a los introducidos en las l\'ineas posteriores a la expresi\'on \eqref{NC:H.cuantico} en el espacio de fases.

Para obtener provecho de la dependencia funcional del potencial, conviene introducir los operadores no herm\'iticos
\begin{equation}\label{NC:central.op.escalera}
    a_{\pm}:=\frac{1}{2\sqrt{\theta}}(x^\pm_1 \mp i x^\pm_2)\,,
\end{equation}
los cuales satisfacen un \'algebra de operadores de creaci\'on y destrucci\'on. Efectivamente, sus \'unicos conmutadores no nulos son
\begin{equation}
    [a_+,a^\dagger_+]=[a_-,a^\dagger_-]=1\,.
\end{equation}
Los beneficios de trabajar con ellos resultan evidentes al introducir los operadores n\'umero $N_{\pm}:=a^\dagger_\pm a_\pm$, puesto que son herm\'iticos y en su base de autovectores los t\'erminos de potencial son diagonales. Por fortuna, el Laplaciano tambi\'en puede ser escrito como combinaci\'on cuadr\'atica de los operadores $a^{\pm}$, resultando
\begin{multline}\label{NC:central.opan}
    \theta\, A=N_++N_-+1-a^\dagger_+a^\dagger_--a_+a_-+\\
    +\frac\theta2\left\{ V(2\theta(2N_++1))+V(2\theta(2N_-+1))\right\}\,.
    \end{multline}

Por otra parte, en t\'erminos de estos operadores el operador $L$ de momento angular se puede escribir como
\begin{equation}
    \label{L}L:=-i\epsilon_{ij}x_i\partial_j=N_+-N_-\,.
\end{equation}
Conviene notar que $L$ es tambi\'en el generador de las rotaciones en las nuevas coordenadas $x^{\pm}$, a saber, $[L,x^\pm_i]=i\epsilon_{ij}\,x^{\pm}_{j}$. Esto sugiere la descomposici\'on del espacio de funciones de cuadrado integrable en el plano como una suma de espacios de Fock $\mathcal{F}^{\pm l}$ con momento angular definido\footnote{De aqu\'i en adelante tomaremos $l\in\mathbb{Z}^{+}$.}
$$L_2(\mathbb{R}^2)=\mathcal{F}^{0}\oplus_{l=1}^\infty\mathcal{F}^{l}\oplus_{m=1}^\infty\mathcal{F}^{-m}.$$ 
Como es habitual, las funciones normalizadas que conforman la base en cada uno de los espacios de Fock $\mathcal{F}^{\pm l}$ se pueden construir en la forma 
\begin{equation}\label{NC:central.basis}
    \phi^{\pm l}_{n}(x)=\frac{1}{\sqrt{(n+l)!\,n!}}\,(a^\dagger_{\pm})^{n+l}(a^\dagger_{\mp})^{n}\phi^0_0(x)\,,
\end{equation}
donde $\phi^0_0(x)$ est\'a definido por las ecuaciones $a_+\,\phi^0_0(x)=a_-\,\phi^0_0(x)=0$. Por supuesto, \'estas resultan ser autofunciones de los operadores n\'umero $N_{\pm}$ con autovalores $n+ l$ y $n$. Empleando la definici\'on \eqref{NC:central.op.escalera} en la expresi\'on \eqref{NC:central.basis} encontramos luego de algunas manipulaciones algebraicas las autofunciones normalizadas
\begin{align}\label{NC:central.basisex}
    \phi^{\pm l}_n(x)=\frac{(-1)^n}{\sqrt{\pi\theta^{l+1}}}\frac{\sqrt{n!}}{\sqrt{(n+l)!}}
    \ r^{l}e^{-\frac{r^2}{2\theta}}L^{l}_n(r^2/\theta)\ e^{\pm il\varphi}\,,
\end{align}
donde $L^{l}_n(\cdot)$ son los polinomios de Laguerre generalizados y $r,\varphi$ son las coordenadas polares de $x\in\mathbb{R}^2$.

La dificultad en la resoluci\'on del problema de autovalores para $A$ depender\'a de la forma expl\'icita del potencial $V$. En el Anexo \ref{NC.disco}, obtendremos el espectro del operador de fluctuaciones cu\'anticas para un potencial central espec\'ifico, en lo que llamaremos el problema del disco NC.

\section{El disco no conmutativo}\label{NC.disco}
Uno de los principales inconvenientes para construir una TCC sobre variedades con borde es la interpretaci\'on misma del t\'ermino borde: en una geometr\'ia granulada como el plano NC, referirnos a un borde como un conjunto de puntos no parece ser la g\'enesis adecuada.  Una opci\'on es asirse del m\'etodo del potencial confinante utilizado en teor\'ias conmutativas, el cual se puede generalizar al caso NC mediante el producto Moyal $\star$. 

A modo de ejemplo, para definir el disco NC en $\mathbb{R}^2$ utilizamos un campo escalar cuya acci\'on es de la forma \eqref{NC:central.action} con un potencial 
\begin{equation}\label{NC:central.theta}
    V(r^2):=\frac{2\Lambda}{\theta}\,\Theta(r^2-R^2)\,,
\end{equation}
donde $\Theta(\cdot)$ es la funci\'on escal\'on definida como uno si su argumento es mayor o igual a cero y se anula en caso contrario. Este potencial de fondo representa una barrera cil\'indrica de radio $R$ y altura $2\Lambda$ (en unidades de $\theta^{-1}$); en el caso conmutativo y en el l\'imite $\Lambda\rightarrow\infty$ es tal que confina al campo a vivir en el disco de radio $R$. Dicho l\'imite, tomado en el caso no conmutativo, es nuestro punto de partida para la definici\'on del disco no conmutativo.

Reemplazando \eqref{NC:central.theta} en  \eqref{NC:central.opan} encontramos la siguiente expresi\'on para $A^\Lambda_N$, la parte espacial del operador de fluctuaciones cu\'anticas:
\begin{align}\label{NC:central.opan2}
\begin{split}    \theta\, A^\Lambda_N=&N_++N_-+1-a^\dagger_+a^\dagger_--a_+a_-
    \\
    &\hspace{1.5cm}+\Lambda\,\left\{ \Theta(N_+-N)+\Theta(N_--N)\right\}\,,
    \end{split}
\end{align}
donde $N$ es el menor entero mayor o igual a $R^2/4\theta-1/2$ o, haciendo uso de la funci\'on techo $\lceil\cdot\rceil$,
\begin{equation}
N:=\lceil R^2/4\theta-1/2\rceil\, .
\end{equation}

Explotando la simetr\'ia rotacional del operador $A^\Lambda_N$, podemos determinar su espectro en cada uno de los subespacios de Fock $\mathcal{F}^{\pm l}$; con este objetivo, proponemos para las autofunciones $\psi^{\pm l}_\lambda(x)$ de $A^\Lambda_N$
con autovalor $\frac{\lambda}{\theta}$ un desarrollo en t\'erminos de las funciones \eqref{NC:central.basisex} con momento angular $l$ definido,
\begin{equation}\label{NC:central.expansion}
    \psi^{\pm l}_\lambda(x)=\sum_{n=0}^\infty c^{l}_n(\lambda)\ \phi^{\pm l}_n(x)\,.
\end{equation}
Empleando esta serie la ecuaci\'on de autovalores para $A^{\Lambda}_N$
se convierte en una relaci\'on de recurrencia entre los coeficientes $c^{l}_n(\lambda)$. No reviste importancia el conocer la forma precisa de la relaci\'on de recurrencia; intuitivamente se puede comprender de \eqref{NC:central.opan2} que para $\Lambda\rightarrow \infty$ las soluciones son posibles s\'olo si los argumentos de las funciones Heaviside son negativos o lo que es lo mismo, recordando c\'omo act\'uan $N_{\pm}$ sobre las funciones $\phi^{\pm l}_n(x)$, si $l<N$ y las componentes $c^{l}_n(\lambda)$ se anulan para $n\geq N-l$. El resultado exacto es \parencite{Falomir:2013vaa}
\begin{align}\label{NC:central.cl}
    c^l_n(\lambda)=\frac{\sqrt{n!}}{\sqrt{(n+l)!}}\,L^l_n(\lambda),\qquad \text{\rm para }l<N \text{\rm\ y\ }n<N-l\,.
\end{align}
Visto que sobre las componentes $c^l_n$ pesa la relaci\'on de recurrencia y deben ser nulas si $n\geq N-l$, necesariamente debe cumplirse
\begin{equation}\label{NC:central.spec}
    L^l_{N-l}(\lambda)=0\,.
\end{equation}
Es consabido que el polinomio $L^l_{N-l}$ posee $N-l$ ra\'ices $\lambda^l_k$ \parencite{A-S}; ello implica que el operador $A^{\infty}_N=\lim_{\Lambda\rightarrow\infty}A^{\Lambda}_N$ est\'a definido en un espacio de Hilbert de dimensi\'on $N^2$. Desde el punto de vista semicl\'asico, esto corresponde a la existencia de una celda de volumen m\'inimo $(4\pi\theta)^2$ fijado por la relaci\'on de conmutaci\'on  $[x^\pm_i,x^\pm_j]=\mp 2i\theta\,\epsilon_{ij}$. Efectivamente, al estar limitado por las condiciones $x_{\pm}^2\leq R^2$, el volumen accesible en el espacio de fases cl\'asico es $(\pi R^2)^2$. El n\'umero de estados posibles se obtiene entonces como el n\'umero de celdas m\'inimas que caben en dicho volumen, esto es $(R^2/4\theta)^2\sim N^2$ en el l\'imite semicl\'asico $\theta\ll R^2$. La existencia de un n\'umero finito de estados es un hecho que hab\'ia sido ya notado para el caso de la esfera difusa \parencite{Madore:1991bw} y el disco difuso \parencite{Lizzi:2005zx,Lizzi:2006bu}. 

\begin{figure}[ht]
\centering
\begin{minipage}{.5\textwidth}
\centering
\includegraphics[width=0.8\textwidth,height=0.8\textwidth]{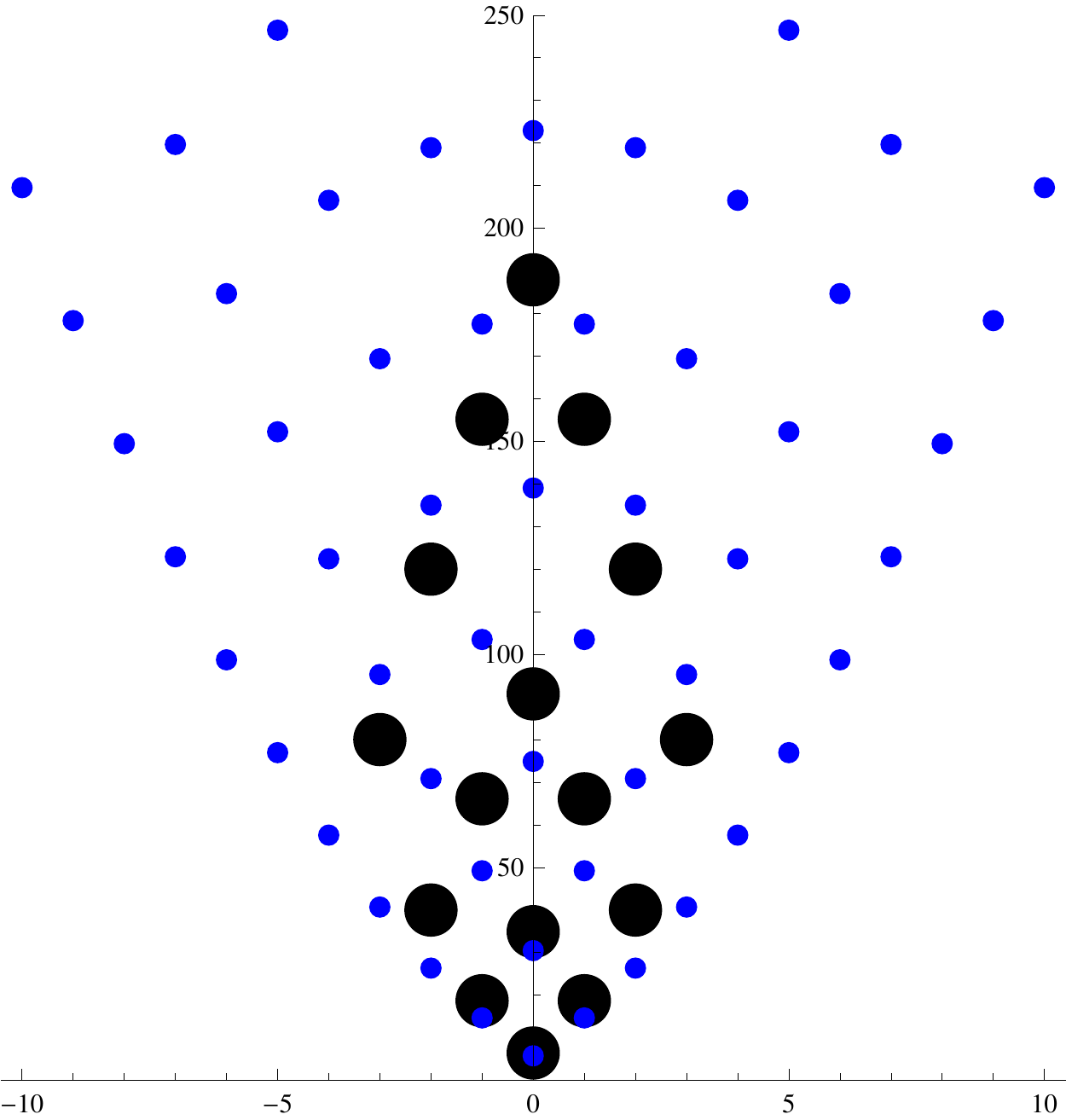}
\captionsetup{width=0.9\textwidth}
\caption{\small Los puntos grandes representan el espectro de $A^\infty_N$ para $\theta=0.05$ ($N=4$) en funci\'on del momento angular. Los puntos peque\~nos corresponden al espectro del disco conmutativo. $R$ es tomado como la longitud unidad.}
\label{fig:n4}
\end{minipage}%
\begin{minipage}{.5\textwidth}
\centering
\includegraphics[width=0.8\textwidth,height=0.8\textwidth]{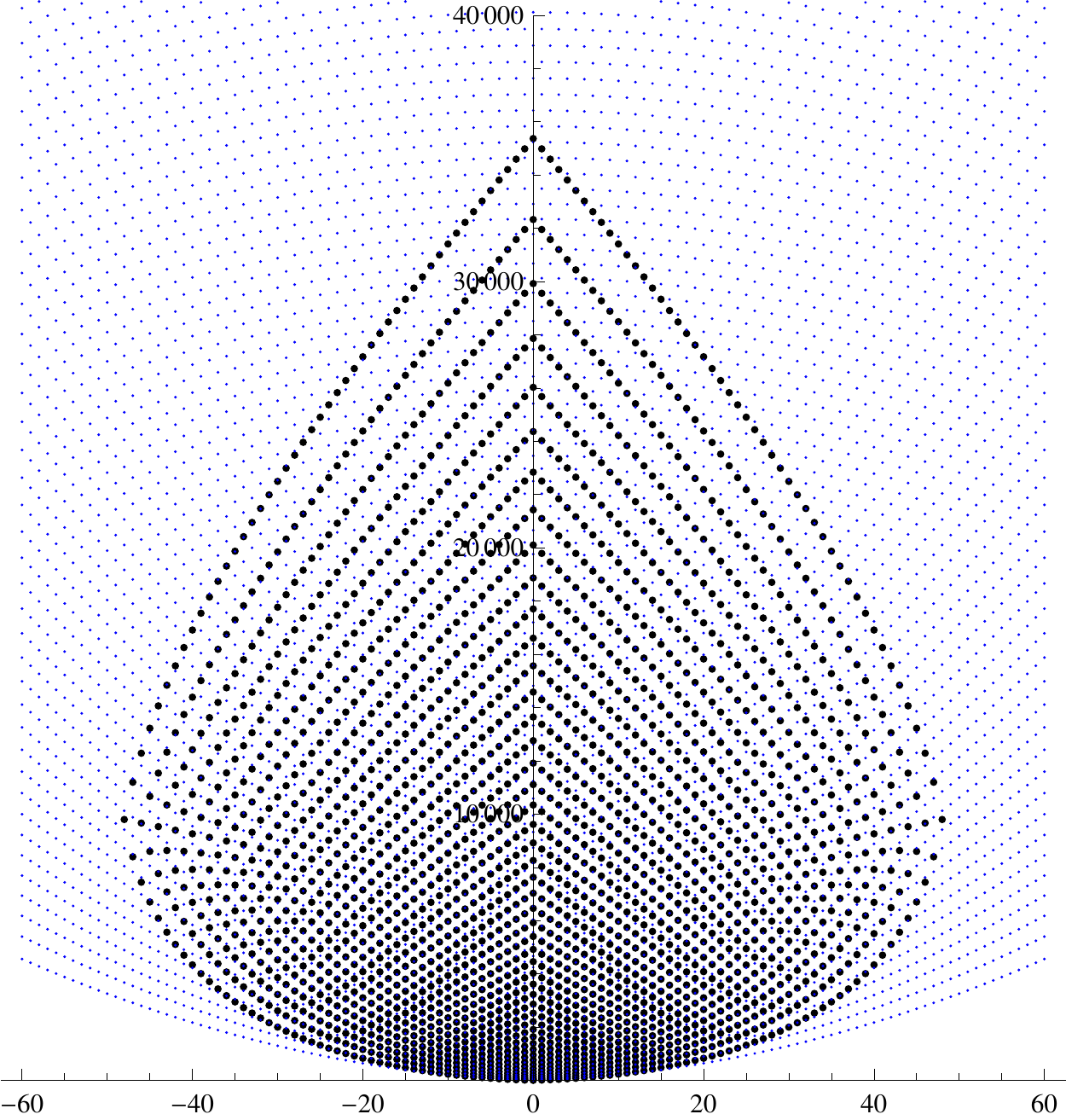}
\captionsetup{width=0.9\textwidth}
\caption{\small Los puntos que conforman la gota representan el espectro de $A^\infty_N$ para $\theta=0.005$ ($N=49$) en funci\'on del momento angular. Los puntos en el fondo corresponden al espectro del disco conmutativo. $R$ es tomado como la longitud unidad.}
\label{fig:n49}
\end{minipage}
\end{figure}

Los autovalores y autofunciones del operador $A^{\infty}_{N}$ presentan varias caracter\'isticas interesantes que pueden ser consultadas en la publicaci\'on de \textcite{Falomir:2013vaa}. A modo de ejemplo, en las Figuras \ref{fig:n4} y \ref{fig:n49} se puede apreciar la comparaci\'on de los espectros del laplaciano en el disco conmutativo y NC para dos valores de $\theta$.

\subsection{Energ\'ia de Casimir para el disco no conmutativo}\label{NC.disco.casimir}
Una de las curiosidades que presenta el disco NC es que, al existir solo un n\'umero finito de autovectores de $A_{N}^{\infty}$, su energ\'ia de Casimir no precisa ser regularizada. Para mostrarlo, tengamos presente que los modos de oscilaci\'on cu\'anticos $\psi_n(t,x)$ para un campo escalar en el disco NC son los modos cero del problema
\begin{equation}
    \delta^2S\cdot \psi_n(t,x)=\left\{-\partial_t^2-m^2-A^\infty_N\right\}\psi_n(t,x)=0\,.
\end{equation}
La soluci\'on a este problema se reduce, realizando la transformada de Fourier en la variable temporal, a la identificaci\'on de las autofunciones $\psi_n(x)$ del operador $A_{N}^{\infty}$:
\begin{equation}
    \psi_n(t,x)=:e^{-i\omega_n t}\psi_n(x)\,.
\end{equation}
Por cuanto conocemos los autovalores de $A^\infty_N$, dados por la expresi\'on \eqref{NC:central.spec}, obtenemos las frecuencias
\begin{equation}
    \omega^{\pm l}_k:=\sqrt{m^2+\frac{\lambda^l_k}{\theta}}\,,
\end{equation}
donde $L^l_{N-l}(\lambda^l_k)=0$, para $l=0,1,\ldots,N-1$ y $k=1,2,\ldots,N-l$.

Como hemos explicado en la secci\'on \ref{DET.casimir} del cap\'itulo 2, la energ\'ia de Casimir $E_{NC}$ es la semisuma de las energ\'ias $\omega_k$ de los modos de oscilaci\'on:
\begin{align}\label{casi}
    E_{NC}&=\frac12 \sum_{k=1}^{N}\omega^0_k+\sum_{l=1}^{N-1}\sum_{k=1}^{N-l}\omega^l_k\\\nonumber
    &=\frac{1}{2\sqrt{\theta}} \sum_{k=1}^{N}\sqrt{\lambda^0_k+\theta\, m^2}+
    \frac{1}{\sqrt{\theta}} \sum_{l=1}^{N-1}\sum_{k=1}^{N-l}\sqrt{\lambda^l_k+\theta\, m^2}\,.
\end{align}
Visto que el espacio de Hilbert de las oscilaciones cu\'anticas es de dimensi\'on finita, la energ\'ia de Casimir del disco NC es correspondientemente finita. En el caso de masa nula, podemos obtener el l\'imite asint\'otico para $N$ grande
\begin{align}\label{casim=0}
    E_{NC}&=\frac{1}{2\sqrt{\theta}} \sum_{k=1}^{N}\sqrt{\lambda^0_k}+
    \frac{1}{\sqrt{\theta}} \sum_{l=1}^{N-1}\sum_{k=1}^{N-l}\sqrt{\lambda^l_k}\\\nonumber
    &\simeq \frac{c}{R}\,N^3+\ldots\,.
\end{align}
Una primera estimaci\'on, la cual otorga el resultado $0.69<c<1.11$, se puede calcular utilizando las siguientes cotas para las ra\'ices de los polinomios de Laguerre \parencite{Gatteschi,IS}:
\begin{equation}
    \frac{2k+l+1}{\sqrt{N-l/2+1/2}}>\sqrt{\lambda^l_k}>
   \frac{\pi k+l-1/2}{\sqrt{4N-2l+2}}\,.
\end{equation}

\end{subappendices}


\chapter[Modelo de GW]{El modelo de Grosse-Wulkenhaar}\label{GW}
\setlength\epigraphwidth{6cm}
\epigraph{\itshape At every single moment of one’s life one is what one is going to be no less than what one has been. Art is a symbol, because man is a symbol.}{-- \textsc{Oscar Wilde}, \textit{De Profundis}.}

Casi un lustro despu\'es de la publicaci\'on del trabajo de \textcite{Minwalla:1999px}, en el cual se pone en relieve el problema inherente la mezcla de divergencias UV-IR en TCC NC, Grosse y Wulkenhaar mostraron a trav\'es de un ejemplo la existencia de teor\'ias NC renormalizables \parencite{Grosse:2003nw,Grosse:2004yu}. Para ello introdujeron, en el modelo del tipo $\lambda\,\phi_{\star}^4$ que recibe su nombre (GW), un t\'ermino harm\'onico de fondo que modifica el propagador libre. Con ello recuperan la dualidad de Langmann-Szabo \parencite{Langmann:2002cc} a costo de romper la invariancia de translaci\'on. El lagrangiano resultante ha sido interpretado a posteriori como fruto de la interacci\'on con la curvatura en un espacio NC \parencite{Buric:2009ss,deGoursac:2010zb}.

Las propiedades sobre este modelo que se van descubriendo d\'ia a d\'ia son cada vez m\'as prometedoras. Primeramente, se mostr\'o su renormalizabilidad a todo orden en teor\'ia de perturbaciones \parencite{Grosse:2003nw,Grosse:2004yu,Rivasseau:2005bh}; luego, encontraron que la constante de acoplamiento pose\'ia una funci\'on $\beta$ que se anulaba en el punto autodual, descartando as\'i la posible presencia del fantasma de Landau \parencite{Disertori:2006nq,Rivasseau:2007qda}. Posteriormente se demostr\'o su equivalencia, en el punto autodual y en el l\'imite de extrema no conmutatividad o infinito volumen,  a un modelo matricial exactamente soluble y no trivial \parencite{Grosse:2012uv}. Esto ha sugerido que podr\'ia dar lugar a la construcci\'on axiom\'atica de una TCC en cuatro dimensiones \parencite{Grosse:2014lxa,Grosse:2014nza}. P\'arrafo aparte, tambi\'en ha sido objeto de estudio en el marco del grupo de renormalizacion funcional, donde puede ser estudiado sin hacer truncamiento alguno \parencite{
Sfondrini:2010zm}.

En contraposici\'on, dos son las principales cr\'iticas que suele recibir este modelo. La primera es que el t\'ermino harm\'onico rompe la invariancia traslacional e introduce un punto preferencial (el origen de coordenadas). A este respecto podemos mencionar que la idea de GW ha sido utilizada para regularizar otros modelos, en los cuales el t\'ermino harm\'onico surge de una elecci\'on de gauge y en consecuencia los resultados f\'isicos respetan la simetr\'ia traslacional \parencite{Rivasseau:2007}. Segundo, suele pensarse que el s\'olo hecho de utilizar un potencial confinante bastar\'ia para obtener una teor\'ia igualmente bien comportada. Que este no es el caso lo demuestra la simetr\'ia de Langmann-Szabo \parencite{Langmann:2002cc} y deber\'ia quedar claro en el an\'alisis de la secci\'on \ref{GW.aniso}.

El cap\'itulo anterior ha sido utilizado para mostrar que el FLM es particularmente conveniente en el estudio de operadores no locales cuyos s\'imbolos son funciones no polinomiales de los momentos, las variables duales de Fourier de las coordenadas. En efecto, utilizando IdC en el espacio de fases los c\'alculos son relativamente sencillos, por cuanto la dependencia no habitual en el momento aparece en pie de igualdad con la dependencia en las coordenadas.

Adaptaremos en este cap\'itulo las t\'ecnicas del FLM al estudio del modelo de GW, modelo que ser\'a introducido en la secci\'on \ref{GW.GW}. Siguiendo este enfoque, en el apartado \ref{GW.hk} lograremos como resultado una f\'ormula magistral para la traza del NdC de operadores no locales con potenciales de fondo cuadr\'aticos. M\'as a\'un, como veremos en la secci\'on \ref{GW.FLM}, la AE al orden de un bucle (a partir de la cual pueden construirse las funciones de Green de $n$ puntos a id\'entico orden) en cierto modo resultar\'a ser un caso particular de esa f\'ormula magistral. Luego de analizar el comportamiento del modelo en el r\'egimen UV mediante el c\'alculo de las funciones $\beta$, en las \'ultimas dos secciones consideraremos dos variaciones: tomaremos una matriz $\Theta$ de no conmutatividad degenerada y un potencial harm\'onico anisotr\'opico.

\section{Definici\'on del modelo de Grosse-Wulkenhaar}\label{GW.GW}
El modelo de GW se encuentra definido en el ET eucl\'ideo Moyal de dimensi\'on arbitraria $d$. Para definir este espacio, tal y como hemos detallado en la secci\'on \ref{NC.tccnc}, debemos introducir el producto Moyal a trav\'es de la matriz de no conmutatividad $\Theta$, vid. (\ref{NC:moyal}). A lo largo de esta secci\'on, salvo que se explicite lo contrario\footnote{En la secci\'on \ref{GW.degenerado} analizaremos la situaci\'on m\'as general, en la que $\Theta$ puede ser singular.}, consideraremos una matriz $\Theta$ no degenerada. Adicionalmente, supondremos por el momento que los par\'ametros $\theta_{\mu=1,\ldots, d}$ que definen los $d/2$ planos de no conmutatividad son todos iguales entre s\'i y a una constante\footnote{Ver el comentario posterior a la ecuaci\'on \eqref{NC:theta}.} $\theta$.

Hechas estas aclaraciones, introduzcamos el modelo de GW para un campo escalar real, defini\'endolo a trav\'es del lagrangiano\footnote{Nuestro par\'ametro $\omega$ est\'a vinculado con el par\'ametro $\Omega$ presente en el lagrangiano original de \textcite{Grosse:2003nw} a trav\'es de la relaci\'on $\omega\,\theta=\Omega$.}
\begin{equation}\label{GW:lagrangian}
    \mathcal{L}_{GW}=\frac{1}{2}(\partial\varphi)^2+\frac{m^2}{2}\varphi^2+
    \frac{\omega^2}{2}\,x^2\varphi^2+\frac{\lambda}{4!}\varphi^4_\star\,,
\end{equation}
donde nuevamente $\varphi^4_\star:=\varphi\star\varphi\star\varphi\star\varphi$. La inclusi\'on del producto $\star$ en el t\'ermino arm\'onico implica s\'olo la aparici\'on de una contribuci\'on adicional t\'ermino cin\'etico, y por ende puede ser obviada. La correspondiente acci\'on cl\'asica es por lo tanto
\begin{equation}\label{GW:action}
    S_{GW}[\varphi]=\int_{\mathbb{R}^d}dx\,\left(\frac{1}{2}\varphi\, G_{GW}\varphi+\frac{\lambda}{4!}\varphi^4_\star\right)\,,
\end{equation}
donde hemos definido el operador $G$ a partir del cual, por motivos que m\'as adelante ser\'an expuestos, realizaremos las perturbaciones:
\begin{equation}\label{GW:G}
  G_{GW}:=-\partial^2+m^2+\omega^2x^2\,.
\end{equation}

Como es usual, podemos escribir la correcci\'on de un bucle a la AE 
en funci\'on del operador de fluctuaciones cu\'anticas $\delta^2S_{GW}$, calculado como la variaci\'on segunda de la acci\'on con respecto al campo cu\'antico $\varphi$ y evaluada en el campo medio $\phi$:
\begin{align}\label{GW:qf-operator}
  \begin{split}\delta^2S_{GW}&=G_{GW}+\frac{\lambda}{3!}\left[L(\phi^2_\star)+R(\phi^2_\star)+L(\phi)R(\phi)\right]\\
           &=G_{GW}+\frac{\lambda}{3!}\left[\phi^2_\star(x+i\Theta \partial)+
  \phi^2_\star(x-i\Theta \partial)\right.\\
  &\left.\hspace{3.5cm}+\phi(x+i\Theta \partial)\phi(x-i\Theta \partial)\right]\,.
\end{split}\end{align}
Ciertamente, como se puede ver comparando las expresiones (\ref{NC:phi4.effactapp}) y (\ref{GW:qf-operator}), la diferencia con el caso considerado en la secci\'on \ref{NC.phi4} reside en el t\'ermino harm\'onico que contiene el operador $G_{GW}$. Utilizando tambi\'en en esta ocasi\'on la representaci\'on del determinante funcional en t\'erminos del tiempo propio de Schwinger, la acci\'on efectiva $\Gamma_{GW}$ al orden de un bucle se puede escribir
\begin{equation}\label{GW:eff-action-pt}
    \Gamma_{GW}[\phi]=S_{GW}[\phi]-\frac{1}{2}\int_{\Lambda^{-2}}^\infty \frac{d\beta}{\beta}\ {\rm Tr}\,
    e^{-\beta\left\{\delta^2S_{GW}\right\}}\,.
\end{equation}
Hemos regularizado la expresi\'on utilizando un par\'ametro $\Lambda$ de corte UV, puesto que  en este marco las divergencias UV de los diagramas de Feynman provienen de las divergencias de la integral en (\ref{GW:eff-action-pt}) para $\beta\rightarrow0$. En la siguiente secci\'on, mostraremos c\'omo calcular la traza del NdC del operador no local $\delta^2S_{GW}$ adaptando la formulaci\'on, estudiada en el cap\'itulo previo, de IdC en el espacio de fases de una part\'icula puntual.

\section[NdC de operadores con un t\'ermino harm\'onico]{N\'ucleo de calor de operadores con un t\'ermino harm\'onico}\label{GW.hk}
Tomemos como punto de partida un operador no local $H$ que act\'ua sobre funciones de $x \in\mathbb{R}^d$ en la forma
\begin{equation}\label{GW:h}
    H:=-\partial^2+m^2+\omega^2x^2+V(x,-i\partial)\,.
\end{equation}
El operador no local $V$ es en este punto arbitrario y, llegado el caso, ser\'a elegido apropiadamente para que $H$ coincida con $\delta^2S_{GW}$.

A esta altura, resultar\'a natural al lector reconocer al NdC $e^{-\beta H}$ como el operador evoluci\'on, en tiempo eucl\'ideo $\beta$, de una part\'icula puntual movi\'endose en $\mathbb{R}^d$ bajo la acci\'on del hamiltoniano no local $H$. Por ende, el NdC puede ser escrito en t\'erminos de IdC en mec\'anica cu\'antica que, dada la no localidad del operador, conviene dejar expresadas como integrales en el espacio de fases. En resumen, siendo nuestro inter\'es el c\'alculo de la traza del NdC, la formulaci\'on usual conduce a la siguiente integral sobre trayectorias peri\'odicas (t.p.)
\begin{multline}\label{GW:pi-h}
    {\rm Tr}\,e^{-\beta H}
        =e^{-\beta m^2}\mathcal{N}(\beta) \int_{t.p.}\mathcal{D}q(t)\mathcal{D}p(t)\,e^{-\mathcal{S}[q(t),p(t)]}\times\\
        \times e^{-\beta\int_0^{1}dt\,V_W(\sqrt{\beta}q(t),p(t)/\sqrt{\beta})}
          \,,
\end{multline}
donde $\mathcal{N}(\beta)$ es una constante de proporcionalidad que ser\'a luego determinada a partir de una traza conocida, y $\mathcal{S}$ es la acci\'on eucl\'idea de un oscilador harm\'onico en el espacio de fases $\mathbb{R}^{2d}$, una funcional cuadr\'atica en los caminos de la posici\'on y el momento:
\begin{equation}\label{GW:action.harmonic}
  \mathcal{S}[q(t),p(t)]:=\int_0^1 dt\,\left\{p(t)^2-ip(t)\dot{q}(t)+\omega^2\beta^2q(t)^2\right\}.
\end{equation}

Antes de proseguir conviene realizar algunas aclaraciones sobre la expresi\'on (\ref{GW:pi-h}). Primero, notemos que hemos realizado el reescaleo habitual en las variables de la IdC, $t\rightarrow \beta t$, $q(t)\rightarrow \sqrt{\beta}q(t)$ y $p(t)\rightarrow p(t)/\sqrt{\beta}$, de modo que las trayectorias $p(t)$, $q(t)$ y el par\'ametro $t$ son ahora adimensionales. En la constante de proporcionalidad $\mathcal{N}(\beta)$ absorberemos eventuales factores provenientes de la medida de integraci\'on. 

Segundo, en concordancia con la prescripci\'on del punto medio para integrales de camino, la funci\'on $V_W$ se obtiene reemplazando los operadores $x$ y $-i\partial$ por $\sqrt{\beta}\,q(t)$ y $p(t)/\sqrt{\beta}$, respectivamente, en la expresi\'on del potencial $V(x,-i\partial)$ ordenada seg\'un Weyl. Esto implica que deber\'iamos simetrizar el operador no local $V(x,-i\partial)$ con respecto a $x$ y $-i\partial$ antes de reemplazarlos por las trayectorias $q(t)$ y $p(t)$. Afortunadamente, de acuerdo al an\'alisis precedente a  (\ref{NC:weyl}), el operador (\ref{GW:qf-operator}) puede ser reexpresado en forma sim\'etrica sin introducir t\'erminos adicionales.

Por \'ultimo, hemos mencionado que la integral de (\ref{GW:pi-h}) es realizada sobre trayectorias $q(t)$ y $p(t)$ peri\'odicas, es decir tales que cumplen $q(0)=q(1)$ y $p(0)=p(1)$. Si nos hubieramos propuesto obtener el NdC $\langle x'|e^{-\beta H}|x\rangle$, podr\'iamos haber simplemente integrado sobre trayectorias cuya proyecci\'on en el espacio de configuraci\'on satisficiera las condiciones de borde\footnote{En el anexo \ref{GW.meh} realizamos el c\'alculo de una cantidad local, el n\'ucleo de la inversa del operador $G_{GW}$ definido en \eqref{GW:G}, utilizando esta t\'ecnica.} $q(0)=x$ y $q(1)=x'$. La traza podr\'ia desde luego obtenerse a partir de la expresi\'on resultante, tomando $x=x'$ e integrando sobre la variable $x$. En el caso que nos respecta resulta suficiente ojear el ap\'endice \ref{GW.cc}, donde consideramos condiciones de contorno tipo Dirichlet,  para convencerse de que el c\'alculo haciendo uso de condiciones peri\'odicas resulta m\'as elegante.

Habiendo compartido estas consideraciones, definimos los valores de expectaci\'on de una funcional $f[q(t),p(t)]$ en la forma habitual
\begin{equation}\label{GW:valormediofunc}
    \left\langle
      f[q(t),p(t)]\right\rangle_{\!\rm per}:=\mathcal{Z}(\omega\beta)^{-1}\int_{t.p.}\mathcal{D}q(t)\mathcal{D}p(t)\,e^{-\mathcal{S}[q(t),p(t)]} f[q(t),p(t)]\,,
\end{equation}
eligiendo la constante de normalizaci\'on $\mathcal{Z}(\omega\beta)$ de modo tal que se cumpla $\langle 1 \rangle=1$. Como veremos en poco, la dependencia funcional de $\mathcal{Z}(\omega\beta)$ con $\omega\beta$ carece de importancia.
Tambi\'en nos ser\'a de utilidad la introducci\'on de la funcional generatriz  $Z_{per}[k,j]$ evaluada en las fuentes $k$ y $j$:
\begin{align}\label{GW:Z}
Z_{\rm per}[k(t),j(t)]:=  \left\langle
    e^{\int_0^1 dt\,\left\{k(t)p(t)+j(t)q(t)\right\}}
    \right\rangle_{\!\rm per}.
\end{align}

En forma concisa, la traza (\ref{GW:pi-h}) puede ahora escribirse en t\'erminos del valor medio de la exponencial del potencial
\begin{equation}\label{GW:pi4}
        {\rm Tr}\,e^{-\beta H}
        =e^{-\beta m^2}\mathcal{N}(\beta)\mathcal{Z}(\omega\beta)\,
        \left\langle e^{-\beta\int_0^{1}dt\,V_W(\sqrt{\beta}q(t),p(t)/\sqrt{\beta})}\right\rangle_{\!\rm per}
          \,.
\end{equation}
Por otro lado, el producto de las constantes de proporcionalidad y normalizaci\'on $\mathcal{N}(\beta)\,\mathcal{Z}(\omega\beta)$ puede ser determinada por el valor conocido de la traza del NdC para el oscilador arm\'onico en un espacio $d$ dimensional. En efecto, tomando $m=V_W=0$ en la expresi\'on (\ref{GW:pi4}) obtenemos 
\begin{equation}\label{GW:normalizacionIdc}
  {\rm Tr}\,e^{-\beta \{-\partial^2+\omega^2x^2\}}
        =\mathcal{N}(\beta)\mathcal{Z}(\omega\beta)=
        \left(\sum_{n=0}^\infty e^{-2\beta\omega(n+1/2)}\right)^d=
        \frac{1}{(2\sinh{\omega\beta})^{d}}
          \,.
\end{equation}

El paso siguiente consiste en hacer el desarrollo de la expresi\'on (\ref{GW:pi4}) en potencias de $V_W$, imitando el c\'alculo perturbativo, en el n\'umero de v\'ertices de los diagramas de Feynman, de una TCC en $0+1$ dimensiones. Asimismo, expresamos el resultado en t\'erminos de la transformada de Fourier $\tilde{V}_W(\cdot,\cdot)$ de la funci\'on $V_W(\cdot,\cdot)$ en sus dos variables, de modo que el \'unico valor medio a calcular es la propia FG; luego de trabajar un poco la expresi\'on obtenemos
\begin{align}{\label{GW:pi3}}
        \begin{split}
{\rm Tr}\,e^{-\beta H}
        &=\frac{e^{-\beta m^2}}{(2\sinh{\omega\beta})^{d}}
        \ \sum_{n=0}^\infty (-\beta)^n
        \int_{\mathbb{R}^{2d}}d\tilde{\sigma}_1d\tilde{\xi}_1\ldots
        \int_{\mathbb{R}^{2d}}d\tilde{\sigma}_nd\tilde{\xi}_n\,\times\\[2mm]
        &\times\,\tilde{V}_W(\sigma_1,\xi_1)\ldots \tilde{V}_W(\sigma_n,\xi_n)
        \times K^{(n)}_\beta(\sigma_1,\ldots,\sigma_n;\xi_1,\ldots,\xi_n)
        \,,
        \end{split}
\end{align}
donde las funciones $K^{(n)}_{\beta}$ que hemos introducido, utilizando la simetr\'ia en las variables $t_i$, pueden ser escritas como integrales de la FG\footnote{Notemos que el t\'ermino $n=0$ de la expresi\'on (\ref{GW:pi3}) debe ser tomado igual a 1.}:
\begin{align}\label{GW:kn00}
   \begin{split}
 K^{(n)}_\beta(\sigma_1,\ldots,\sigma_n;\xi_1,\ldots,\xi_n)
    &:=\int_0^1dt_1\ldots\int_0^{t_{n-1}}dt_n\, Z_{\rm per}[k_n(t),j_n(t)].
\end{split}
\end{align}
Por su parte, las fuentes $k_n$ y $j_n$ en las cuales debe evaluarse la FG, dependen de los tiempos $t_i$ y est\'an dadas por 
\begin{align}\label{GW:kj}
 \begin{split}
    k_n(t)&:=i\,\beta^{-1/2}\,\sum_{i=1}^{n}\delta(t-t_i)\,\xi_i\,,\\
    j_n(t)&:=i\,\beta^{1/2}\,\sum_{i=1}^{n}\delta(t-t_i)\,\sigma_i\,.
\end{split}
\end{align}

El c\'alculo a partir de este momento resulta directo si notamos que la acci\'on de la ec. (\ref{GW:action.harmonic}) para una part\'icula puntual es una funcional cuadr\'atica en los caminos $p$ y $q$, y puede ser reescrita como
\begin{equation}\label{GW:qufo}
  \mathcal{S}[q(t),p(t)]=\frac12\int_0^1dt\,
  \begin{pmatrix}
    p(t)&q(t)
  \end{pmatrix}
        D_{\rm per}
  \begin{pmatrix}
    p(t)\\q(t)
  \end{pmatrix}\,,
\end{equation}
vali\'endonos del operador diferencial $D_{\rm per}$ herm\'itico que act\'ua sobre trayectorias peri\'odicas tanto en $p$ como en $q$ de acuerdo a 
\begin{equation}
        D_{\rm per}:=\begin{pmatrix}
                    2&-i\partial_t\\
                    i\partial_t&2\omega^2\beta^2
        \end{pmatrix}\,.
\end{equation}
Basta por consiguiente con completar cuadrados en la IdC asociada a la FG para deshacernos de las funcionales lineales de la trayectoria presentes en el exponente y obtener la siguiente expresi\'on cerrada:
\begin{align}\label{GW:z}
    Z_{\rm per}[k,j]=
    \exp{\left\{\frac{1}{2}\int_0^1\!\!\int_0^1 \!\! dtdt'\, \begin{pmatrix}
    k(t)&j(t)
    \end{pmatrix}
    G^{\rm (per)}(t-t')
    \begin{pmatrix}
    k(t')\\j(t')
    \end{pmatrix}\right\}}\,;
\end{align}
el n\'ucleo del operador inverso $D_{\rm per}^{-1}$, $G^{\rm (per)}(t-t')$, est\'a dado por la f\'ormula
\begin{align}\label{GW:dkernel}
\begin{split}    G^{\rm (per)}(\Delta)&=\frac{1}{2\sinh{\omega\beta}}
    \begin{pmatrix}
     G^{\rm (per)}_{pp}        &  G^{\rm (per)}_{px} \\
     G^{\rm (per)}_{xp}        &  G^{\rm (per)}_{xx} 
    \end{pmatrix}\,,
    \end{split}
\end{align}
donde sus componentes est\'an definidas como\footnote{La funci\'on signo $\epsilon(\cdot)$ es $\pm1$ si su argumento es positivo (negativo).}
\begin{align}
\begin{split}
    G^{\rm (per)}_{pp}:&=\omega\beta\,\cosh{[\omega\beta(2|\Delta|-1)]},\\
    G^{\rm (per)}_{px}:&=i\,\epsilon(\Delta)\,\sinh{[\omega\beta(2|\Delta|-1)]},\\
    G^{\rm (per)}_{xp}:&=-i\,\epsilon(\Delta)\,\sinh{[\omega\beta(2|\Delta|-1)]},\\
    G^{\rm (per)}_{xx}:&=\frac{1}{\omega\beta}\,\,\cosh{[\omega\beta(2|\Delta|-1)]}.
\end{split}
\end{align}
Entre otras propiedades, podemos mencionar que $G^{\rm (per)}(t-t')$ es sim\'etrico y, como es de esperar teniendo en cuenta que su dominio de definici\'on es el de las trayectorias peri\'odicas, depende s\'olo de la diferencia $\Delta:=t-t'$.

Finalmente, evaluando la FG en las fuentes $k_n(t)$ y $j_n(t)$, dadas por (\ref{GW:kj}), y utilizando nuevamente la simetr\'ia en las variables temporales $t_i$, podemos escribir para las funciones $K_{\beta}^{(n)}$ la f\'ormula 
\begin{align}\label{GW:kn}
\begin{split}
    &K^{(n)}_\beta(\sigma_1,\ldots,\sigma_n;\xi_1,\ldots,\xi_n)=
    e^{-\frac{1}{4\omega\tanh{\omega\beta}}\, \stackrel[i]{}{\sum}(\omega^2\xi^2_i+\sigma^2_i)}
    \!\int_0^1\!dt_1\ldots\!\int_0^{t_{n-1}}\!dt_n\ \times\\[2mm]
    &\!\!\times\!
    e^{-\frac{1}{2\omega\sinh{\omega\beta}}\, \stackrel[i<j]{}{\sum}
    \left\{\cosh{\left[\omega\beta(2|t_i-t_j|-1)\right]}\,(\omega^2\xi_i\xi_j+\sigma_i\sigma_j)
    +i\omega\sinh{\left[\omega\beta(2|t_i-t_j|-1)\right]}\,(\xi_i\sigma_j-\xi_j\sigma_i)\right\}}
    \!.
\end{split}
\end{align}
La combinaci\'on de esta expresi\'on con la ecuaci\'on (\ref{GW:pi3}) da el resultado deseado para el desarrollo, exacto en el potencial arm\'onico y perturbativo en $V$, de la traza del NdC del operador no local (\ref{GW:h}).

\section{FLM en el modelo de Grosse-Wulkenhaar}\label{GW.FLM}
En la secci\'on anterior hemos encontrado un desarrollo para la traza del NdC de un operador no local con un potencial arm\'onico, el cual nos servir\'a para calcular a continuaci\'on la AE a un bucle en el modelo de GW. Con ese prop\'osito principiamos uniendo las f\'ormulas (\ref{GW:eff-action-pt}) y (\ref{GW:pi3}); de este modo obtenemos la siguiente expresi\'on para la AE a un bucle correspondiente a un campo escalar cuyo operador de fluctuaciones cu\'anticas posee un potencial harm\'onico y uno arbitrario $V_W$:
\begin{multline}\label{GW:eff-action-pt2}
    \Gamma_{1-\text{\rm bucle}}[\phi]=\frac{1}{2}\sum_{n=1}^\infty
        \int \prod_{i=1}^n \left\{d\tilde{\sigma}_id\tilde{\xi}_i\ \tilde{V}_W(\sigma_i,\xi_i)\right\}\times\\
        \times\int_{\Lambda^{-2}}^{\infty} d\beta\,
        \frac{e^{-\beta m^2}\,(-\beta)^{n-1}}{(2\sinh{\omega\beta})^{d}}
        \ K^{(n)}_\beta
        \,,
\end{multline}
donde las funciones $K^{(n)}_\beta$ que figuran en esta f\'ormula han sido plasmadas en (\ref{GW:kn}).

Sin m\'as, la identificaci\'on del potencial $V_W$ asociado al modelo de GW puede realizarse mediante la simple comparaci\'on de las ecuaciones para el operador de fluctuaciones y H, viz. (\ref{GW:qf-operator}) y (\ref{GW:h}):
\begin{equation}\label{GW:qf-operator2}
    V_W(x,p)=\frac{\lambda}{3!}\left[\phi^2_\star(x-\Theta p)+
    \phi^2_\star(x+\Theta p)+
    \phi(x-\Theta p)\phi(x+\Theta p)\right]\,.
\end{equation}
Como hemos mostrado en la secci\'on \ref{NC.intcaminofase}, al escribir (\ref{GW:qf-operator2}) no es que nos hemos olvidado de realizar el ordenamiento de Weyl sino que nos sustentamos en el hecho de que cualquier operador escrito en la forma $f(x\pm i\Theta\partial)$, o como el producto $f(x+ i\Theta\partial)g(x+ i\Theta\partial)$ puede ser llevado a dicho orden sin la introducci\'on de t\'erminos adicionales\footnote{Un resultado quiz\'as poco intuitivo es que un operador escrito como $f(x+ i\Theta\partial)g(x- i\Theta\partial)$ en general no corresponde al ordenamiento de Weyl.}. Por su parte, la transformada de Fourier $\tilde{V}_W(\cdot,\cdot)$ del potencial (\ref{GW:qf-operator2}) en sus dos variables es
\begin{align}\label{GW:qf-operator2-ft}
    \tilde{V}_W(\sigma,\xi)&
    =\frac{\lambda}{3!}\,(2\pi)^d\left[\delta(\xi-\Theta\sigma)\ \mathcal{F}\left\{\phi^2_\star\right\}(\sigma)+
    \delta(\xi+\Theta\sigma)\ \mathcal{F}\left\{\phi^2_\star\right\}(\sigma)+\right.\nonumber\\
    &\left.\mbox{}+
    {\rm det}^{-1}(4\pi\Theta)\ \mathcal{F}\{\phi\}(\sigma/2-\Theta^{-1} \xi/2)\ \mathcal{F}\{\phi\}(\sigma/2+\Theta^{-1} \xi/2)\right]\,,
\end{align}
al emplear el s\'imbolo $\mathcal{F}$ para representar la transformada de Fourier de una funci\'on en $\mathbb{R}^d$.

Previo a analizar cada t\'ermino de la ecuaci\'on (\ref{GW:eff-action-pt2}) por separado para el campo escalar de GW, conviene mencionar algunos detalles a ella concernientes. Ante todo, deberemos mantener el par\'ametro $\Lambda$  de corte UV para regularizar las divergencias de las integrales en el l\'imite $\beta\rightarrow 0$. Segundo, hemos eliminado el t\'ermino $n=0$, en concordancia con una adaptaci\'on al presente caso de los argumentos que llevan a \eqref{DET:w1}: en esta ocasi\'on, el lugar del operador $G$ es ocupado por $G_{GW}=-\partial^2+m^2+x^2$. Tercero, las divergencias IR de la AE que podr\'ian aparecer en la integraci\'on (para $\beta\rightarrow \infty$), est\'an ausentes incluso en el caso sin masa; el responsable de este comportamiento ben\'evolo es el factor exponencialmente decreciente $(\sinh{\omega\beta})^{-d}$ que tiene por su parte origen en el potencial de fondo harm\'onico. 

Por \'ultimo, visto que el potencial $V_W$ es cuadr\'atico en los campos, el t\'ermino indexado con $n$ en la f\'ormula (\ref{GW:eff-action-pt2}) da la contribuci\'on de un bucle a la funci\'on de correlaci\'on de $2n$ puntos de la AE. En los sucesivos par\'agrafos centraremos nuestra atenci\'on en los t\'erminos $n=1$ y $n=2$ para estudiar las correcciones a un bucle de las funciones de dos (propagador) y cuatro puntos, respectivamente\footnote{Dado que todos los resultados enunciados a continuaci\'on corresponden al orden de un bucle, omitiremos la inclusi\'on del sub\'indice ``$1-\text{\rm bucle}$'' en lo que resta del cap\'itulo.}.

\subsection{La funci\'on de dos puntos}\label{GW.2p}
Las contribuciones a la acci\'on efectiva que son cuadr\'aticas en el campo y, en consecuencia, corrigen el propagador, est\'an dadas por el t\'ermino $n=1$ de (\ref{GW:eff-action-pt2})
\begin{equation}\label{GW:2p-gf}
    \Gamma^{(2)}[\phi]=
        \frac{1}{2}
        \int_{\mathbb{R}^{2d}}d\tilde\sigma d\tilde\xi
        \ \tilde{V}_W(\sigma,\xi)
        \int_{\Lambda^{-2}}^{\infty} d\beta\,
        \frac{e^{-\beta m^2}}{(2\sinh{\omega\beta})^d}
        \ K^{(1)}_\beta(\sigma;\xi)
    \,,
\end{equation}
donde, luego de realizar la integral respecto a $t_1$, la expresi\'on (\ref{GW:kn}) arroja el valor: 
\begin{align}\label{GW:K1beta.nosingular}
    K^{(1)}_{\beta}(\sigma;\xi)=e^{-\frac{1}{4\omega\tanh{\omega\beta}}\,(\omega^2\xi^2+\sigma^2)}\,.
\end{align}

Espec\'ificamente, la f\'ormula (\ref{GW:2p-gf}) posee tres contribuciones, a saber una por cada t\'ermino de la transformada $\tilde{V}_W$ del potencial. El lector recordar\'a de la secci\'on \ref{NC.phi4} que dos de estos t\'erminos, aquellos que no mezclan la multiplicaci\'on Moyal a izquierda y a derecha, corresponden a contribuciones de diagramas de Feynman planares. El restante,  aquel que mezcla el producto Moyal a izquierda y a derecha, es resultado de diagramas no planares. Analizaremos estos casos por separado.

\subsubsection{Contribuciones planares}\label{GW.2pp}

Como era de esperar, las dos contribuciones planares a la expresi\'on (\ref{GW:2p-gf}) coinciden. Esto resulta evidente del hecho de que difieran solamente en el signo del argumento de las funciones delta $\delta(\xi\pm\Theta \sigma)$, mientras que la funci\'on $K_{\beta}^{(1)}(\sigma,\xi)$ depende de $\xi^2$. 

Por consiguiente, la suma de estas dos correcciones planares $\Gamma_P^{(2)}[\phi]$ al t\'ermino cuadr\'atico de la acci\'on efectiva puede ser escrita como
\begin{align}
\label{GW:2p-gf-p2}
\begin{split}
        \Gamma^{(2)}_{{\rm P}}[\phi]
        &=\frac{\lambda}{6}
        \int_{\mathbb{R}^{d}}d\tilde\sigma
        \ \mathcal{F}\left\{\phi^2_\star\right\}(\sigma)
        \int_{\Lambda^{-2}}^{\infty} d\beta\,
        \frac{e^{-\beta m^2}}{(2\sinh{\omega\beta})^d}
        \ e^{-\frac{1}{4\omega\tanh{\omega\beta}}\,(1+\omega^2\theta^2)\,\sigma^2}
        \\[2mm]
        &=\frac{1}{2}\int_{\mathbb{R}^{d}}dx
        \ \Gamma^{(2)}_{{\rm P}}(x)\star\phi(x)\star\phi(x)\,,
\end{split}
\end{align}
donde, debido al comportamiento del integrando para peque\~nos valores de $\beta$, hemos introducido la funci\'on $\Gamma^{(2)}_{{\rm P}}(x)$ conservando la regularizaci\'on mediante el par\'ametro de corte $\Lambda$:
\begin{align}
  \Gamma^{(2)}_{{\rm P}}(x):=\frac{\lambda}{3}\left\{\frac{\omega}{2\pi(1+\omega^2\theta^2)}\right\}^{d/2}
        \int_{\Lambda^{-2}}^{\infty} d\beta\,
        \frac{e^{-\beta m^2}}{(\sinh{2\omega\beta})^{d/2}}
        \ e^{-\frac{\omega\tanh{\omega\beta}}{1+\omega^2\theta^2}\,x^2}\,.
        \label{GW:eff-back}
\end{align}
En el caso $\omega=0$, que corresponde al modelo invariante frente a traslaciones, $\Gamma^{(2)}_{{\rm P}}(x)$ resulta ser independiente de $x$ y de $\theta$, y su divergencia puede ser eliminada a trav\'es de la renormalizaci\'on usual de la masa. En contraste, para $\omega\neq0$ y dependiendo de la dimensi\'on del ET, la regularizaci\'on de esta divergencia podr\'ia precisar la redefinici\'on de las constantes del t\'ermino cin\'etico, del t\'ermino harm\'onico e incluso de otras que acompa\~nan potencias mayores de $x$. 

A modo de ejemplo, consideremos las correcciones planares en un ET bidimensional. Reemplazando $d=2$ en la f\'ormula (\ref{GW:eff-back}) obtenemos
\begin{equation}\label{GW:2p-gf-c-2}
        \Gamma^{(2)}_{{\rm P}}[\phi]=\frac{1}{2}\int_{\mathbb{R}^{2}}dx\,
        \left\{ m_{2,\rm ren}^2\,\phi^2(x)+U_2(x)\star\phi(x)\star\phi(x)\right\}
        \,.
\end{equation}
Al igual que en el caso conmutativo, la expresi\'on (\ref{GW:2p-gf-c-2}) posee una \'unica divergencia logar\'itmica en t\'erminos de $\Lambda$, presente en el t\'ermino de masa
\begin{equation}
    m_{2,\rm ren}^2:=\frac{\lambda}{12\pi(1+\omega^2\theta^2)}
        \log{\left(\Lambda^{2}/\omega\right)}+\text{t\'erminos finitos}\,.
\end{equation}
El t\'ermino remanente, correspondiente al potencial de fondo $U_2(x)$, es regular en el r\'egimen UV y puede ser representado como
\begin{equation}
    U_2(x):=\frac{\lambda}{12\pi(1+\omega^2\theta^2)}
        \int_{0}^{1}\frac{dt}{t}\,\left(\frac{1-t}{1+t}\right)^{m^2/2\omega}\,
        \left\{e^{-\frac{\omega}{1+\omega^2\theta^2}\,t\,x^2}-1\right\}\,.
\end{equation}
Esta expresi\'on se comporta como $\sim x^2$ para peque\~nos valores de $x$, de modo que renormaliza la frecuencia $\omega$ en una cantidad finita, cuyo valor depende de la prescripci\'on de renormalizaci\'on a utilizar. En particular, para el caso sin masa se puede obtener la expresi\'on cerrada definida a trav\'es de $\gamma$ y $\Gamma(0,\cdot)$, la constante de Euler-Mascheroni y la funci\'on gamma incompleta respectivamente:
\begin{equation}
    U_2(x)\Big\rvert_{m=0}=-\frac{\lambda}{12\pi(1+\omega^2\theta^2)}
            \left\{
            \Gamma\left(0,\tfrac{\omega}{1+\omega^2\theta^2}\,x^2\right)+
            \log{\left(\tfrac{\omega}{1+\omega^2\theta^2}\,x^2\right)}+
            \gamma\right\}.
\end{equation}

Como hemos adelantado, en dimensiones mayores, la AE precisa al menos de la renormalizaci\'on de la masa, el campo y la frecuencia. Efectivamente, si tomamos $d=4$, la expresi\'on (\ref{GW:eff-back}) se reduce a 
\begin{multline}\label{GW:f4}
  \Gamma^{(2)}_{{\rm P}}(x)=\frac{\lambda}{48\pi^2}\frac{\omega}{(1+\omega^2\theta^2)^2}
         \times\\ \times \int_{\tanh{\frac\omega{\Lambda^2}}}^{1}\ \frac{dt}{t^2}
        \ \frac{(1-t)^{\frac{m^2}{2\omega}+1}}{(1+t)^{\frac{m^2}{2\omega}-1}}
        \ \left\{1-\tfrac{\omega}{1+\omega^2\theta^2}\,t\,x^2\right\}+O(x^4)\,,
\end{multline}
la cual diverge cuadr\'aticamente para $\Lambda\rightarrow\infty$. Sin lugar a dudas, el t\'ermino $O(x^4)$ es finito en el r\'egimen UV y positivo, a la vez que el primero entre llaves es independiente de $x$ y puede ser removido absorbiendo en la masa un t\'ermino proporcional a $\Lambda^2$. El segundo t\'ermino entre llaves, en cambio, es proporcional a $x^2$ y resulta logar\'itmicamente divergente; m\'as a\'un, reemplazado en la segunda l\'inea de (\ref{GW:2p-gf-p2}) da lugar a la renormalizaci\'on del campo y la frecuencia, ya que $\int dx\,\{x^2\star\phi\star\phi\}=\int dx\,\{x^2\,\phi^2+\theta^2(\partial\phi)^2\}$. La constante $Z_4$ de renormalizaci\'on del campo a primer orden en las perturbaciones resulta
\begin{equation}\label{GW:z.renor}
    Z_4=1+\frac{\lambda}{48\pi^2}\,\frac{\omega^2\theta^2}{(1+\omega^2\theta^2)^3}
    \left[\log{\left(\Lambda^2/\omega\right)}+\text{(t. f.)}\right]\,,
\end{equation}
donde $\phi_{\rm ren}(x):=Z_4^{-1/2}\,\phi(x)$. Es importante remarcar que, en forma opuesta a lo que sucede en el caso conmutativo $\theta=0$, la renormalizaci\'on del campo ya posee contribuciones a este orden.

De igual manera, usando (\ref{GW:f4}) y teniendo en cuenta la renormalizaci\'on de los campos, podemos escribir para la frecuencia $\omega_{\rm ren}$ y la masa $m_{\rm ren}$ renormalizadas
\begin{align}\label{GW:w.renor}
\omega^2_{\rm ren}&=\omega^2\left\{1-\frac{\lambda}{48\pi^2}\,\frac{1-\omega^2\theta^2}{(1+\omega^2\theta^2)^3}
    \left[\log{\left(\Lambda^2/\omega\right)}+{\rm (t.\ f.)}\right]\right\}\,,\\
\label{GW:m.renor}
\begin{split}
m_{\rm ren}^2&=m^2\left\{1+\frac{\lambda}{48\pi^2}\frac{1}{(1+\omega^2\theta^2)^2\,m^2}\Lambda^2+\right.\\
&\hspace{3cm}+\left.\frac{\lambda}{48\pi^2}\,\frac{\omega^2\theta^2}{(1+\omega^2\theta^2)^3}
    \left[\log{\left(\Lambda^2/\omega\right)}+\text{(t. f.)}\right]\right\}.
\end{split}
\end{align}
Las cantidades renormalizadas (\ref{GW:z.renor}), (\ref{GW:w.renor}) y \eqref{GW:m.renor} se encuentran en conformidad con las calculadas a un bucle por \textcite{Grosse:2004by}. 

Es de remarcar el hecho de que para el punto autodual $\omega\,\theta=1$, de acuerdo  a la f\'ormula (\ref{GW:w.renor}), no es necesario absorber en la frecuencia $\omega$ t\'erminos que divergen para $\Lambda$ grande; esto no ser\'ia posible de no ser por el valor preciso \eqref{GW:z.renor} que toma la constante $Z_4$ de renormalizaci\'on del campo y es consecuencia de la invariancia del modelo ante la dualidad de \textcite{Langmann:2002cc}. 

En pocas palabras, dicha dualidad corresponde a observar que, tanto a nivel cl\'asico como cu\'antico, para $\theta\omega=1$ la acci\'on de GW tiene el mismo aspecto escrita en t\'erminos del campo o de su transformada de Fourier. Teniendo en cuenta que bajo esta transformaci\'on el t\'ermino cin\'etico se convierte en el t\'ermino harm\'onico y que la 
renormalizaci\'on de la teor\'ia puede realizarse por igual en el espacio de coordenadas o de momentos, es plausible el resultado $\omega_{\rm ren}=\omega$ para el punto autodual. Ahondaremos sobre esta cuesti\'on en la secci\'on \ref{GW.4p}; llegado el momento precisaremos las expresiones de las funciones $\beta$ para las diversas constantes:
\begin{align}\label{GW:funcionesbeta}
 \begin{split}\beta_{\omega^2}&=\Lambda \partial_{\Lambda} \omega^2=\frac{\lambda\omega^2}{24\pi^2}\,\frac{1-\omega^2\theta^2}{(1+\omega^2\theta^2)^3}\\
\beta_{m^2}&=\Lambda \partial_{\Lambda} m^2 =-\frac{\lambda}{24\pi^2}\frac{\Lambda}{(1+\omega^2\theta^2)^2}-\frac{\lambda\omega^2}{24\pi^2}\,\frac{\omega^2\theta^2}{(1+\omega^2\theta^2)^3}\\
\gamma&=\Lambda \partial_{\Lambda} Z_4= \frac{\lambda}{24\pi^2}\,\frac{\omega^2\theta^2}{(1+\omega^2\theta^2)^3}.
\end{split}
\end{align}
Para completar la lista, hemos incluido tambi\'en la funci\'on $\gamma$ que indica el flujo de $Z_4$, la renormalizaci\'on del campo, con la escala de energ\'ia~$\Lambda$.

\subsubsection{Contribuciones no planares}\label{GW.2pnp}
La contribuci\'on no planar de un bucle a la AE, a saber, la correspondiente al tercer t\'ermino de (\ref{GW:qf-operator2-ft}), es
\begin{equation}\label{GW:2p-gf-np}
        \Gamma^{(2)}_{{\rm NP}}[\phi]=\frac12
        \int_{\mathbb{R}^{2d}}d\tilde{p}d\tilde{p}'
        \,\tilde{\phi}(p)\tilde{\phi}(p')
        \ \Gamma^{(2)}_{{\rm NP}}(p,p')
        \,,
\end{equation}
donde  $\Gamma^{(2)}_{{\rm NP}}(p,p')$ est\'a definida en t\'erminos de la funci\'on  hipergeom\'etrica confluente de Kummer $U(\cdot,\cdot,\cdot)$ \parencite{A-S}:
\begin{align}\label{GW:2p-gamma}
  \begin{split} 
   \Gamma^{(2)}_{{\rm NP}}(p,p')
            &=\frac{\lambda}{6}
        \int_{0}^{\infty} \frac{d\beta}{({2}\sinh{\omega\beta})^d}\,
        e^{-\beta m^2-\frac{(p+p')^2+\omega^2\theta^2(p-p')^2}{4\omega\tanh{\omega\beta}}}
        \\[2mm]
            &=\frac{\lambda}{12\omega}\,
        \Gamma(\tfrac{d}2+\tfrac{m^2}{2\omega})
        \ e^{-\frac{1}{4\omega}\,\left\{(p+p')^2+\omega^2\theta^2(p-p')^2\right\}}\times
        \\ &\hspace{3cm}\times U\left(\tfrac{d}2+\tfrac{m^2}{2\omega},d;\tfrac{(p+p')^2+\omega^2\theta^2(p-p')^2}{2\omega}\right)
        \,.
  \end{split}\end{align}
 Como forma de verificar el c\'alculo, podemos tomar el l\'imite $\omega\rightarrow 0^+$ de la expresi\'on (\ref{GW:2p-gamma}). Este resultado, escrito en t\'erminos de la funci\'on modificada de Bessel $K_{n}(\cdot)$ de orden $n$ \parencite{A-S}, es id\'entico al ya expresado en la f\'ormula (\ref{NC:phi4.bess}) y muestra que, para $\omega=0$, la correcci\'on no planar diverge para momentos entrantes peque\~nos\footnote{Recordemos que este efecto recibe el nombre de mezcla UV-IR \parencite{Minwalla:1999px}.}:
\begin{align}\label{GW:2p-gf-np-ti}
        \Gamma^{(2)}_{\rm NP}(p,p')
        \overset{\omega\rightarrow 0^+}{\rightarrow}
        \frac{\lambda}{3}\,\pi^{\frac{d}2}m^{\frac{d}2-1}
        \ \frac{K_{d/2-1}(2m\theta |p|)}{(\theta |p|)^{\frac{d}2-1}}\ \delta(p+p').
\end{align}

\begin{figure}[t]
\centering
\begin{minipage}{.8\textwidth}
\centering
\includegraphics[height=45mm]{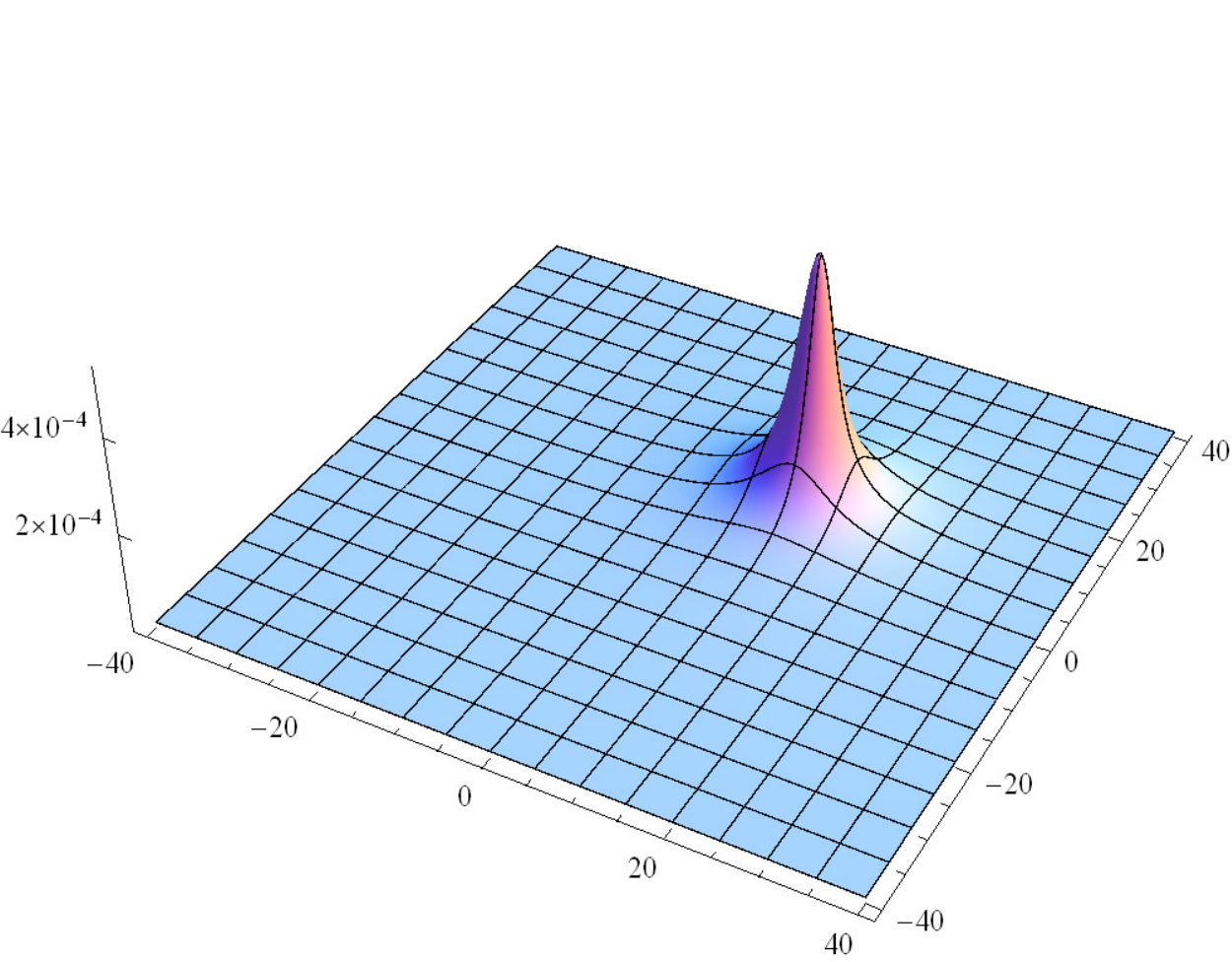}
\caption{\small Potencial de fondo no local $U_{NP}(x,x')$ como funci\'on de $x\in\mathbb{R}^2$ para $x'=(10,10)\in\mathbb{R}^2$, $\lambda=1$ y $\omega=0.1$ (la unidad de longitud es $\sqrt{\theta}$).}
\label{fig:GW:1}
\end{minipage}
\end{figure}

\begin{figure}[t]
\begin{minipage}{.8\textwidth}
\centering
\includegraphics[height=45mm]{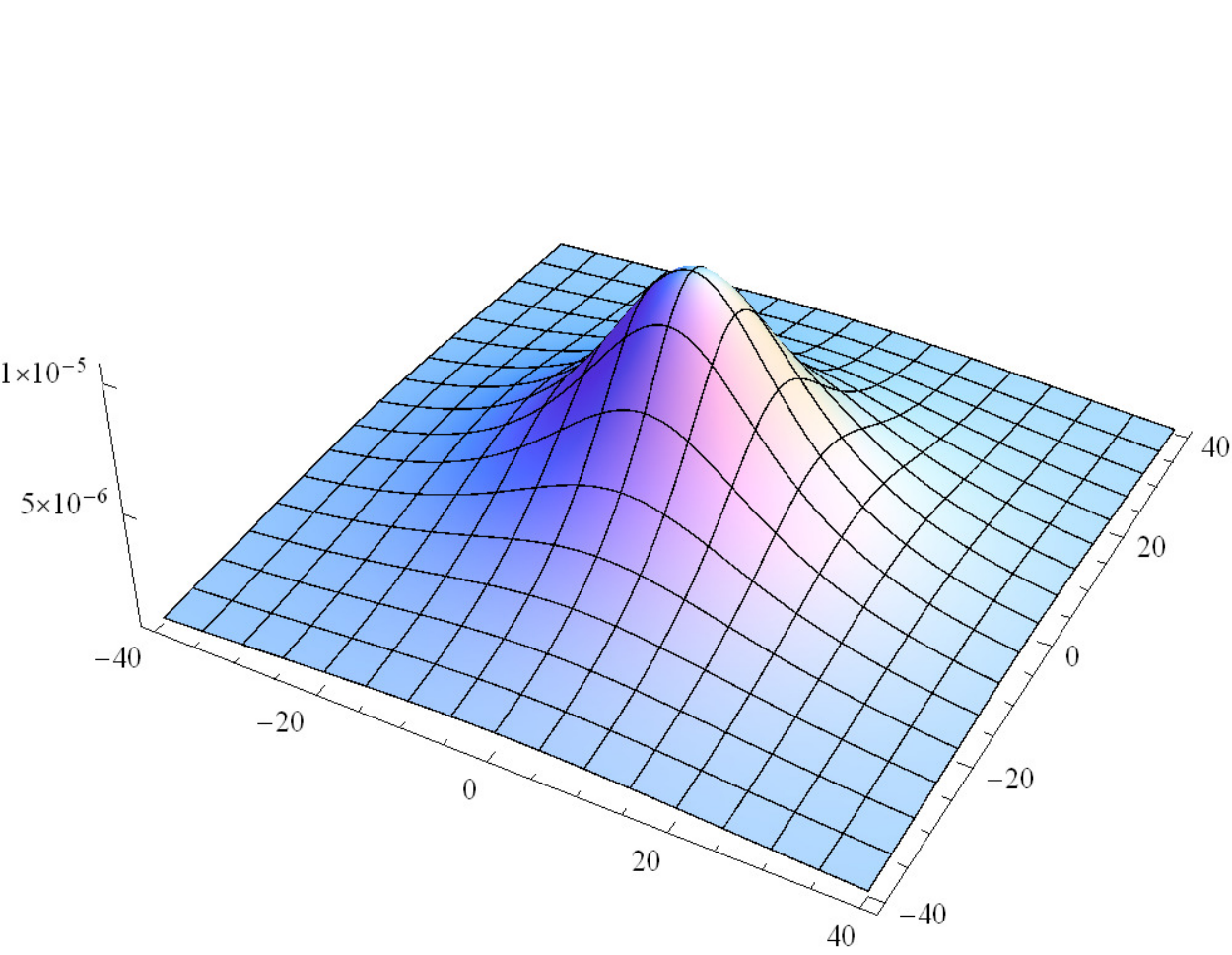}
\caption{\small Potencial de fondo no local $U_{NP}(x,x')$ como funci\'on de $x\in\mathbb{R}^2$ para $x'=(10,10)\in\mathbb{R}^2$, $\lambda=1$ y $\omega=1$ (la unidad de longitud es$\sqrt{\theta}$).}
\label{fig:GW:2}
\end{minipage}
\end{figure}

Retomando la ecuaci\'on (\ref{GW:2p-gamma}), conviene notar que hemos eliminado el pa\-r\'a\-metro de corte $\Lambda$, puesto que la integral es convergente en el r\'egimen UV. Por este motivo e independientemente de la dimensi\'on $d$ del espacio, la contribuci\'on no planar representa una correcci\'on no local finita, que puede escribirse como
\begin{equation}
        \Gamma^{(2)}_{{\rm NP}}[\phi]
        =\frac12\int_{\mathbb{R}^{2d}}dxdx'\, \phi(x)\phi(x')\  U_{\rm NP}(x,x')
        \,,
\end{equation}
donde el potencial de fondo $U_{\rm NP}(x,x')$ inducido por la autointeracci\'on es, a orden $\hbar$, 
\begin{equation}\label{GW:polpot}
  U_{\rm NP}(x,x')=
  \frac{\lambda}{6}\,\frac{1}{(4\pi\theta)^d}\int_0^\infty d\beta\,\frac{e^{-\beta m^2}}{(\cosh{\omega\beta})^d}
    \ e^{-\frac{\tanh{\omega\beta}}{4\omega\theta^2}\left[(x-x')^2+\omega^2\theta^2(x+x')^2\right]}
  \,.
\end{equation}
Para el caso particular de un campo sin masa ($m=0$) y en un ET bidimensional ($d=2$), la expresi\'on (\ref{GW:polpot}) se puede escribir en forma cerrada como
\begin{equation}
  U_{\rm NP}(x,x')\rvert_{m=0,d=2}=
  \frac{\lambda}{24\pi^2}\ \frac{\left(1-e^{-\frac{1}{4\omega\theta^2}\,\left[(x-x')^2+\omega^2\theta^2(x+x')^2\right]}\right)}%
  {\left[(x-x')^2+\omega^2\theta^2(x+x')^2\right]}
  \,.
\end{equation}
En las Figuras \ref{fig:GW:1} y \ref{fig:GW:2} mostramos los gr\'aficos de $U_{NP}(x,x')$ como funci\'on de $x\in\mathbb{R}^2$ para dos valores diversos de $\omega\theta$. Observamos que el potencial de fondo muestra un pico agudo alrededor de $x=x'$ para peque\~nos valores de $\omega\theta$, mostrando una gran dependencia en la distancia $\lvert x-x'\lvert$ (ver Figura \ref{fig:GW:1}). Por el contrario, cerca del punto autodual $\omega\theta=1$, el fondo depende \'unicamente de $\lvert x\lvert$  y $\lvert x'\lvert$ por separado, repeliendo el campo del origen (ver Figura \ref{fig:GW:2}). Por \'ultimo, a medida que $\omega\theta$ se incrementa, la funci\'on $U_{NP}$ se concentra cada vez m\'as en la regi\'on alrededor de $x=-x'$.

\subsection{La funci\'on de cuatro puntos}\label{GW.4p}
Luego de haber estudiado en el par\'agrafo previo la renormalizaci\'on del t\'ermino cuadr\'atico de la AE, consideraremos a continuaci\'on la del t\'ermino cu\'artico, que involucra la constante de acoplamiento $\lambda$. Con tal fin, examinemos la contribuci\'on de un bucle a la funci\'on de cuatro puntos $\Gamma^{(4)}$, indexada con $n=2$ en la f\'ormula (\ref{GW:eff-action-pt2}):
\begin{multline}\label{GW:eff-action-pt4}
    \Gamma^{(4)}[\phi]=-\frac{1}{2}
        \int d\tilde{\sigma}_1d\tilde{\sigma}_2d\tilde{\xi}_1d\tilde{\xi}_2
        \ \tilde{V}_W(\sigma_1,\xi_1)
        \,\tilde{V}_W(\sigma_2,\xi_2)\,\times
        \\
        \times \int_{\Lambda^{-2}}^{\infty} d\beta\,\beta\,
        \frac{e^{-\beta m^2}}{(2\sinh{\omega\beta})^{d}}
        \ K^{(2)}_\beta
        \,.
\end{multline}

Las divergencias de la expresi\'on (\ref{GW:eff-action-pt4}) en el l\'imite $\Lambda\rightarrow\infty$ pueden aislarse considerando los primeros t\'erminos del desarrollo de $K^{(2)}_\beta$ para peque\~nos valores de $\beta$. Como veremos, para ET de dimensiones $d\leq4$, bastar\'a con tomar el t\'ermino principal de la expansi\'on para $\beta$ peque\~no de la f\'ormula (\ref{GW:kn}) para $n=2$:
\begin{multline}\label{GW:k2b0}
    K^{(2)}_\beta(\sigma_1,\sigma_2;\xi_1,\xi_2)\sim
    \,e^{-\frac{1}{4\omega^2\beta}\,\left\{\omega^2(\xi_1+\xi_2)^2+(\sigma_1+\sigma_2)^2\right\}}\times\\
    \times  \int_{-1/2}^{1/2}dt\,\left(\tfrac12-t\right)\,e^{-it\,\left(\xi_1\sigma_2-\xi_2\sigma_1\right)}
    \,.
\end{multline}

Esta expresi\'on nos permite realizar el siguiente razonamiento: debido a la presencia del prefactor exponencial, bajo su s\'imbolo integral las variables $\xi_1,\sigma_1\sim\sqrt{\beta}$; por ello, podemos reemplazar el desarrollo (\ref{GW:k2b0}) en (\ref{GW:eff-action-pt4}) aproximando la exponencial imaginaria a su valor en cero. Adem\'as, dado que los factores restantes dependen solo de las sumas $\xi_1+\xi_2$ y $\sigma_1+\sigma_2$, las integrales en (\ref{GW:eff-action-pt4}) dan como resultado la convoluci\'on de las transformadas de Fourier $\tilde{V}_W$. En definitiva, como funcional del potencial $V_W$ la contribuci\'on cu\'artica toma la forma
\begin{multline}\label{GW:eff-action-4p}
    \Gamma^{(4)}[\phi]=-\frac{1}{4}\frac{1}{(2\pi)^d}
        \int_{\mathbb{R}^{2d}} dxdp\ V^2_W(x,p)\times \\
        \times \int_{\Lambda^{-2}}^{\infty}d\beta\,\beta\,e^{-\beta m^2}\,\left(\frac{\omega\beta}{\sinh{\omega\beta}}\right)^d
        \,e^{-\beta\omega^2x^2-\beta p^2}
        +\ldots\,,
\end{multline}
donde los puntos indican contribuciones de orden superior en $\beta$. 

Notemos que si la integral de $V^2_W(x,p)$ en el espacio de fases  fuera finita, podr\'iamos acotar las exponenciales $e^{-\beta\omega^2x^2-\beta p^2}\leq 1$ de la f\'ormula (\ref{GW:eff-action-pt4})  y mostrar que la contribuci\'on resulta finita en el l\'imite $\Lambda\rightarrow\infty$. Este es el caso de las correcciones no planares, para las cuales el factor $V_W^2(x,p)$ depende de ambas combinaciones independientes $(x\pm\Theta p)$. 

A la inversa, el t\'ermino planar $(\phi^2_\star(x-\Theta p))^2$ presente  en $V^2_W(x,p)$, dependiente solo de $(x-\Theta p)$, no ser\'ia integrable si removi\'eramos los factores exponenciales de la expresi\'on (\ref{GW:eff-action-4p}). En efecto, en esta ocasi\'on la exponencial $e^{-\beta\omega^2x^2-\beta p^2}$ garantiza la convergencia en la direcci\'on $(x+\Theta p)$, bajo el precio de agregar un factor $\beta^{-d/2}$; como resultado, para $d\geq4$, la contribuci\'on se vuelve divergente UV. Por 
supuesto, un argumento similar vale para el t\'ermino $(\phi^2_\star(x+\Theta p))^2$. A fin de cuentas, las contribuciones divergentes para la funci\'on de cuatro puntos pueden obtenerse del desarrollo 
\begin{multline}\label{GW:4p-div}
    \Gamma^{(4)}[\phi]=-\frac{\lambda^2}{72}
        \int_{\mathbb{R}^d} dx
        \,\frac{\left\{\phi^2_\star(x)\right\}^2}{(4\pi)^{d/2}}\times\\
        \times
        \int_{\Lambda^{-2}}^{\infty}d\beta\,\beta^{1-\tfrac d2}\left(\frac{\omega\beta}{\sinh{\omega\beta}}\right)^d
        \frac{e^{-\beta\left(m^2+\frac{\omega^2x^2}{1+\omega^2\theta^2}\right)}}{(1+\omega^2\theta^2)^{d/2}}
        +\ldots
\end{multline}
En dos dimensiones, esta expresi\'on es regular en el r\'egimen UV y por lo tanto implica solo una renormalizaci\'on finita de la constante de acoplamiento $\lambda$. En cambio, para $d\geq4$, las contribuciones de un bucle a la funci\'on de cuatro puntos son divergentes.

En particular, para $d=4$ la divergencia en (\ref{GW:4p-div}) proviene del \'unico t\'ermino escrito, visto que los puntos suspensivos indican contribuciones $O(\beta^0)$; reescribi\'endola en la forma 
\begin{equation}\label{GW:4p-div-d=4}
    \Gamma^{(4)}[\phi]=-\frac{\lambda^2}{1152\,\pi^2}
    \,\frac{1}{(1+\omega^2\theta^2)^2}
        \int_{\mathbb{R}^4} dx
        \ \phi^4_\star(x)
        \int_{\Lambda^{-2}}^{\omega^{-1}}\frac{d\beta}{\beta}\ +{\rm (\text {t. f.})}\,,
\end{equation}
observamos que  puede ser removida introduciendo la constante de acoplamiento renormalizada, $\lambda_R$; su c\'alculo, recordando la apropiada renormalizaci\'on de los campos, es directo:
\begin{equation}\label{GW:lambdar}
    \lambda_{\rm ren}=Z_4^2\,\lambda\left\{1-\frac{\lambda}{48\pi^2}\,\frac{1}{(1+\omega^2\theta^2)^2}
    \left[\log{\left(\Lambda^2/\omega\right)}+{\rm (t.\ f.)}\right]\right\}\,.
\end{equation}
Ahora, haciendo uso de la ecuaci\'on (\ref{GW:lambdar}) y del desarrollo para $Z_4$ dado por la f\'ormula (\ref{GW:z.renor}),  podemos calcular la correspondiente funci\'on $\beta_{\lambda}$ para la constante de acoplamiento, 
\begin{equation}\label{GW:betalambda}
    \beta_\lambda:=\Lambda\,\partial_\Lambda\lambda=\frac{\lambda^2}{24\pi^2}
    \ \frac{\left(1-\theta^2\omega^2\right)}{\left(1+\theta^2\omega^2\right)^3}\,,
\end{equation}
la cual se anula en el punto autodual $\omega\theta=1$. El conjunto de las expresiones \eqref{GW:funcionesbeta} y \eqref{GW:betalambda} implica que el flujo del grupo de renormalizaci\'on tiene, al orden de un bucle, un punto fijo para $\omega\theta=1$ y un valor finito de $\lambda$; ello justifica la aseveraci\'on $\omega\sim\theta^{-1}$ y el hecho de que el potencial arm\'onico deba ser tenido en cuenta no perturbativamente. No obstante la idea original de GW de obtener un modelo renormalizable tomando en cuenta la simetr\'ia de LS no intu\'ia la presencia de un punto fijo, este resultado no es sorpresivo si tenemos en consideraci\'on las similitudes se\~naladas por \textcite{Langmann:2003if} entre teor\'ias autoduales y cierto tipo de modelos matriciales. Notemos asimismo que de \eqref{GW:funcionesbeta} y \eqref{GW:betalambda} se desprende la igualdad
\begin{align}\label{GW:landau.beta}
\frac{\beta_{\lambda}}{\lambda}=\frac{\beta_{\omega^2}}{\omega^2}
\end{align}
y, a su vez, de la integraci\'on de \eqref{GW:landau.beta} resulta inmediato que, al considerar el flujo de la escala de energ\'ia $\Lambda$, las constantes de acoplamiento $\lambda$ y $\omega^2$ se mantienen proporcionales entre s\'i.

\begin{figure}[t]
\begin{minipage}{.9\textwidth}
\centering
\hspace{-0.5cm}\includegraphics[width=.65\textwidth]{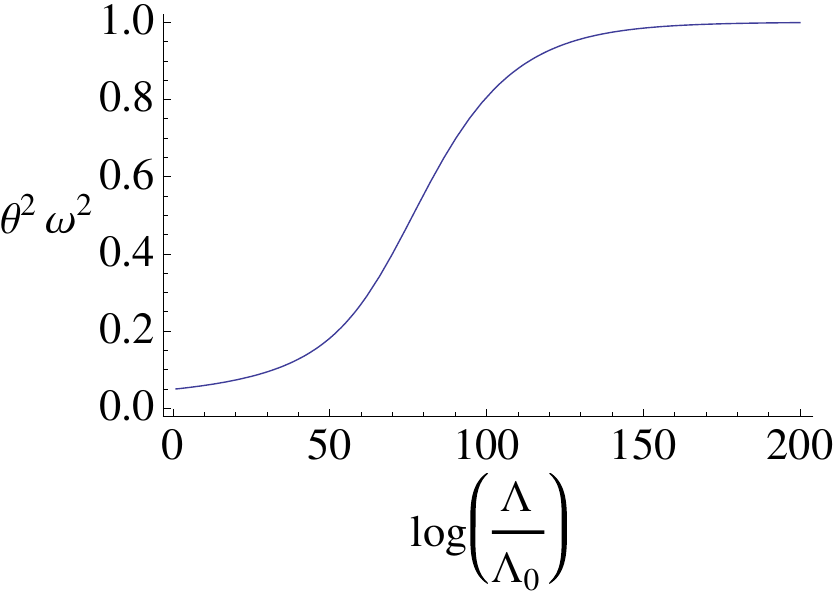}
\caption{\small Soluci\'on num\'erica para $\theta\omega$, la frecuencia $\omega$ en unidades de $\theta^{-1}$, como funci\'on de $\log\left(\frac{\Lambda}{\Lambda_0}\right)$, el logaritmo de la escala de energ\'ia $\Lambda$ relativa a la escala inicial $\Lambda_0$. Las condiciones iniciales impuestas son $\omega^2(\Lambda_0)=0.05\,\theta^{-2}$ y $\lambda(\Lambda_0)=1$.}
\label{fig:GW:flujow1}
\end{minipage}
\end{figure}

\begin{figure}
\begin{minipage}{.9\textwidth}
\centering
\hspace{-0.5cm}\includegraphics[width=.65\textwidth]{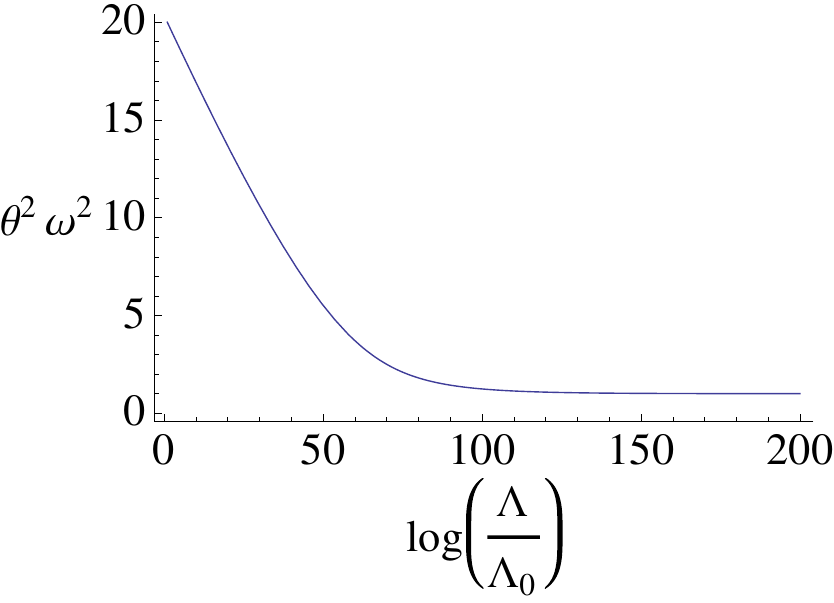}
\caption{\small Soluci\'on num\'erica para $\theta\omega$, la frecuencia $\omega$ en unidades de $\theta^{-1}$, como funci\'on de $\log\left(\frac{\Lambda}{\Lambda_0}\right)$, el logaritmo de la escala de energ\'ia $\Lambda$ relativa a la escala inicial $\Lambda_0$. Las condiciones iniciales impuestas son $\omega^2(\Lambda_0)=20\,\theta^{-2}$ y $\lambda(\Lambda_0)=1$.}
\label{fig:GW:flujow2}
\end{minipage}
\end{figure}

Para exhibir cualitativamente el flujo de las constantes con la escala de energ\'ia, resolvemos num\'ericamente el sistema de ecuaciones diferenciales que plantean las funciones $\beta$, viz. \eqref{GW:funcionesbeta} y \eqref{GW:betalambda}. En las Figuras \ref{fig:GW:flujow1} y \ref{fig:GW:flujow2}  inclu\'imos  los gr\'aficos que obtenemos para $\theta\omega(\Lambda)$, la frecuencia $\omega$ en unidades de la inversa del par\'ametro de no conmutatividad $\theta$, en funci\'on de $\log(\Lambda/\Lambda_0)$, el logaritmo de la escala de energ\'ia $\Lambda$ relativa a la escala $\Lambda_0$ que utilizamos para imponer las condiciones iniciales $\omega(\Lambda_0)=\omega_0$ y $\lambda(\Lambda_0)=\lambda_0$. 

En particular, las Figuras \ref{fig:GW:flujow1} y \ref{fig:GW:flujow2} corresponden a $\omega_0<\theta^{-1}$ y $\omega_0>\theta^{-1}$ respectivamente. De ellas se puede inferir que, efectivamente, sin importar su valor inicial la frecuencia tiende al valor l\'imite $\omega(\infty)=\theta^{-1}$ para grandes escalas de energ\'ias. Recordemos que de acuerdo a \eqref{GW:landau.beta}, el comportamiento de $\lambda$ es proporcional a $\omega^2$ y en consecuencia podemos presumir que el modelo de GW no presenta el problema del polo de Landau. En el anexo \ref{GW.landau} demostraremos esta aseveraci\'on al orden de un bucle, a la par que mostraremos que la elecci\'on de las condiciones iniciales s\'olo modifica la velocidad con la que los par\'ametros tienden a sus valores l\'imites al tomar $\Lambda$ grande.

Todos los c\'alculos hasta aqu\'i desarrollados pertenecen al estudio de una matriz $\Theta$ de no conmutatividad no degenerada. Como veremos en la secci\'on entrante, los resultados referentes a los par\'ametros de renormalizaci\'on del modelo de GW se trasladan, al menos cualitativamente, al caso en que $\Theta$ es singular.

\section[An\'alisis para una matriz $\Theta$ degenerada]{An\'alisis para una matriz de no conmutatividad degenerada}\label{GW.degenerado}
Para realizar el estudio del modelo de GW con una matriz $\Theta$ que posee subespacios conmutativos tenemos, entre varias, las siguientes dos opciones para utilizar las t\'ecnicas del FLM: la primera, que podr\'ia ser explorada siguiendo los lineamientos de la secci\'on \ref{GW.aniso}, consiste en tomar par\'ametros de no conmutatividad $\theta_{\mu}$ diversos, de tal manera que al final del d\'ia podr\'iamos imponer la conmutatividad en algunos planos tomando el l\'imite $\theta_{{\mu}}$ yendo a cero. La segunda, la elegida para desarrollar a continuaci\'on, plantea como punto de partida la existencia de los subespacios conmutativos.

Tomemos como ejemplo el ET eucl\'ideo Moyal de cuatro dimensiones, parametrizado con coordenadas $(\bar{x},x)\in\mathbb{R}^4$; cambiando ligeramente la notaci\'on de la secci\'on \ref{NC.tccnc}, $\bar{x}\in\mathbb{R}^2$ son coordendas conmutantes y $x\in\mathbb{R}^2$ describe un espacio de dos dimensiones donde la no conmutatividad est\'a caracterizada por la matriz
\begin{equation}\label{GW:deg-theta}
    \Theta=\left(\begin{array}{cccc}
      0&&&\theta\\
      -\theta&&&0
    \end{array}
        \right)\,,\qquad \theta\in\mathbb{R}\ .
\end{equation}
En este espacio, tomemos un campo escalar $\varphi(\bar{x},x)$ cuyo lagrangiano sea similar al de GW en las coordenadas no conmutativas,
\begin{equation}\label{GW:lagrangiandeg}
    \mathcal{L}=\frac{1}{2}(\bar{\partial}\varphi)^2+\frac{1}{2}(\partial\varphi)^2+\frac{m^2}{2}\varphi^2+
    \frac{\omega^2}{2}\,x^2\varphi^2+\frac{\lambda}{4!}\varphi^4_\star\,;
\end{equation}
hemos dividido el t\'ermino cin\'etico acordemente, llamando $\bar{\partial}$ y $\partial$ a los gradientes con respecto a $\bar{x}$ y $x$. Por supuesto, el producto $\star$ del t\'ermino cu\'artico est\'a definido como en la expresi\'on (\ref{NC:moyal}), recordando que en este caso las variables involucradas son $x\in\mathbb{R}^2$ y la matriz $\Theta$ est\'a dada por la ec. (\ref{GW:deg-theta}): este producto $\star$ no supone derivadas $\bar{\partial}$ con respecto a las variables conmutativas $\bar{x}$.

Buscando obtener las correcciones de un bucle a la AE de este modelo, prosigamos con el c\'alculo del operador de fluctuaciones cu\'anticas. De acuerdo a los pasos detallados en la secci\'on \ref{GW.GW}, es evidente que este operador tendr\'a la forma 
\begin{equation}\label{GW:hdeg}
    H:=-\bar{\partial}^2-\partial^2+m^2+\omega^2x^2+V(\bar{x},x,-i\partial)\,.
\end{equation}
Podemos por ende investigar la traza del NdC del operador general H; para obtener el resultado requerido, bastar\'a reemplazar posteriormente $V$ por un potencial que, salvo por la distinci\'on entre coordenadas conmutantes y no conmutantes, es equivalente al dado en (\ref{GW:qf-operator2}). 

Como primer paso, escribimos la representaci\'on usual para la traza del NdC  del operador H:
\begin{equation}\label{GW:pi-h-deg}
    {\rm Tr}\,e^{-\beta H}
        =e^{-\beta m^2}\mathcal{N}(\beta) \int_{\mathbb{R}^2} d\bar{x}
        \int\mathcal{D}\bar{q}\mathcal{D}q\mathcal{D}p\ e^{-\mathcal{S}[\bar{q},q,p]}
        \ e^{-\beta\int_0^{1}dt\,V_W}
          \,.
\end{equation}
Esta f\'ormula merece varias aclaraciones. Entre ellas, notemos que hemos realizado los reescaleos necesarios en las variables de los caminos y el tiempo propio para que $V_W:=V_W(\bar{x}+\sqrt{\beta}\,\bar{q}(t),\sqrt{\beta}\,q(t),p(t)/\sqrt{\beta})$. 
Igualmente, al escribirla hemos aprovechado que $V_W$ no depende de $\bar{\partial}$ y hemos integrado el momento conjugado $\bar{p}(t)$: resta efectuar solo la integral en las trayectorias $\bar{q}(t)$ del espacio de configuraci\'on. Por su parte, estas trayectorias $\bar{q}(t)$ satisfacen condiciones de contorno Dirichlet $\bar{q}(0)=\bar{q}(1)=0$, mientras que las trayectorias correspondientes al plano no conmutante, $q(t)$ y $p(t)$, satisfacen condiciones de contorno peri\'odicas. Esto explica la aparici\'on de la integral sobre la variable $\bar{x}$ para obtener la traza\footnote{El uso de trayectorias peri\'odicas $\bar{q}(t)$ no simplificar\'ia el c\'alculo de la traza; ello se debe a la existencia de un modo cero para la acci\'on de la part\'icula libre bajo \'estas condiciones de contorno.}. Finalmente, los t\'erminos en la acci\'on de la part\'icula puntual que no dependen del potencial $V$ admiten una separaci\'on  $\mathcal{S}[\bar{q},q,p]=\bar{\mathcal{S}}[\bar{q}]+\mathcal{S}[q,p]$, con
\begin{align}\label{GW:actionpoint-deg}
\begin{split}
  \bar{\mathcal{S}}[\bar{q}]&:=\frac14\int_0^1 dt\ \dot{\bar{q}}^2(t)\,,\\
  \mathcal{S}[q,p]&:=\int_0^1 dt\,\left\{p(t)^2-ip(t)\dot{q}(t)+\omega^2\beta^2q(t)^2\right\}\,.
\end{split}
\end{align}
Alternativamente, podemos aseverar que ser\'a el t\'ermino potencial $V$ el que decidir\'a si se puede desacoplar el problema en uno conmutativo y otro NC.

En la forma habitual, la traza \eqref{GW:pi-h-deg} puede ser entendida como un valor de expectaci\'on cuya definici\'on involucra la medida gaussiana dada por la IdC de las acciones (\ref{GW:actionpoint-deg}).
El paso siguiente consiste en realizar un desarrollo en potencias de la transformada de Fourier $\tilde{V}_W$; si como en los casos previos determinamos la normalizaci\'on de manera que la expresi\'on \eqref{GW:pi-h-deg} contemple el resultado conocido para $V_W=0$, llegamos a la f\'ormula
\begin{multline}{\label{GW:pi3-deg}}
        {\rm Tr}\,e^{-\beta H}
        =\frac\pi4\,\frac{e^{-\beta m^2}}{\beta\sinh^2{\omega\beta}}
        \ \sum_{n=1}^\infty (-\beta)^n
        \int_{\mathbb{R}^{6}}d\tilde{\bar{\sigma}}_1d\tilde{\sigma}_1d\tilde{\xi}_1\ldots
        \int_{\mathbb{R}^{6}}d\tilde{\bar{\sigma}}_n\tilde{\sigma}_nd\tilde{\xi}_n\,\times\\[2mm]
        \times\,\delta\left(\bar{\sigma}_1+\ldots+\bar{\sigma}_n\right)\,\tilde{V}_W(\bar{\sigma}_1,\sigma_1,\xi_1)\ldots \tilde{V}_W(\bar{\sigma}_n,\sigma_n,\xi_n)
        \times K^{(n)}_\beta
        \,.
\end{multline}
Estas funciones $K^{(n)}_\beta$ quedan completamente definidas por las FG para la part\'icula libre con condiciones de contorno Dirichlet ($Z_D[j]$) y con condiciones de contorno peri\'odicas bajo la acci\'on de un potencial arm\'onico ($Z_{per}[k,j]$), escritas en \eqref{FLM:zetafunction} y \eqref{GW:z} respectivamente:
\begin{equation}\label{GW:kn00-deg}
    K^{(n)}_\beta:=
    \int_0^1dt_1\ldots\int_0^{t_{n-1}}dt_n\,
  Z_{per}[k_n,j_n]\,Z_D[\bar{j}_n];
\end{equation}
las fuentes que deben ser empleadas involucran las variables temporales intermedias $t_i$ y las coordenadas de Fourier $\xi_i$, $\sigma_i$ y $\bar{\sigma}_i$ en la forma
\begin{align}\label{GW:kdeg}
    \begin{split}
k_n(t):&=i\,\beta^{-\frac12}\,\sum_{i=1}^{n}\delta(t-t_i)\,\xi_i\,,\\
j_n(t):&=i\,\beta^{\frac12}\,\sum_{i=1}^{n}\delta(t-t_i)\,\sigma_i\,,\\
\bar{j}_n(t):&i\,=\beta^{\frac12}\,\sum_{i=1}^{n}\delta(t-t_i)\,\bar{\sigma}_i\,.\quad
\end{split}
\end{align}

Conviene notar que hemos eliminado en (\ref{GW:pi3-deg}) el t\'ermino dado por $n=0$, ya que corresponde a la normalizaci\'on de la FG de los campos\footnote{Cfr. \eqref{DET:w1} y los comentarios posteriores.}. En tanto, la funci\'on delta con soporte en $\sum_i \bar{\sigma}_i$ es consecuencia de la integraci\'on hecha en $\bar{x}$.  Adicionalmente, la descomposici\'on \eqref{GW:actionpoint-deg} se ve reflejada en la aparici\'on del producto de funciones generatrices en (\ref{GW:kn00-deg}).

Retomando el hilo de la discusi\'on, las contribuciones $\Gamma^{(2)}$ a la AE, cuadr\'aticas en el campo,  est\'an indexadas por $n=1$ en la ecuaci\'on (\ref{GW:pi3-deg}), debido a la estructura  tambi\'en cuadr\'atica del potencial $V_W$:
\begin{equation}\label{GW:2p-gf-deg}
    \Gamma^{(2)}[\phi]=
        \frac{1}{32\pi}
        \int_{\mathbb{R}^{4}}d\tilde\sigma d\tilde\xi\ \tilde{V}_W(0,\sigma,\xi)
        \int_{\Lambda^{-2}}^{\infty} d\beta\,
        \frac{e^{-\beta m^2}}{\beta\sinh^2{\omega\beta}}
        \ K^{(1)}_\beta
    \,.
\end{equation}
En esta expresi\'on hemos empleado la delta de Dirac para eliminar la integral en $\bar{\sigma}$, y, considerando \eqref{FLM:zetafunction} y \eqref{GW:K1beta.nosingular}, la funci\'on $K^{(1)}_\beta$ es sencillamente
\begin{equation}
    K^{(1)}_\beta=e^{-\frac{1}{4\omega\tanh{\omega\beta}}\,(\omega^2\xi^2+\sigma^2)}\,.
\end{equation}
Para facilitar la exposici\'on, nuevamente separaremos el estudio de los casos planares y no planares.

\subsection{Contribuciones planares}
Si consideramos los t\'erminos de $\tilde{V}_W$ que no mezclan productos Moyal a izquierda y derecha, obtenemos la contribuci\'on planar a la funci\'on de dos puntos
\begin{align}\label{GW:2p-gf-p2-deg}
        \Gamma^{(2)}_{{\rm P}}[\phi]
        =\frac{1}{2}\int_{\mathbb{R}^{4}}d\bar{x}dx\ \Gamma^{(2)}_{{\rm P}}(x)\star\phi(\bar{x},x)\star\phi(\bar{x},x)\,,
\end{align}
efectuando la regularizaci\'on de $\Gamma^{(2)}_{{\rm P}}(x)$ por medio de un par\'ametro $\Lambda$  de corte  UV:
\begin{equation}
  \Gamma^{(2)}_{{\rm P}}(x)=\frac{\lambda}{24\pi^2}\,\frac{\omega}{(1+\omega^2\theta^2)}
        \int_{\Lambda^{-2}}^{\infty} \frac{d\beta}{\beta}\, \frac{e^{-\beta m^2}}{\sinh{2\omega\beta}}
        \ e^{-\frac{\omega\tanh{\omega\beta}}{1+\omega^2\theta^2}\,x^2}\,.\label{GW:eff-action-deg}
\end{equation}
Este resultado es cualitativamente igual al obtenido para las contribuciones planares en el caso no degenerado, cfr. (\ref{GW:eff-back}). Desde el punto de vista de la renormalizaci\'on, $\Gamma^{(2)}_{{\rm P}}(x)$ posee una divergencia UV independiente de $x$ y proporcional a $\Lambda^2$ que puede ser removida renormalizando la masa, y una divergencia logar\'itmica que se elimina redefiniendo la normalizaci\'on de los campos y la frecuencia del t\'ermino harm\'onico. 

En forma precisa, las correcciones de un bucle al t\'ermino cin\'etico ahora s\'olo introducen divergencias en las direcciones NC. Como consecuencia, para curarlas debemos introducir no s\'olo un par\'ametro de renormalizaci\'on del campo $Z$, sino tambi\'en uno nuevo $a$ que tenga en cuenta esta asimetr\'ia entre direcciones conmutativas y no conmutativas \parencite{Grosse:2012my}; en consecuencia el t\'ermino cin\'etico conmutativo del lagrangiano deber\'ia leerse $a(\bar{\partial}\phi)^2$. La lista completa de par\'ametros involucrados en la renormalizaci\'on de la funci\'on de dos puntos es
\begin{align}
\begin{split}
    Z&=1+\frac{\lambda}{48\pi^2}\,\frac{\omega^2\theta^2}{(1+\omega^2\theta^2)^2}
    \left[\log{\left(\Lambda^2/\omega\right)}+{\rm (t.\ f.)}\right]\,,\label{GW:zdeg}\\
    a_{ren}&=a\,\left\lbrace 1-\frac{\lambda}{48\pi^2}\,\frac{\omega^2\theta^2}{(1+\omega^2\theta^2)^2}
    \left[\log{\left(\Lambda^2/\omega\right)}+{\rm (t.\ f.)}\right]\right\rbrace\,,\\
    \omega^2_{\rm ren}&=\omega^2\left\{1-\frac{\lambda}{48\pi^2}\,\frac{1-\omega^2\theta^2}{(1+\omega^2\theta^2)^2}
    \left[\log{\left(\Lambda^2/\omega\right)}+{\rm (t.\ f.)}\right]\right\}\,,\\
m_{\rm ren}^2&=m^2\left\{1+\frac{\lambda}{48\pi^2}\frac{1}{(1+\omega^2\theta^2)\,m^2}\Lambda^2+\right.\\
&\hspace{2cm}+\left.\frac{\lambda}{48\pi^2}\,\frac{\omega^2\theta^2}{(1+\omega^2\theta^2)^2}
    \left[\log{\left(\Lambda^2/\omega\right)}+\text{(t. f.)}\right]\right\}.
\end{split}
\end{align}
Estos resultados, comparados con (\ref{GW:z.renor}) y (\ref{GW:w.renor}), muestran una potencia menor de $(1+\omega^2\theta^2)$ en los t\'erminos divergentes; esta diferencia es entendible observando la dependencia con la dimensi\'on en \eqref{GW:eff-back}.  

\subsection{Contribuciones no planares}
La contribuci\'on no planar se calcula insertando en (\ref{GW:2p-gf-deg}) el t\'ermino del potencial que mezcla productos Moyal a izquierda y a derecha; el resultado es 
\begin{equation}\label{GW:2p-gf-np2-deg}
        \Gamma^{(2)}_{{\rm NP}}[\phi]
        =\frac{1}{2}\int_{\mathbb{R}^{6}}d\bar{x}dxdx'\,\phi(\bar{x},x)\,\phi(\bar{x},x')\ U_{{\rm NP}}(x,x')\,,
\end{equation}
en t\'erminos del potencial de fondo $U_{{\rm NP}}(x,x')$ que debe ser regularizado en el regimen UV,
\begin{equation}
  U_{{\rm NP}}(x,x')=\frac{\lambda}{384\pi^3}\,\frac{1}{\theta^2}
        \int_{\Lambda^{-2}}^{\infty} \frac{d\beta}{\beta}\, \frac{e^{-\beta m^2}}{\cosh^2{\omega\beta}}
        \ e^{-\frac{\tanh{\omega\beta}}{4\omega\theta^2}\,\left[(x-x')^2+\omega^2\theta^2(x+x')^2\right]}\,.\label{GW:eff-back-deg}
\end{equation}
Las diferencias con el caso no degenerado son en esta oportunidad evidentes: 
mientras que cuando la matriz $\Theta$ es no degenerada, de acuerdo a (\ref{GW:2p-gf-np-ti}), las contribuciones no planares son finitas  sin importar la dimensi\'on de la variedad de base, tomando un subespacio degenerado de dimensi\'on\footnote{La divergencia de la contribuci\'on no planar se se incrementa con el n\'umero de variables conmutativas, pues se ganan  potencias $\beta^{-1/2}$ en perjuicio de  factores $(\cosh^2{\omega\beta})^{-1}$ dentro del integrando de la expresi\'on \eqref{GW:eff-back-deg}.} 2 obtenemos una divergencia UV logar\'itmica
\begin{equation}\label{GW:lognonpladeg}
  U_{{\rm NP}}(x,x')=\frac{\lambda}{384\pi^3}\,\frac{1}{\theta^2}\,\log{\Lambda^2/\omega}+{\rm (t.\ f.)}\,.
\end{equation}
Este tipo de divergencia no puede ser eliminada mediante la redefinici\'on de los par\'ametros del lagrangiano (\ref{GW:lagrangiandeg}), dado que el mismo no posee un t\'ermino con la forma de (\ref{GW:2p-gf-np2-deg}). Para lograr la renormalizaci\'on, se vuelve necesario introducir una nueva interacci\'on no local al lagrangiano, en conformidad con lo descripto en el trabajo de \textcite{Grosse:2008df}:
\begin{equation}
        \frac{1}{2}\,\frac{\kappa^2}{\theta^2}\int_{\mathbb{R}^{6}}d\bar{x}dxdx'\,\phi(\bar{x},x)\,\phi(\bar{x},x')\,.
\end{equation}
De este modo, la divergencia logar\'itmica encontrada en la f\'ormula (\ref{GW:lognonpladeg}) se puede remover por v\'ia de la renormalizaci\'on del par\'ametro $\kappa$. Las funciones $\beta$ que se obtienen sin dificultad a partir de  (\ref{GW:zdeg}) y (\ref{GW:lognonpladeg}) coinciden con los resultados de  \textcite{Grosse:2012my}.

\section{Anisotrop\'ia en el t\'ermino harm\'onico}\label{GW.aniso}
Finalizaremos este cap\'itulo analizando qu\'e sucede cuando el t\'ermino harm\'onico del modelo de GW es anis\'otropo, a saber, cuando el lagrangiano toma la forma\footnote{De aqu\'i en adelante, escribiremos expl\'icitamente todas las sumas sobre \'indices para evitar confusiones.}
\begin{equation}\label{GW:anisolagrangian}
    \mathcal{L}= \overset{d}{\underset{{\mu}=1}{\sum}} \left[\frac{1}{2}(\partial_{\mu}\phi)^2+
    \frac{\omega_{\mu}^2}{2}\,x_{\mu}^2\phi^2\right]+\frac{m^2}{2}\phi^2+\frac{\lambda}{4!}\phi^4_\star\,.
\end{equation}
En adici\'on, tomaremos una matriz de no conmutatividad $\Theta$ que en estas coordenadas se encuentra expresada en $d/2$ bloques de la forma (\ref{GW:deg-theta}) con constantes $\theta_{\mu}$ en principio diversas entre s\'i. Tomando en cuenta que el desarrollo de los c\'alculos involucrados en la obtenci\'on de las correcciones de un bucle a la AE es visiblemente semejante al realizado en las secciones previas, no nos detendremos en c\'alculos intermedios sino que mencionaremos los resultados m\'as relevantes. 

El operador de fluctuaciones de este modelo posee la forma de un operador $H$ como el de la ecuaci\'on (\ref{GW:h}); para este tipo de operadores, la anisotrop\'ia implica un peque\~no cambio en la f\'ormula (\ref{GW:pi3}), que ahora se lee
\begin{multline}{\label{GW:anisopi3}}
{\rm Tr}\,e^{-\beta H}
        =\frac{e^{-\beta m^2}}{\underset{{\mu}=1}{\overset{d}{\prod}}2\sinh{\omega_{\mu}\beta}}
        \ \sum_{n=0}^\infty (-\beta)^n
        \int_{\mathbb{R}^{2d}}d\tilde{\sigma}_1d\tilde{\xi}_1\ldots
        \int_{\mathbb{R}^{2d}}d\tilde{\sigma}_nd\tilde{\xi}_n\,\times\\[2mm]
        \times\,\tilde{V}_W(\sigma_1,\xi_1)\ldots \tilde{V}_W(\sigma_n,\xi_n)
        \times K^{(n)}_\beta(\sigma_1,\ldots;\xi_1,\ldots)
        \,.
\end{multline}
Como era de esperar, tambi\'en las funciones $K^{(n)}_\beta$ deben modificarse en concordancia con el nuevo t\'ermino arm\'onico. Llamando $\xi^{\mu}_{i}$ a la ${\mu}$-\'esima componente de la coordenada $\xi_i$, la expresi\'on (\ref{GW:kn}) se ve modificada seg\'un
\begin{multline}\label{GW:anisokn}
    K^{(n)}_\beta(\sigma_1,\ldots;\xi_1,\ldots)=
    e^{-\overset{d}{\underset{{\mu}=1}{\sum}}\frac{1}{4\omega_{\mu}\tanh{\omega_{\mu}\beta}}\, \stackrel[i]{}{\sum}(\omega_{\mu}^2{\xi^{{\mu}}_{i}}^2+{\sigma^{{\mu}}_{i}}^2)}
    \!\!\int_0^1\!\!dt_1\ldots\!\int_0^{t_{n-1}}\!\!dt_n\ \times\\[2mm]
    \times\,
    e^{-\overset{d}{\underset{{\nu}=1}{\sum}}\frac{1}{2\omega_{\nu}\sinh{\omega_{\nu}\beta}}\, \stackrel[i<j]{}{\sum}
    \left\{\cosh{\left[\omega_{\nu}\beta(2|t_i-t_j|-1)\right]}\,\left(\omega_{\nu}^2\xi_{i}^{{\nu}}\xi_{j}^{{\nu}}+\sigma_{i}^{{\nu}}\sigma_{j}^{{\nu}}\right)
    \right\}}\times\\
    \times e^{-\overset{d}{\underset{{\rho}=1}{\sum}}\frac{1}{2\sinh{\omega_{\rho}\beta}}\, \stackrel[i<j]{}{\sum}
    \left\{i\sinh{\left[\omega_{\rho}\beta(2|t_i-t_j|-1)\right]}\,\left(\xi_{i}^{{\rho}}\sigma_{j}^{{\rho}}-\xi_{j}^{{\rho}}\sigma_{i}^{{\rho}}\right)\right\}}
    \,.
\end{multline}

Para analizar como entra en juego la no conmutatividad en este nuevo modelo, tomemos por caso la contribuci\'on de un bucle a la funci\'on de 2 puntos,
\begin{multline}\label{GW:aniso2p-gf}
    \Gamma^{(2)}[\phi]=
        \frac{1}{2}
        \int_{\mathbb{R}^{2d}}d\tilde\sigma d\tilde\xi
        \ \tilde{V}_W(\sigma,\xi)\times\\
        \times\int_{\Lambda^{-2}}^{\infty} d\beta\,
        \left({\underset{\mu=1}{\overset{d}{\prod}} 2\sinh{\omega_{\mu}\beta}}\right)^{-1} {e^{-\beta m^2}}        \ K^{(1)}_\beta(\sigma;\xi)
    \,,
\end{multline}
donde, visto lo expresado en (\ref{GW:anisokn}), $K^{(1)}_\beta$ es una ligera variaci\'on de \eqref{GW:K1beta.nosingular}:
\begin{equation}
    K^{(1)}_\beta(\sigma;\xi)=e^{-\underset{{\mu}=1}{\overset{d}{\sum}}\frac{1}{4\omega_{\mu}\tanh{\omega_{\mu}\beta}}\,(\omega_{\mu}^2\xi_{{\mu}}^2+\sigma_{{\mu}}^2)}\,.
\end{equation}
Haciendo uso de estas ecuaciones, es posible mostrar que la contribuci\'on planar a la funci\'on de dos puntos puede a\'un ser escrita en la forma
\begin{align}
\label{GW:aniso2p-gf-p2}
\begin{split}
        \Gamma^{(2)}_{{\rm {P}}}[\phi]
        &=\frac{1}{2}\int_{\mathbb{R}^{d}}dx
        \ \Gamma^{(2)}_{{\rm {P}}}(x)\star\phi(x)\star\phi(x)\,,
\end{split}
\end{align}
definiendo adecuadamente $\Gamma^{(2)}_{{\rm {P}}}$; esta funci\'on posee una estructura de divergencias id\'entica a la del caso analizado en la secci\'on \ref{GW.2p}, lo cual es claro de la siguiente f\'ormula:
\begin{multline}
  \Gamma^{(2)}_{{\rm {P}}}(x):=\frac{\lambda}{3}\;\left\{\underset{{\mu}=1}{\overset{d}{\prod}} \;\frac{\omega_{\mu}}{2\pi(1+\omega_{\mu}^2\theta_{\mu}^2)}\right\}^{1/2}\times\\
          \times\int_{\Lambda^{-2}}^{\infty} d\beta\,\left(\underset{{\nu}=1}{\overset{d}{\prod}} \sinh{2\omega_{\nu}\beta}\right)^{-1/2}
        e^{-\beta m^2}
        \ e^{-\underset{{\rho}=1}{\overset{d}{\sum}}\frac{\omega_{\rho}\tanh{\omega_{\rho}\beta}}{1+\omega_{\rho}^2\theta_{\rho}^2}\,x_{\rho}^2}\,.\label{GW:anisoeff-back}
\end{multline}

Por cierto, dichas contribuciones a la funci\'on de dos puntos tienen su contraparte no planar $\Gamma^{(2)}_{{\rm NP}}$, proveniente siempre del t\'ermino con productos Moyal cruzados en el potencial $V_W$:
\begin{equation}
        \Gamma^{(2)}_{{\rm NP}}[\phi]
        =\frac12\int_{\mathbb{R}^{2d}}dxdx'\, \phi(x)\phi(x')\  U_{\rm NP}(x,x')
        \,.
\end{equation}
El potencial de fondo $U_{\rm NP}(x,x')$ es sin duda regular en el r\'egimen UV, excluyendo los casos $\theta_{\mu}\neq0$ y $\omega_{\mu}\neq0$, visto que 
\begin{align}\label{GW:anisopolpot}
  \begin{split}
U_{\rm NP}(x,x')&=
  \frac{\lambda}{6}\,\frac{1}{(4\pi)^d\,\det\Theta}\int_0^\infty d\beta\left(\underset{{\mu}=1}{\overset{d}{\prod}} \cosh{\omega_{\mu}\beta}\right)^{-1}
    \ e^{-\beta m^2}\times\\
    &\hspace{3cm}\times e^{-\underset{{\nu}=1}{\overset{d}{\sum}}\frac{\tanh{\omega_{{\nu}}\beta}}{4\omega_{\nu}\theta_{\nu}^2}\left[\left(x-x'\right)_{{\nu}}^2+\omega_{\nu}^2\theta_{\nu}^2\left(x+x'\right)_{{\nu}}^2\right]}
  \,.
  \end{split}
\end{align}
Cuando la matriz $\Theta$ es degenerada, i.e. cuando alg\'un $\theta_{\mu}$ se anula, hay que actuar con cierta cautela: el l\'imite en cuesti\'on da como resultado una delta de Dirac con soporte en $(x-x')_{{\mu}}$, e introduce un prefactor $(\tanh\omega_{\mu}\beta)^{-1/2}$ que puede afectar la convergencia de la integral para peque\~nos valores de $\beta$. Esto es lo que sucede, por ejemplo, en el caso particular $d=4$ al tomar un plano conmutativo: el resultado que se obtiene a partir de (\ref{GW:anisopolpot}) es id\'entico al de \eqref{GW:eff-back-deg}.

Tras haber analizado el caso degenerado en la secci\'on \ref{GW.degenerado}, no resulta extra\~no que las componentes del t\'ermino cin\'etico deban ser renormalizadas independientemente y, consecuentemente, debamos escribir el lagrangiano renormalizado como
\begin{equation}\label{GW:anisolagrangian.renor}
    \mathcal{L_{\rm ren}}=\underset{{\mu}=1}{\overset{4}{\sum}} \left[\frac{Z_{\mu}}{2}(\partial_{\mu}\phi)^2+
    \frac{\omega_{{\mu},\rm ren}^2}{2}\,x_{\mu}^2\phi^2\right]+\frac{m_{\rm ren}^2}{2}\phi^2+\frac{\lambda_{\rm ren}}{4!}\phi^4_\star\,.
\end{equation}
Para ser precisos, si tomamos el ET eucl\'ideo Moyal de dimensi\'on $d=4$, las constantes intervenientes en la renormalizaci\'on se pueden escribir como 
\begin{align}
   \label{GW:anisoz.renor} Z_{\mu}&=1+\frac{\lambda}{48\pi^2}\,\left[\underset{{\nu}=1}{\overset{4}{\prod}}(1+\omega_{\nu}^2\theta_{\nu}^2)^{-1/2}\right]\times\\
\nonumber &\hspace{4.3cm}\times
    \frac{\omega_{\mu}^2\theta_{\mu}^2}{(1+\omega_{\mu}^2\theta_{\mu}^2)}\left[\log{\left(\Lambda^2/\omega\right)}+\text{(t. f.)}\right]\,,\\
\nonumber     \omega^2_{{\mu},\rm ren}&=\omega_{\mu}^2\left.\Biggl\{1-\frac{\lambda}{48\pi^2}\,\left[\underset{{\nu}=1}{\overset{4}{\prod}}(1+\omega_{\nu}^2\theta_{\nu}^2)^{-1/2}\right]\left(1+\omega_{\mu}^2\theta_{\mu}^2\right)^{-1}\times\right.\\
\nonumber    &\hspace{6cm}\left.\times \left[\log{\left(\Lambda^2/\omega\right)}+{\rm (t. \ f.)}\right]\right.\Biggr\}\,,\\
\nonumber  m_{\rm ren}^2&=m^2\left\{1+\frac{\lambda}{48\pi^2}\left[\underset{{\nu}=1}{\overset{4}{\prod}}(1+\omega_{\nu}^2\theta_{\nu}^2)^{-1/2}\right]\times\right.\\
\nonumber &\hspace{1.5cm}\times\left.\,\Biggl[\frac{\Lambda^2}{m^2}+\sum_{\rho}\frac{\omega^2_{\rho}\theta_{\rho}^2}{(1+\omega_{\rho}^2\theta_{\rho}^2)^2}
    \left[\log{\left(\Lambda^2/\omega\right)}+\text{(t. f.)}\right]\Biggr]\right\}.
\end{align}

A fin de cuentas, cabe destacar la importancia de la anisotrop\'ia en el potencial arm\'onico de fondo como opci\'on para eliminar las divergencias UV en un modelo no conmutativo caracterizado por $d$ par\'ametros $\theta_{\mu}$ diversos, con ${\mu}=1,\ldots,d$. Si eligi\'eramos $\omega_{{\mu}}=\Omega/\theta_{\mu}$, con $\Omega$ un nuevo par\'ametro adimensional, encontrar\'iamos que bastar\'ia la sola introducci\'on de un par\'ametro $Z$ para renormalizar el t\'ermino cin\'etico. A su vez, el c\'alculo de las funciones $\beta$ mostrar\'ia que se anulan para el valor especial $\Omega=1$. En otras palabras, la anisotrop\'ia del potencial arm\'onico cancelar\'ia la de los par\'ametros de no conmutatividad y, por consiguiente, las bondades del modelo no dependen de una elecci\'on particular de los par\'ametros de no conmutatividad.  

\begin{subappendices}
 \renewcommand{\theequation}{\Roman{chapter}.\Alph{section}.\arabic{equation}}
 
 \section{El n\'ucleo de Mehler}\label{GW.meh}
 El FLM puede ser utilizado para calcular no s\'olo la traza del NdC, sino tambi\'en sus propiedades locales. En este anexo calcularemos, a modo de ejemplo, el propagador libre para el modelo de GW,
 \begin{equation}
  \Delta(p,p'):=\int_{\mathbb{R}^{2d}}dxdx'\,e^{-ipx-ip'x'}\,\langle x'|G^{-1}|x\rangle\,,
\end{equation}
es decir la transformada de Fourier del n\'ucleo del operador $G^{-1}$, cfr. \eqref{GW:G}.  Evidentemente, este propagador satisface la ecuaci\'on diferencial
\begin{equation}\label{GW:meh:ed}
  (-\omega^2\partial^2+m^2+p^2)\Delta(p,p')=(2\pi)^d\,\delta(p+p')\,.
\end{equation}

Para comenzar el c\'alculo deseado, expresamos el operador $G^{-1}$ en funci\'on del n\'ucleo de calor $e^{-\beta\, G}$, obteniendo como resultado
\begin{equation}
  \Delta(p,p')=\int_{\mathbb{R}^{2d}}dxdx'\,e^{-ipx-ip'x'}\,
  \int_{0}^\infty d\beta
  \,\langle x'|e^{-\beta G}|x\rangle
  \,.
\end{equation}
A continuaci\'on, introducimos la representaci\'on de dicho n\'ucleo en t\'erminos de una IdC:
\begin{multline}\label{GW:meh:a1}
\Delta(p,p')=\int_{\mathbb{R}^{2d}}dxdx'\,e^{-ipx-ip'x'}
  \ \times\\
  \times
  \int_{0}^\infty d\beta\,e^{-\beta m^2}\int_{q(0)=x}^{q(\beta)=x'}\mathcal{D}q(t)\mathcal{D}p(t)
  \,e^{-\int_0^{\beta}
          dt\,\left\{p^2(t)-ip(t)\dot{q}(t)+\omega^2q^2(t)\right\}}\,.
  \end{multline}
Es preciso notar que en la IdC \eqref{GW:meh:a1} las trayectorias $p(t)$ no est\'an sujetas a condici\'on de contorno alguna, mientras que las trayectorias $q(t)$ satisfacen $q(0)=x$ y $q(\beta)=x'$. Como es frecuente, es conveniente introducir las perturbaciones alrededor de las soluciones cl\'asicas $q_0(t),p_0(t)$, a saber, las trayectorias que minimizan la acci\'on\footnote{Notar que no usaremos los reescaleos en $\beta$ de $p(t),q(t)$ y $t$ que hemos utilizado frecuentemente.}
\begin{equation}
    \mathcal{S}_0[q(t),p(t)]=\int_0^{\beta}dt\,\left\{p^2(t)-ip(t)\dot{q}(t)+\omega^2q^2(t)\right\}
\end{equation}
con las condiciones de contorno recientemente mencionadas. Las soluciones a este problema son
\begin{align}\begin{split}
  q_0(t)&=\frac{1}{\sinh{2\omega\beta}}\left[x'\,\sinh{2\omega t}+x\,\sinh{2\omega (\beta-t)}\right]\,,\\
  p_0(t)&=\frac{i\omega}{\sinh{2\omega\beta}}\left[x'\,\cosh{2\omega t}-x\,\cosh{2\omega (\beta-t)}\right]\,;
\end{split}
\end{align} 
estas nos permiten realizar las translaciones $q(t)\rightarrow q_0(t)+q(t)$ y $p(t)\rightarrow p_0(t)+p(t)$ en la expresi\'on \eqref{GW:meh:a1} para obtener
\begin{multline}\label{GW:meh:a30}
\Delta(p,p')=\int_{0}^\infty d\beta\,e^{-\beta m^2}\int_{\mathbb{R}^{2d}}dxdx'\,e^{-ipx-ip'x'}
  \     \,
  e^{-\mathcal{S}_0[q_0(t),p_0(t)]}\,\times\\
  \times
\int_{q(0)=0}^{q(\beta)=0}\mathcal{D}q(t)\mathcal{D}p(t)
  \,e^{-\left.ip_0q\right\rvert_{0}^{\beta}-\int_0^{\beta}
  dt\,\left\{p^2(t)-ip(t)\dot{q}(t)+\omega^2q^2(t)\right\}}
  \,.
\end{multline}
De esta forma hemos logrado separar $\mathcal{S}_0[q_0(t),p_0(t)]$, la contribuci\'on cl\'asica, 
del aporte de las fluctuaciones cu\'anticas, cifrado en la IdC en $q(t)$ y $p(t)$. Asimismo, luego del cambio de variables, la integral debe ser realizada sobre trayectorias que satisfacen las condiciones de contorno tipo Dirichlet $q(0)=q(\beta)=0$. 

Al utilizar el FLM es habitual interpretar las integrales de camino resultantes como amplitudes de transici\'on; este caso no es la excepci\'on. Pensando que corresponde a una transici\'on entre puntos $x=x'=0$ a tiempo eucl\'ideo $\beta$ para un oscilador arm\'onico en $d$ dimensiones, se pueden utilizar los polinomios  de Hermite $H_n(\cdot)$, autofunciones del hamiltoniano del oscilador arm\'onico, para calcular
\begin{align}\label{GW:meh:n2}
  \begin{split}
 \int_{q(0)=0}^{q(\beta)=0}\mathcal{D}q(t)\mathcal{D}p(t)
  \,e^{-\int_0^{\beta}
  dt\,\left\{p^2(t)-ip(t)\dot{q}(t)+\omega^2q^2(t)\right\}}=&\\
  &\hspace{-3cm}=\langle x'=0|\,e^{-\beta\left\{-\partial^2+\omega^2x^2\right\}}\,|x=0\rangle\\
  &\hspace{-3cm}=\left(\sum_{n=0}^\infty e^{-2\beta\omega(n+\frac12)}
  \frac{\sqrt{\omega}}{2^n n!\sqrt{\pi}}\,H^2_n(0)\right)^d\\
  &\hspace{-3cm}=
  \left(\frac{\omega}{2\pi\sinh{2\omega\beta}}\right)^{d/2}\,.
\end{split}
\end{align}
Finalmente, uniendo los resultados \ \eqref{GW:meh:a30} y \eqref{GW:meh:n2} encontramos el n\'ucleo de Mehler en el espacio de Fourier \parencite{mehler}:
\begin{equation}\label{a3}
  \Delta(p,p')
  =\left(\frac{2\pi}{\omega}\right)^{d/2}
  \!\!\int_{0}^\infty \!\!d\beta\,\frac{e^{-\beta m^2}}{(\sinh{2\omega\beta})^{d/2}}
  e^{-\frac{1}{2\omega\sinh{2\omega\beta}}\left\{
  (p^2+p'^2)\cosh{2\omega\beta}+2pp'
  \right\}}
  \,.
\end{equation}

\section[Sobre las condiciones de contorno...]{Sobre las condiciones de contorno en las integrales de camino para el modelo Grosse-Wulkenhaar}\label{GW.cc}
En la secci\'on \ref{GW.hk} hemos visto que es posible calcular la traza del NdC de operadores no locales de la forma \begin{equation}\label{GW:cc.H}
    H:=-\partial^2+m^2+\omega^2x^2+V(x,-i\partial),
\end{equation}
definidos sobre $\mathbb{R}^d$, implementando el FLM sobre trayectorias con condiciones de contorno peri\'odicas. En este anexo mostraremos que id\'entico resultado se obtiene, no obstante el c\'alculo sea m\'as intrincado, considerando IdC con condiciones de contorno tipo Dirichlet. 

Efectivamente, siguiendo los pasos que nos llevaron a obtener la ecuaci\'on \eqref{DET:efectivafinal}, es posible mostrar que el operador $H$ puede ser interpretado, en mec\'anica cu\'antica y en el espacio de fases, como la suma de todas las transiciones de un punto a otro con la misma componente espacial, sin restricciones sobre la variable de momento que surge en la IdC\footnote{Llamaremos a esta traza ${\rm Tr}_{(D)}$ hasta tanto no demostremos su igualdad con la traza calculada utilizando trayectorias peri\'odicas.}:
\begin{multline}
 {\rm Tr}_{(D)}\,    e^{-\beta H}=\\
=\beta^{d/2} \mathcal{N}(\beta) e^{-\beta m^2}\int_{\mathbb{R}^d}\! dx \!\int\limits_{q(0)=x}^{q(1)=x}	\! \dd p \dd q\, e^{-\mathcal{S}_{0}[q(t),p(t)]-\int^1_0 dt V_W(\sqrt{\beta} q,p/\sqrt{\beta})}\,.
 \end{multline}
En esta expresi\'on hemos utilizado la escala $\sqrt{\beta}$ para trabajar con variables $q$ y $p$ sin dimensi\'on, hemos introducido una normalizaci\'on $\mathcal{N}(\beta)$ que luego ser\'a determinada a partir de un caso conocido y hemos separado el potencial $V_W$ ordenado seg\'un la prescripci\'on de Weyl de la acci\'on cl\'asica del oscilador harm\'onico,
\begin{align}\label{GW:cc.accionharmonico}
 \mathcal{S}_{0}[q(t),p(t)]:=\int^{1}_0 dt \left\lbrace p^2-i\dot{q}p+\beta^2\omega^2q^2\right\rbrace.
\end{align}
Por otro lado, a partir de este momento y sin p\'erdida de generalidad\footnote{Basta con hacer la transformada de Fourier de la exponencial del potencial para llegar a expresiones similares a aquella con la cual trabajaremos.}, podemos considerar que el potencial corresponde s\'olo a un par de fuentes $j(t)$ y $k(t)$ en la forma 
\begin{align}\label{GW:cc.Vw}
V_W(\sqrt{\beta}q,p/\beta)=j(t)\,q(t)+k(t)\,p(t).                                                                                                                                                                                                                                                                                                                                              
 \end{align}

Como es usual al calcular IdC, simplificar\'a los c\'alculos el considerar la trayectoria cl\'asica $P_0^{(x)}$ en el espacio de fases, con componentes espacial $q_0^{(x)}$ y de momento $p_0^{(x)}$, que minimiza la acci\'on $\mathcal{S}_{0}$ al partir desde un punto dado $x$ del espacio de configuraci\'on y volver a \'el:
\begin{align}
\label{GW:cc.trayectorias}
\begin{split}
P_0^{(x)}(t)&:=
\begin{pmatrix}
p_0^{(x)}(t) \\[0.1cm]
q_0^{(x)}(t)
\end{pmatrix}\\[0.1cm]
&:=\begin{pmatrix}
\frac{i\omega\beta\,x}{\sinh{2\omega\beta}}\left[\cosh{2\omega\beta t}-\cosh{2\omega\beta (1-t)}\right]\\[0.2cm]
\frac{x}{\sinh{2\omega\beta}}\left[\sinh{2\omega\beta t}+\sinh{2\omega\beta (1-t)}\right]
\end{pmatrix}
\,.\end{split}	
\end{align} 
Si asimismo definimos las fluctuaciones cu\'anticas $\Psi$ de los caminos en el espacio de fase y el vector fuente $K(t)$ como
\begin{align}
\begin{split}P(t)&=P_0^{(x)}(t)+\Psi(t)\\
&=P_0^{(x)}(t)+\begin{pmatrix}
                \psi(t)\\
\phi(t)
\end{pmatrix},\\[0.1cm]
K(t)&=\begin{pmatrix}
      k(t)\\
j(t)
     \end{pmatrix},
\end{split}
\end{align}
la expresi\'on para la traza se simplifica notablemente pues la acci\'on del oscilador harm\'onico se separa en un t\'ermino de borde correspondiente a la trayectoria cl\'asica y la acci\'on harm\'onica original \eqref{GW:cc.accionharmonico} para las fluctuaciones cu\'anticas:
\begin{multline}\label{GW:cc.traza}
 {\rm Tr}_{(D)}\,    e^{-\beta H}=\beta^{d/2}e^{-\beta m^2} \mathcal{N}(\beta) \int_{\mathbb{R}^d} dx\, e^{-\omega\beta\tanh(\omega\beta)x^2+\int_0^1\beta^{1/2}K^t(t) P_0^{(x)}(t)}\\
\times \underset{\phi(0)=\phi(1)=0}{\int} \dd \psi \dd \phi\, e^{-\mathcal{S}_0[\phi(t),\psi(t)]+K^t(t)\Psi(t)}.
\end{multline}

De esta manera, hemos logrado separar el problema en el c\'alculo de dos integrales: una  en la variable $x$ y otra en los caminos $\phi(t)$ y $\psi(t)$ con condiciones de contorno Dirichlet. La primera no es m\'as que una integral gaussiana cuyo resultado puede ser escrito como
\begin{multline}\label{GW:cc.integralx}
 \int_{\mathbb{R}^d} dx\, e^{-\omega\beta\tanh(\omega\beta)x^2+\int_0^1\beta^{1/2}K^t(t) P_0^{(x)}(t)}=\\
 = \left(\frac{\pi}{\omega\beta\tanh(\omega\beta)}\right)^{d/2}{\rm exp}\left\lbrace\frac{1}{2}\int^1_0\int_0^1 dtdt'\, K^t(t) I(t,t') K(t')\right\rbrace,
\end{multline}
donde, en t\'erminos del vector de trayectorias $P_0^{(1,\cdots,1)}(t)$ dado por la expresi\'on \eqref{GW:cc.trayectorias} con $x=(1,\ldots,1)$, el n\'ucleo $I(t,t')$ es
\begin{align}
I(t,t')=\frac{1}{2\omega\beta\tanh(\omega\beta)} P_0^{(1,\cdots,1)}(t)\left(P_0^{(1,\cdots,1)}\right)^t(t').
\end{align}

La restante integral tambi\'en es gaussiana y puede ser trabajada como en los dem\'as casos estudiados a lo largo de este trabajo. Para ello, observemos que la acci\'on $\mathcal{S}_0$ puede ser interpretada como el producto escalar
\begin{align}\label{GW:cc.accion.vectores}
 \mathcal{S}_0[\phi,\psi]=-\frac{1}{2}\int^{1}_0 \Psi^t
        D_D  \Psi,
 \end{align}
en t\'erminos del operador herm\'itico $D_D$ que act\'ua sobre trayectorias cuyas componentes espaciales satisfacen condiciones de contorno tipo Dirichlet, 
\begin{align}
D_D= 
\begin{pmatrix}
 2 & -i\partial_t\\
 i\partial_t &2\beta^2\omega^2
\end{pmatrix},
\end{align}
y cuyo inverso posee un n\'ucleo de componentes $G^{(D)}_{ab}$ que pueden ser escritas como
\footnotesize\begin{align}
 G^{(D)}_{pp}(t,t')&=\frac{\beta\omega}{2\sinh(2\beta\omega)}
  \bigl[\cosh\bigl(2\beta\omega(t'+t-1)\bigr)+\cosh\bigl(2\beta\omega(\vert t'-t\vert -1)\bigr)\bigr], \\
\nonumber G^{(D)}_{px}(t,t')&=\frac{-i}{2\sinh(2\beta\omega)}\bigl[\sinh\bigl(2\beta\omega(t'+t-1)\bigr)-\epsilon(t'-t) \sinh\bigl(2\beta\omega(\vert t'-t\vert -1)\bigr)\bigr],\\
\nonumber G^{(D)}_{xp}(t,t')&=\frac{-i}{2\sinh(2\beta\omega)} \bigl[\sinh\bigl(2\beta\omega(t'+t-1)\bigr)+\epsilon(t'-t) \sinh\bigl(2\beta\omega(\vert t'-t\vert -1)\bigr)\bigr], \\
 \nonumber G^{(D)}_{xx}(t,t')&=\frac{1}{2\beta\omega\sinh(2\beta\omega)}\bigl[\cosh\bigl(2\beta\omega(t'+t-1)\bigr)-\cosh\bigl(2\beta\omega(\vert t'-t\vert -1)\bigr)\bigr].
\end{align}
\normalsize

Definamos tambi\'en el valor medio $  \left\langle f[q(t),p(t)]\right\rangle_{\!\rm D}$ 
de una funci\'on $f(q,p)$ con una medida gaussiana dada por la IdC
con la acci\'on \eqref{GW:cc.accion.vectores},
\begin{align}
  \left\langle
      f[q(t),p(t)]\right\rangle_{\!\rm D}:=\mathcal{Z}(\omega\beta)^{-1}  \underset{\begin{subarray}{c} 
 \phi(0)=\phi(1)=0\end{subarray}}{\int} \dd \psi \dd \phi\, e^{-\frac{1}{2}\int^{1}_0 \Psi^t
        D_D  \Psi}
  f[q(t),p(t)]\,,
\end{align}
normalizada de forma tal que $  \left\langle1\right\rangle_{\!\rm D}=1$ gracias a la presencia de la funci\'on $\mathcal{Z}(\omega\beta)$.
Haciendo uso de estos resultados, obtenemos para la funci\'on generatriz $Z_D[k,j]$  la siguiente expresi\'on que surge de completar cuadrados:
\begin{align}\label{GW:cc.Z}
 Z_D[k(t),j(t)]&:=\left\langle e^{\int^{1}_0 K^t \Psi}\right\rangle_{\!\rm D}\\
 &={\rm exp}\left\lbrace\frac{1}{2}\int^1_0\int_0^1 dtdt'\, K^t(t) G^{(D)}(t,t') K(t')\right\rbrace.
\end{align}

Resta por determinar a\'un las funciones de normalizaci\'on $\mathcal{N}(\beta)$ y $\mathcal{Z}(\omega\beta)$. Ello puede realizarse, tal y como hemos hecho en \eqref{GW:normalizacionIdc}, tomando el resultado conocido $m=V_W=0$; en este caso llegamos a
\begin{align}\label{GW:cc.normalizacion}
\begin{split}
  {\rm Tr}_{(D)}\,e^{-\beta \{-\partial^2+\omega^2x^2\}}
        &=\beta^{d/2}\,\mathcal{N}(\beta)\mathcal{Z}(\omega\beta)\int dx\, e^{-\omega\beta\tanh(\omega\beta)x^2}\left\langle1\right\rangle_D\\
        &=\frac{1}{(2\sinh{\omega\beta})^{d}}
          \,.
\end{split}
\end{align}

Volviendo al c\'alculo de \eqref{GW:cc.traza}, si reunimos los resultados \eqref{GW:cc.integralx}, \eqref{GW:cc.Z} y \eqref{GW:cc.normalizacion}, obtenemos finalmente la expresi\'on cerrada
\begin{align}
\begin{split}{\rm Tr}_{(D)}\,    e^{-\beta H}&=\left(\frac{\pi}{\omega\tanh(\omega\beta)}\right)^{d/2}e^{-\beta m^2} \\
&\hspace{.5cm}\times {\rm exp}\left\lbrace\frac{1}{2}\int^1_0\int_0^1 dtdt'\, K^t(t) \left(G^{(D)}(t,t')+I(t,t')\right) K(t')\right\rbrace .
\end{split}\end{align}
Esta expresi\'on para la traza del NdC del operador H coincide con la que se obtiene de combinar \eqref{GW:Z},\eqref{GW:pi4}, \eqref{GW:normalizacionIdc} y \eqref{GW:z} para el caso especial del potencial $V_W$ dado por \eqref{GW:cc.Vw}, 
\begin{align}
\begin{split} {\rm Tr}\,    e^{-\beta H}&=\left(\frac{\pi}{\omega\tanh(\omega\beta)}\right)^{d/2}e^{-\beta m^2} \\
&\hspace{2cm}\times {\rm exp}\left\lbrace\frac{1}{2}\int^1_0\int_0^1 dtdt'\, K^t(t) G^{(per)}(t-t') K(t')\right\rbrace,
\end{split}
\end{align}
si tomamos en cuenta que a partir de la definici\'on \eqref{GW:dkernel} vale
\begin{align}
G^{(per)}(t-t')= G^{(D)}(t,t')+I(t,t').
\end{align}

\section[El polo de Landau...]{El polo de Landau y el modelo de Grosse-Wulkenhaar al orden de un bucle}\label{GW.landau}

En este anexo nos dedicaremos a probar que, al orden de un bucle, el modelo de GW no presenta el conocido problema del polo de Landau. Para ello, mostraremos que la constante de acoplamiento $\lambda$, al igual que el cuadrado de la frecuencia $\omega^2$, se mantiene finita para cualquier valor del par\'ametro de corte $\Lambda$.

Para comenzar, podemos utilizar el resultado \eqref{GW:landau.beta} con el objetivo de reescribir la funci\'on $\beta_{\omega^2}$, cfr. \eqref{GW:funcionesbeta}, en t\'erminos de $\omega^2$ y los valores $\lambda_0$ y $\omega_0^2$ de la constante de acoplamiento y la frecuencia al cuadrado correspondientes a cierta escala $\Lambda_0$:
\begin{align}\label{GW:landau.betaparaomega}
\begin{split}
\beta_{\omega^2}(\Lambda,\omega^2)&=\frac{\lambda_0}{24\pi^2\omega_0^2}\,\omega^4\frac{1-\omega^2\theta^2}{(1+\omega^2\theta^2)^3}.
\end{split}
\end{align}
Visto que la funci\'on $\beta_{\omega^2}(t,x)$ es continua y acotada en todo el plano $(t,x)$, regi\'on donde tambi\'en satisface la condici\'on de Lipschitz en la variable $x$, el teorema de Picard-Lindel\"of nos garantiza la existencia y la unicidad de la soluci\'on $\omega^2(\Lambda)$ a la ecuaci\'on diferencial \eqref{GW:landau.betaparaomega} para todo valor de $\Lambda$ y condici\'on inicial. Si bien encontrar una expresi\'on cerrada para $\omega^2(\Lambda)$ no es sencillo, veremos que nos bastar\'a obtener funciones que la acoten por arriba y por debajo para lograr nuestro cometido. 

Con el fin de simplificar la demostraci\'on, consideremos $\Lambda>\Lambda_0$ y $0<\omega_0\leq\theta^{-1}$ ya que el caso $\omega_0\geq\theta^{-1}$ se puede analizar en forma similar. Puesto que $\omega^2(\Lambda)$ es continua, existir\'a un valor $\Lambda_1$ hasta el cual ser\'a cierto que $\omega(\Lambda)\leq\theta^{-1}$ y ser\'an v\'alidas las cotas
\begin{align}\label{GW:landau.desigualdadbeta}
 a_{m}(1-\omega^2\theta^2) \leq\beta_{\omega^2}\leq a_{M}(1-\omega^2\theta^2),
\end{align}
donde hemos definido las constantes $a_{m}$ y $a_{M}$ a trav\'es de las f\'ormulas 
\begin{align}\label{GW:landau.am.y.aM}
\begin{split}a_{m}:&=\frac{\lambda_0\,\omega_0^2}{48\pi^2\,(1+\omega_0^2\theta^2)^2},\\
a_{M}:&=\frac{\lambda_0}{96\pi^2\,\omega_0^2\,\theta^4}.
\end{split}
\end{align}

El punto crucial es notar que las desigualdades impuestas por las expresiones \eqref{GW:landau.desigualdadbeta} pueden emplearse para construir otras en las que entre en juego expl\'icitamente la funci\'on $\omega(\Lambda)$; efectivamente, es posible demostrar que en el intervalo $\Lambda_0<\Lambda<\Lambda_1$ se cumple 
\begin{align}\label{GW:landau.cotas.omega}
\omega_{min}^2(\Lambda)\leq \omega^2(\Lambda)\leq\omega^2_{max}(\Lambda),
\end{align}
llamando $\omega_{i=m,M}^2$ a las funciones cuyas ecuaciones diferenciales involucran las cotas superior e inferior escritas en \eqref{GW:landau.cotas.omega}:
\begin{align}\label{GW.landau.ec.dif}
 \Lambda\partial_{\Lambda}{\omega^2_{i}}=a_{i}(1-\omega_{i}^2\theta^2),\qquad \omega^2_{i}(\Lambda_0)=\omega_0^2, \qquad i=m,M.
\end{align} 
Por cierto, las soluciones a \eqref{GW.landau.ec.dif} se pueden obtener sin dificultades y son funciones crecientes, acotadas superiormente por $\theta^{-2}$:
\begin{align}\label{GW:landau.soluciones}
 \omega_{i}^2(\Lambda)={\theta^{-2}} + \left(\omega_0^2-\theta^{-2}\right) \left(\frac{\Lambda}{\Lambda_0}\right)^{-a_{i}\theta^2}, \qquad i=m,M.
\end{align}
A dicho valor $\theta^{-2}$ tienden cuando $\Lambda$ se vuelve grande, con una potencia dada por los valores de $a_{m}$ y $a_{M}$, relacionados por su parte con los par\'ametros iniciales seg\'un \eqref{GW:landau.am.y.aM}.

 Ahora, observando \eqref{GW:landau.cotas.omega} deber\'ia ser claro que $\omega^2(\Lambda)$ no puede llegar a tomar el valor $\theta^{-2}$, pues si lo tomara por primera vez para $\Lambda=\Lambda_2$ valdr\'ia $\omega_{M}(\Lambda_2)\geq \theta^{-2}$, lo cual se contradice con \eqref{GW:landau.soluciones}. 
En definitiva, \eqref{GW:landau.cotas.omega} es v\'alida para cualquier valor $\Lambda_0<\Lambda<\infty$ y, por ende, la frecuencia $\omega$ resulta acotada superior e inferiormente por el valor l\'imite 
\begin{align}
 \theta^{-2}=\lim_{\Lambda\rightarrow\infty}\omega_{min}^2(\Lambda)\leq \lim_{\Lambda\rightarrow\infty} \omega^2(\Lambda)\leq\lim_{\Lambda\rightarrow\infty} \omega^2_{max}(\Lambda)=\theta^{-2}.
\end{align}

Recordando que seg\'un \eqref{GW:landau.betaparaomega} la constante de acoplamiento $\lambda(\Lambda)$ y la frecuencia al cuadrado $\omega^2(\Lambda)$ son proporcionales, $\lambda$ debe ser una funci\'on acotada tal que 
\begin{align}
\lim_{\Lambda\rightarrow\infty} \lambda(\Lambda)=\frac{\lambda_0}{\omega_0^2\theta^2}.
\end{align}
Esto termina por demostrar que el modelo de GW no presenta un polo de Landau al orden de un bucle.

\end{subappendices}


\chapter{Conclusiones}\label{CONC}

\setlength\epigraphwidth{8cm}
\epigraph{\itshape Ex parte enim cognoscimus, et ex parte prophetamus; cum autem venerit quod perfectum est, evacuabitur quod ex parte est... Videmus nunc per Speculum et in Aenigmate, tunc autem facie ad faciem. Nunc cognosco ex parte, tunc autem cognoscam sicut et cognitus sum.}{-- I Corinthios 13:12.}

El problema de la gravedad cu\'antica ha estado presente en el mundo de la f\'isica fundamental desde la misma concepci\'on de las teor\'ias cu\'antica y de la gravedad, a juzgar por las palabras de Einstein en uno de sus trabajos fundadores de la teor\'ia general de la gravedad\footnote{``\emph{Gleichwohl m\"u{\ss}ten die Atome zufolge der inneratomischen Elektronenbewegung nicht nur elektromagnetische, sondern auch Gravitationsenergie ausstrahlen, wenn auch in winzigem Betrage. Da dies in Wahrheit in der Natur nicht zutreffen d\"urfte, so scheint es, da\ss die Quantentheorie nicht nur die Maxwellsche Elektrodynamik, sondern auch die neue Gravitationstheorie wird modifizieren m\"ussen}'', \parencite{Einstein}.}. Si bien han sido reformuladas a lo largo de los a\~nos, las dos principales l\'ineas de trabajo quedaron establecidas a mediados del siglo pasado \parencite{Rovelli:2004tv}. 
En concreto, nos referimos a las que suelen mencionarse como l\'inea de investigaci\'on covariante, en la cual se intenta construir una teor\'ia cu\'antica de campos que considere las fluctuaciones de la m\'etrica alrededor de una cierta m\'etrica de fondo, y l\'inea de investigaci\'on can\'onica, seg\'un la cual el espacio de Hilbert a considerar debe contener toda la m\'etrica, sin necesidad de incurrir en la introducci\'on de una m\'etrica de fondo; claro est\'a, en la actualidad corresponden a la teor\'ia de cuerdas y la teor\'ia de gravedad cu\'antica de bucles, 
respectivamente. Sin poner en tela 
de juicio la posible validez de estas teor\'ias, no se puede negar ni la existencia de dificultades t\'ecnicas en los c\'alculos que involucran, ni que probablemente estas dificultades no sean resueltas por la actual generaci\'on de f\'isicos.

La teor\'ia cu\'antica de campos no conmutativa ofrece una soluci\'on de compromiso: existe una estrecha relaci\'on entre ella y el l\'imite de bajas energ\'ias de la teor\'ia de cuerdas, a la vez que ofrece posibilidades ciertas de c\'alculo. Su principal caracter\'istica es la presencia de una longitud m\'inima, que surge del par\'ametro de no conmutatividad $\theta$ de los operadores posici\'on en la teor\'ia. Como si eso no bastara, no es impensado que en conjunto con las ideas de fenomenolog\'ia surgidas a fines del siglo pasado en gravedad cu\'antica, ofrezcan en un futuro cercano posibles comparaciones entre predicciones te\'oricas y valores experimentales \parencite{AmelinoCamelia:2008qg}.

En el escenario que hemos planteado, esta tesis ha propuesto una nueva herramienta de c\'alculo, la aplicaci\'on del Formalismo de L\'inea de Mundo a la TCC NC. Para ello, luego de haber ofrecido una breve rese\~na hist\'orica en el primer cap\'itulo, en el cap\'itulo \ref{DET} hemos introducido las nociones generales de c\'alculos perturbativos a un bucle y renormalizaci\'on en TCC con m\'etodos funcionales, los cuales nos llevaron en forma natural a la consideraci\'on de t\'ecnicas espectrales. 

Dichas t\'ecnicas espectrales encontraron su formulaci\'on matem\'atica en el cap\'itulo \ref{FE}. El principal teorema de ese apartado establece la existencia de un desarrollo asint\'otico, en potencias del tiempo propio, del n\'ucleo de calor para operadores en derivadas parciales que cumplan con las siguientes hip\'otesis: 
\begin{itemize}
 \item estar definido en un fibrado vectorial $V$ suave, sobre $\mathcal{M}$, una variedad Riemaniana   $d$-dimensional, suave, compacta y sin borde;
 \item ser de orden $g>0$, el\'iptico y autoadjunto;
 \item y poseer un s\'imbolo principal definido positivo para las variables duales $\xi\neq0$.
\end{itemize}
En ese caso, para el NdC $K(t,x,y)$, vale el desarrollo asint\'otico de su diagonal
\begin{align}\label{CONC:desarrollo.heat-local}
  K(t,x,x)\sim \sum_{n=0}^{\infty} t^{\frac{n-d}{k}}\,e_n(x,P) ,\qquad \text{para}\;\;\; t\!\downarrow\!0 ,
\end{align}
 donde los coeficientes locales e invariantes $e_n(x,P)$ dependen de un n\'umero finito de lo que en la jerga de la comunidad f\'isica llamar\'iamos t\'erminos de potencial, de la conexi\'on $A_{\mu}$ y de la m\'etrica, y se anulan para $n$ impar. El caso de variedades con borde es an\'alogo, salvo por el hecho que en general los $e_n$ con $n$ impar dejan de ser nulos.
 
Partiendo de esa base, nos hemos dedicado en el transcurso del cap\'itulo \ref{FLM} a desarrollar el FLM para la TCC usual. El FLM supone la utilizaci\'on de IdC en mec\'anica cu\'antica para el c\'omputo de trazas de NdC, y resulta aplicable al estudio de correcciones cu\'anticas en TCC. La eficiencia de este m\'etodo qued\'o plasmada al analizar modelos escalares con potenciales regulares arbitrarios. Para el caso $\lambda\phi^4$ hemos discutido su renormalizaci\'on al orden de un bucle, en la cual surgen dos caras del mismo problema: el polo de Landau y la trivialidad cu\'antica, seg\'un los cuales el modelo degenera en $\lambda\rightarrow\infty$ o $\lambda=0$ respectivamente. Este problema dej\'o perpleja a la comunidad f\'isica hasta el descubrimiento de la libertad asint\'otica en el marco de las teor\'ias de gauge no abelianas.

Adem\'as, hemos considerado el caso de operadores con potenciales singulares tipo delta de Dirac, los cuales generan condiciones de pegado que suelen ser interpretados en la literatura como condiciones de borde semitransparentes. Esta interpretaci\'on permite aseverar que, por m\'as que este tipo de potenciales no sean suaves, los resultados est\'an en concordancia con las conclusiones del teorema resumido en \eqref{CONC:desarrollo.heat-local}; en efecto, es de destacar que el  desarrollo del NdC posee potencias semienteras del tiempo propio. Una aplicaci\'on de estas condiciones de contorno es el estudio de densidades de carga el\'ectrica confinadas a capas muy delgadas, por ejemplo, en mol\'eculas gigantes de carbono; vista la relevancia que adquiere la fuerza de Casimir a escalas nanom\'etricas, hoy en d\'ia al alcance de la mano, hemos cre\'ido apropiada la determinaci\'on de su valor para este problema de condiciones de contorno semitransparentes. El resultado muestra que las fuerzas involucradas para 
esta geometr\'ia son atractivas.

El cap\'itulo \ref{NC} ha cumplido con el objetivo de implementar el FLM a las TCC NC definidas sobre el ET eucl\'ideo Moyal. Teniendo en cuenta que la NC pone en pie de igualdad coordenadas y momentos, decidimos encarar el problema en el espacio de fases; en este espacio calculamos la FG y el desarrollo de la traza del NdC de operadores no locales. Estos resultados, m\'as all\'a de su aplicaci\'on f\'isica, son de por s\'i importantes en el estudio formal de las FE. Un punto crucial fue el reconocimiento de t\'erminos no locales en el desarrollo del NdC, los cuales en el estudio de dos modelos, uno sobre el toro NC y otro el de un potencial $\lambda\varphi^4_{\star}$, sugirieron la no renormalizabilidad de los mismos. Un efecto relacionado es  la mezcla UV-IR, que justamente vincula las divergencias de los regimenes UV e IR. Otros efectos propios de TCC NC son estudiados en el modelo del disco NC.

Por \'ultimo, a lo largo del cap\'itulo \ref{GW}, hemos modificado el FLM para emplearlo en el estudio del campo escalar de GW, cuyo lagrangiano es:
\begin{equation}\label{CONC:lagrangian}
    \mathcal{L}_{GW}=\frac{1}{2}(\partial\varphi)^2+\frac{m^2}{2}\varphi^2+
    \frac{\omega^2}{2}\,x^2\varphi^2+\frac{\lambda}{4!}\varphi^4_\star\,.
\end{equation}
El t\'ermino arm\'onico debe ser tratado exactamente, debido al gran valor que \emph{a posteriori} vimos que toma su constante de acoplamiento, mientras que el potencial $\lambda\phi_{\star}^4$ puede ser tratado perturbativamente. Fruto del an\'alisis de la renormalizabilidad al orden de un bucle, obtuvimos constantes renormalizadas que coinciden con  \textcite{Disertori:2006nq} y muestran la existencia de un punto fijo en el flujo del grupo de renormalizaci\'on para el valor $\omega\theta=1$. De esta manera, la constante de acoplamiento se mantiene acotada sin necesidad de introducir la libertad asint\'otica. Tambi\'en investigamos dos variaciones del modelo de GW, en las cuales la matriz de no conmutatividad es singular o el potencial arm\'onico de fondo es anisotr\'opico. Demostramos para ambos que la estructura de divergencias es similar a la del modelo original, con algunas nuevas complicaciones. Conviene mencionar que para ninguno de estos dos casos ha sido probada la renormalizabilidad a todo orden en 
teor\'ias de perturbaciones. 

\section{Trabajo a futuro}
Para concluir esta tesis, proponemos algunas l\'ineas de investigaci\'on que ser\'ian una continuaci\'on l\'ogica de los resultados hasta aqu\'i obtenidos. 

Una primera opci\'on consiste en implementar el FLM para obtener cantidades a ordenes superiores en el n\'umero de bucles. En la obra de \textcite{Schubert:2001he,Schubert:1996jj,Sato:1998sf,Sato:1999xy}, el c\'alculo de diagramas de m\'ultiples bucles es realizado para campos escalares, espinores y de gauge en TCC usual. La ventaja que en este caso ofrece el FLM surge de la ausencia de variables de momento virtuales que oscurezcan la distinci\'on  de los t\'erminos involucrados en la acci\'on efectiva.

Por otro lado, la aplicaci\'on del FLM al estudio de variedades chatas con borde es de reciente data \parencite{Bastianelli:2006hq,Bastianelli:2007jr,Bastianelli:2008vh,Bastianelli:2009mw}. El estudio de modelos con bordes resulta de por s\'i interesante en el contexto de TCC NC, en cuanto como punto de partida debe definirse qu\'e se entiende por borde en una geometr\'ia donde es imposible localizar puntos. En esta direcci\'on, hemos analizado el disco NC \parencite{Falomir:2013vaa} y actualmente estamos investigando dos problemas concretos:
\begin{itemize}
 \item el problema de Casimir de dos placas paralelas sobre las que el campo satisface condiciones Dirichlet, en un ET eucl\'ideo Moyal de 2+1 dimensiones;
 \item la termodin\'amica de part\'iculas bos\'onicas y fermi\'onicas en el problema ya estudiado del disco NC.
\end{itemize}
La aplicaci\'on del FLM a este tipo de problemas con borde ser\'ia un objetivo por alcanzar.

Tercero y \'ultimo, un problema interesante a analizar es el de las teor\'ias de gauge NC, las cuales podr\'ian dar paso a la generalizaci\'on NC del modelo est\'andar. Los desaf\'ios que surgen en este caso son varios. Por una parte, el tipo y el n\'umero de \'algebras que puede ser utilizado para la construcci\'on de estos modelos no abelianos queda restringido \parencite{Matsubara:2000gr,Chaichian:2001mu}.
Por otro lado, tal y como sucede en el modelo escalar, la no localidad implica la presencia del efecto de mezcla UV-IR \parencite{Matusis:2000jf, Grosse:2000yy,Bichl:2002wb}. Las divergencias IR originadas por este efecto provienen de t\'erminos que no est\'an presentes en el lagrangiano original y son de naturaleza diversa a las encontradas por la inclusi\'on de mediadores no masivos en las teor\'ias de gauge conmutativas. 

Luego del \'exito del modelo escalar de GW, sus t\'ecnicas se han propuesto en varios modelos de campos de gauge con el fin de eliminar las divergencias IR   \parencite{Grosse:2007dm, deGoursac:2007gq,Blaschke:2008yj,Blaschke:2009aw, Blaschke:2013gha}. Sin embargo, la inclusi\'on de un t\'ermino ``arm\'onico'' a\'un no ha dado sus frutos en este tipo de teor\'ias. Uno de los principales escollos que surge a la hora de incluir este tipo de t\'erminos es la preservaci\'on de la simetr\'ia de gauge; \'esta resulta ser asimismo un obst\'aculo al intentar demostrar la renormalizaci\'on de 
estos modelos mediante las t\'ecnicas 
usadas para el campo de GW. La implementaci\'on del enfoque de BPHZ
 \parencite{Blaschke:2014vda} podr\'ia ser una salida elegante de esta encrucijada. 
Tambi\'en se ha evaluado la posibilidad de que la regularizaci\'on de la mezcla UV-IR provenga de la interacci\'on con la curvatura del espacio sobre el cual el campo est\'a definido \parencite{Buric:2010xs,deGoursac:2010zb}. Asimismo, consideramos que las ventajas t\'ecnicas del FLM pueden ser de ayuda en el estudio perturbativo de los modelos actualmente en desarrollo.


\chapter*{Agradecimientos}
Si acaso esta tesis resultare en un m\'inimo progreso para esa cosa intangible que hemos denominado humanidad, las miradas no deber\'an posarse sobre m\'i sino sobre Pablo A. G. Pisani, encomiable Virgilio que me ha deparado el destino. La sinceridad me obliga a reconocer que no poco han contribuido Horacio Falomir, Mariel Santangelo, Gabriela Beneventano, Mariela Nieto y la comunidad de cuerdas y altas energ\'ias del Dto. de F\'isica de la UNLP, en particular a Nicol\'as Grandi y Mercedes Mosquera.\\[0.5cm]

P\'arrafo aparte, agradezco a Mauricio Leston, Gerardo Rossini y \foreignlanguage{russian}{Дмитрий Василевич} por su predisposici\'on; a Olindo Corradini, Roberto Bonezzi, Mariano Salvay, Giovanni Amelino-Camelia, Lorenzo Orifici y Giacomo Rosati por sus horas.\\[0.5cm]

El CONICET ha financiado este trabajo a trav\'es de una beca de postgrado, en conjunto con la UNLP a trav\'es del programa de retenci\'on de recursos humanos del Dto. de F\'isica.\\[0.5cm]

El interesado lector y aquellos a quienes no hago expl\'icito mi agradecimiento sepan disculparme.\\[0.5cm]

Juani, Juancho, Facundo, Francisco, Charly y Eugenia, no dudo que ustedes habr\'an sabido reconocerse en estas l\'ineas.\\[0.5cm]

\begin{flushright}

\begin{minipage}{0.6\textwidth}
\vspace{0.1\textheight}
\normalsize\emph{No habr\'a una sola cosa que no sea\\[0.1cm]
una nube. Lo son las catedrales \\[0.1cm]
de vasta piedra y b\'iblicos cristales \\[0.1cm]
que el tiempo allanar\'a. Lo es la Odisea,\\[0.1cm]
que cambia como el mar. Algo hay distinto\\[0.1cm]
cada vez que la abrimos. El reflejo\\[0.1cm]
de tu cara ya es otro en el espejo\\[0.1cm]
y el d\'ia es un dudoso laberinto.\\[0.1cm]
Somos los que se van. La numerosa\\[0.1cm]
nube que se deshace en el poniente\\[0.1cm]
es nuestra imagen. Incesantemente\\[0.1cm]
la rosa se convierte en otra rosa.\\[0.1cm]
Eres nube, eres mar, eres olvido.\\[0.1cm]
Eres tambi\'en aquello que has perdido.}\\[0.2cm]
\begin{flushright}
\textsc{Jorge Luis Borges},\\[0.1cm] \emph{Los Conjurados}, Nube (I).
\end{flushright}
\end{minipage}
 
\end{flushright}

{\sloppy
\printbibliography}

\end{document}